\documentclass[usenatbib,useAMS]{mnras}
\usepackage{tablefootnote}
\usepackage{multirow}
\usepackage{xcolor}
\usepackage{graphicx}	
\usepackage{amsmath}	
\usepackage{amssymb}	
\usepackage{multicol}   
\usepackage{bm}		
\usepackage[normalem]{ulem} 
\usepackage{changes}
\usepackage{caption} 
\usepackage{url}
\usepackage{longtable}
\usepackage{pdfpages}
\usepackage{subcaption}
\usepackage{hyperref}
\usepackage{pdflscape}
\usepackage{orcidlink}

\title[Quasi-periodic variability of Ton\,599]
{Multiwavelength quasi-periodic variability of the blazar Ton\,599}

\author[Sotnikova et al.]{%
Yu.~V.~Sotnikova,$^{1}\orcidlink{0000-0001-9172-7237}$\thanks{E-mail: lacerta999@gmail.com} 
T.~V.~Mufakharov,$^{1,12}\orcidlink{0000-0001-9984-127X}$ 
A.~E.~Volvach,$^{2}\orcidlink{0000-0002-3839-3466}$
V.~V.~Vlasyuk,$^{1}\orcidlink{0009-0002-6596-7274}$
\newauthor
M.~L.~Khabibullina,$^{1}\orcidlink{0000-0001-9515-4552}$
A.~G.~Mikhailov,$^{1}\orcidlink{0000-0002-0279-0777}$
T.~An,$^{3,4}\orcidlink{0000-0003-4341-0029}$
D.~O.~Kudryavtsev,$^{1}\orcidlink{0000-0003-2461-1570}$
Yu.~A.~Kovalev,$^{5,6}\orcidlink{0000-0002-8017-5665}$
\newauthor
Y.~Y.~Kovalev,$^{7}\orcidlink{0000-0001-9303-3263}$
A.~V.~Popkov,$^{8,5}\orcidlink{0000-0002-0739-700X}$
S.~S.~Savchenko,$^{9,10}\orcidlink{0000-0003-4147-3851}$
A.~K.~Erkenov,$^{1}\orcidlink{0000-0002-6086-9299}$
D.~A.~Morozova,$^{9}\orcidlink{0000-0002-9407-7804}$
\newauthor
T.~A.~Semenova,$^{1}\orcidlink{0000-0002-2902-5426}$
O.~I.~Spiridonova,$^{1}\orcidlink{0009-0007-7315-3090}$
M.~A.~Kharinov,$^{11}\orcidlink{0000-0002-0321-8588}$
I.~A.~Rakhimov,$^{11}\orcidlink{0000-0002-9185-6239}$
T.~S.~Andreeva,$^{11}\orcidlink{0000-0003-3613-6252}$
\newauthor
L.~Cui,$^{12}\orcidlink{0000-0003-0721-5509}$
X.~Wang,$^{12}\orcidlink{0000-0001-8221-9601}$
N.~Chang,$^{12}\orcidlink{0000-0002-8684-7303}$
R.~Yu.~Udovitskiy,$^{1}$
P.~G.~Zhekanis,$^{1}$
\newauthor
G.~A.~Borman,$^2\orcidlink{0000-0002-7262-6710}$
T.~S.~Grishina,$^{9}\orcidlink{0000-0002-3953-6676}$
E.~N.~Kopatskaya,$^{9}$
E.~G.~Larionova,$^{9}\orcidlink{0000-0002-2471-6500}$
I.~S.~Troitskiy,$^{9}\orcidlink{0000-0002-4218-0148}$
\newauthor
Yu.~V.~Troitskaya,$^{9}\orcidlink{0000-0002-9907-9876}$
A.~A.~Vasilyev,$^{9}\orcidlink{0000-0002-8293-0214}$
A.~V.~Zhovtan,$^{2}$
D.~V.~Kratov,$^{1}\orcidlink{0000-0002-2506-5746}$
L.~N.~Volvach,$^{2}\orcidlink{0000-0001-6157-003X}$
\newauthor
E.~V.~Shishkina,$^{9}\orcidlink{0009-0002-2440-2947}$
A.~I.~Dmytrotsa,$^{2}$
V.~I.~Zharov$^{1}\orcidlink{0000-0002-2841-8017}$
\\
$^{1}$ Special Astrophysical Observatory of the Russian Academy of Sciences, Nizhny Arkhyz, 369167, Russia\\
$^{2}$ Crimean Astrophysical Observatory of the Russian Academy of Sciences, 298409, Nauchny, Russia \\
$^{3}$ State Key Laboratory of Radio Astronomy and Technology, Shanghai Astronomical Observatory, \\
Chinese Academy of Sciences, 80 Nandan Road, Shanghai 200030, P. R. China\\
$^{4}$ Guizhou Radio Astronomical Observatory, Guizhou University, 550000, Guiyang, China \\
$^{5}$ Lebedev Physical Institute of Russian Academy of Sciences, Leninsky prosp. 53, Moscow 119991, Russia \\
$^{6}$ Institute for Nuclear Research, Russian Academy of Sciences, 60th October Anniversary Prospect 7a, Moscow 117312, Russia\\
$^{7}$ Max-Planck-Institut f\"ur Radioastronomie, Auf dem H\"ugel 69, Bonn 53121, Germany\\
$^{8}$ Moscow Institute of Physics and Technology, Institutsky per. 9, Dolgoprudny 141700, Russia\\
$^{9}$ Saint Petersburg State University, 7/9 Universitetskaya nab., St. Petersburg, 199034 Russia\\
$^{10}$ Pulkovo Observatory, St.Petersburg, 196140, Russia\\
$^{11}$ Institute of Applied Astronomy of the Russian Academy of Sciences, Kutuzova Embankment 10, St. Petersburg 191187, Russia\\
$^{12}$ Xinjiang Astronomical Observatory, CAS, 150 Science-1 Street, Urumqi 830011, China\\
}
\date{Accepted 2026 February 10. Received 2026 January 4; in original form 2025 August 3}
\pubyear{2026}

\begin{document}
\label{firstpage}
\pagerange{\pageref{firstpage}--\pageref{lastpage}}
\maketitle 

\begin{abstract} 
During the last 40--50 years, the TeV blazar Ton\,599 has systematically experienced major outbursts detected in a wide wavelength range from radio to $\gamma$-rays. In this work, we present an analysis of Ton\,599 quasi-periodic variability across multiple wavelengths using an observing baseline from 1983 to 2025. The $\gamma$-ray, optical, and radio emissions are found to be highly correlated with time lags $\sim0$--$360$ days, which indicates that they are triggered by the same population of particles. Using the Lomb--Scargle periodogram and the Weighted Wavelet Z-transform, we revealed several periodic components with characteristic periods of 1.4, 1.7, 2.3, 6.5, and 7.5 yrs. The result is consistent with the detection of periodic components in the 1997--2011 light curves, which means that we observe the same mechanism causing long-term periodic variability. A model of a binary supermassive black hole (SMBH) with a precessing jet, applied to the radio light curves of Ton\,599, yields frequency-dependent best-fitting parameters with orbital periods ranging from $\sim$1.2 to 1.7 yrs and precession periods from $\sim$5.8 to 7.7~yrs. This result implies the existence of an SMBH system modulating emission through both the orbiting and jet precession effects, with differing observed periods possibly reflecting frequency-dependent emission regions along a structured, stratified jet. Nonetheless, the short-term periodicity and exceptionally strong flares likely arise from internal jet shocks, which aligns with typical blazar behavior. We suggest that the multiband quasi-periodicity of Ton\,599 is reasonably described by a combination of geometric effects (orbital motion and precession) and stochastic processes. 
\end{abstract}

\maketitle 
\section{Introduction}

Multiwavelength variability in active galactic nuclei (AGNs), particularly blazars, provides crucial insights into the structure and dynamics of relativistic jets, accretion processes, and the possible presence of supermassive binary black holes (SMBBHs). However, distinguishing genuine quasi-periodic signals from stochastic variability remains one of the most challenging aspects of AGN studies. Blazars exhibit violent variability across the entire
electromagnetic spectrum, often dominated by stochastic processes such as internal shocks and turbulent magnetic reconnection \citep{1985ApJ...298..114M, 2019ARA&A..57..467B}. Despite this predominantly stochastic nature, a subset of blazars exhibits long-term quasi-periodic oscillations (QPOs) in flux densities, which may reflect geometric or dynamical mechanisms on parsec scales.

The periodic or quasi-periodic behavior in AGNs has been observed at various time-scales, from months to decades, and across a broad range of wavelengths from radio to $\gamma$-rays \citep{2018Galax...6....1G, 2014MNRAS.443...58W, 2016AJ....151...54S,2021MNRAS.501.5478A, 2023NatAs...7.1368E, 2023A&A...672A..86R}. The statistical significance and physical reality of many claimed periodicities are still being debated due to limited observation periods, gaps in the data, and the red-noise characteristics inherent in AGN variability. These modulations have been tentatively interpreted through several physical scenarios, including jet precession induced by 
the Lense--Thirring torque \citep[e.g.][]{2013MNRAS.428..280C}, orbital motion in a close SMBBH system \citep[e.g.][]{1996ApJ...460..207L, 2004ApJ...615L...5R}, quasi-helical jet structures \citep[e.g.][]{2018MNRAS.478.3199B}, and periodic injection of energetic particles in the jet base \citep[e.g.][]{2017MNRAS.471.3036P}.

Statistically robust evidence of persistent periodicity has been claimed in only a limited number of AGNs, such as OJ\,287, J0735$+$178, J2202.7$+$4216, PG\,1553$+$113, TXS\,0506$+$056, 3C\,273, 3C\,279, 3C\,345, J1308$+$326, AO\,0235$+$164, and PKS\,1502$+$106 \citep{2010SCPMA..53S.252Z,2013ARep...57...34V, 2015ApJ...813L..41A,2016ApJ...819L..37V, 2017A&A...602A..29B,2018MNRAS.478.3199B, 2021PASP..133g4101Y,2022ApJ...941L..25B, 2023MNRAS.523L..52B,2024MNRAS.527.6970K, 2024MNRAS.535.2775V,2025MNRAS.537.2380Z, 2025arXiv250703967P}. 
Even in these cases, the detected periodicities often span only a few cycles within available observing baselines, raising questions about their long-term stability and statistical significance. Despite growing interest, only a limited number of blazars show statistically robust QPO signals, and the physical interpretation of these phenomena remains uncertain due to data gaps, red noise contamination, and methodological limitations.

The flat-spectrum radio quasar (FSRQ) Ton\,599 
(\mbox{$\rm RA_{2000.0}=11$:59:31.82}, $\rm Dec_{2000.0}=+29$:14:43.7) at \mbox{$z = 0.725$} \citep{2010MNRAS.405.2302H}) is a bright TeV blazar \citep{2017ATel11075....1M} with a history of violent multiwavelength flaring behavior and has been previously identified as a QPO candidate. \cite{2006PASJ...58..797F} reported possible periods of 1.58 and 3.55 years in the optical band from 1974 to 2002. \cite{2007A&A...469..899H, 2008A&A...488..897H} found 3.3--3.5-year radio periodicities at 22 and 37 GHz using the Mets{\"a}hovi data. Later, \cite{2014Ap&SS.351..281W} applied the wavelet and Lomb--Scargle (LS) techniques to the University of Michigan Radio Astronomy Observatory (UMRAO) radio light curves and discovered a harmonic set of four periodicities (1.7, 2.3, 3.4, and 6.8 years), while \cite{2021ARep...65.1233H} found in the R band only one period of $\sim2.1$ years.

Nevertheless, the prior studies of Ton\,599 suffer from several limitations. Most analyses focus on a single wavelength range (optical or radio) or relatively short time intervals ($\leq15$ years), and frequently neglect rigorous red noise modeling. Furthermore, the potential coherence of variability across the bands, e.g., lagged correlations between the optical and radio emissions, has not been systematically investigated. In particular, 
a comprehensive multifrequency study of the QPOs, covering a multidecade baseline and supported by robust statistical techniques, has not been conducted for Ton\,599 before.

In this work, we perform a systematic multiwavelength analysis of Ton\,599 using an extensive dataset spanning \mbox{$\sim40$~years} (1983--2025) in the radio band at 22 and 37 GHz, over 20 years in the optical R band and in the radio band at 2, 5, 8, 11, and 230 GHz, and nearly 17 years of the $\gamma$-ray data from Fermi-LAT. We apply two independent methods: the LS periodogram and Weighted Wavelet Z-transform (WWZ) to detect and characterize quasi-periodic variability in these light curves. To explore possible common origin for the observed variability, we calculate time lags between flares at different wavelengths using the discrete correlation function (DCF) technique. We examine the physical plausibility of several scenarios, including the jet precession and binary SMBH models, in explaining the long-term variability patterns. The results provide strong support for a mixed scenario in which both geometric and stochastic processes drive the observed quasi-periodicity of Ton\,599.

\section{Observed data}
We have collected long-term datasets from five radio and five optical instruments and the $\gamma$-ray Fermi-LAT telescope (Table~\ref{tab:faci}). Most of the measurements are presented in this paper for the first time (1-22 GHz RATAN-600 measurements, 5 and 8 GHz data from three RT-32, R-band Zeiss-1000 and AS-500/2 data). The flux densities at 230 GHz (1.3 mm), obtained in the period from January 2003 to March 2025, were taken from the Submillimeter Array archive \footnote{\url{http://sma1.sma.hawaii.edu/callist/callist.html}} (SMA; \citealt{2007ASPC..375..234G}).  

\begin{table}
\caption{The instruments used in this work.} 
\label{tab:faci}
\centering
\begin{tabular}{cccc}
\hline
Telescope & Institute & Epoch & Band \\
\hline
 RATAN-600 &  SAO RAS  & 1997--2025 & 1--22 GHz \\ 
 RT-32     & IAA RAS  & 2018--2025 & 5.0, 8.6 GHz\\ 
 RT-22     & CrAO RAS  & 1983--2025 & 22.2, 36.8 GHz \\ 
  NSRT     & XAO CAS  & 2007--2024 & 4.8 GHz \\ 
 SMA &  SAO \& ASIAA & 2003--2025  & 230 GHz \\ 
 Zeiss-1000 & SAO RAS  & 2023--2025 & $R$ band \\ 
 AS-500/2 & SAO RAS  & 2023--2024 & $R$ band \\ 
 LX-200 & SPbU  & 2006--2025 & $R$ band \\ 
 AZT-8 & CrAO RAS & 2005--2023 & $R$ band \\ 
 48$''$ Schmidt & ZTF Palomar & 2018--2023 & $r$ band \\ 
 Fermi-LAT & NASA  & 2008--2025 & $\gamma$-ray \\ 
\hline
\end{tabular}
\end{table} 

\subsection{RATAN-600}

The broadband radio spectra at 0.96/1.2, 2.25, 4.7, 7.7/8.2, 11.2, and 21.7/22.3 GHz were measured with the RATAN-600 radio telescope in 1997--2025. The spectra were obtained quasi-simultaneously, on the time-scale of 3--5 minutes. The RATAN-600 angular resolution depends on the antenna elevation angle and is presented in Table~\ref{tab:radiometers}. The resolution along declination ${\rm FWHM}_{\rm Dec.}$ is three to five times worse than that along right ascension ${\rm FWHM}_{\rm RA}$. The flux density detection limit is 8 mJy at 4.7 GHz (integration time is about 3 s) under good weather conditions and at the average antenna elevation angle (${\rm Dec.} \sim 0\degr$). It also depends on the atmospheric extinction and effective area for different antenna elevation angles $H$ (from 10$\degr$ up to 90$\degr$) at corresponding frequencies \citep{1993IAPM...35....7P,1997ASPC..125...46V,1999A&AS..139..545K,2011AstBu..66..109T,2016AstBu..71..496U,2018AstBu..73..494T,2020gbar.conf...32S}. 

The data reduction was made using the automated data reduction system \citep{1999A&AS..139..545K,2011AstBu..66..109T,2016AstBu..71..496U,2018AstBu..73..494T} and the Flexible Astronomical Data Processing System (\textsc{FADPS}) standard package modules \citep{1997ASPC..125...46V} for the broadband RATAN continuum radiometers. We used the following flux density secondary calibrators: 3C\,48, 3C\,147, 3C\,161, 3C\,286, and NGC\,7027  \citep{1977A&A....61...99B,2013ApJS..204...19P,2017ApJS..230....7P,1994A&A...284..331O,1980A&AS...39..379T}. The flux densities at 1-22 GHz $S_{\nu}$, their errors $\sigma$, and average observing epochs (yyyy.dd.mm, JD and yyyy.yyyy) are presented in Table~\ref{TableA1_part}.

\begin{table}
\caption{RATAN-600 continuum radiometer parameters: the central frequency $f_0$, the bandwidth $\Delta f_0$, and the detection limit for point sources per transit $\Delta F$. ${\rm FWHM}_{\rm {RA} \times \rm {Dec.}}$ is the angular resolution along RA and Dec., calculated for the average angles.} 
\label{tab:radiometers}
\centering
\begin{tabular}{cccr@{$\,\times\,$}l}
\hline
$f_{0}$ & $\Delta f_{0}$ & $\Delta F$ &  \multicolumn{2}{c}{FWHM$_{\rm {RA} \times \rm{Dec.}}$}\\
(GHz)    &   (GHz)           &  (mJy beam$^{-1}$)   &   \multicolumn{2}{c}{}  \\
\hline
 $21.7/22.3$ & $2.5$  &  $50$ & $0\farcm17$ & $1\farcm6$  \\ 
 $11.2$ & $1.4$  &  $1 5$ & $0\farcm34$ & $3\farcm2$ \\ 
 $7.7/8.2$  & $1.0$  &  $10$ & $0\farcm47$ & $4\farcm4$   \\ 
 $4.7$  & $0.6$  &  $8$  & $0\farcm81$ & $7\farcm6$   \\ 
 $2.25$  & $0.08$  &  $40$ & $1\farcm7$ & $16\arcmin$  \\ 
 $0.96/1.25$  & $0.08$ &  $200$ & $3\farcm1$ & $27\arcmin$ \\ 
\hline
\end{tabular}
\end{table} 

\begin{table*}
\caption{\label{TableA1_part} A fragment of the RATAN-600 measurements in 1997--2025: epochs in yyyy.mm.dd (Col.~1), Julian Date (JD) (Col.~2), yyyy.yyyy (Col.~3), flux densities at 22, 11.2, 8, 5, 2, and 1 GHz and their errors in Jy (Cols.~4--15). The full version is available as online supplementary material.} 
\begin{tabular}{|l|c|c|c|c|c|c|c|c|c|c|c|c|c|c|c|}
\hline
yyyy.mm.dd & JD  &  yyyy.yyyy & $S_{22}$ & $\sigma$ & $S_{11.2}$ & $\sigma$ & $S_{8}$ & $\sigma$ & $S_{5}$ & $\sigma$ & $S_{2}$ & $\sigma$ & $S_{1}$ & $\sigma$  \\
1 & 2 & 3 & 4 & 5 & 6 & 7 & 8 & 9  & 10 & 11 & 12 & 13 & 14 & 15 \\
\hline
1997.03.18 & 2450526 &	1997.2082 &	1.09 &	0.02 &	1.40 &	0.02 &	1.46 &	0.06 &	1.82 &	0.02 &	2.08 &	0.03 &	2.55 &	0.16 \\
1997.06.21 & 2450621 &	1997.4684 &	0.78 &	0.06 &	0.98 &	0.05 &	1.20 &	0.03 &	1.59 &	0.02 &	1.77 &	0.10 &	2.54 &	0.11 \\
1997.12.10 & 2450793 &	1997.9397 &	0.95 &	0.07 &	0.87 &	0.05 &	0.91 &	0.02 &	1.17 & 	0.04 &	1.46 &	0.01 &	2.06 &	0.25 \\
1998.04.16 & 2450920 &	1998.2876 &	2.01 &	0.02 &	1.30 &	0.18 &	1.05 &	0.01 &	1.06 &	0.03 &	1.23 &	0.03 &	2.00 &	0.04 \\
1999.04.13 & 2451282 &	1999.2793 &	2.38 &	0.07 &	2.25 &	0.02 &	2.09 &	0.04 &	1.85 &	0.06 &	1.62 &	0.02 &	1.96 &	0.03 \\
1999.09.28 & 2451450 &	1999.7396 &	3.13 &	0.08 &	2.93 &	0.04 &	2.69 &	0.04 &	2.39 &	0.03 &	1.93 &	0.01 &	1.67 &	0.11 \\
\hline
\end{tabular}
\end{table*}

\subsection{RT-32}

The flux densities at frequencies of 5.05 and 8.63~GHz were measured using three RT-32 stations -- Svetloe~(Sv), Zelenchukskaya~(Zc), and Badary~(Bd) -- in several different epochs \citep{2019..VLBI..Quasar}. All the antennas and receivers have similar parameters: a bandwidth $\Delta f_{0}=900$~MHz for both receivers with the central frequencies specified above; beamwidths at the half-power level $\rm HPBW=7\arcmin$ and 3\farcm9, respectively; a flux density limit $\Delta F$ reaching about 20~mJy per scan with the 1-s time constant for both frequencies under optimal observing conditions.

The almost weekly single-dish monitoring was performed with drift scan mode by elevation (Bd, Zc) and azimuth (Sv) from January 2018 to March 2025. One drift scan lasted about 1 minute at 8.63 GHz and 1.5 minutes at 5.05 GHz with a cadence of 1 second. To accumulate the signal, the scanning cycle was repeated on average 15 times, forming a continuous observing set.

The radio data at both 5.05 and 8.63~GHz for the period of 2018--2023 were obtained by the single RT-32 telescope from the Svetloe station. The data for 2024--2025 were measured separately at 5.05~GHz at the Badary station and at 8.63~GHz at Zelenchukskaya.
3C\,48, 3C\,147, 3C\,295, and 3C\,309.1 were used as reference sources. The flux density scales were calculated similarly to the RATAN-600 observations \citep{1977A&A....61...99B,2013ApJS..204...19P}.

The observed data were processed with the original program package CV \citep{kharinov2012} and the Database of Radiometric Observations. The drift scans were filtered, rejected if spoiled considerably by weather or industrial noise, averaged, and fitted with a Gaussian curve. Before averaging, the zero level of each scan was approximated by a parabola. Correction for pointing offsets was applied to the scans by the peak values of the Gaussian fittings. The antenna temperature and its error were estimated from the Gaussian analysis of the averaged scan. 
The flux densities $S_{\nu}$ at 5.05~GHz, their errors $\sigma$, and average observing epochs are presented in Table~\ref{TableA2_part}. The RT-32 measurements at 8.63~GHz are presented in the online supplementary material in the same form.

The observing frequencies of the RT-32 (5.05, 8.63 GHz) and RATAN-600 (4.7, 8.2~GHz) are close, 
therefore we use their rounded values, 8 and 5 GHz, in the further analysis. Frequencies of 21.7/22.3, 11.2, 2.3, and 0.96/1.2 GHz are rounded as 22, 11, 2, and 1 GHz.

\begin{table}
\caption{\label{TableA2_part} A fragment of the RT-32 measurements in 2018--2025: epoch in yyyy.mm.dd (Col.~1), JD (Col.~2), epoch in yyyy.yyyy (Col.~3), flux densities at 5 GHz and their errors in Jy (Cols.~4--5), and the telescope (Col.~6), in accordance with RT-32 station -- Svetloe (RT-32\_Sv), Badary (RT-32\_Bd), and Zelenchukskaya (RT-32\_Zc). The full version is available as online supplementary material} 
\begin{tabular}{|c|c|c|c|c|c|}
\hline
yyyy.mm.dd & JD  &  yyyy.yyyy & $S_{5}$ & $\sigma$ & Telescope \\
1 & 2 & 3 & 4 & 5 & 6 \\
\hline
2018.01.23 & 2458142 & 2018.0603 & 2.98	& 0.07 & RT-32\_Sv \\
2018.01.24 & 2458143 & 2018.0630 & 3.08	& 0.03 & RT-32\_Sv \\
2018.02.06 & 2458156 & 2018.0986 & 3.26 & 0.02 & RT-32\_Sv \\
2025.03.22 & 2460757 & 2025.2245 & 3.39	& 0.11 & RT-32\_Bd \\
2025.03.23 & 2460758 & 2025.2272 & 3.37	& 0.08 & RT-32\_Bd \\
\hline
\end{tabular}
\end{table}

\subsection{RT-22}

The observations at 22.2 and 36.8 GHz were carried out with the \mbox{22-m} \mbox{RT-22} radio telescope (CrAO RAS). The beamwidth is 160\arcsec\ and 100\arcsec\ for the both frequencies respectively, with a pointing accuracy of several arcseconds. The \mbox{0.2-mm} surface accuracy allows observations down to 2-mm wavelengths.
The effective antenna area is 220 m$^{2}$ at 36.8~GHz, varying slightly with temperature and elevation (within \mbox{5--7}~per~cent).

The data were collected using diagram-modulated receivers in on-on mode, with 10--30 measurements per source. Antenna temperatures were averaged and converted to flux densities using elevation-dependent effective area, approximated by a trigonometric fit. Typical measurement errors were \mbox{2--4}~per~cent (signal) and 3--6 per cent (calibration). Other
details are described in \cite{2021A&A...648A..27V,2024A&A...691L...9V} and \cite{2023AstBu..78..105S}.

\subsection{NSRT}

The observations at 4.8 GHz were carried out with the Nanshan 26-m radio telescope (NSRT)
operated by the Xinjiang Astronomical Observatory (XAO), CAS. Ton 599 is part of an ongoing long-term monthly flux density monitoring program for a sample of about 130 $\gamma$-ray AGNs at 4.8 and 23.6 GHz using the XAO-NSRT. The flux density was measured in cross-scan mode, with each scan comprised of eight 
subscans (four in azimuth and four in elevation) over the source position. 
More details are described in \cite{2011A&A...530A.129M} and \cite{2015A&A...578A..34L}.

\subsection{Optical data}

We have collected the optical R-band data from three several sources: the SAO RAS 1-metre Zeiss-1000 and \mbox{0.5-metre} \mbox{AS-500/2} optical reflectors (since January 2023), the data provided by the Saint Petersburg State University team (since May 2005), and the data from the Zwicky Transient Facility (ZTF; \cite{2019PASP..131a8002B}) that has been operating since 2018. 

All observations with the optical telescope Zeiss-1000 were obtained using a CCD photometer in the Cassegrain focus, equipped with a $2048\times2048$ back-illuminated E2V~CCD~\mbox{42-40}. 

In order to reduce the readout noise and dark current of the FLI camera used at the AS-500/2 telescope, we installed in July 2023 a more compact but less noisy detector: a back-illuminated electron-multiplying
CCD camera \mbox{Andor~iXon~897}. It has a capability of single-photon detection, combined with high quantum efficiency, which is above 90 per cent in the wavelength range of 450--700~nm. This CCD camera, mounted in the Cassegrain focus of the \mbox{AS-500/2}, achieves a 7$'$ field of view with a $0\farcs82$/pixel angular resolution (1~pixel~$=$~16~microns). The system readout noise is about 6 $e^-$ with a gain equal to 3 $e^-$/ADU in 
conventional CCD mode, while the electron-multiplying mode provides a readout noise below 1 $e^-$. The air cooling regime of the CCD system was chosen to maintain a temperature of $-70^\circ$C and allows minimization of dark current down to 0.003 $e^-$ per second. The details of the instrumental setup are described in  \cite{2023AstBu..78..464V}. 

Both CCD photometers are equipped with similar sets of filters that are close to the standard broadband Johnson--Cousins filters, given the sensitivity of both CCDs.
The typical integration time in the Ton\,599 observations was 300~s for Zeiss-1000 and 120~s for the AS-500/2. During the periods of object high activity, the integration time was less: down to 30~s for higher time cadence.
To estimate the flux density, we performed the standard reduction steps: dark current subtraction, image flat-fielding, integration of the individual object signal within rings of increasing size, etc. All details of these reduction steps were described earlier \citep{1993BSAO...36..107V}.

The observations of the SPbU team were performed with two telescopes: the 40 cm LX-200 telescope located at Svetloe near Saint Petersburg and the 70 cm AZT-8 telescope of the Crimean Observatory. Initially, both telescopes were equipped with identical photopolarimeters based on the SBIG ST-7b CCD cameras with the standard Johnson--Cousins filters and Savart plates. Since 2019, the camera of the \mbox{LX-200} telescope 
has been replaced with a more sensitive FLI ML4710 CCD camera. The observing strategy and the main steps of data reduction are described in \cite{2002A&A...390..407V}.

In order to provide a joint analysis of the optical and radio data, we averaged our measurements over individual nights and transformed the resulting values into fluxes according to the constant from \cite{1990A&AS...83..183M}. The $R$-band flux densities $R_{\rm flux}$, their errors $\sigma$, and average observing epochs for SAO RAS and SPbU team data are presented in Table~\ref{TableA3_part}.

\begin{table}
\caption{\label{TableA3_part} 
A fragment of the $R$-band measurements
in 2005--2025: epoch in yyyy.mm.dd (Col.~1), JD (Col.~2), epoch in yyyy.yyyy (Col.~3), flux densities and their errors in mJy (Cols.~4--5) and the telescope (Col.~6). The full version is available as online supplementary material.} 
\begin{tabular}{|c|c|c|c|c|c|}
\hline
yyyy.mm.dd & JD  &  yyyy.yyyy & $R_{\rm flux}$ & $\sigma$ & Telescope \\
1 & 2 & 3 & 4 & 5 & 6 \\
\hline
2024.05.22	& 2460453	& 2024.3891	& 1.46	& 0.03	& AS-500 \\
2024.05.23	& 2460454	& 2024.3914	& 1.40	& 0.01	& Z-1000 \\
2024.05.23	& 2460454	& 2024.3917	& 1.39	& 0.03	& AS-500 \\
2024.05.24	& 2460455	& 2024.3942	& 1.39	& 0.03	& AS-500 \\
2024.05.24	& 2460455	& 2024.3945	& 1.36	& 0.03	& Z-1000 \\
\hline
\end{tabular}
\end{table}

\subsection{Fermi-LAT}
We utilized the $\gamma$-ray light curves available in the Fermi Large Area Telescope (Fermi-LAT) public Light Curve Repository\footnote{\url{https://fermi.gsfc.nasa.gov/ssc/data/access/lat/LightCurveRepository/about.html}} (LCR, \citealt{2023ApJS..265...31A}). The LCR is a database of multi-cadence flux-calibrated light curves for over 1500 sources considered as variable in the 10-yr Fermi-LAT point source (4FGL-DR2) catalogue \citep{2020arXiv200511208B}. The analysis has been performed with the standard Fermi-LAT science tools (version \verb|v11r5p3|), which use maximum likelihood analysis \citep{2009ApJS..183...46A}, considering a region with a radius of $12^{\circ}$ centered at the location of the source of interest and the \mbox{$\gamma$-rays} energy range between 100 MeV and 100 GeV. 
We allowed the photon index to vary for a better representation of the spectrum evolution during quiescent and flaring periods.
A limit of Test Statistics (TS) ${\rm TS}\geq 4$ (approximately $2\sigma$) was applied to compute the fluxes for each bin of the light curve, while only the measurements with ${\rm TS}\geq 10$ were selected for further MW analysis. We adopted the weekly binned light curves to have more evenly-sampled time series with higher time cadence in the period from August 2008 to March 2025. The $\gamma$-ray flux uncertainties are smaller than the plotted symbol size in Fig.~\ref{fig:fig1}. The typical 2$\sigma$ statistical errors on the weekly photon fluxes are about 4 $\times$ 10$^{-8}$ photons cm$^{-2}$ s$^{-1}$ corresponding to about 20 per cent of fractional uncertainty.

The $\gamma$-ray light curves used in this work were obtained from the default LCR products, which include only convergent likelihood fits and exclude unconstrained solutions where the photon index reaches the limits of the allowed parameter range.

To ensure that the extracted $\gamma$-ray light curve is not contaminated by nearby sources, we verified the field isolation around Ton\,599 (4FGL J1159.5$+$2914) using the \texttt{astroquery} interface to the 4FGL-DR4 catalog \citep{2022ApJS..260...53A}. A search within a $2^{\circ}$ radius centered on the target revealed no additional $\gamma$-ray sources, confirming that Ton\,599 is isolated within the LAT point spread function and that the retrieved light curve represents the intrinsic variability of the source. These checks confirm that the timing and periodicity results reported in this work are not affected by known LCR caveats.

\section{Long-term light curves}
\label{sec:lc}
\begin{figure*}
\centerline{\includegraphics[width=0.8\linewidth]{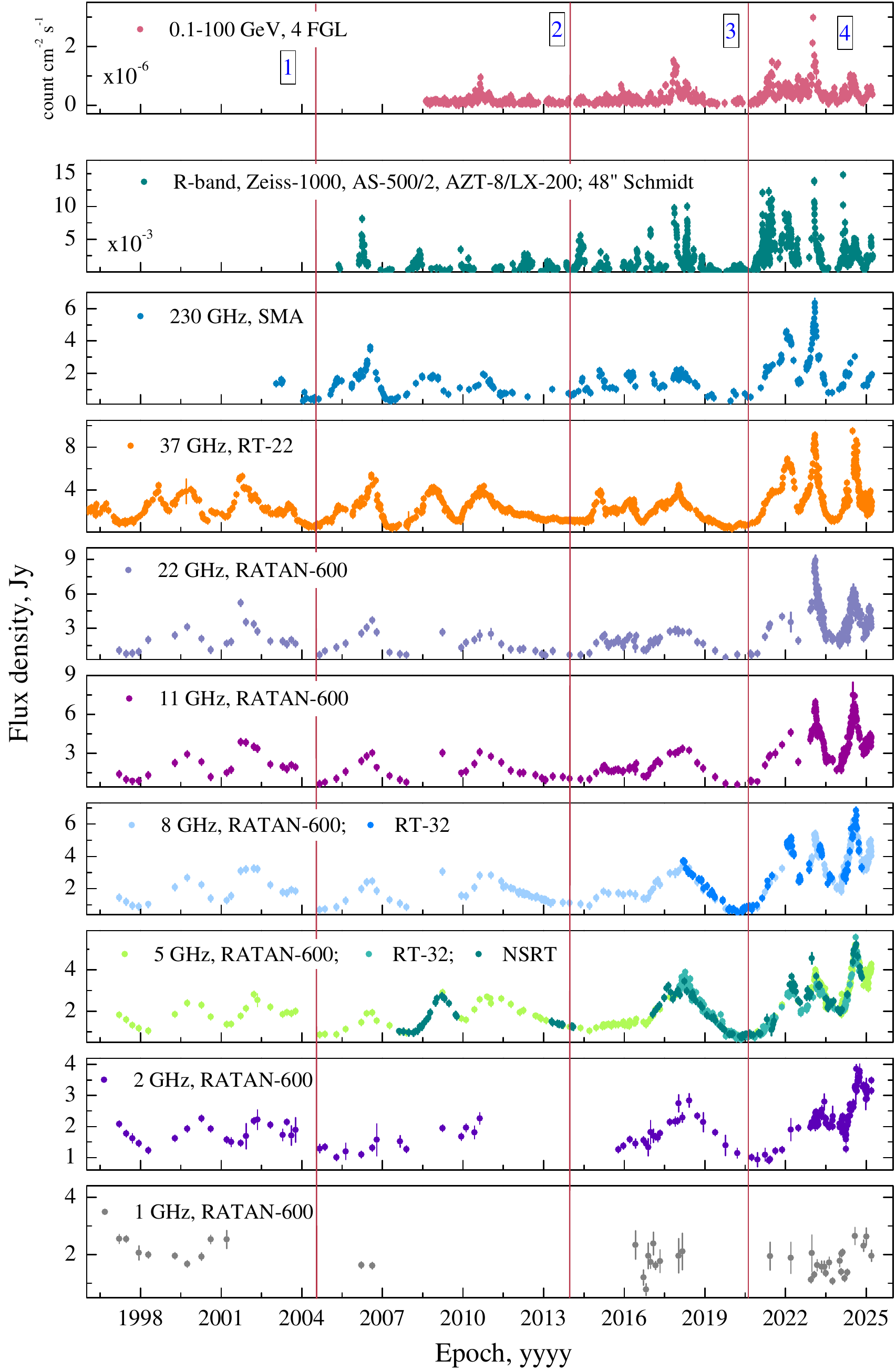}}
\caption{Multiband light curves of Ton\,599 in 1997--2025.} 
\label{fig:fig1}
\end{figure*}

\begin{figure*}
\centerline{\includegraphics[width=1.8\columnwidth]{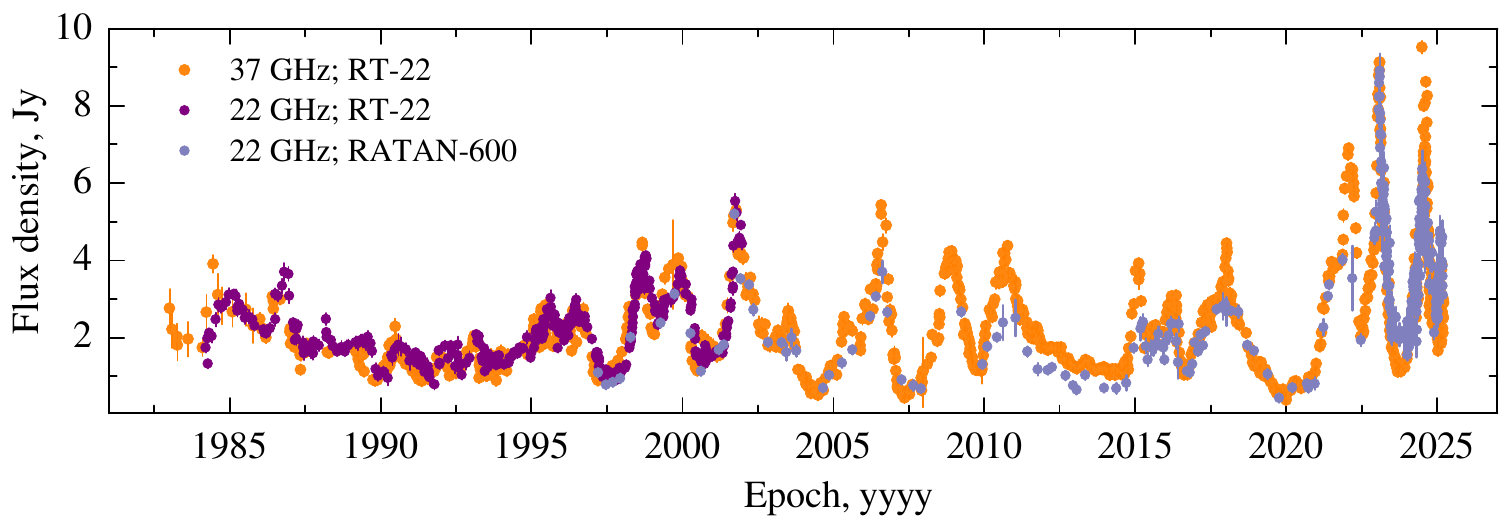}}
\caption{The 22 and 37 GHz radio light curves of Ton\,599 measured with the RT-22 and RATAN-600 in 1983--2025.} 
 \label{fig:fig2}
\end{figure*}

The long-term multiwavelength light curves for Ton\,599 are presented in Fig.~\ref{fig:fig1}. The individual 22 and 37 GHz light curves, measured with the RT-22 and covering the time span from 1983 to 2025, are presented in Fig.~\ref{fig:fig2}. Since the year 1997, a pattern has been observed: series of three major flares separated by short time intervals of low activity. 

In accordance with this, we separated the light curves into four epochs of activity 1, 2, 3, and 4. The epochs are denoted in Fig.~\ref{fig:fig1} by vertical lines; the corresponding dates are given in Table~\ref{tab:epochs}. As is shown further in Section~\ref{sec:dcf}, the time lags between flares at different frequencies decrease from epoch~1 to epoch~4. 
Although this does not affect the investigation of quasi-periodicity conducted in this paper, it is essential for the significance of the observed cross-correlations between flares at different frequencies and is, as well, an observed phenomenon that is worth noting.

\begin{table}
\caption{Four epochs with flaring activity} 
\label{tab:epochs}
\centering
\begin{tabular}{ccc}
\hline
Epoch &  Period, yyyy.yyyy  & Dates \\
\hline
1  &  1997.2082--2004.4959 & Mar 1997 -- Jun 2004 \\
2  &  2004.4959--2013.9671 & Jun 2004 -- Dec 2013 \\
3  &  2013.9671--2020.6000 & Dec 2013 -- Aug 2020 \\
4  &  2020.6000--2025.2475 & Aug 2020 -- Apr 2025 \\
\hline
\end{tabular}
\end{table} 

During the last 30--40 years, Ton\,599 has systematically experienced major outbursts detected in the entire electromagnetic spectrum. The variability level both for the whole period and within each of the epochs 1--4 was calculated using the expression for fractional variability $F_{\rm var}$ \citep{2003MNRAS.345.1271V}:
\begin{equation}
\label{eq:frac}
F_{\rm var}=\sqrt{\frac{V^2-\bar\sigma^{2}_{\rm err}}{\bar S^2}}
\end{equation}
where $V^2$ is variance, $\bar S$ is the mean flux density, and $\sigma_{\rm err}$ is the root mean square error. The uncertainty of $F_{\rm var}$ is determined as
\begin{equation}
\label{eq:frerr}
\bigtriangleup F_{\rm var}=\sqrt{\left(\sqrt{\frac{1}{2N}}\frac{\bar\sigma^{2}_{\rm err}}{F_{\rm var}\,\bar S^2}\right)^2 + \left(\sqrt{\frac{\bar \sigma^{2}_{\rm err}}{N}}\frac{1}{\bar S}\right)^2}.
\end{equation}

\begin{figure}
\centerline{\includegraphics[width=1.0\columnwidth]{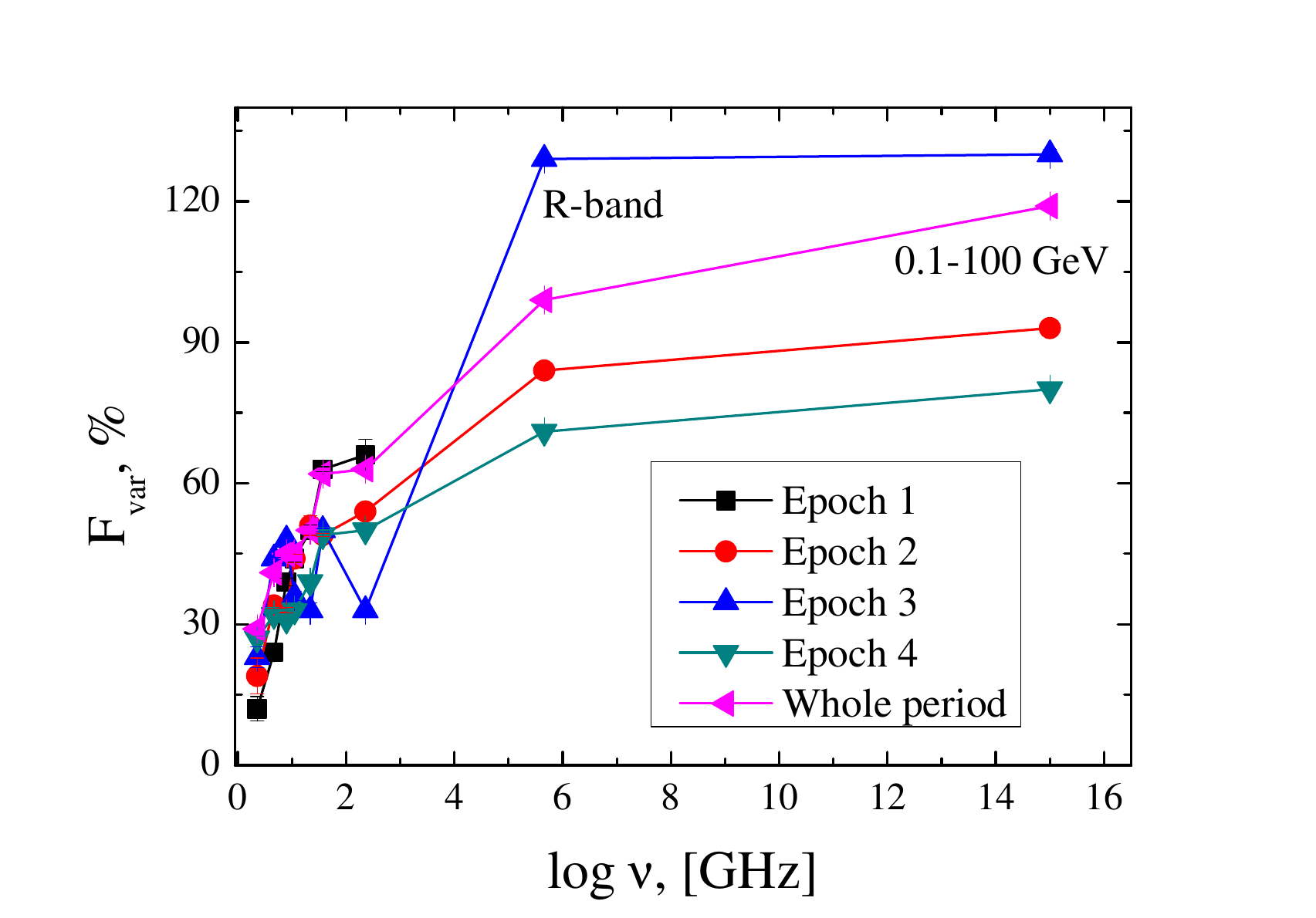}}
\caption{Multiband fractional variability in different epochs.}
\label{fig:var-band}
\end{figure}

The $F_{\rm var}$--$\log \nu$ plot is given in Fig.~\ref{fig:var-band}. The $F_{\rm var}$ errors are from 0.1 to 3.8 per cent in different wavelength bands. 
The fractional variability increases with increasing frequency both for the whole time period and for most of the individual epochs. The exception is epoch 3: it is clearly seen that $F_{\rm var}$ decreases with frequency in the 8--22 GHz range. This fact can be explained by strong opacity at these frequencies. 

\section{Cross-correlation between the light curves}
\label{sec:dcf}

The flares observed in the light curves at different frequencies correlate. To describe these correlations numerically, we used the method of discrete correlation functions (DCFs) suggested by \cite{1988ApJ...333..646E}, which allows analyzing unevenly sampled variability data without interpolation in the temporal domain. We used the Python software package {\tt pydcf}\footnote{\url{https://github.com/astronomerdamo/pydcf}} \citep{2015MNRAS.453.3455R}. A specificity of DCF calculation is the use of bins of time lags, therefore DCF values are intrinsically discrete, with the size of the increment (bin width) to be defined manually, which introduces both the observational and human biases in the derived lag values. Moreover, the uncertainties of the lags cannot be strictly estimated in this approach. To overcome this issue, \cite{1998PASP..110..660P} suggested the method of flux redistribution and random subset selection (FR/RSS). On the step of flux redistribution, white noise with a standard deviation equal to the uncertainties of the measured flux densities is added to the original observations. On the second step of random subset selection, a subset of the flux densities is randomly selected with replacement (bootstrapping). Then, a DCF is derived from the modified data, and a lag can be obtained at the maximum of the DCF or at a centroid of several DCF values near the maximum. By repeating these steps many times in Monte Carlo simulations, a distribution of lags can be obtained, from which a realistic observation-based lag can be derived along with its uncertainty. In our implementation of the method, we also varied the bin size in the DCF calculations to make the best-fitting lag bin-independent.

The significance levels for the DCFs were calculated by the Monte Carlo method, where we simulated a large number of light curves (5000 per frequency) randomized by amplitude and phase but with power spectral densities (PSDs) and probability density functions (PDFs) similar to the observed light curves at particular frequencies. Flux densities in the simulated curves were then sampled so that only the flux densities corresponding to the time of actual observations were taken into account to calculate the significance levels. The $1\sigma$, $2\sigma$, and $3\sigma$ levels were estimated based on the obtained normal distributions of DCF values at each particular lag. 

The method of light curve simulation is described in \cite{2013MNRAS.433..907E}, we used the software tool {\tt DELightcurveSimulation}\footnote{ \url{https://github.com/samconnolly/DELightcurveSimulation/}} \citep{2015arXiv150306676C}. We have additionally modified the original code to fit the PDFs derived from the observed light curves using the nested sampling Monte Carlo algorithm MLFriends \citep{2019PASP..131j8005B} using the UltraNest\footnote{\url{https://johannesbuchner.github.io/UltraNest/}} package 
\citep{2021JOSS....6.3001B}. The approximations of the PSDs and PDFs for the observed light curves at each frequency, along with the parameters of the distributions, are given in Appendix~\ref{app:stat}.

The PSD and PDF can be estimated from an observed light curve using either only the actual measurements or by imputing missing measurements, e.g., via interpolation. In our analysis we attempted to derive the PDF shape from both the real and interpolated data. As well, the Lomb--Scargle periodogram was explored to obtain the PSD without interpolation. The most realistic simulated light curves were obtained in the case of interpolation. The significance levels calculated by the two approaches are consistent in most cases. We find that the less realistic simulated light curves in the first (real data) approach is due to the uneven cadence of the observed light curves in different time intervals.

In the interpolation approach, the light curve
first must be made evenly distributed in the temporal domain, this stage was done using the Steffen spline interpolation\footnote{ \url{https://gvar.readthedocs.io/en/latest/gvar_other.html}} \citep{1990A&A...239..443S}. The method performs 1D series interpolation. In each interval the interpolation function is assumed to be a third-order polynomial passing through the data points. The slope at each grid point is determined in such a way as to guarantee a monotonic behavior of the interpolating function. The result is a smooth curve with continuous first-order derivatives that passes through any given set of data points without spurious oscillations. Local extrema can occur only at grid points where they are given by the data, but not in between two adjacent grid points. The method is suitable for the AGN light curves with their possible abrupt flares. To make the interpolated light curves even more realistic, we add to the interpolated points white noise with a standard deviation equal to the median uncertainty of the measured flux densities.

We calculated pair-wise time lags between light curves at different frequencies for each of the epochs mentioned above. Table~\ref{tab:ccf-lags} summarizes the results for the four epochs (columns) and each pair of frequencies (rows). The time lags and their significances are given. We present only those values where DCF significances obtained using Monte Carlo simulations are higher than $2\sigma$. 

The frequency pairs for which a correlation cannot be estimated because of the absent data are indicated by the N/A abbreviation. In epoch~2, the time span of the $\gamma$-ray observations is significantly shorter than time spans for the other frequencies, therefore in the pairwise DCF calculations with the $\gamma$-ray light curve in epoch~2 we limited the measurements of a paired frequency within the $\gamma$-ray observation time span. We also note that using the full time span does not actually change the results: the lags remain the same within measurement errors, and the significance does not improve or degrade.

A large fraction of frequency pairs exhibit correlations at significance levels greater than $2\sigma$, for a number of pairs correlations are greater than $3\sigma$. The DCFs corresponding to the measurements from Table~\ref{tab:ccf-lags} are presented in Figs.~\ref{fig:dcf_ep1}--\ref{fig:dcf_ep4} along with the derived lags and their uncertainties estimated by the FR/RSS method mentioned above.

\begin{table*}
\centering
\caption{\label{tab:ccf-lags} 
Time lags between the light curves at different frequencies with significances greater than the $2\sigma$ level. The dashes are where no significant correlation has been found for a frequency pair; the N/A abbreviation indicates the absence of measurements.
}
\begin{tabular}{ccccccccc}
\hline
\multirow{2}{*}{Bands}  & Lag, days & Signif. & Lag, days & Signif. & Lag, days & Signif. & Lag, days & Signif.\\
  & \multicolumn{2}{c}{epoch 1} & \multicolumn{2}{c}{epoch 2} & \multicolumn{2}{c}{epoch 3} & \multicolumn{2}{c}{epoch 4}\\
\hline
$\gamma$ vs $R$     & N/A & N/A & -- & -- & $0\pm20$ & $3\sigma$ & $10\pm10$ & $2\sigma$ \\
$\gamma$ vs 230 GHz & N/A & N/A & -- & -- & -- & -- & $10\pm5$ & $2\sigma$\\
$\gamma$ vs 37 GHz  & N/A & N/A & -- & -- & $40\pm30$ & $2\sigma$ & $20\pm10$ &  $2\sigma$ \\
$\gamma$ vs 22 GHz   & N/A & N/A & $25\pm110$ & $2\sigma$ & $0\pm50$ & $2\sigma$ & $30\pm5$ & $3\sigma$ \\
$\gamma$ vs 11 GHz   & N/A & N/A & $70\pm100$ & $2\sigma$ & $20\pm50$ & $2\sigma$ & $20\pm10$ & $2\sigma$ \\
$\gamma$ vs 8 GHz   & N/A & N/A & -- & -- & $80\pm30$ & $2\sigma$ & -- & -- \\
$\gamma$ vs 5 GHz   & N/A & N/A & -- & -- & $75\pm30$ & $2\sigma$ & -- & -- \\
$\gamma$ vs 2 GHz   & N/A & N/A & -- & -- & $120\pm70$ & $2\sigma$ & -- & -- \\
\hline
$R$ vs 230 GHz & N/A & N/A & $90\pm25$ & $3\sigma$ & $70\pm35$ & $2\sigma$ & $20\pm10$ & $2\sigma$\\
$R$ vs 37 GHz  & N/A & N/A & $120\pm25$ & $2\sigma$ & $25\pm45$ & $2\sigma$ & -- & -- \\
$R$ vs 22 GHz   & N/A & N/A & $115\pm65$ & $2\sigma$ & $-15\pm65$ & $2\sigma$ & $10\pm10$ & $2\sigma$ \\
$R$ vs 11 GHz   & N/A & N/A & $100\pm95$ & $2\sigma$ & $30\pm55$ & $2\sigma$ & -- & -- \\
$R$ vs 8 GHz   & N/A & N/A & -- & -- & $105\pm25$ & $2\sigma$ & -- & -- \\
$R$ vs 5 GHz   & N/A & N/A & -- & -- & $95\pm20$ & $2\sigma$ & -- & -- \\
$R$ vs 2 GHz   & N/A & N/A & -- & -- &  $175\pm80$ & $2\sigma$ & -- & -- \\
\hline
230 GHz vs 37 GHz  & N/A & N/A & $40\pm10$ & $3\sigma$ & $-10\pm20$ & $2\sigma$ & $0\pm5$ & $3\sigma$ \\
230 GHz vs 22 GHz   & N/A & N/A & $85\pm40$ & $3\sigma$ & $20\pm40$ & $2\sigma$ & $10\pm5$ & $3\sigma$\\
230 GHz vs 11 GHz   & N/A & N/A & $115\pm50$ & $2\sigma$ & $15\pm50$ & $2\sigma$ & $10\pm5$ & $3\sigma$\\
230 GHz vs 8 GHz   & N/A & N/A & $160\pm65$ & $2\sigma$ & $70\pm40$ & $2\sigma$ & $10\pm5$ & $2\sigma$ \\
230 GHz vs 5 GHz   & N/A & N/A & -- & -- & $80\pm30$ & $2\sigma$ & -- & -- \\
\hline
37 GHz vs 22 GHz   & $150\pm55$ & $2\sigma$ & $30\pm25$ & $3\sigma$ & $15\pm25$ & $3\sigma$ & $10\pm5$ & $2\sigma$\\
37 GHz vs 11 GHz   & $195\pm45$ & $3\sigma$ & $55\pm30$ & $3\sigma$ & $20\pm30$ & $2\sigma$ & $10\pm5$ & $2\sigma$ \\
37 GHz vs 8 GHz   & $220\pm40$ & $3\sigma$ & $90\pm25$ & $3\sigma$ & $90\pm25$ & $3\sigma$ &  $0\pm5$ & $2\sigma$\\
37 GHz vs 5 GHz   & $315\pm45$ & $2\sigma$ & $120\pm35$ & $2\sigma$ & $105\pm20$ & $3\sigma$ & -- & -- \\
37 GHz vs 2 GHz   & $365\pm80$ & $3\sigma$ & -- & -- &  $95\pm55$ & $2\sigma$ & -- & -- \\
\hline
22 GHz vs 11 GHz   & -- & -- & $20\pm45$ & $3\sigma$ & $-5\pm45$ & $2\sigma$ & $-10\pm5$ & $3\sigma$\\
22 GHz vs 8 GHz   & $90\pm70$ & $3\sigma$ & $75\pm70$ & $3\sigma$ & $60\pm40$ & $3\sigma$ & $0\pm5$ & $2\sigma$ \\
22 GHz vs 5 GHz   & $180\pm60$ & $3\sigma$ & -- & -- & $110\pm35$ & $3\sigma$ & -- & -- \\
\hline
11 GHz vs 8 GHz   & $30\pm70$ & $3\sigma$ & $45\pm55$ & $3\sigma$ & $40\pm35$ & $3\sigma$ & $0\pm5$ & $3\sigma$\\
11 GHz vs 5 GHz   & $130\pm75$ & $3\sigma$ & $65\pm80$ & $2\sigma$ & $75\pm35$ & $3\sigma$ & $10\pm5$ & $2\sigma$ \\
11 GHz vs 2 GHz   & $185\pm90$ & $3\sigma$ & -- & -- & -- & -- & $60\pm10$ & $2\sigma$ \\
\hline
8 GHz vs 5 GHz   & $95\pm70$ & $3\sigma$ & $15\pm60$ & $3\sigma$ & $20\pm20$ & $3\sigma$ & $20\pm5$ & $3\sigma$\\
8 GHz vs 2 GHz   & $185\pm70$ & $3\sigma$ & -- & -- & $100\pm60$ & $3\sigma$ & $60\pm10$ & $2\sigma$\\
\hline
5 GHz vs 2 GHz   & $95\pm70$ & $3\sigma$ & -- & -- & $90\pm50$ & $3\sigma$ & $40\pm5$ & $3\sigma$\\
\hline
\end{tabular}
\end{table*}

\begin{figure}
\includegraphics[width=\columnwidth]{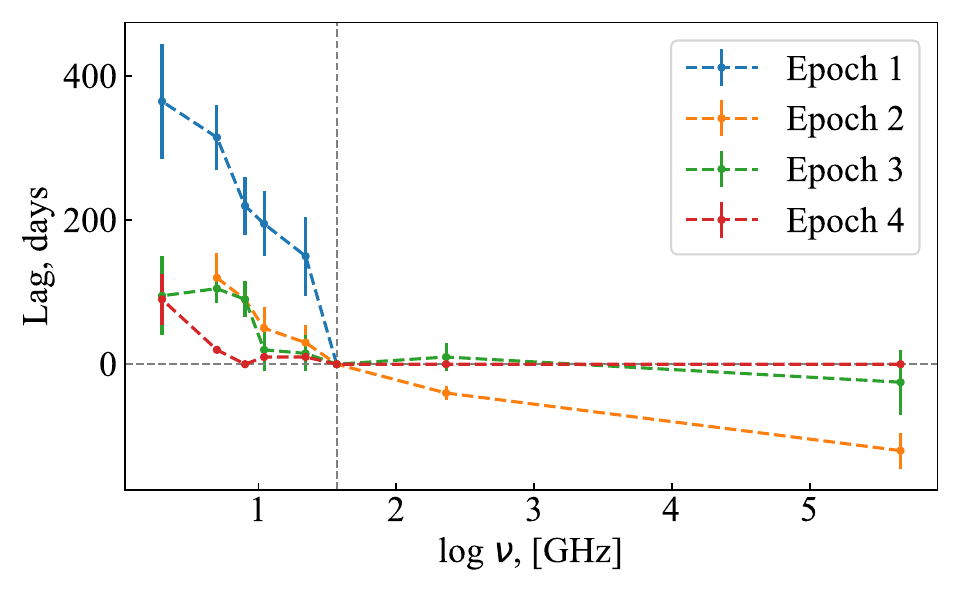}
\caption{Lags of the light curves at various frequencies ($\log_{10} \nu$) with respect to the light curve at a frequency of 37~GHz (the zero lag marked by the grey lines). The frequency range is from 2~GHz (the leftmost points) to the optical $R$ band (on the right). The epochs are shown by different colors. 
}
\label{fig:lags}
\end{figure}

There is strong evidence that the observed lags decrease with time. Figure~\ref{fig:lags} presents the overall behavior of the time lags with respect to the lower/higher frequencies and for different observing epochs. In the figure we took the 37~GHz light curve as the reference and showed the lags relative to this light curve for frequencies from 1~GHz to the optical $R$~band during the four epochs. We can see that (i)~the light curves at lower frequencies lag those at higher frequencies and (ii)~the absolute values of the lags decrease from the earliest epoch~1 to the latest epoch~4; actually the lags are zero in epoch~4 between all the light curves except the lag at the lowest frequency of 2~GHz (the lags for the lower frequency of 1~GHz are not measurable because of the scarcity of data). Similar behavior of time lags can be observed if we take other frequencies as the reference.

The lags for the $\gamma$-ray band are omitted in Fig.~\ref{fig:lags}. It is seen from Table~\ref{tab:ccf-lags}, though, that in epoch~3 the lags with respect to $\gamma$-rays resemble those for the optical $R$ band.
Also, the lags between the $\gamma$-ray and $R$-band light curves in epochs~3 and~4 are zero with significances greater than $3\sigma$ and $2\sigma$, respectively. It is evident that the $R$ band and $\gamma$-rays correlate in epochs~3 and~4, where both curves have good cadence. Similar results for the $\gamma$-ray~-- optical band correlation was noted in \cite{2024MNRAS.52711900R}.

The systematic time delays between the high-energy (\mbox{$\gamma$-rays}, optical emission) and low-energy (radio) spectral ranges in Fig.~\ref{fig:lags} may be explained within the common paradigm of a flare expanding along a relativistic jet outside an active nucleus. The decrease of time delays from the first epoch to the last one may indicate the evolution of the physical conditions in the medium through which the jet propagates. In the case of an expanded volume, the optical depth varies comparatively slowly, providing higher time delays. For the last epoch, we can see very short time intervals (below 100 days) between the flares across all spectral bands, which indicates a more compact volume of jet propagation.

\section{Search for quasi-periodicity}

\subsection{Lomb--Scargle periodogram}
To investigate quasi-periodic behavior in unevenly sampled astronomical time series, we employed the Lomb--Scargle (LS) periodogram method \citep{1976Ap&SS..39..447L,1982ApJ...263..835S}. This technique is particularly well-suited for data with gaps or irregular cadence. It provides a periodogram resembling a classical Fourier power spectrum, where peaks indicate potential periodic signals present in the data. For the implementation, we used the {\tt astropy} library’s built-in LS algorithm,\!\footnote{\url{https://docs.astropy.org/en/stable/timeseries/lombscargle.html}} which is grounded in the computational framework presented by \cite{2014sdmm.book.....I}. This tool allows robust handling of heterogeneously sampled data and offers enhanced frequency resolution over traditional methods.

To ensure that detected periodicities are not artifacts of red noise contamination, commonly present in AGN light curves, we estimated the significance of periodogram peaks by comparing them with red-noise simulations. Following the methodology of \cite{2022MNRAS.513.5238R}, we generated synthetic light curves with power-law noise characteristics matching the observed data. By computing periodograms for a large number of these red-noise realizations, we constructed a statistical distribution of power values at each frequency, which allowed us to assess the false-alarm probability for the observed peaks and identify truly significant quasi-periodic components.

\begin{figure*}
\centerline{\includegraphics[height=3.2cm]{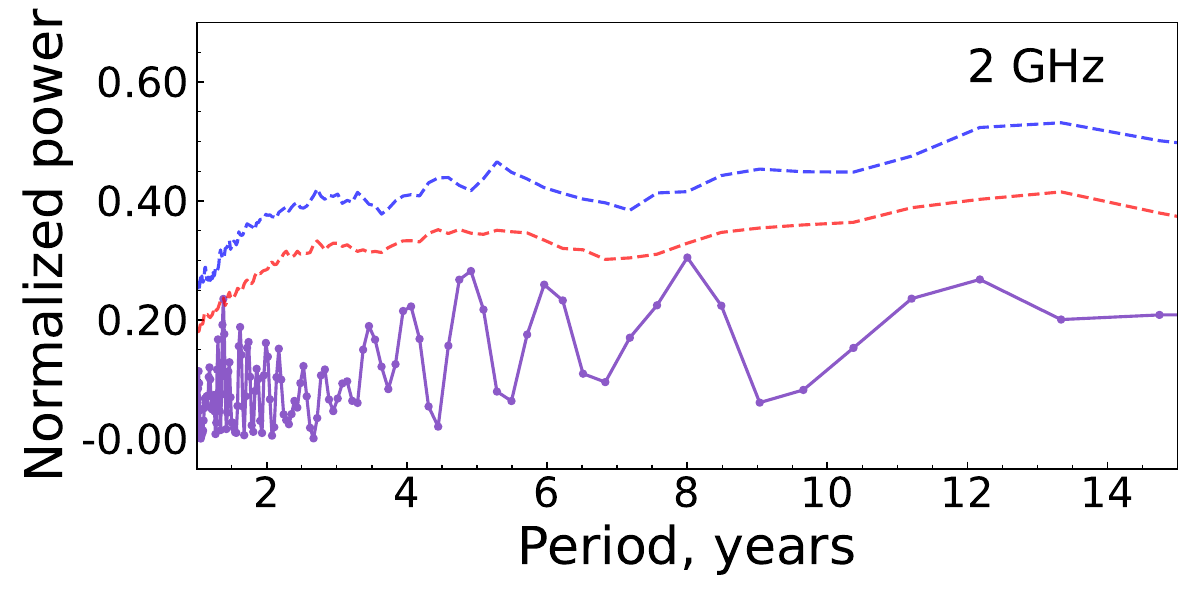} \includegraphics[height=3.2cm]{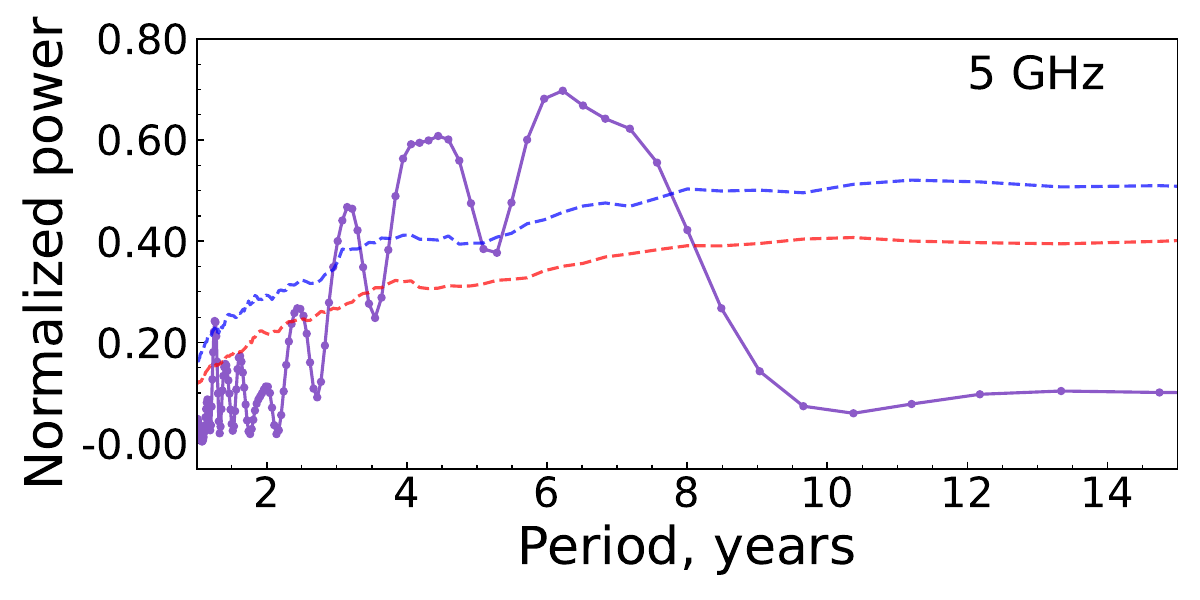}}
\centerline{\includegraphics[height=3.2cm]{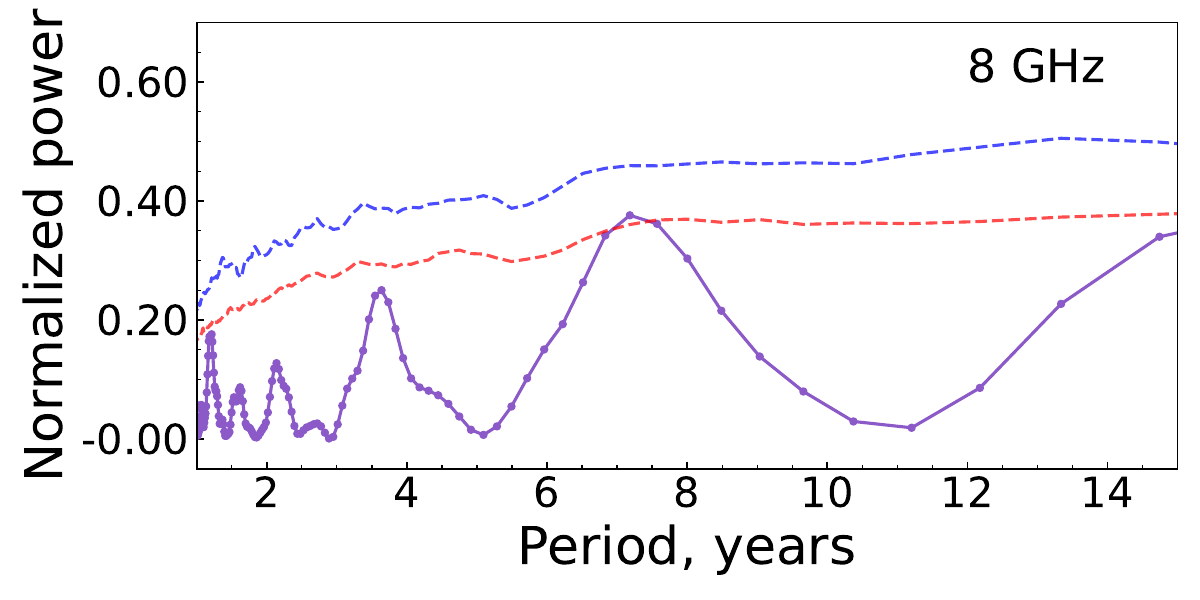} \includegraphics[height=3.2cm]{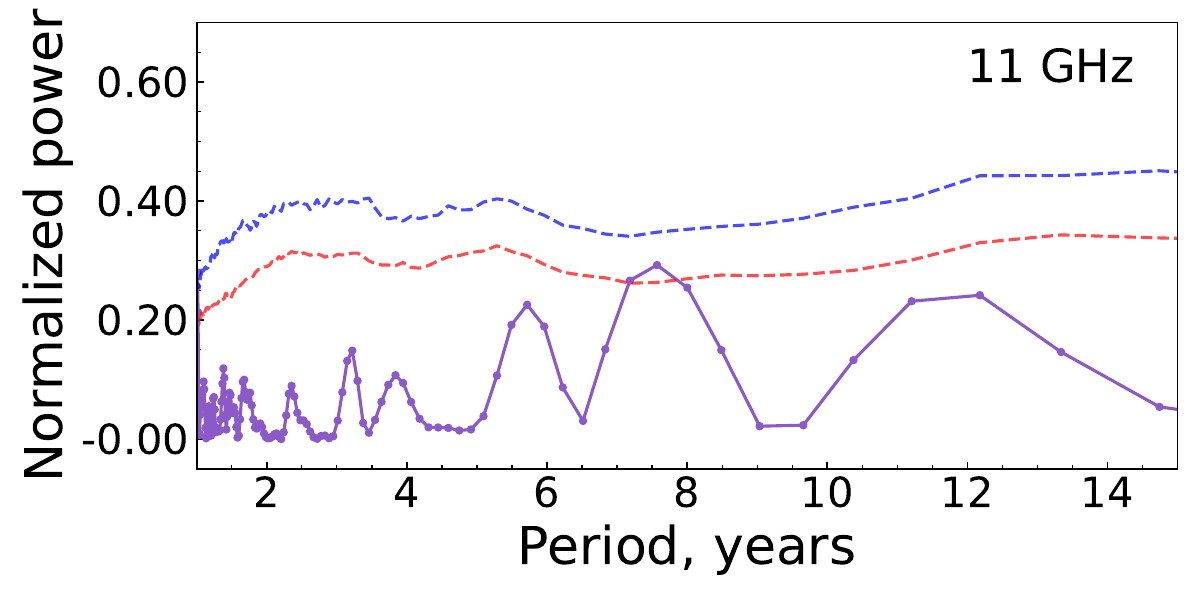}}
\centerline{\includegraphics[height=3.2cm]{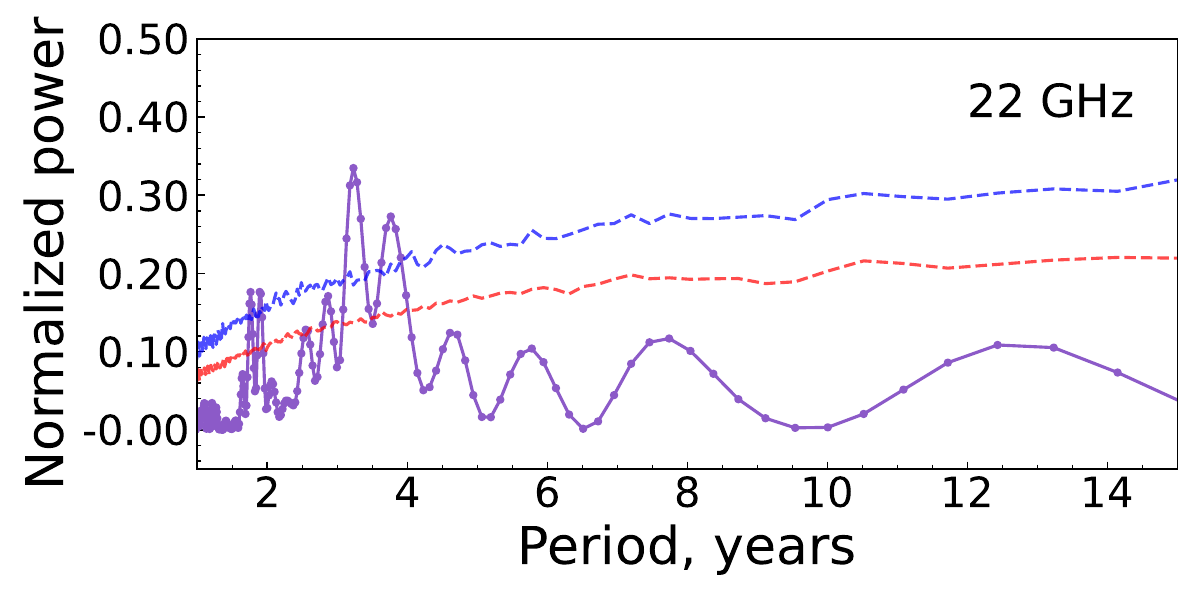} \includegraphics[height=3.2cm]{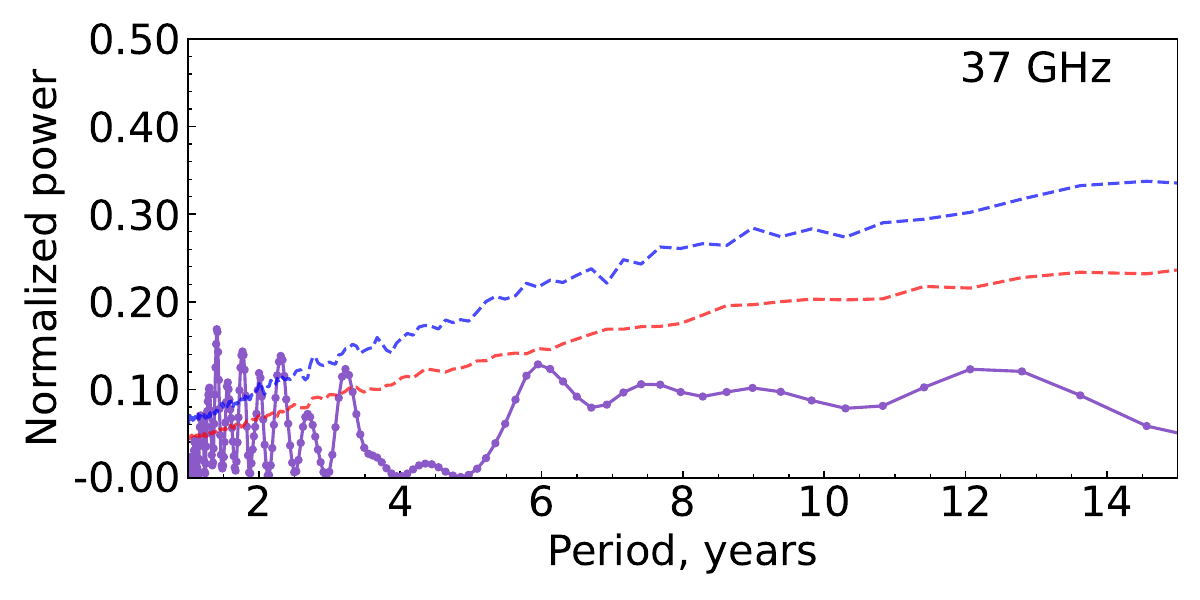}}
\centerline{\includegraphics[height=3.2cm]{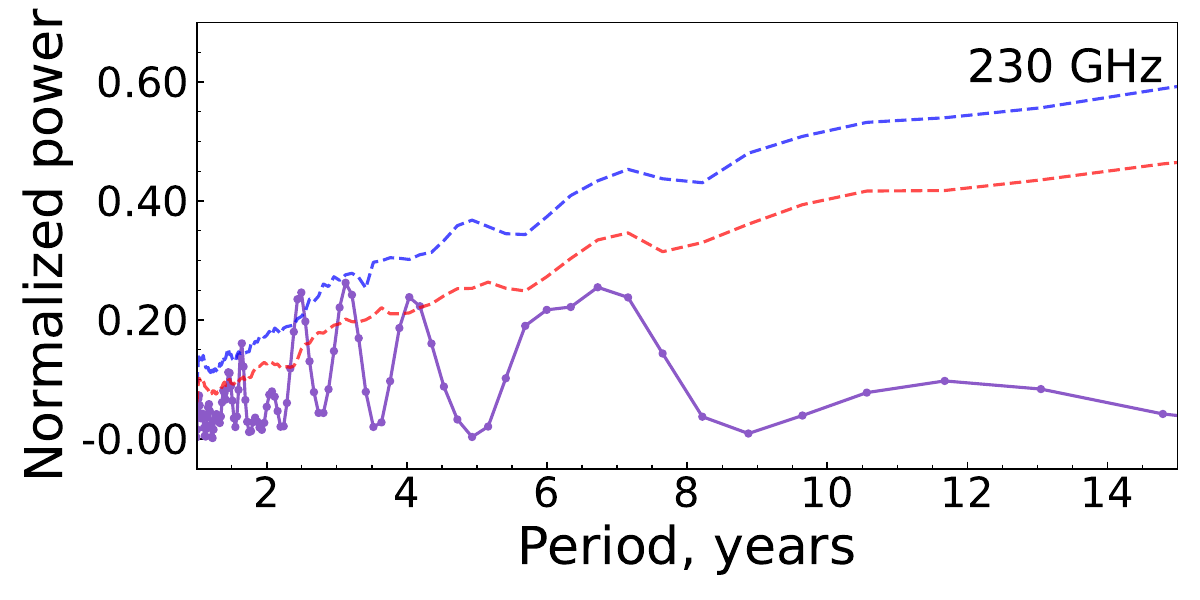} \includegraphics[height=3.2cm]{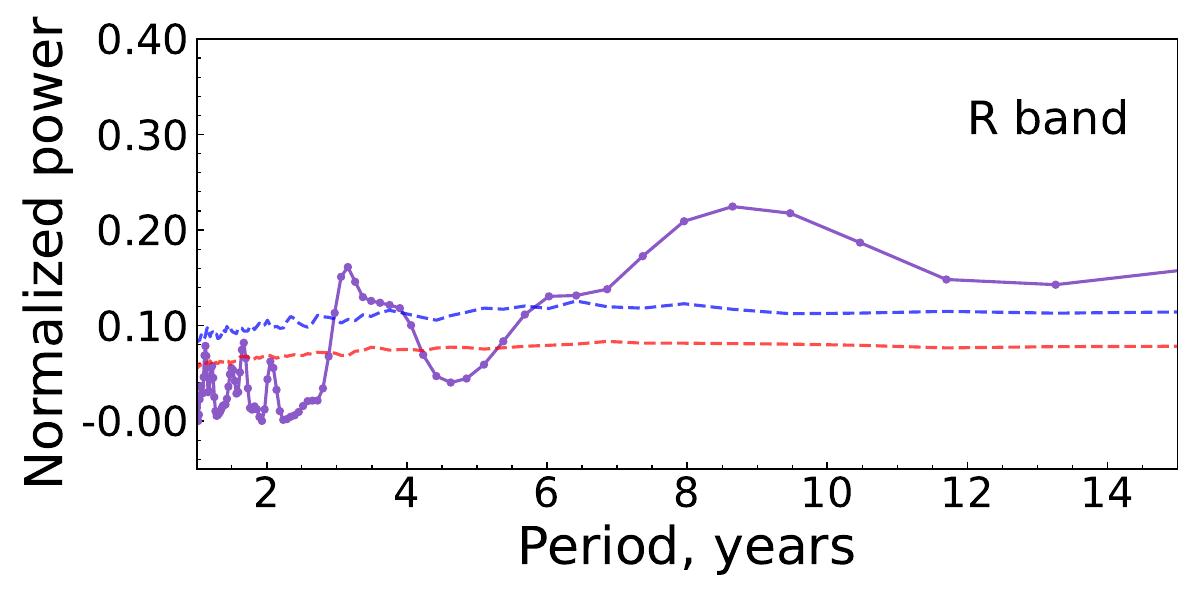}}
\centerline{\includegraphics[height=3.2cm]{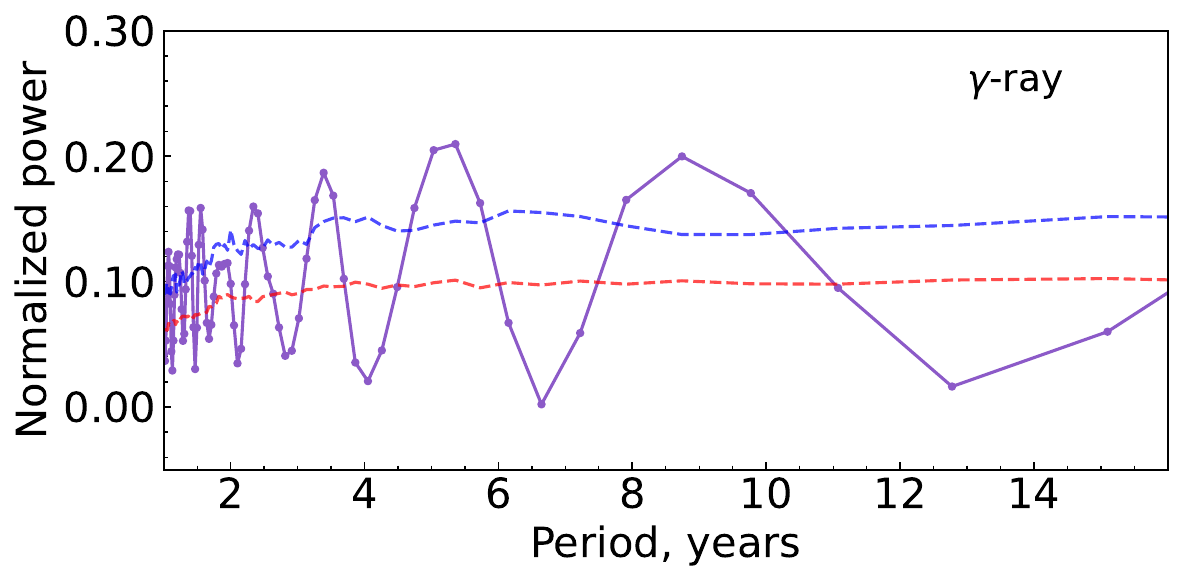}}
\caption{The LS periodograms for the total flux variations at 2, 5, 8, 11, 22, 37, 230~GHz, in the optical $R$ band, and in $\gamma$-rays. The dashed blue and red lines show the false alarm probability levels ${\rm FAP} =1$ and $5$ per cent.}
\label{fig:LS1}
\end{figure*}

The detected quasi-periods across all bands range from 1.0 to 12.2~years, with varying levels of statistical significance (Fig.~\ref{fig:LS1} and Table~\ref{tab:wwz-ls}).

\subsection{Weighted wavelet Z-transform}
\label{sec:wwz}

To investigate possible quasi-periodic oscillatory (QPO) behavior of the source, we employed the Weighted Wavelet Z-transform (WWZ) technique, which is particularly well-suited for unevenly sampled time series. It provides time-frequency localization while allowing for gaps in the data and for non-stationary signals.

The WWZ method was introduced by \citet{1996AJ....112.1709F}. Rather than relying on a classical wavelet integral transform, the WWZ
fits a local model to the data of the form
\begin{equation}
x(t) \approx A \cos(\omega t) + B \sin(\omega t) + C
\end{equation}
using weighted least squares, where each observation at a time \( t \) is weighted by a Gaussian:
\begin{equation}
w(t) = \exp\left[- c\, \omega^2 (t - \tau)^2\right].
\end{equation}
Here \( \omega \) is the angular frequency, \( \tau \) is the time shift, and \( c \) is a decay parameter controlling the temporal resolution. The resulting WWZ power quantifies how strongly a sinusoidal oscillation of frequency \( \omega \) is present in the data around the time \( \tau \).

We employed an open-source implementation of the WWZ algorithm, developed by Sebastian Kiehlmann and available at {\tt github},\!\footnote{\url{https://github.com/skiehl/wwz}} which closely follows the algorithmic framework described in \citet{1996AJ....112.1709F}. The implementation handles:
\begin{itemize}
\item unevenly spaced data without resampling;
\item Gaussian-weighted projections onto the basis functions \( \{1, \cos(\omega t), \sin(\omega t)\} \);
\item optional weighting by the signal-to-noise ratio (SNR) derived from flux uncertainties;
\item output of the WWZ power \( Z(\tau, \omega) \) on a dense grid in \((\tau, \omega)\).
\end{itemize}

For our analysis, we set the decay constant \( c = 1/(8\pi^2) \), following the recommendation of \citet{1996AJ....112.1709F}. We explored 1000 logarithmically spaced trial periods between \( P_{\rm min} = 0.05 \) yr and \( P_{\rm max} = 15 \) yr and computed WWZ power over 1000 uniformly spaced time steps across the observation span. The SNR weighting option was enabled.

The global WWZ power spectrum was computed by averaging the WWZ power over all time shifts:
\begin{equation}
Z_{\rm global}(\omega) = \frac{1}{N_\tau} \sum_{i=1}^{N_\tau} {\rm WWZ}(\omega, \tau_i).
\end{equation}

To assess the significance of peaks in the WWZ power spectrum, we performed Monte Carlo simulations following the procedure of \citet{2013MNRAS.433..907E}, implemented in the \texttt{DELightcurveSimulation} package \citep{2015arXiv150306676C}. 

For each simulated light curve, we computed the WWZ power and derived the global WWZ spectrum. We produced \( N_{\rm sim} = 1000 \) such realizations to build empirical distributions of WWZ power at each frequency. Following the approach described by \citet{2013MNRAS.433..907E} and adopted by \citet{2022MNRAS.513.5238R}, we determined the significance threshold at the 95\%\ confidence level as
\begin{equation}
Z_{95\%}(\omega) = \text{95$^\text{th}$ percentile of } \left\{ Z^{(k)}_{\rm global}(\omega) \right\}_{k=1}^{N_{\rm sim}},
\end{equation}
with similar thresholds computed at 99 per cent and higher levels.

\begin{figure*}
\centerline{\includegraphics[height=5cm]{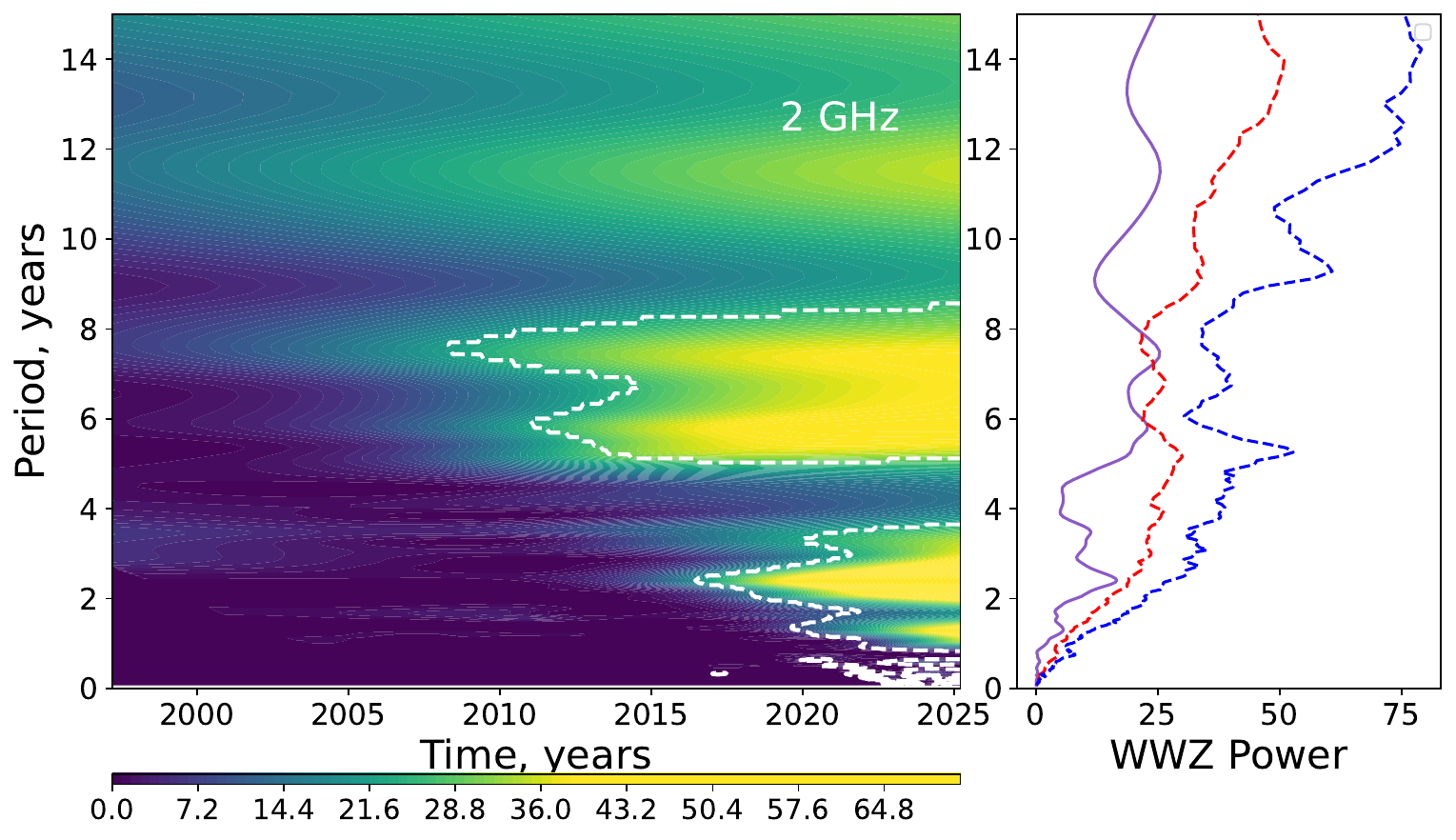}\includegraphics[height=5cm]{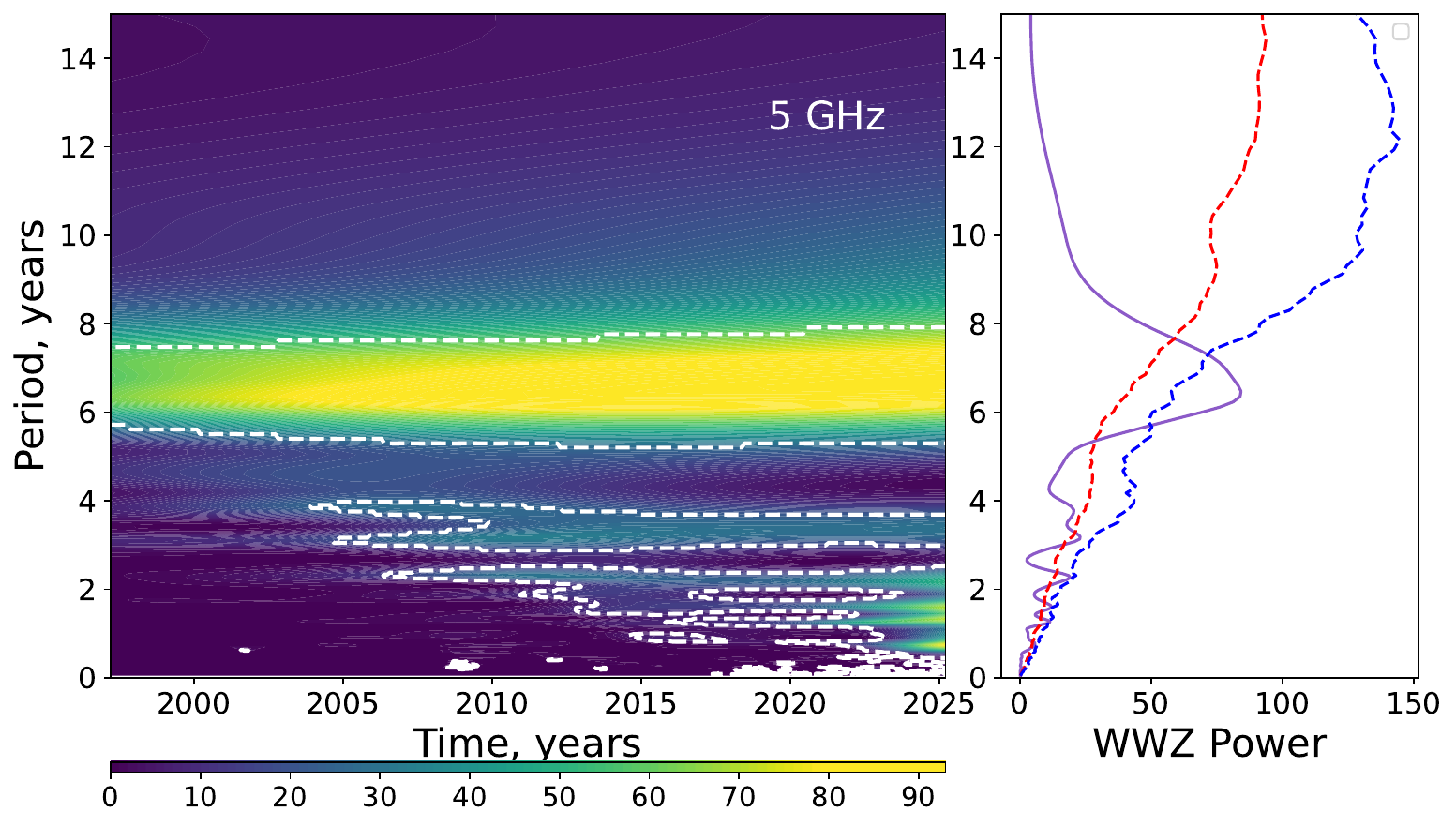}}\centerline{\includegraphics[height=5cm]{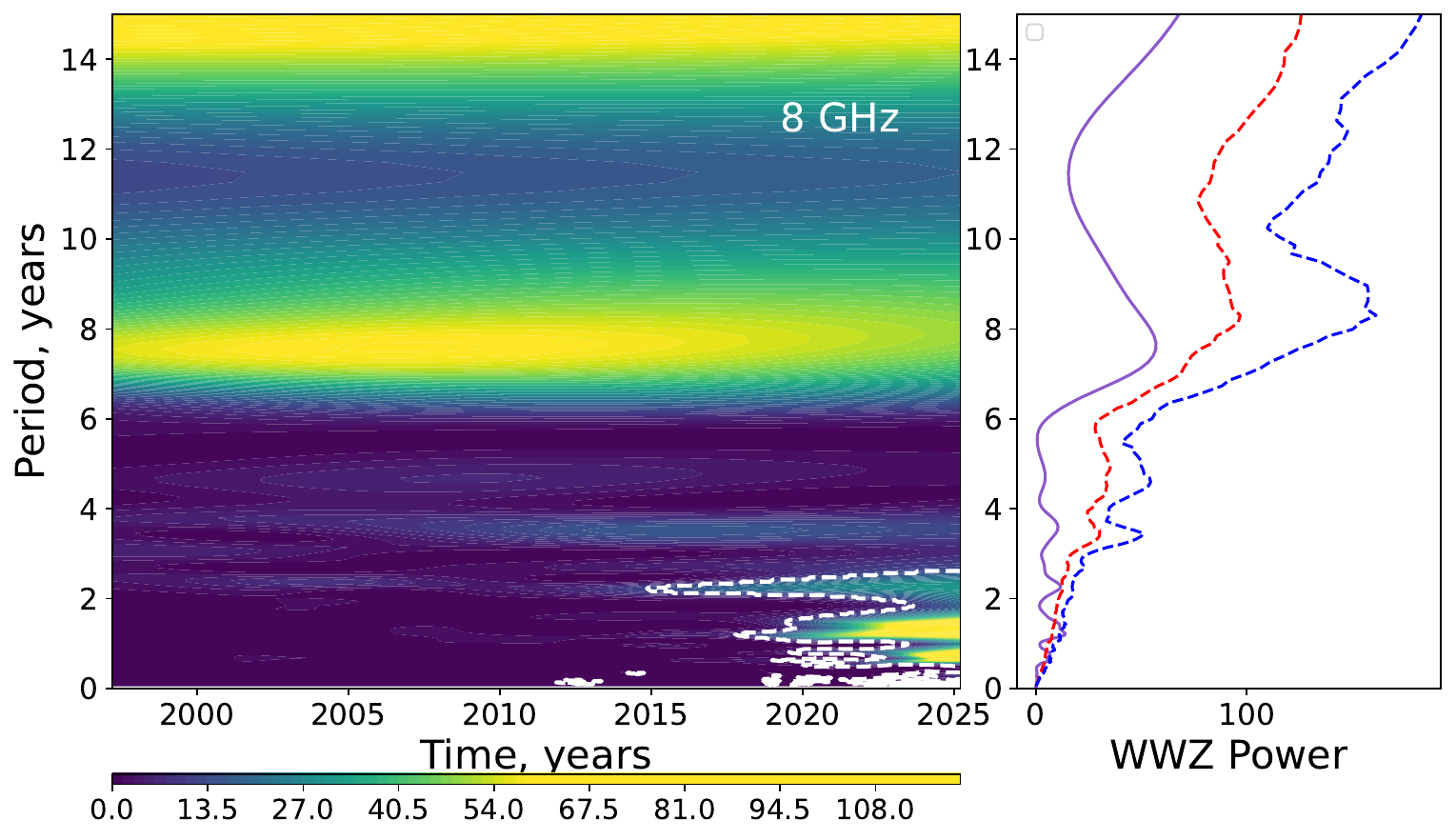}\includegraphics[height=5cm]{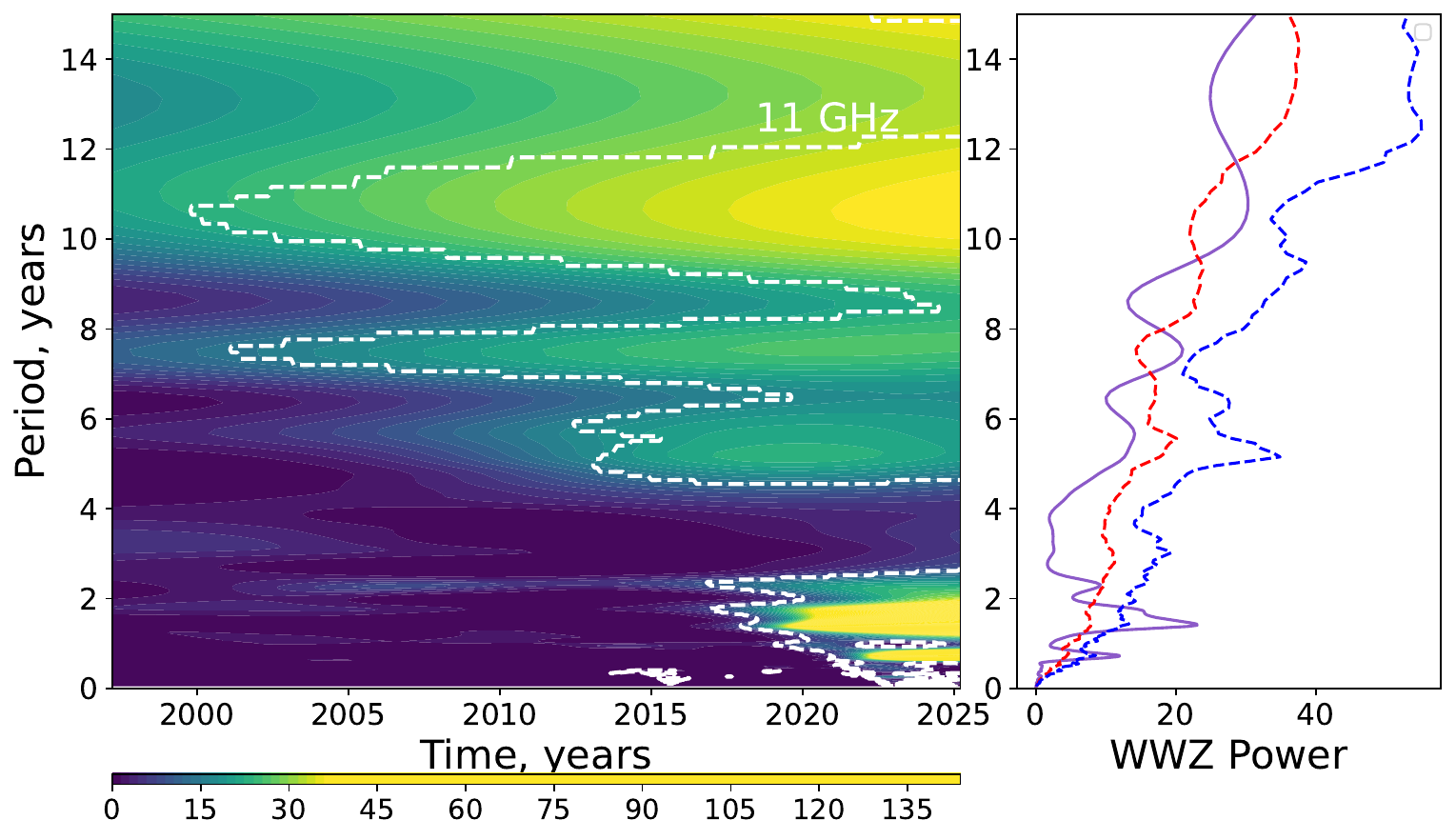}}
\centerline{\includegraphics[height=5cm]{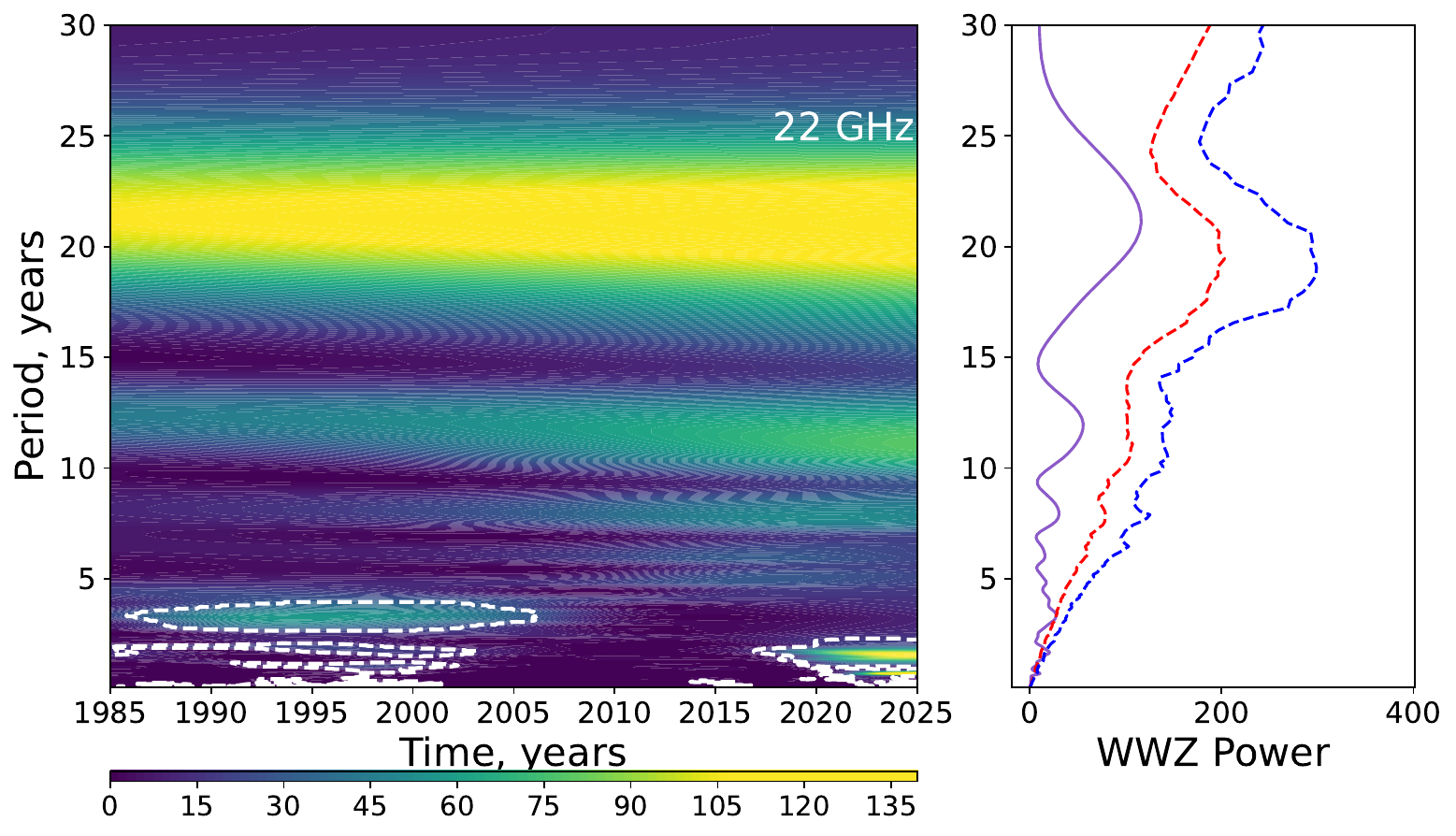}\includegraphics[height=5cm]{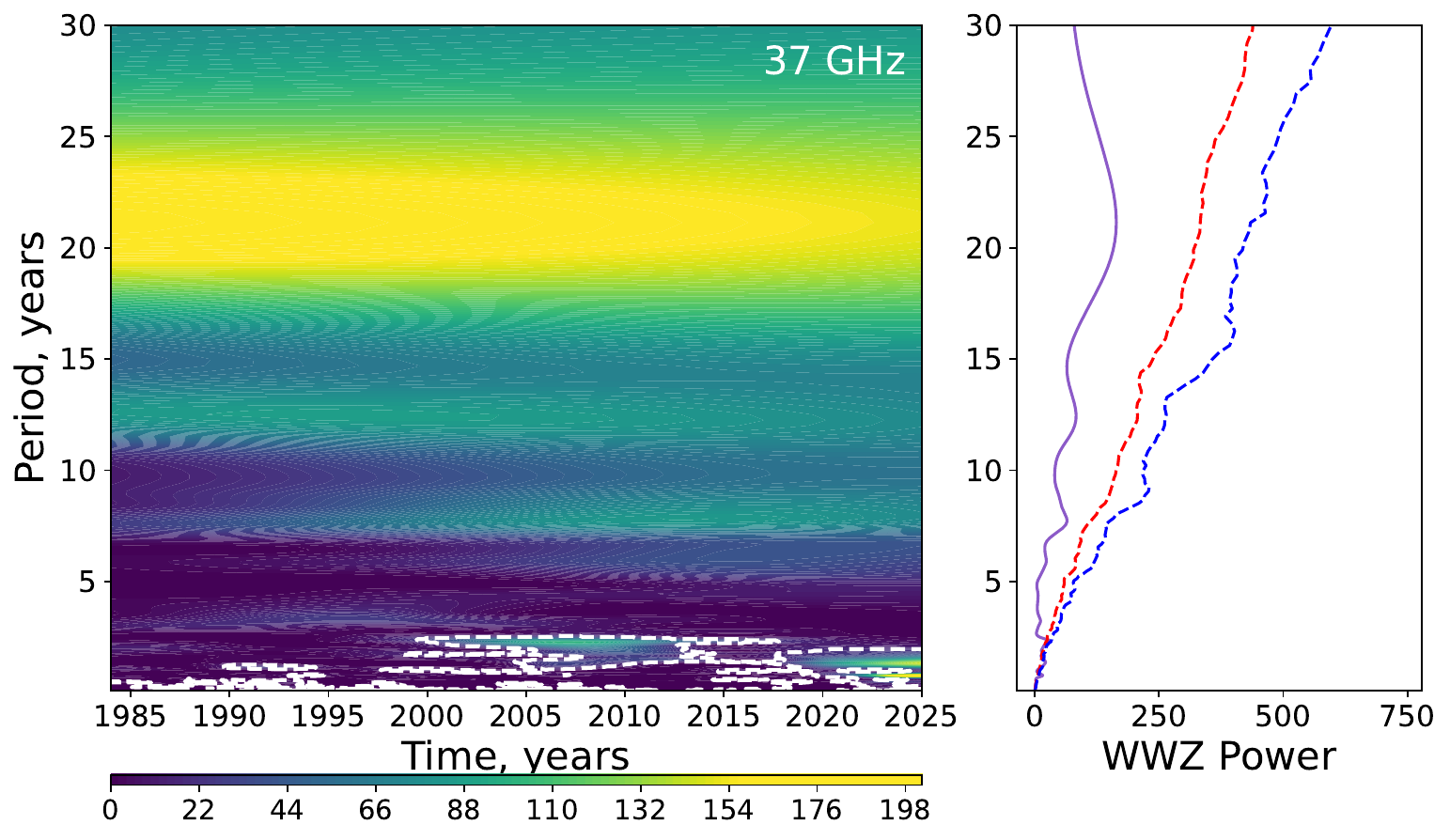}}
\centerline{\includegraphics[height=5cm]{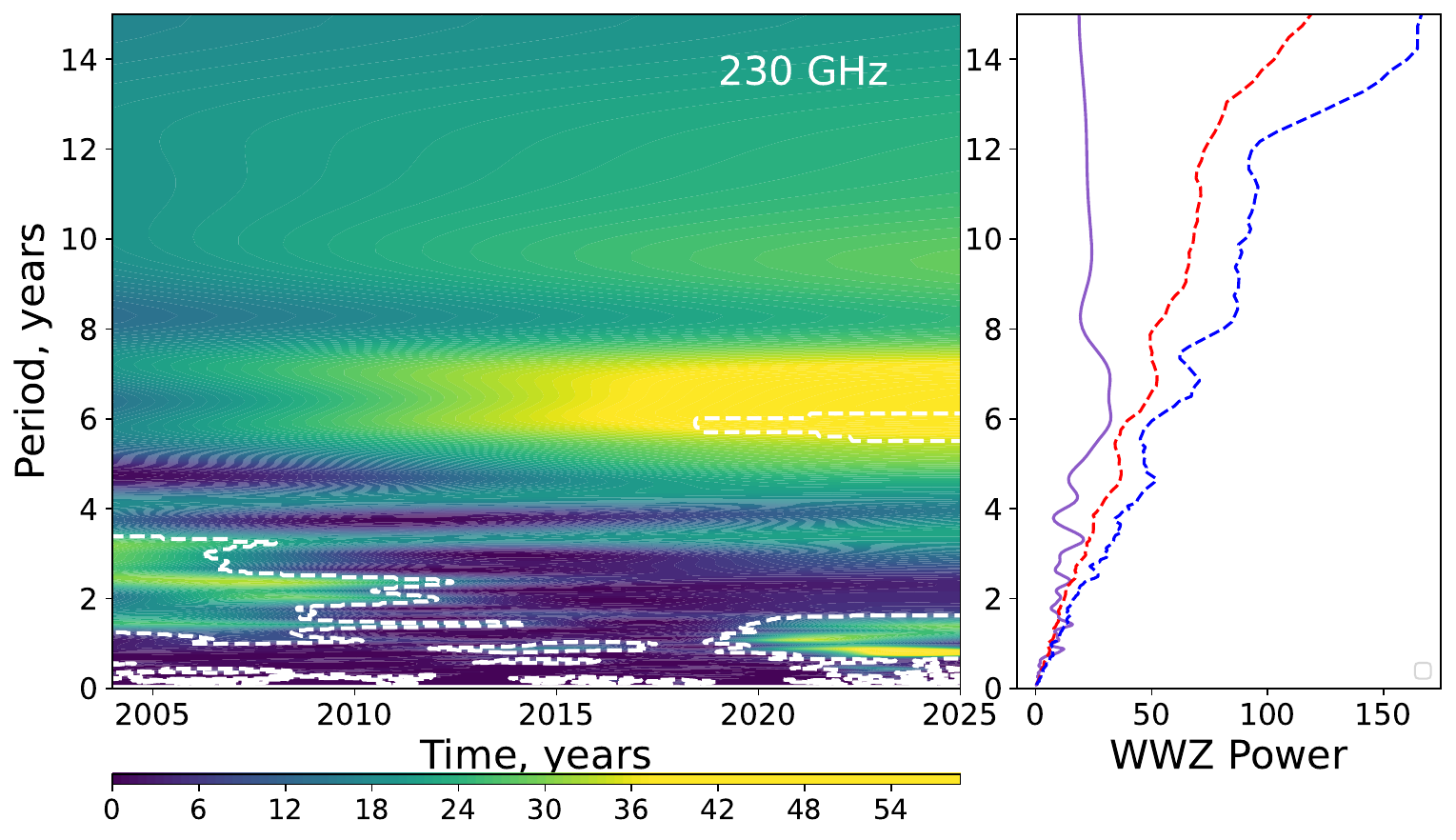}\includegraphics[height=5cm]{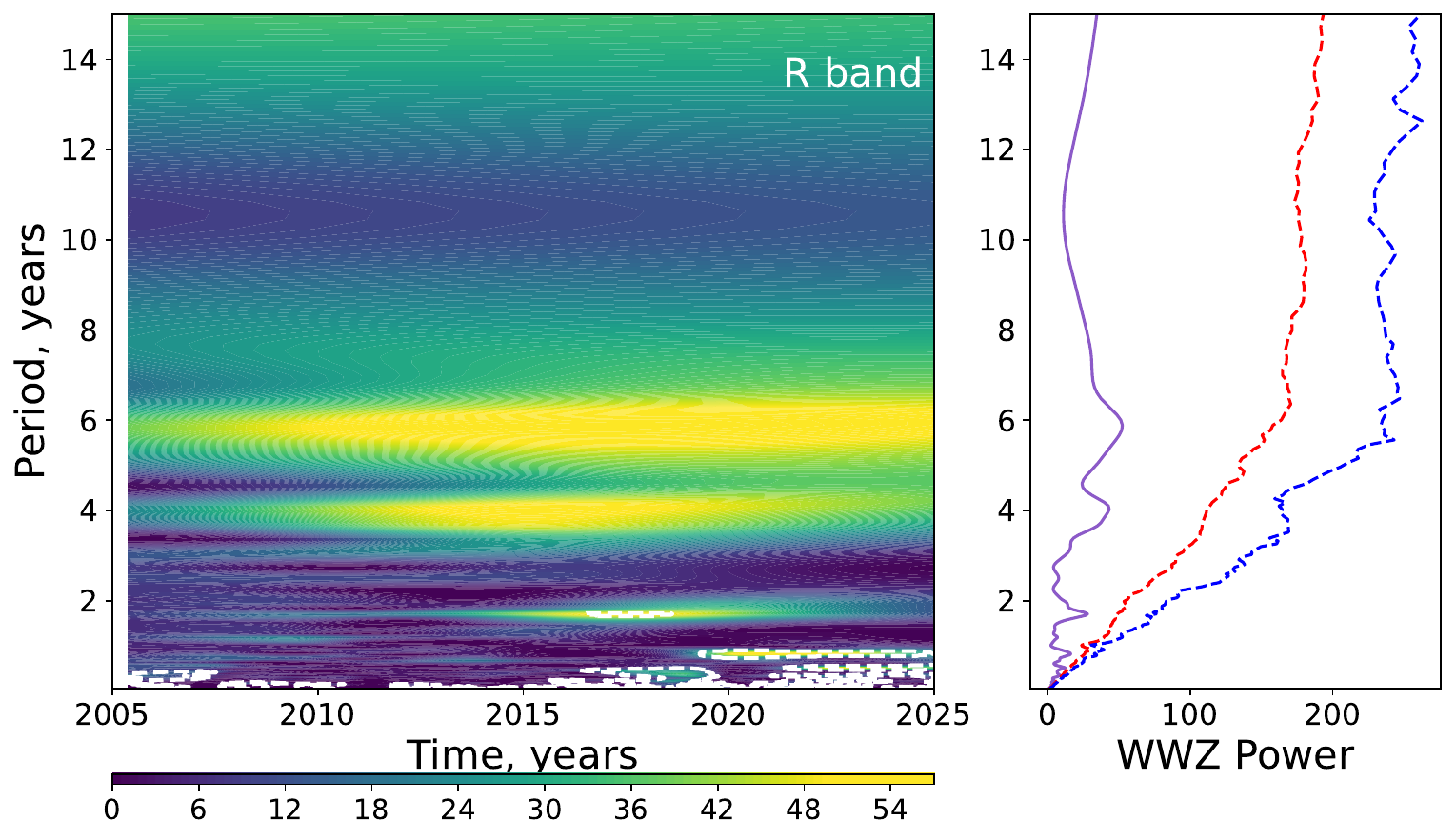}}
\caption{
WWZ power maps and global spectra for Ton\,599 in the radio (5--230 GHz) and optical R band. Each left panel shows the WWZ power as a function of time and trial period, where the horizontal axis indicates time in years and the vertical axis represents trial periods in years. The colour scale reflects the WWZ power, with brighter regions corresponding to stronger time-localised periodicities. White dashed contours mark areas where the local WWZ power exceeds the 95th percentile of the red-noise distribution estimated from 1000 light curve simulations using the method of \citet{2013MNRAS.433..907E}, indicating statistically significant features. The corresponding right panels display the global WWZ power spectra calculated as the time-averaged WWZ power at each period. The solid curves show the observed power spectra, while the red and blue dashed lines denote the 95 and 99 per cent significance thresholds, respectively, derived from the simulations. Significant local peaks may not appear in the global spectrum if the signal is not persistent throughout the full observing interval. The colour bars indicate the WWZ power scale used in each time--period panel.}
\label{fig:WT1}
\end{figure*}

\begin{figure}
\centerline{\includegraphics[width=1.0\columnwidth]{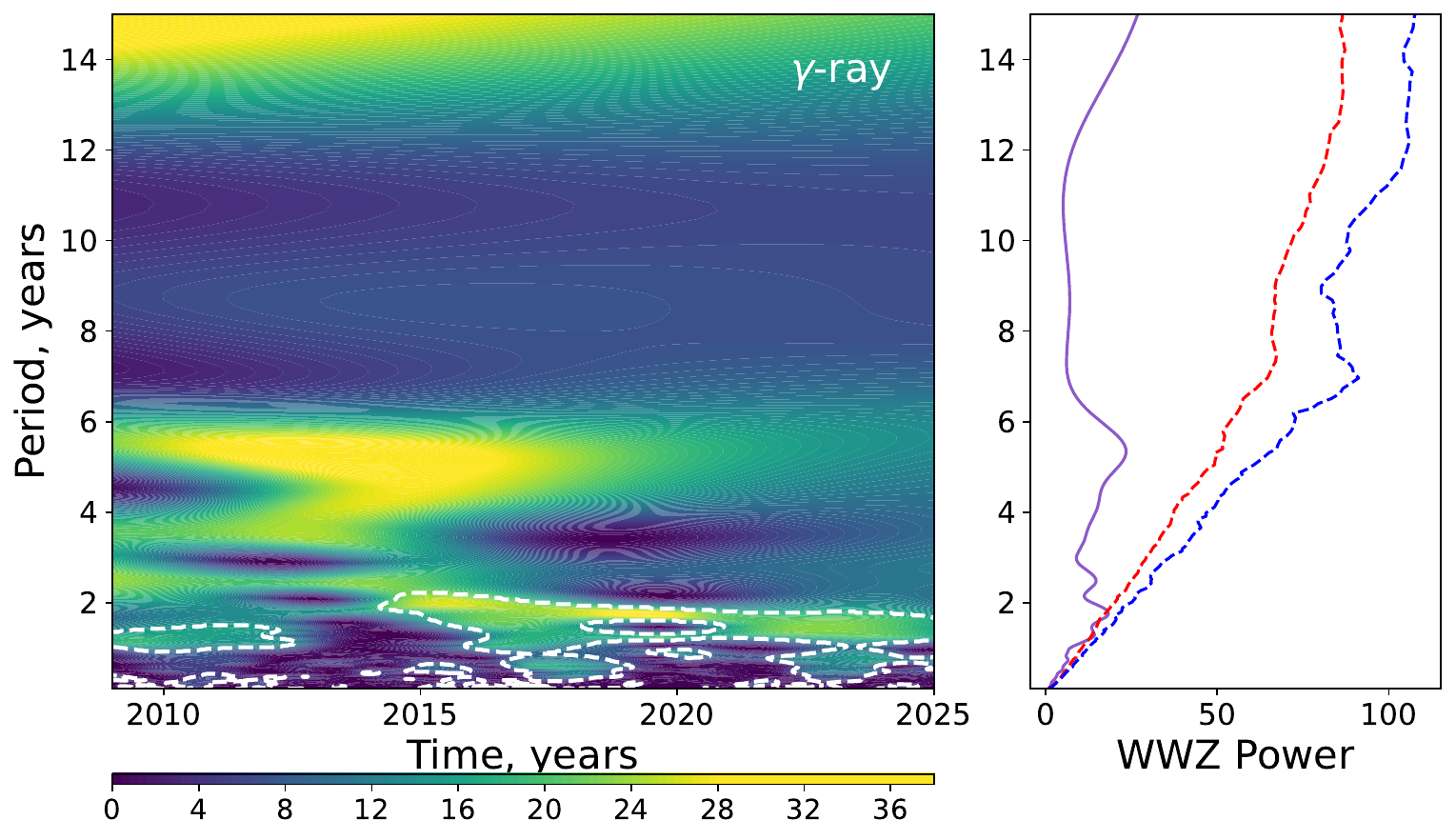}}
\caption{Same as Fig.~\ref{fig:WT1} but for $\gamma$-rays. The analysis was performed using the WWZ method in $\gamma$-rays, with significance levels computed from 1000 red-noise simulations as described in Section~\ref{sec:wwz}.}
\label{fig:WT2}
\end{figure}

Figures~\ref{fig:WT1} and ~\ref{fig:WT2} (right panel) show the global WWZ power \( Z_{\rm global}(\omega) \) as a function of period, overlaid with the 95 per cent and 99 per cent significance thresholds derived from these simulations. In the time--period WWZ map (Figs.~\ref{fig:WT1} and ~\ref{fig:WT2}, left panel), we delineated regions where the local WWZ power exceeds the 95 per cent level by plotting contours based on the boolean mask:
\begin{equation}
\texttt{sig\_mask} = \left\{ {\rm WWZ}(\omega, \tau) > Z_{95\%}(\omega) \right\}.
\end{equation}
This approach facilitates the detection of localized, transient QPO features beyond the expectations of red-noise variability.

The results of the quasi-periodicity analysis, performed using both the WWZ and the LS periodogram, are presented in Table~\ref{tab:wwz-ls}. A global examination of the detected periods reveals several systematic features as well as distinct differences between the methods, attributable to their inherent sensitivity to temporal localization
and the stationarity of the signal.

A consistent agreement is observed between the two methods in the detection of long-term 
quasi-periodicities across several radio bands (Table~\ref{tab:wwz-ls}). In the 5--22~GHz range, the LS periodogram identifies significant periods spanning both long and short time-scales. At 5~GHz, prominent peaks appear at $P = 6.2$~yr ($3.1\sigma$) and $P = 3.1$~yr ($3.1\sigma$), while shorter periodicities are also found near $P = 1.3$--$2.4$~yr with significances of $1.6$--$2.5\sigma$. At 22~GHz, the strongest component occurs at $P = 3.2$~yr ($6.4\sigma$), with additional short-term signals around $P = 1.1$--$1.8$~yr, the most significant being $P = 1.8$~yr ($2.8\sigma$). The WWZ method supports these LS detections by identifying corresponding time-localised signals at similar periods, with significances reaching up to $2.9\sigma$. 

At 37\,GHz, the variability is dominated by short-term periodicities. The LS periodogram identifies highly significant peaks at $P = 1.4$~yr ($6.4\sigma$) and $P = 1.8$~yr ($3.3\sigma$) as well as a longer period at $P = 2.3$~yr ($2.8\sigma$). These periodicities are independently traced by the WWZ method, which highlights corresponding time-localised features with significance levels reaching $2.9\sigma$. This reinforces the interpretation that the short-term components at this frequency are not merely statistical fluctuations but likely represent genuine, albeit temporally variable, quasi-periodic processes associated with internal jet dynamics.

At higher frequencies (230~GHz), the detected periods tend to cluster around $P = 1.0$--$1.5$~yr. The WWZ analysis identifies localised periodicities in this range with significances up to $2.6\sigma$, exceeding those found in the LS periodogram
\mbox{($1.1$--$1.8\sigma$)}, and additionally confirms longer-term components at $P = 2.4$--$3.3$~yr.

In the $R$ band, the LS periodogram reveals several formally significant peaks, with the strongest at $P \sim 8.6$~yr ($6.4\sigma$). However, no corresponding features are found in the WWZ results, suggesting that these periodicities may represent transient or harmonic structures rather than persistent oscillations.

In the $\gamma$-ray band, the WWZ analysis reveals two persistent quasi-periodic components at $\sim2.0$ and $\sim1.3$ years, with global significance levels of $1.8\sigma$ and $1.3\sigma$, respectively. The $\sim2$-year signal spans over a decade and aligns well with the features found in the optical and radio domains, indicating a likely common origin across the wavebands.

\begin{table*}
\centering
\caption{Periods detected in the light curves of Ton\,599 using the LS periodogram and the WWZ method. The columns ``Band'' and ``No.'' indicate the observing frequency and the period label. ``Period (yr)'' and ``Sign.'' list the period in years and its significance in $\sigma$ from the LS analysis. ``Epoch'' shows the time interval where the WWZ signal is strongest, while ``Period$_{\rm max}$ (yr)'' and ``Sign.'' give the period and significance from the WWZ global power spectrum. Dashes indicate the absence of significant detections.}
\label{tab:wwz-ls}
\begin{tabular}{ccccccccc}
\hline
\multirow{2}{*}{Band}  & No. & \multicolumn{2}{c}{LS} & \multicolumn{3}{c}{WWZ} \\
  & & Period (yr) & Sign. & Epoch & Period$_{\rm max}$ (yr) & Sign.\\
\hline
2 GHz  & P1 & 8.0 & 1.5 & 2008.4 - 2025.2 &  7.5 (2025.2) & 1.9\\
  & P2 & 6.0 & 1.1 & 2011.2 - 2025.2 & 5.9 (2025.2) & 1.6 \\
  & P3 & 1.7 &0.9 &2016.5 - 2025.2 & 2.4 (2025.2) & 1.4\\
  & P4 & 1.4 & 1.6 &2019.8 - 2025.2 & 1.3 (2025.2) & 1.3 \\
\hline
5 GHz  & P1 & 6.2 & 3.1 &1997.2 - 2025.2 & 6.5 (2020.9) & 2.9 \\
  & P2 & 1.3 & 2.5 & 2016.1 - 2025.2 & 1.3 (2025.2) & 2.4 \\
  & P3 &2.4 & 1.8 &2007.1 - 2025.2 & 2.3 (2025.2)& 2.2 \\
  & P4 &1.6 &1.6 & 2013.2 - 2025.2 & 1.6 (2025.2)& 2.2 \\
  & P5 & 3.1 & 3.1  &2005.9 - 2025.2 & 3.2 (2011.5)& 1.9 \\
  & P6 &4.4 & 6.4 &2004.9 - 2012.3 &3.8 (2008.2) & 1.4 \\  
\hline
8 GHz  & P1 &1.2 &1.5 & 2017.8 - 2025.2 &1.2 (2025.2) & 2.7 \\
  & P2 & --& -- & 2019.7 - 2025.2& 1.4 (2025.2)& 2.0 \\
\hline
11 GHz  & P1 &1.0 & 2.5 & 2018.0 - 2025.2& 1.4 (2024.3)& 3.1 \\
  & P2 & 7.6 & 1.9 & 2001.1 - 2025.2& 7.6 (2021.7)& 2.1 \\
  & P3 & 12.2 & 1.0 &2001.3 - 2025.2 & 10.9 (2025.2)& 1.9 \\
  & P4 & --& -- &2017.0 - 2025.2 &2.3 (2023.9) & 1.5 \\
  & P5 & 5.7 & 1.0 &2014.1 - 2025.2 & 5.7 (2020.0)&  1.3 \\
\hline
22 GHz    & P1 & 1.8 & 2.8 &2021.8 – 2025.2 &1.7 (2024.0)  &2.9 \\
  & P2 & 1.6& 1.3 &2018.3 - 2025.2 & 1.4 (2024.5)& 2.2 \\
  & P3 & 3.2& 6.4 &1990.8 – 2000.4 &3.3 (1995.0)  & 1.6 \\
  & P4 & 1.1& 0.6 &2021.4 – 2025.2 &1.1 (2025.2)  & 1.6 \\
 \hline
37 GHz   & P1 & 1.4 & 6.4 &2017.8 – 2025.2 & 1.4 (2025.2) &2.9 \\
  & P2 & 2.3 & 2.8 & 2000.1 – 2017.7& 2.3 (2006.0)& 2.2 \\
  & P3 & 1.8 & 3.3 & 1998.8 – 2025.2& 1.8 (2010.0) & 1.9 \\
\hline
230 GHz    & P1 & 1.5& 1.8 & 2004.7 – 2025.2& 1.4 (2025.2)& 2.6 \\
  & P2 & 1.0 & 1.1 & 2018.5 – 2025.2& 1.1 (2021.1)& 2.5 \\
  & P3 & 1.2 & 1.1 & 2019.1 – 2025.2& 1.2 (2022.3)& 2.1 \\
  & P4 & 2.5 & 2.7 &2003.0 - 2012.0& 2.4 (2006.9)& 1.5 \\
  & P5 & 3.1& 2.2 & 2003.0 - 2006.9& 3.3 (2003.0)& 1.4 \\
  & P6 & 2.1 & 0.9 & 2003.4 - 2011.9& 2.0 (2008.5)& 1.4 \\
\hline
R band  & P1 & 8.6 & 6.4&-- &-- &--\\
   & P2 & 3.2 & 2.9 &-- &-- &--\\
   & P3 & 2.1 & 1.5 &-- &-- &--\\
    & P4 & 1.7 & 2.0 &-- &-- &--\\
\hline
$\gamma$-ray  & P1 & 2.0 & 2.1 & 2014.3 – 2024.6 & 2.0 (2019.3) & 1.8 \\
  & P2 & 1.4 & 3.3 & 2021.8 – 2024.5 & 1.3 (2023.4) & 1.3 \\
\hline
\end{tabular}
\end{table*}

\section{Physical driving mechanisms}

The long-term light curves of Ton\,599, covering the monitoring period of 1997--2025 (Fig.~\ref{fig:fig1}), show two different types of variability. The first type is associated with groups of triple flares that last for 7--9 years period of time. The second type of variability is represented by 2--3 year flares within the triple groups. Another notable feature of the flares is that in triple groups there is at least one double flare. This feature is clearly visible at high frequencies of 22, 37, and 230 GHz. In the period of monitoring 1984--1997, the light curves at 22 and 37 GHz demonstrate variable behavior with relatively less amplitudes.  

Such a complex behavior suggests different scenarios of quasi-periodic variability, which can be caused by several mechanisms: (i) the disc-driven precession due to the Lense--Thirring effect, proposed originally for SS\,433 by \citet{1980ApJ...238L.129S} and applied to extragalactic radio jets by \citet{1990A&A...229..424L,2005ApJ...635L..17L}; (ii) a supermassive binary black hole (SMBBH) in which the jet precesses due to the torque arising from the misalignment of an accretion disc and the orbital plane of a binary SMBH; (iii) the precession model of 
a Doppler-beamed jet ejected by a single SMBH \citep{1999A&A...344...61A,2023ApJ...951..106B}; (iv) observable manifestations of an inhomogeneous, curved, and twisting jet \citep{2017Natur.552..374R}; (v) Kelvin–Helmholtz (K-H) (\cite{2005ApJ...620..646H}) or Magneto-Hydrodynamic (MHD) (\cite{2020ApJ...901..149Z}) jet instabilities: helical K–H modes triggered at the jet–sheath interface can imprint multi-year cycles. 

To estimate models parameters, one should adopt several basic parameters of Ton\,599 based on the published values and estimates from Table~\ref{tab:par}. 
We take the SMBH total mass as $5\times10^{8}~M_{\odot}$, which is close to a geometric mean between the masses calculated by \cite{2006ApJ...637..669L} and \cite{2022ApJ...926..180H}. The absolute magnitude of Ton\,599 in the $B$ band is estimated as $M_{\rm abs}=-27\fm3$ using its distance modulus from the NED database\footnote{\url{https://ned.ipac.caltech.edu/}} $m_B-M_B=43\fm3$ and adopting the mean apparent magnitude and K correction in this band as $m_B=16\fm5$ and $-0\fm5$, respectively.
The Doppler factor depends on the jet viewing angle, the frequency, and the state of the blazar. For example, \cite{2002PASJ...54..159Z} analysed averaged historical broadband data for Ton\,599 and obtained $\delta_{\rm radio}=9.42$, $\delta_{\rm optical}=10.65$, \mbox{$\delta_{\rm X-ray}=4.22$}, and $\delta_{\rm \gamma-ray}=12.5$. \cite{2018ApJ...866..102P} estimated 
the Doppler factors for two prominent $\gamma$-ray flares (in 2010 and 2017) as 10.75 and 13.45, while \cite{2009A&A...494..527H} reported $\delta=28.5$ calculated using the radio data at 22~GHz. 
The Doppler factor in the highest-frequency radio band (43~GHz), reported by \cite{2014MNRAS.445.1636R} and measured during the 2011 multiband flare, is 45.95.
We adopted the Doppler factor $\delta=18$ \citep{2004A&A...417..887H,2024MNRAS.52711900R} derived from the radio interferometric observations and consistent with the reported values of other authors in different bands. The viewing angle of the jet in Ton\,599 has been estimated to be within the range from $2\degr$ to $8\degr$ (e.g., \citealt{1999ApJ...521..493L,2004A&A...417..887H,2014MNRAS.445.1636R}). We adopted its value as $\theta=2\degr$ \citep{2009A&A...507L..33P}.

\begin{table}
\centering
\caption{\label{tab:par}Parameters used in the SMBBH model} 
\begin{tabular}{|l|l|}
\hline
Parameter & Value\\
\hline
Jet viewing angle $\theta$       & 2$\degr$--8$\degr$ \\
Redshift $z$  & 0.725 \\
Doppler factor $\delta$  & 18 \\
Lorentz factor $\Gamma$  & 6, 13, 28 \\
Total mass $M_{\rm sys}$ & $5\times10^{8}~M_{\odot}$ \\
\hline
\end{tabular}
\end{table}

\subsection{Lense--Thirring precession}
\label{sec:Lense-Thirring}

The misalignment between the accretion disk and the SMBH spin and the Lense–Thirring effect leads to jet precession around the central black hole. The precession period is estimated as \citep{2005ApJ...635L..17L}

\begin{multline*}
\log P(\rm yr)=0.48M_{\rm abs}+20.06+\frac{48}{35}\log\alpha+\frac{5}{7}\log a\\
+\frac{1}{7}\log \left(\frac{M}{10^8~M_\odot}\right)+\frac{6}{5}\log \eta,
\end{multline*}
where $M_{\rm abs}$ is the absolute magnitude of the source in the $B$ band, $\alpha$ is the viscosity parameter of the accretion disk, $a$ is the spin, $M$ is the black hole mass, and $\eta$ is the radiative efficiency. 

There is much evidence that central black holes in AGNs are strongly rotating (see, e.g., 
the review of \citealt{2021ARA&A..59..117R} about constraints on the SMBH spins based on observed data). 
\citealt{2007MNRAS.376.1740K} argue that the viscosity parameter ranges  between
0.1--0.4, these values fit the observed data better. However, the authors also pointed out that numerical simulations provided $\alpha$ values that were an order of magnitude smaller. Thus, we assume that $\alpha=0.1$. Assuming that Ton\,599 contains a Kerr black hole (i.e., $a=0.998$, $\eta=0.42$), we have $P \approx 1.7\times10^5$ years at the source rest frame. It is impossible to find any features of this period in any existing AGN data series (several dozen yrs). If we consider a slowly rotating black hole with $a=0.1$ and $\eta=0.07$, then we have $P \approx4\times10^3$ yr. A slowly rotating black hole is not effective in production of relativistic jets, according to the spin paradigm (e.g., \citealt{1995ApJ...438...62W,1998MNRAS.301..142M}), thus we consider $P \approx4\times10^3$ yr as a lower limit. We conclude that it is difficult to detect the features of these periods at $\alpha=0.1$ having the available time series. If we consider $\alpha>0.1$, the corresponding $P$ are greater, up to $10^6$~yr. We estimate $P \approx7\times10^3$ yr for the Kerr black hole and $\alpha=0.01$. 

Taking into account the most probable parameters, it is impossible to detect the Lense--Thirring precession in the available observations. All the estimations were made for the prograde rotation, when the angle between the angular momenta of the black hole and the accretion disk is less than $90\degr$. However, it is possible to have a retrograde rotation, when this angle exceeds $90\degr$. The radiative efficiency for the retrograde case lies within the range of 0.03--0.06, and this fact leads to a decreasing precession period compared to the prograde rotation. At the same time, according to the flux-trapping model \citep{2010MNRAS.406..975G}, the retrograde configuration is more effective in producing relativistic jets compared to the prograde rotation. We have $P \approx 6\times 10^3 \div 4\times 10^4$ yr for $\alpha=0.1 \div 0.4$ and a retrograde strongly rotating black hole with $a=-0.9$ and $\eta=0.03$. If we assume $\alpha=0.01$, then $P\approx300$ yr for a retrograde strongly rotating black hole. Moreover,
we need to convert the $P$ value estimates to the observer’s frame of reference according to the relation $P_{\rm obs}=P(1+z)/\gamma^2$ (for details see the next section), and we expect $P_{\rm obs} \approx P/200 \div P/20$ for Ton\,599. 

Thus, our main conclusion is the following: if the reason for the observed periodicity in Ton\,599 was the Lense--Thirring precession, this system would have an extremely small viscosity parameter ($\alpha\leq0.01$) and/or a retrograde rotating black hole. This condition seems implausible. Moreover, an accretion rate of tens of solar masses per year is necessary in the retrograde rotation case to provide
the bolometric luminosity $L_{\rm bol}\sim10^{47}$ erg s$^{-1}$ of Ton\,599.
  
\subsection{SMBBH model}

One of the explanations for the observed quasi-periodicity involves a SMBBH model. In such a system, jet precession arises due to the misalignment of an accretion disc and the orbital plane of a binary SMBH \citep{1997ApJ...478..527K,2000A&A...355..915A,2000A&A...360...57R,2018MNRAS.478.3199B,2023ApJ...951..106B}. We found that the $\gamma$-ray and optical periods are generally statistically consistent with the radio periods. In addition, significant correlation was found between $\gamma$-ray, optical, and radio emissions
with time delays varying from 100--200 days in the earlier epochs to the zero lags at present (Section~\ref{sec:dcf}). This suggests that the driving mechanism of the periodicity for the $\gamma$-ray and optical emission regions is similar to that in the radio region, although there is a spatial separation between them. In this case the jet precession can be described as for a rigid body.

The SMBBH model allows estimating the distance between the primary and secondary BHs \citep{2021A&A...648A..27V,2023Galax..11...96V}.
The system total mass $m+M$, the orbital period $P_{\rm orb}$ of the central SMBH, and the radius of the companion’s orbit $r$ can be expressed via 
generalized third Kepler's law:
\begin{equation}
m + M = \frac{4\pi^2r^3}{G\,P_{\rm orb}^2},
\label{eq:7}
\end{equation} 
where $M$ and $m$ are the masses of the central BH and the companion. 
Structured variability, characterized by quasi-periodic components, state-dependent multi-band time lags, and significant polarization swings \citep{2018ApJ...866..102P,2004A&A...417..887H,2011A&A...529A.113Z} supports a jet geometry consistent with a non-ballistic helical flow in Ton 599. Hence the observed period is related to the rest frame period as
\begin{equation}
P_{\rm source} = \frac{P_{\rm obs} \Gamma^2}{1+z},
\end{equation}
where $z$ is the redshift, and $\Gamma$ is the Lorentz factor. 

The angular velocity $\Omega_{pr}$ for the primary component precession can be estimated from the standard equation:
\begin{equation}
\Omega_{\rm pr} = \frac{3G\, m \cos i}{4r^{3} \, \omega}, 
\label{eq:9}
\end{equation} 
where $i$ is the half angle of the precession cone,
and $\omega$ is the primary component spin angular speed. By replacing \mbox{$\Omega_{\mathrm{pr}}=2\pi/P_{\rm pr}$}, $\omega=2\pi/P_{\mathrm{rot}}$, $\cos i\approx 1$ ($i$ does not exceed $10^{\circ}$ for blazars), and $P_{\mathrm{rot}}=P_{\mathrm{orb}}$, the previous relationship can be written as

\begin{equation}
P_{\rm orb} \, P_{\rm pr} = \frac{16\pi^{2}r^3}{3G \, m}.
\label{eq:10}
\end{equation} 

In order to determine $m$, $M$, and $r$, we combined relations (\ref{eq:7}) and (\ref{eq:10}) in the following way:
\begin{equation}
\frac{m+M}{m} = 0.75 \frac{P_{\mathrm{pr}}}{P_{\mathrm{orb}}}.
\end{equation} 
Thus, the binary mass function depends on the ratio of the orbital and precession periods, and the masses can be calculated as

\begin{equation}
m = \frac{16\pi^2r^3}{3G \, P_{\rm orb} \, P_{\rm pr}},\quad 
M = \frac{16\pi^2r^3 \, (0.75P_{\rm pr} - P_{\rm orb})}{3G \, {P_{\rm orb}}^2 \, P_{\rm pr}}.
\label{eq:12}
\end{equation}

To describe the observed quasi-periodic variability  in the framework of the binary SMBH model, we chose the following pairwise periods [$P_{\mathrm{orb}}$; $P_{\mathrm{prec}}$] from Table~\ref{tab:wwz-ls}: \mbox{[1.4; 6.5/7.5]}, \mbox{[1.4; 11]}, \mbox{[1.7; 6.5/7.5]}, [1.7; 11], [2.3; 6.5/7.5], \mbox{[2.3; 11]}, \mbox{[1.4; 2.3]}, [6.5/7.5; 11]. The longest calculated period \mbox{$P_{\mathrm{prec}} = 11$--$12$}~yr is found by both the WWZ and LS methods at 11 GHz ($>2\sigma$) and at 2 and 37 GHz with a lower significance level $>1\sigma$.

\begin{table}
\centering
\caption{\label{tab:smbh} Estimates of the primary ($M$) and companion ($m$) SMBH masses and 
the distance ($r$) between them for different orbital and precession period pairs.} 
\begin{tabular}{|c|c|c|c|c|c|}
\hline
\multirow{2}{*}{$\Gamma$} & $P_{\mathrm{orb}}$ & $P_{\mathrm{pr}}$ & $m$ & $M$ & $r$ \\
& (yr) & (yr) & ($10^{8}$$M_{\odot}$) & ($10^{8}$$M_{\odot}$) & (pc) \\
\hline
6  & 1.4 & 7.5 & 1.2 & 3.8 & 0.04 \\
13 & 1.4 & 7.5 & 1.2 & 3.8 & 0.10 \\
28 & 1.4 & 7.5 & 1.2 & 3.8 & 0.28 \\
\hline
6  & 1.4 & 6.5 & 1.4 & 3.6 & 0.04 \\
13 & 1.4 & 6.5 & 1.4 & 3.6 & 0.10 \\
28 & 1.4 & 6.5 & 1.4 & 3.6 & 0.28 \\
\hline
6  & 1.4 & 11 & 0.8 & 4.2 & 0.04 \\
13 & 1.4 & 11 & 0.8 & 4.2 & 0.10 \\
28 & 1.4 & 11 & 0.8 & 4.2 & 0.28 \\
\hline
\hline
6  & 1.7 & 7.5 & 1.5 & 3.5 & 0.04 \\
13 & 1.7 & 7.5 & 1.5 & 3.5 & 0.12 \\
28 & 1.7 & 7.5 & 1.5 & 3.5 & 0.32 \\
\hline
6  & 1.7 & 6.5 & 1.7 & 3.3 & 0.04 \\
13 & 1.7 & 6.5 & 1.7 & 3.3 & 0.12 \\
28 & 1.7 & 6.5 & 1.7 & 3.3 & 0.32 \\
\hline
6  & 1.7 & 11 & 1.0 & 4.0 & 0.04 \\
13 & 1.7 & 11 & 1.0 & 4.0 & 0.12 \\
28 & 1.7 & 11 & 1.0 & 4.0 & 0.32 \\
\hline
\hline
6  & 2.3 & 7.5 & 2.0 & 3.0 & 0.05 \\
13 & 2.3 & 7.5 & 2.0 & 3.0 & 0.14 \\
28 & 2.3 & 7.5 & 2.0 & 3.0 & 0.40 \\
\hline
6  & 2.3 & 6.5 & 2.4 & 2.6 & 0.05 \\
13 & 2.3 & 6.5 & 2.4 & 2.6 & 0.14 \\
28 & 2.3 & 6.5 & 2.4 & 2.6 & 0.40 \\
\hline
6  & 2.3 & 11 & 1.4 & 3.6 & 0.05 \\
13 & 2.3 & 11 & 1.4 & 3.6 & 0.14 \\
28 & 2.3 & 11 & 1.4 & 3.6 & 0.40 \\
\hline
\end{tabular}
\label{SMBBH_param}
\end{table}

Assuming the presence of a binary SMBH in the blazar Ton\,599, we obtained several reasonable paired values \mbox{[$P_{\rm orb}$; $P_{\rm pr}$]} for the orbital and precession periods that physically correspond to the SMBBH conditions. These values are listed in Table~\ref{tab:smbh}: [1.4/1.7; 6.5/7.5], [2.3; 6.5/7.5], \mbox{[1.4/1.7; 11]}, and [2.3; 11]. The distance between the two components varies from 0.04 to 0.4 pc.

\subsubsection{Estimation of SMBBH merging time}

A system of two SMBHs generates gravitational waves (GW), and as a consequence, the orbit of this binary system will progressively decrease and the orbital period will be shortening until the collapse of the system. The merging time-scale, which represents the transfer time from an initial orbit until the system collapses, can be expressed as a function of the masses of the system and the components and the size of the orbit (see, for example, \citealt{2018MNRAS.478.3199B,2024iSci...27j9427V,2024A&A...691L...9V}).

The calculation of the time necessary for the system to merge, based on expression (17) from \cite{2018MNRAS.478.3199B}, gives system merging times between $\sim 60$ Myrs and hundreds of billion years for the set of the binary system orbit sizes within the 0.04--0.4 pc interval presented in Table~\ref{SMBBH_param}. Thus, this result cannot provide reasonable constraints on the SMBBH model since the energy losses caused by GW emission are unobservable.

\subsection{Modeling the light curves}
\label{sec:jet_precession_orbital}

In a binary SMBH system two independent jets may emerge if both BHs are spinning and accreting significantly and also if each retains its own accretion disk within the
shared circumbinary environment. However, typically only the primary SMBH is considered as active, with the secondary being either too low-mass or lacking sufficient inflow to launch its own jet; for example, the cases of OJ\,287 (e.g., \cite{2018ApJ...866...11D}) and PKS\,2131-021 (e.g., \cite{2022ApJ...926L..35O}). There is an example of a possible binary SMBH system with two jets in 3C\,279 \citep{2019A&A...621A..11Q}. However, confirming dual jets observationally is challenging because it requires clear spatial or kinematic separation. We do not have such evidence to 
the existence of a pair of jets in Ton\,599, and hence in our model we consider one jet for simplicity.

\begin{figure}
\centerline{\includegraphics[width=0.9\columnwidth]{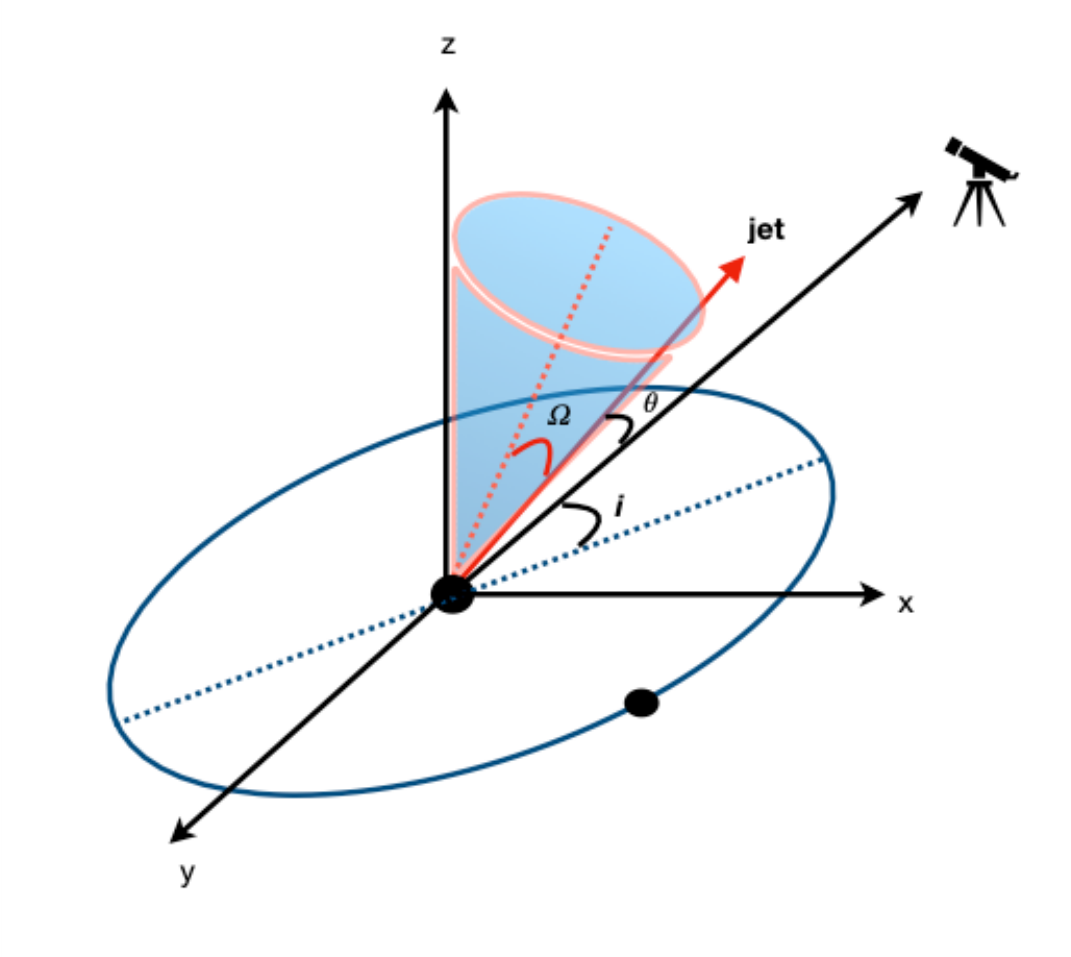}}
\caption{A sketch showing jet orbital motion and precession. The elements are not in scale. The binary SMBH is shown as two black dots, the jet is the red arrow, the angle between it and the observer's line of sight is $\theta$, the orbit inclination angle with respect to the observer is $i$, and the precessing cone angle is $\Omega$.}
\label{fig:scheme}
\end{figure}

As it is shown in the previous section, we have found plausible pairs of periods for the orbital-motion and precession-caused variability: the orbital periods within the range
from 1.4 to 2.3 yr and the precession periods from 6.5 to 11 yr. Having this result, we 
have adopted a jet model in which the observed flux is modulated by both jet precession (e.g., \cite{1992A&A...255...59C,1999A&A...344...61A,2023ApJ...951..106B}) and orbital Doppler boosting from a binary SMBH system (e.g., \cite{1999A&A...347...30V,2022ApJ...926L..35O}). 
We assume that Ton\,599 is an SMBH binary with a subparsec separation, where a combination of two processes takes place: (i) the primary black hole's jet precesses due to the torque from the binary companion, and this causes periodic variation in the jet
viewing angle, amplifying variability via relativistic beaming; (ii) the secondary black hole orbital motion modulates the jet Doppler boosting, producing periodic flux 
variation (an illustration of the basic geometry of the model is shown in Fig.~\ref{fig:scheme}). The total flux is computed as the sum of the Doppler-boosted components, following the formalism outlined in \cite{2022ApJ...926L..35O}.

The viewing angle changes due to the precession as
\begin{equation}
\cos\theta(t) = \cos\theta_0 \cos\Omega + \sin\theta_0 \sin\Omega \cos\left( \frac{2\pi t}{P_{\rm pr}} + \phi_p \right),
\end{equation}
where $\theta_0$ is the average viewing angle of the precessing jet, $\Omega$ is the half-opening angle of the precession cone, $\phi_p$ is the phase offset of precession, and
$P_{\rm pr}$ is the precession period. 

The flux density from the orbital Doppler modulation is relativistically enhanced as

\begin{equation}
S_{\rm orb}(t) = S_0 \, \delta_{\rm orb}(t)^{p} \label{eq:s_orb},
\end{equation}
and the flux density from the precessing jet is expressed in the same way:

\begin{equation}
S_{\rm pr}(t) = S_0 \, \delta_{\rm pr}(t)^{p}, \label{eq:s_pr}
\end{equation}
where $p = 2 + \alpha$ (e.g., \citealt{1979ApJ...232...34B,2010A&A...512A..24S}), $\alpha$ is the spectral index, $S_0$ is the internal flux density, and 
$\delta_{\rm orb}$, $\delta_{\rm pr}$ are the Doppler factors. The Doppler factor for orbital motion
\begin{equation}
\delta_{\rm orb}(t) = \frac{1}{\Gamma_{\rm o}} \left(1 - \beta_o \sin i \cos\left(\frac{2\pi t}{P_{\rm orb}} + \phi_o \right)\right)^{-1},
\end{equation}
where $i$ is the inclination angle of the orbit, and $\phi_o$ is the orbital phase offset. The Doppler factor for the precessing jet
\begin{equation}
\delta_{\rm pr}(t) = \frac{1}{\Gamma_{\rm p}} \left(1 - \beta_p \cos\theta(t)\right)^{-1}.
\end{equation}
The jet and orbital speeds are defined through the Lorentz factor
\begin{equation}
\beta = \sqrt{1-\frac{1}{\Gamma^2}} \label{eq:beta}.
\end{equation}
The combination of equations \ref{eq:s_orb} and \ref{eq:s_pr} gives the total observed flux density:
\begin{equation}
S_{\rm total}(t) = S_0 \, (\delta_{\rm orb}\,\delta_{\rm pr})^{p}.
\end{equation}

For simplicity we adopted for the both processes the same initial values of the parameters and the same ranges of their variation based on the observed data, see Table~\ref{tab:precession_orbit_params}. We assumed $\theta_0=2\degr$ \citep{2009A&A...507L..33P}, the cone angle $\Omega$ varying from 0.1\degr to 10\degr (adopting typical values as in, e.g., \citealt{2018MNRAS.478.3199B,2023ApJ...951..106B}), the bulk Lorentz factor of the jet plasma $\Gamma$ varying from 6 to 28 (based on the values obtained in \citealt{2014MNRAS.445.1636R}), and the phase $\phi$ from $-\pi$ to $+\pi$.

We varied the parameters within the defined ranges to obtain the best fitting by the least squares method. The model fitting results are shown in Figs.~\ref{fig:Ton599_model_2GHz(1.4-6)}--\ref{fig:Ton599_model_230GHz(1.3-7.7)}, and the obtained parameters are summarized in Table~\ref{tab:precession_orbit_results}. We found the best-fitting precession periods to lie in the range from 5.8 to 7.7 yrs and the best-fitting orbital periods to locate in an interval from 1.2 to 1.7 yrs. Close values (within 10 per cent) of pairs of the periods are found for 2 and 11 GHz (6/1.5 yrs) and for 8 and 37 GHz (6.5/1.4~yrs).

By analysing the best-sampled light curve (43-yr time span at 37 GHz), we have found that the periodic emission is consistent with a jet orbiting in a binary SMBH system with a period of 1.5 yr, enhancing the emission by a bulk Lorentz factor of 8, the jet is also precessing with a period of 6.4~yr in the observer's frame of reference, which in turn causes emission boost with a bulk Lorentz factor of 22. The model, in general, adequately reproduces positions of radio flares, indicating that the intrinsic jet flux density may indeed be modulated by the orbital and precession motions. While the long-term quasi-periodic trends in the Ton\,599 multifrequency emission can be explained by orbital motion and jet precession, the most prominent flares (e.g., the latest historically highest activity state in \mbox{2024--2025}) cannot be fully accounted for by Doppler boosting within our model. We propose that the strongest flares require a shock-in-jet explanation. A comprehensive analysis of the light curves from $\gamma$-rays to 2 GHz, along with the measured time delays (Section~\ref{sec:dcf}), suggests that shock propagation in the jet is a likely mechanism behind the strongest flares (e.g., \citealt{1985ApJ...298..114M,2001MNRAS.325.1559S,2004ApJ...613..725S,2013ApJ...768...40L}). The lower-frequency emission arises downstream relative to the high-energy emission zone, reflecting  the opacity and light-travel-time effects. Additionally, the observed decrease in the fractional variability parameter with decreasing frequency (Section~\ref{sec:lc}) is consistent with synchrotron self-absorption (SSA) and the smoothing of variability as disturbances propagate downstream the jet.

We tested a precession-only periodicity model and compared it with the results of a combined orbital-precessional model (see an example of the application at 37 GHz in Fig.~\ref{fig:Ton599_compare_model}). In the case of a precessing jet, the fitted peaks appear smoother and broader due to long-term modulations. As our analysis shows, orbital motion is necessary to explain the shorter time-scale variations in the Doppler factor and to better fit the sharper flares. The $\chi^2$ statistics confirm that the two-component (orbital and precessional) periodicity model provides a better fitting in the case of Ton\,599.

\subsubsection{ARIMA and GARCH methods applied to the model residuals}

We modeled the stochastic variability that left after subtracting the periodical precession and orbital model, taking it as a linear combination of the autoregressive (AR) and moving average (MA) terms. For that purpose we applied the \textsc{ARIMA} (Auto Regressive Integrated Moving Average) function\footnote{\url{https://www.statsmodels.org/stable/index.html}} \citep{seabold2010statsmodels}. We first interpolated the residuals onto a uniform time grid with time steps of 0.05~yr (37 and 22 GHz) and 0.1 yr (other bands) and then applied ARIMA ($p,q$) with different combinations of the $p$ and $q$ parameters: ($1,1$), ($2,1$), ($1,2$), and ($2,2$). The model with $p=2$, $q=2$ showed the best fitting, its parameters are listed in Table~\ref{tab:arma}. We have found strong autocorrelation indicating structured variability at all frequencies, which shows strong memory (measured by the heteroskedasticity parameter), possibly tied to physical modulation or a coherent structure (e.g., shock dynamics, turbulence modulation). The residuals are non-Gaussian and do not show autocorrelation, the residual variances after model fitting are quite low ($\sigma^2=0.01$--$0.10$). 

We further implemented the \textsc{GARCH} (Generalized Autoregressive Conditional Heteroskedasticity, \citealt{BOLLERSLEV1986307}) modeling\footnote{\url{https://github.com/bashtage/arch}} of the residuals to capture time-varying variance. Its results indicate a strong response of volatility to recent residuals ($\alpha\simeq1$) and its low persistence (Table~\ref{tab:arma}). The plots in Figs.~\ref{fig:Ton599_model_2GHz_GARCH}--\ref{fig:Ton599_model_230GHz_GARCH} show how the ARIMA model gives the filtered, smoothed, model-based approximation of the residuals left from the orbital-precessional model and together with it quite well captures the deterministic physics (precession and orbital Doppler periodicity) and the structured stochastic component (e.g., short-term variability from shocks or turbulence in the jet); the shaded regions are uncertainty bands from the GARCH-estimated conditional volatility that show how the uncertainty in residuals evolves over time.

\begin{table}
\centering
\caption{Parameters used for the orbital and jet precession modeling.}
\begin{tabular}{|l|c|c|c|}
\hline
Parameter & Symbol & Unit & Range \\
\hline
Lorentz factor & $\Gamma$ & --  & 6--28 \\
Viewing angle & $\theta$ & \degr & 0.1\degr--10\degr  \\
Orbit inclination angle & $i$ & \degr & 0.1\degr--10\degr  \\
Cone angle & $\Omega_i$ & \degr  & 0.1\degr--10\degr  \\
Precession period & $P_{\rm pr}$ & yr & 5.0--11 \\
Orbital period & $P_{\rm orb}$ & yr & 1.0--2.5 \\
Phase offset & $\phi_i$ & rad & -$\pi$--$\pi$ \\
Flux density scaling & $S_{\text{int}}$ & Jy & 0--10 \\
Spectral index ($S \propto \nu^{\alpha}$) & $\alpha$ & --  & $-$1--1 \\
\hline
\end{tabular}
\label{tab:precession_orbit_params}
\end{table}

\subsection{Alternative mechanisms for quasi-periodic variability}

While the SMBBH framework provides a compelling explanation for the observed quasi-periodic oscillations in Ton~599, particularly through the interplay of orbital motion and jet precession modulating Doppler boosting, it is essential to consider alternative physical origins that do not invoke a binary system. These mechanisms, rooted in intrinsic instabilities or geometric effects within a single black hole environment, can also generate apparent periodicity on multi-year timescales, often mimicking the signatures we observe. 

Accretion disk instabilities represent a prominent non-binary pathway to QPOs. For instance, magneto-rotational instability (MRI) or thermal-viscous cycles in the disk can lead to quasi-periodic mass accretion fluctuations, injecting plasma into the jet base and producing recurrent flares \citep[e.g.][]{2013MNRAS.436L.114K}. Such processes have been invoked to explain harmonic oscillations in blazars like Ton~599, where radio variability analyses revealed periods of $\sim$1.7, 2.4, 3.4, and 7.5 years, attributed to disk-driven perturbations \citep{2014MNRAS.443...58W}. Similarly, in NRAO~530, periodic radio modulations were linked to a jet-disk connection, where disk instabilities modulate the jet launching, yielding cycles on scales comparable to those in Ton 599 \citep{2013MNRAS.434.3487A}. These instabilities arise naturally from the turbulent, viscous nature of accretion flows, where small perturbations amplify into limit cycles, potentially aligning with the observed year-scale QPOs. However, such mechanisms often produce less coherent, more stochastic periodicity, evolving with disk conditions like viscosity parameter $\alpha$ or magnetic field strength, and may be difficult to sustain the multi-decade stability seen across our 40-year baseline.

Another class involves localized ``hot spots'' in the accretion disk or jet base---regions of enhanced density or magnetic flux, such as magnetic islands formed via reconnection events \citep[e.g.][]{2001A&A...377..396R}. These could orbit or precess around the black hole, leading to beamed emission that appears quasi-periodic when viewed at small angles. In blazars, hotspots might explain short-term optical-$\gamma$-ray flares, as proposed for sources like PKS 1510-089 \citep{2022MNRAS.510.3641R}, where Doppler-boosted hotspots yield cycles of $\sim$1–2 years without requiring a companion black hole. Whereas, this scenario struggles with the frequency-dependent lags observed in Ton~599, as hotspots would primarily affect compact, high-frequency emission zones, leaving longer-wavelength radio variability less modulated.

Distinguishing these alternatives from the SMBBH model hinges on their different implications for jet structure and emission coherence. Disk instabilities or hotspots predict more erratic period evolution, potentially tied to stochastic accretion events, whereas single-jet precession might show frequency-independent phase shifts due to uniform geometric effects. In contrast, the SMBBH scenario in Ton~599, with orbital periods of 1.2--1.7 yr and precession of 5.8--7.7 yr, uniquely forecasts a hierarchical modulation: short cycles from binary-induced Doppler variations superimposed on longer precessional wobbles, with frequency dependence reflecting stratified emission regions along the jet.

\section{Discussion}

The blazar Ton\,599 is one of the relatively rare AGNs showing signatures of structured long-term variability. Earlier studies based on long-term radio monitoring data (e.g., \citealt{2008A&A...488..897H,2014MNRAS.443...58W,2014Ap&SS.352..215L}) reported the presence of quasi-periodic components $\sim1.4$, 2.3, 3.4, and 7.5 years 
at 4.8, 8.0, 14.5, 22, and 37 GHz, and about 1.6, 2.1, and 3.6~yrs in the optical band \citep{2006PASJ...58..797F,2021ARep...65.1233H}.

\subsection{Cross-correlations}
The $\gamma$-ray radiation is highly variable, showing multiple strong flaring events, its variability is most strongly correlated with the optical band emission in epoch 3, the near-zero lag implies cospatial emission regions for the synchrotron (optical) and inverse Compton ($\gamma$-ray) components.

The mm-band (230 GHz) light curve, reflecting the emission from compact jet regions, is smoother, with the flares continuing on a longer time-scale compared to the higher-energy bands. The delays relative to the higher-energy bands may imply the opacity effects or the propagation of disturbances down the jet. The high DCF values and small time lags between the 230, 37, and 22 GHz emissions, especially in epoch 4, imply compact emission zones for the high frequencies.

The cm-band (2--37 GHz) variability amplitude decreases and the time lags increase with decreasing frequency, indicating that the lower-frequency emission originates farther downstream the jet emitting by synchrotron self-absorption. The large time delays (hundreds of days) for the low-frequency bands in combination with strong correlation indicate the synchrotron self-absorption in larger, more extended jet regions at 2--11 GHz.

Notably, there is a systematic trend of decreasing time lag from epoch 1 to epoch 4 (an example of time lags relative to 37 GHz is given in Fig.~\ref{fig:lags}). This trend likely reflects significant changes in the Ton\,599 jet structure and flaring dynamics over time: (i) it implies reduced synchrotron self-absorption in the later epochs; (ii) shocks might be slower and weaker in epoch 1, later they became faster and stronger, affecting the jet across a broader range almost simultaneously in epoch 4; (iii) possible changes in the jet environment (e.g., expansion, magnetic field weakening, or jet instabilities). The latter could be inspected with VLBI observations of jet evolution during the latest series of flares in 2021--2025.

\subsection{Quasi-periodicity}
While the detection of similar periods across multiple wavebands may indicate a common physical origin, caution remains necessary due to statistical limitations. Although the formal significance of several LS peaks is above $3\sigma$ in some bands, the uncertainties in period estimates are substantial. In particular, the longest detected periods in the range of $7$--$12$~yr have been observed for only 2--4 cycles within the observing baseline (1983--2025), which inherently limits the statistical robustness of such detections. Furthermore, red-noise processes intrinsic to AGN light curves can produce power enhancements that mimic quasi-periodic signals \citep{2005A&A...431..391V}, requiring careful evaluation when interpreting such periodicities as physically meaningful.


The two methods employed---LS and WWZ---provide complementary perspectives on variability. The LS periodogram is more sensitive to globally coherent signals, whereas the WWZ allows the localisation of quasi-periodic components in time. In many bands both methods consistently detect short-term oscillations in the range of $\sim0.5$--$2.5$~yr, particularly at higher radio frequencies such as 37~GHz, where the LS significance reaches up to $6.4\sigma$. These consistent detections strengthen the interpretation of these signals as genuine, though possibly transient, quasi-periodic processes linked to jet-internal dynamics, such as shocks or turbulence \citep{2010ApJ...719L.153R, 2007ApJ...664L..71A, 2009A&A...506L..17L}.

The fundamental period of $\sim$6.8 yr reported for the radio bands in \cite{2014MNRAS.443...58W} matches our best-fitting precession period for most of the light curves. \cite{2014Ap&SS.352..215L} examined the 4.8--14.5 GHz radio light curves over a 32-yr time span and identified a quasi-periodic oscillation with a period of about 2.3 yr and a characteristic variability time-scale of $1.15\pm0.05$ yr. This period is present at 8 and 230 GHz in our data according to the WWZ analysis.

A comprehensive optical study by \cite{2006PASJ...58..797F} analysed photometric data spanning from 1974 to 2002. They identified two potential periods in the $R$-band light curve: approximately 3.55 and 1.58 yrs.  \cite{2021ARep...65.1233H} reported a quasi-periodic component of 2.1 yrs in the analysis of an optical dataset covering the time period of 1997--2002. We found close three periodic components in the $R$-band light curve (2005--2025) using the LS method: $\sim1.7$, 2.1, and 3.2~yrs.


In the $\gamma$-ray band two significant quasi-periodic components are identified: a $\sim2.0$~yr signal appears consistently in both methods, with a significance of $2.1\sigma$ (LS) and $1.8\sigma$ (WWZ), and a $\sim1.3$~yr component is also detected, albeit with a lower WWZ significance. These values are consistent with the short-term periodicities seen in the radio and optical bands, implying a common variability timescale potentially associated with recurring structures in the relativistic jet.

\subsection{Comparison with OJ\,287 and 3C\,345}

According to our model assumptions, Ton\,599 contains a tightly bound asymmetric binary with a short orbital and precession periods. The derived masses may imply that the system has been formed via hierarchical merging or gas-driven migration. The small separation ($\sim 0.1$ pc) may indicate that this system is in advanced evolutionary stage. Here we compare the parameters obtained with our model with those of other authors derived for two SMBH binaries: one with a roughly similar mass (3C\,345) and the other with unequal masses (OJ\,287).

\cite{2005A&A...431..831L} found for 3C\,345 two equal SMBHs with  masses $M_1 \simeq M_2 \simeq 7.1 \times10^{8}~M_{\odot}$, a separation $a \simeq0.33$~pc, an orbital period $P_{\rm orb} \simeq$ 480 yr, and an accretion disk precession period $ P_{\rm pr} \simeq 2570$ yr. These periods are much longer than the $\sim4$ yr period obtained from the observed variability of the radio nucleus in \cite{1999ApJ...521..509L}, which may indicate that some other periodic process affects the accretion disk. \cite{2022arXiv220201915Q} revised this with a VLBI analysis of the jet, applying the nozzle precession models for a number of blazars including 3C\,345, and obtained an orbital period $P_{\rm orb} \simeq$ 4.6 yr in the source frame of reference, a jet precession period of $ P_{\rm pr} \simeq$ 7.3 yr, and a mass ratio of $q \simeq 0.87$.

A frequently discussed example of a binary black hole system is OJ\,287, exhibiting major flares with a quasi-period of about 12 yrs and subflares separated by 1--2 yrs. The system has estimated masses of $1.8\times10^{10}~M_{\odot}$ for the primary BH and $1.4\times10^{8}~M_{\odot}$ for the secondary \citep{2010CeMDA.106..235V,2012MNRAS.427...77V}. Alternatively, an explanation with a single SMBH and a helical jet has been offered for the optical periodicity observed in OJ\,287 (e.g., \citealt{2020Univ....6..191B,2024APh...16002965G,2024A&A...683A.248C}). The model introduced in \cite{2020Univ....6..191B} proposes that jet geometry and instabilities can naturally explain both the 12-yr optical outbursts and the longer-term structural variations. It sidesteps the need for a secondary SMBH and instead emphasizes relativistic boost effects and magnetohydrodynamic processes. The jet components exhibit non-radial motions, naturally arising from the Kelvin--Helmholtz instability, which establishes and maintains the helical geometry. 

For Ton\,599 we suggest that the orbital and precession modulations affect jet direction and Doppler boosting, producing observable quasi-periodic flux variations, while OJ\,287 is reported as likely a binary SMBH with two disk impacts during a 12-yr orbital period, and the quasi-periodicity in 3C\,345 arises from geometric variation of the jet. Our estimate of BH separation ($\sim0.04$--$0.40$~pc) is comparable to both 3C\,345 (0.33 pc) and OJ\,287 ($\sim0.05$--$0.10$~pc). The orbital period $P_\mathrm{orb}\sim1$ yr in Ton\,599, which is shorter than in OJ\,287 and 3C\,345, implies a tight binary: such short periods are expected only after the dynamical friction and circumbinary disk torques have shrunk the orbit. The precession period ($P_\mathrm{pr}\sim7$ yr), which is also shorter than in the other two blazars, might arise from the Lense--Thirring precession of an inner accretion flow or a torqued disk with a very small viscosity parameter ($\alpha\leq0.01$, see Section~\ref{sec:Lense-Thirring}), especially in the case of a system with the misaligned BH spin and orbit. Ton\,599 likely occupies a rare transitional class between systems like 3C\,345 (long-period binaries with stable jets) and OJ\,287 (strong-impact binaries with large mass disparity). A summary of the comparison between the three blazars is shown in Table~\ref{tab:compare}.

The SMBBH model yields testable predictions. It anticipates gravitational wave emission from the tight binary (separation $\sim$0.04-0.4 pc), detectable in the nHz band by pulsar timing arrays (PTAs) like NANOGrav or future space-based interferometers such as LISA, expected to launch in the 2030s. The waveform would encode the system's mass ratio ($\sim$0.39) and eccentricity (assumed low), allowing direct comparison with electromagnetic periods. Future observations will be crucial for distinguishing these models. Extended radio monitoring could track period stability over another decade, while CTA's $\gamma$-ray sensitivity might resolve intra-cycle substructure, favoring binary-induced boosts over disk noise. Integrating these with numerical simulations (e.g., GRMHD codes capturing binary torques) will refine predictions, potentially elevating Ton\,599 as a benchmark for the emerging field of multi-messenger AGN astrophysics.

Our results confirm the presence of the earlier-found variability time-scales using independent datasets and methods and extending the analysis to a broader time and frequency range, including the optical and $\gamma$-ray domains.


\section{Summary}

We have conducted a multiwavelength analysis of the long-term variability of the flat-spectrum radio quasar Ton\,599 using the observed data spanning more than four decades (1983--2025). The results may be summarized as follows.

\begin{enumerate}

\item 
During the last 30--40 years, Ton\,~599 has systematically experienced major outbursts detected in the entire electromagnetic spectrum with a high variability level that reaches 130 per cent. The multiwavelength light curves show complex temporal structure variability. The $\gamma$-ray, optical, and radio emissions are found to be strongly correlated with time lags $\sim 0$--$360$ days, which implies that they are generated by the same population of relativistic particles. The time lags decrease from epoch 1 to epoch 4, reflecting the variation in the jet structure and flaring dynamics, such as reduced synchrotron self-absorption and stronger shocks in the later epochs.

\item 
A search for periodicity reveals several quasi-periodic components across multiple bands. The detected periods cluster around $\sim$1.3--2.3, 3.2, 6.5--7.5, and 11 years, implying the presence of variability processes acting on different time-scales. In the $\gamma$-ray domain, two short-term components (1.3 and 2.0~yr) are detected with moderate significance and show good agreement with those found in the radio and optical bands, supporting a common origin related to processes within the jet.

\item 
Assuming that the observed Ton\,599 quasi-periodic variability is caused by jet precession due to a SMBBH system, we estimated the best-fitting orbital and precession periods of 5.8--7.7 and 1.2--1.7 yrs for the total system mass of $5 \times 10^{8}~M_{\odot}$ with the distance between the two components from 0.04 to 0.4 pc. Although the orbital-precessional model satisfactorily describes  the observed light curves, it cannot fully account for the extreme flux peaks solely through the Doppler boosting caused by the orbital motion and jet precession in an SMBBH system. This discrepancy implies the presence of additional jet-internal processes, likely flaring events consistent with the shock-in-jet scenario. Thus, this study demonstrates that the Ton\,599 emission arises from both ordered and stochastic processes, reflecting the superposed observational effects presented in the object.

\end{enumerate}

\section*{Acknowledgements}

We thank the referee for the valuable comments and suggestions that helped improve the paper and derive significantly more accurate parameter estimations.

We would like to thank Prof.~Eduardo Ros from the Max Planck Institute for Radio Astronomy for his valuable comments on the draft.

Observations with the SAO RAS telescopes are supported by the Ministry of Science and Higher Education of the Russian Federation. The observations were carried out with the RATAN-600 scientific facility, Zeiss-1000 and AS-500/2 optical reflectors of SAO RAS, and RT-22 of CrAO RAS. The observations at 5.05 and 8.63 GHz were performed with the Svetloe, Badary, and Zelenchukskaya RT-32 radio telescopes operated by the Center of Shared Research Facility for the Quasar VLBI Network of IAA RAS (\url{https://iaaras.ru/cu-center/}). 

This research is supported by the International Partnership Program of the Chinese Academy of Sciences under Grant No. 018GJHZ2024025GC.

YYK was supported by the MuSES project, which has received funding from the European Union (ERC grant agreement No.~101142396). The views and opinions expressed are, however, those of the author(s) only and do not necessarily reflect those of the European Union or ERCEA. Neither the European Union nor the granting authority can be held responsible for them.

VAE and VLN are grateful to the staff of the Radio Astronomy Department of CrAO RAS for their participation in the observations. 

TA thanks the support from the National SKA Program of China (2022SKA0120102, 2022SKA0130103), Xinjiang Tianchi Talent Program, the FAST Special Program (NSFC 12041301). 

The observations at 4.8 GHz were performed with the XAO-NSRT 26m radio telescope operated by the Xinjiang Astronomical Observatory, CAS. XAO-NSRT is supported by the Urumqi Nanshan Astronomy and Deep Space Exploration Observation and Research Station of Xinjiang (XJYWZ2303). LC acknowledges the support from the Tianshan Talent Training Program (grant No. 2023TSYCCX0099) and the CAS 'Light of West China' Program (No. 2021-XBQNXZ-005).

We thank M.~Gurwell for the Submillimeter Array (SMA) flux density measurement time series for Ton\,599. The Submillimeter Array is a joint project between the Smithsonian Astrophysical Observatory and the Academia Sinica Institute of Astronomy and Astrophysics and is funded by the Smithsonian Institution and the Academia Sinica. We recognize that Maunakea, the site of the SMA, is culturally important for the indigenous Hawaiian people; we are privileged to study the cosmos from its summit.

The development of the Fermi-LAT Light Curve Repository has been funded in part through the Fermi Guest Investigator Program (NASA Research Announcements NNH19ZDA001N and NNH20ZDA001N).

\section*{Data Availability}
The Zeiss-1000, AS-500/2, AZT-8, LX-200, RT-32 and RATAN-600 data underlying this article are available in the article and in its online supplementary material. The RT-22 data underlying this article were provided by AV by permission. Data will be shared on request to the corresponding author with permission of AV. The NSRT data underlying this article were provided by LC by permission. Data will be shared on request to the corresponding author with permission of LC. The SMA data are available at \url{http://sma1.sma.hawaii.edu/callist/callist.html}. 
The Fermi-LAT data are presented in public Light Curve Repository at \url{https://fermi.gsfc.nasa.gov/ssc/data/access/lat/LightCurveRepository/about.html}
The RATAN-600 data is partly available in the BLcat on-line catalogue at \url{https://www.sao.ru/blcat/}.

\textit{Facilities}: RATAN-600, Zeiss-1000, AS-500/2, AZT-8, LX-200, RT-32, RT-22, XAO-NSRT, SMA, \textit{Fermi}-LAT.

\bibliographystyle{mnras}
\bibliography{sotnikova}

@ARTICLE{1998PASP..110..660P,
       author = {{Peterson}, Bradley M. and {Wanders}, Ignaz and {Horne}, Keith and {Collier}, Stefan and {Alexander}, Tal and {Kaspi}, Shai and {Maoz}, Dan},
        title = "{On Uncertainties in Cross-Correlation Lags and the Reality of Wavelength-dependent Continuum Lags in Active Galactic Nuclei}",
      journal = {\pasp},
         year = 1998,
        month = jun,
       volume = {110},
       number = {748},
        pages = {660-670},
          doi = {10.1086/316177},
archivePrefix = {arXiv},
       eprint = {astro-ph/9802103},
 primaryClass = {astro-ph},
       adsurl = {https://ui.adsabs.harvard.edu/abs/1998PASP..110..660P}
}

@ARTICLE{2018ApJ...866..102P,
       author = {{Patel}, S.~R. and {Chitnis}, V.~R. and {Shukla}, A. and {Rao}, A.~R. and {Nagare}, B.~J.},
        title = "{Temporal Variability and Estimation of Jet Parameters for Ton 599}",
      journal = {\apj},
         year = 2018,
        month = oct,
       volume = {866},
       number = {2},
          eid = {102},
        pages = {102},
          doi = {10.3847/1538-4357/aae1fc},
archivePrefix = {arXiv},
       eprint = {1809.05269},
 primaryClass = {astro-ph.HE},
       adsurl = {https://ui.adsabs.harvard.edu/abs/2018ApJ...866..102P}
}

@ARTICLE{2022MNRAS.510.3641R,
       author = {{Roy}, Abhradeep and {Sarkar}, Arkadipta and {Chatterjee}, Anshu and {Gupta}, Alok C. and {Chitnis}, Varsha and {Wiita}, P.~J.},
        title = "{Transient quasi-periodic oscillations at {\ensuremath{\gamma}}-rays in the TeV blazar PKS 1510-089}",
      journal = {\mnras},
         year = 2022,
        month = mar,
       volume = {510},
       number = {3},
        pages = {3641-3649},
          doi = {10.1093/mnras/stab3701},
archivePrefix = {arXiv},
       eprint = {2112.08955},
 primaryClass = {astro-ph.HE},
       adsurl = {https://ui.adsabs.harvard.edu/abs/2022MNRAS.510.3641R}
}

@ARTICLE{2013MNRAS.436L.114K,
       author = {{King}, O.~G. and {Hovatta}, T. and {Max-Moerbeck}, W. and {Meier}, D.~L. and {Pearson}, T.~J. and {Readhead}, A.~C.~S. and {Reeves}, R. and {Richards}, J.~L. and {Shepherd}, M.~C.},
        title = "{A quasi-periodic oscillation in the blazar J1359+4011.}",
      journal = {\mnras},
         year = 2013,
        month = nov,
       volume = {436},
        pages = {L114-L117},
          doi = {10.1093/mnrasl/slt125},
archivePrefix = {arXiv},
       eprint = {1309.1158},
 primaryClass = {astro-ph.HE},
       adsurl = {https://ui.adsabs.harvard.edu/abs/2013MNRAS.436L.114K}
}

@ARTICLE{2004A&A...417..887H,
       author = {{Hong}, X.~Y. and {Jiang}, D.~R. and {Gurvits}, L.~I. and {Garrett}, M.~A. and {Garrington}, S.~T. and {Schilizzi}, R.~T. and {Nan}, R.~D. and {Hirabayashi}, H. and {Wang}, W.~H. and {Nicolson}, G.~D.},
        title = "{A relativistic helical jet in the {\ensuremath{\gamma}}-ray AGN 1156+295}",
      journal = {\aap},
         year = 2004,
        month = apr,
       volume = {417},
        pages = {887-904},
          doi = {10.1051/0004-6361:20031784},
archivePrefix = {arXiv},
       eprint = {astro-ph/0401627},
 primaryClass = {astro-ph},
       adsurl = {https://ui.adsabs.harvard.edu/abs/2004A&A...417..887H}
}

@ARTICLE{2015A&A...578A..34L,
       author = {{Liu}, X. and {Mi}, L. -G. and {Liu}, J. and {Cui}, L. and {Song}, H. -G. and {Krichbaum}, T.~P. and {Kraus}, A. and {Fuhrmann}, L. and {Marchili}, N. and {Zensus}, J.~A.},
        title = "{Intra-day variability observations and the VLBI structure analysis of quasar S4 0917+624}",
      journal = {\aap},
         year = 2015,
        month = jun,
       volume = {578},
          eid = {A34},
        pages = {A34},
          doi = {10.1051/0004-6361/201424854},
archivePrefix = {arXiv},
       eprint = {1504.04141},
 primaryClass = {astro-ph.GA},
       adsurl = {https://ui.adsabs.harvard.edu/abs/2015A&A...578A..34L}
}

@ARTICLE{2011A&A...530A.129M,
       author = {{Marchili}, N. and {Krichbaum}, T.~P. and {Liu}, X. and {Song}, H. -G. and {Anderson}, J.~M. and {Witzel}, A. and {Zensus}, J.~A.},
        title = "{On the influence of the Sun on the rapid variability of compact extragalactic sources}",
      journal = {\aap},
         year = 2011,
        month = jun,
       volume = {530},
          eid = {A129},
        pages = {A129},
          doi = {10.1051/0004-6361/201016033},
archivePrefix = {arXiv},
       eprint = {1103.4245},
 primaryClass = {astro-ph.SR},
       adsurl = {https://ui.adsabs.harvard.edu/abs/2011A&A...530A.129M}
}

@ARTICLE{2013ARep...57...34V,
       author = {{Vol'vach}, A.~E. and {Kutkin}, A.~M. and {Vol'vach}, L.~N. and {Larionov}, M.~G. and {Lakhteenmaki}, A. and {Tornikoski}, M. and {Nieppola}, E. and {Tammi}, J. and {Savolainen}, P. and {Leon-Tavares}, J. and {Aller}, M.~F. and {Aller}, H.~D.},
        title = "{Results of long-term monitoring of 3C 273 over a wide range of wavelengths}",
      journal = {Astronomy Reports},
         year = 2013,
        month = jan,
       volume = {57},
       number = {1},
        pages = {34-45},
          doi = {10.1134/S1063772912050083},
       adsurl = {https://ui.adsabs.harvard.edu/abs/2013ARep...57...34V}
}

@ARTICLE{2010SCPMA..53S.252Z,
       author = {{Zhang}, HaoJing and {Zhao}, Gang and {Zhang}, Xiong and {Bai}, JinMing},
        title = "{The periodicity of 3C 273's radio light curve at 15 GHz found by the wavelet method}",
      journal = {Science China Physics, Mechanics, and Astronomy},
         year = 2010,
        month = jan,
       volume = {53},
        pages = {252-255},
          doi = {10.1007/s11433-010-0053-x},
       adsurl = {https://ui.adsabs.harvard.edu/abs/2010SCPMA..53S.252Z}
}

@ARTICLE{2017Natur.552..374R,
       author = {{Raiteri}, C.~M. and {Villata}, M. and {Acosta-Pulido}, J.~A. and {Agudo}, I. and {Arkharov}, A.~A. and {Bachev}, R. and {Baida}, G.~V. and {Ben{\'\i}tez}, E. and {Borman}, G.~A. and {Boschin}, W. and {Bozhilov}, V. and {Butuzova}, M.~S. and {Calcidese}, P. and {Carnerero}, M.~I. and {Carosati}, D. and {Casadio}, C. and {Castro-Segura}, N. and {Chen}, W. -P. and {Damljanovic}, G. and {D'Ammando}, F. and {di Paola}, A. and {Echevarr{\'\i}a}, J. and {Efimova}, N.~V. and {Ehgamberdiev}, Sh. A. and {Espinosa}, C. and {Fuentes}, A. and {Giunta}, A. and {G{\'o}mez}, J.~L. and {Grishina}, T.~S. and {Gurwell}, M.~A. and {Hiriart}, D. and {Jermak}, H. and {Jordan}, B. and {Jorstad}, S.~G. and {Joshi}, M. and {Kopatskaya}, E.~N. and {Kuratov}, K. and {Kurtanidze}, O.~M. and {Kurtanidze}, S.~O. and {L{\"a}hteenm{\"a}ki}, A. and {Larionov}, V.~M. and {Larionova}, E.~G. and {Larionova}, L.~V. and {L{\'a}zaro}, C. and {Lin}, C.~S. and {Malmrose}, M.~P. and {Marscher}, A.~P. and {Matsumoto}, K. and {McBreen}, B. and {Michel}, R. and {Mihov}, B. and {Minev}, M. and {Mirzaqulov}, D.~O. and {Mokrushina}, A.~A. and {Molina}, S.~N. and {Moody}, J.~W. and {Morozova}, D.~A. and {Nazarov}, S.~V. and {Nikolashvili}, M.~G. and {Ohlert}, J.~M. and {Okhmat}, D.~N. and {Ovcharov}, E. and {Pinna}, F. and {Polakis}, T.~A. and {Protasio}, C. and {Pursimo}, T. and {Redondo-Lorenzo}, F.~J. and {Rizzi}, N. and {Rodriguez-Coira}, G. and {Sadakane}, K. and {Sadun}, A.~C. and {Samal}, M.~R. and {Savchenko}, S.~S. and {Semkov}, E. and {Skiff}, B.~A. and {Slavcheva-Mihova}, L. and {Smith}, P.~S. and {Steele}, I.~A. and {Strigachev}, A. and {Tammi}, J. and {Thum}, C. and {Tornikoski}, M. and {Troitskaya}, Yu. V. and {Troitsky}, I.~S. and {Vasilyev}, A.~A. and {Vince}, O.},
        title = "{Blazar spectral variability as explained by a twisted inhomogeneous jet}",
      journal = {\nat},
         year = 2017,
        month = dec,
       volume = {552},
       number = {7685},
        pages = {374-377},
          doi = {10.1038/nature24623},
archivePrefix = {arXiv},
       eprint = {1712.02098},
 primaryClass = {astro-ph.HE},
       adsurl = {https://ui.adsabs.harvard.edu/abs/2017Natur.552..374R}
}

@ARTICLE{1996ApJ...460..207L,
       author = {{Lehto}, Harry J. and {Valtonen}, Mauri J.},
        title = "{OJ 287 Outburst Structure and a Binary Black Hole Model}",
      journal = {\apj},
         year = 1996,
        month = mar,
       volume = {460},
        pages = {207},
          doi = {10.1086/176962},
       adsurl = {https://ui.adsabs.harvard.edu/abs/1996ApJ...460..207L}
}

@ARTICLE{2010MNRAS.406..975G,
       author = {{Garofalo}, D. and {Evans}, D.~A. and {Sambruna}, R.~M.},
        title = "{The evolution of radio-loud active galactic nuclei as a function of black hole spin}",
      journal = {\mnras},
         year = 2010,
        month = aug,
       volume = {406},
       number = {2},
        pages = {975-986},
          doi = {10.1111/j.1365-2966.2010.16797.x},
archivePrefix = {arXiv},
       eprint = {1004.1166},
 primaryClass = {astro-ph.CO},
       adsurl = {https://ui.adsabs.harvard.edu/abs/2010MNRAS.406..975G}
}

@ARTICLE{1998MNRAS.301..142M,
       author = {{Moderski}, R. and {Sikora}, M. and {Lasota}, J. -P.},
        title = "{On the spin paradigm and the radio dichotomy of quasars}",
      journal = {\mnras},
         year = 1998,
        month = nov,
       volume = {301},
       number = {1},
        pages = {142-148},
          doi = {10.1046/j.1365-8711.1998.02009.x},
archivePrefix = {arXiv},
       eprint = {astro-ph/9804140},
 primaryClass = {astro-ph},
       adsurl = {https://ui.adsabs.harvard.edu/abs/1998MNRAS.301..142M}
}

@ARTICLE{1995ApJ...438...62W,
       author = {{Wilson}, A.~S. and {Colbert}, E.~J.~M.},
        title = "{The Difference between Radio-loud and Radio-quiet Active Galaxies}",
      journal = {\apj},
         year = 1995,
        month = jan,
       volume = {438},
        pages = {62},
          doi = {10.1086/175054},
archivePrefix = {arXiv},
       eprint = {astro-ph/9408005},
 primaryClass = {astro-ph},
       adsurl = {https://ui.adsabs.harvard.edu/abs/1995ApJ...438...62W}
}

@ARTICLE{2021JOSS....6.3001B,
       author = {{Buchner}, Johannes},
        title = "{UltraNest - a robust, general purpose Bayesian inference engine}",
      journal = {The Journal of Open Source Software},
         year = 2021,
        month = apr,
       volume = {6},
       number = {60},
          eid = {3001},
        pages = {3001},
          doi = {10.21105/joss.03001},
archivePrefix = {arXiv},
       eprint = {2101.09604},
 primaryClass = {stat.CO},
       adsurl = {https://ui.adsabs.harvard.edu/abs/2021JOSS....6.3001B}
}

@ARTICLE{2019PASP..131j8005B,
       author = {{Buchner}, Johannes},
        title = "{Collaborative Nested Sampling: Big Data versus Complex Physical Models}",
      journal = {\pasp},
         year = 2019,
        month = oct,
       volume = {131},
       number = {1004},
        pages = {108005},
          doi = {10.1088/1538-3873/aae7fc},
archivePrefix = {arXiv},
       eprint = {1707.04476},
 primaryClass = {stat.CO},
       adsurl = {https://ui.adsabs.harvard.edu/abs/2019PASP..131j8005B}
}

@ARTICLE{2018MNRAS.478.3199B,
       author = {{Britzen}, S. and {Fendt}, C. and {Witzel}, G. and {Qian}, S. -J. and {Pashchenko}, I.~N. and {Kurtanidze}, O. and {Zajacek}, M. and {Martinez}, G. and {Karas}, V. and {Aller}, M. and {Aller}, H. and {Eckart}, A. and {Nilsson}, K. and {Ar{\'e}valo}, P. and {Cuadra}, J. and {Subroweit}, M. and {Witzel}, A.},
        title = "{OJ287: deciphering the `Rosetta stone of blazars}",
      journal = {\mnras},
         year = 2018,
        month = aug,
       volume = {478},
       number = {3},
        pages = {3199-3219},
          doi = {10.1093/mnras/sty1026},
       adsurl = {https://ui.adsabs.harvard.edu/abs/2018MNRAS.478.3199B}
}

@ARTICLE{2023ApJ...951..106B,
       author = {{Britzen}, Silke and {Zaja{\v{c}}ek}, Michal and {Gopal-Krishna} and {Fendt}, Christian and {Kun}, Emma and {Jaron}, Fr{\'e}d{\'e}ric and {Sillanp{\"a}{\"a}}, Aimo and {Eckart}, Andreas},
        title = "{Precession-induced Variability in AGN Jets and OJ 287}",
      journal = {\apj},
         year = 2023,
        month = jul,
       volume = {951},
       number = {2},
          eid = {106},
        pages = {106},
          doi = {10.3847/1538-4357/accbbc},
archivePrefix = {arXiv},
       eprint = {2307.05838},
 primaryClass = {astro-ph.HE},
       adsurl = {https://ui.adsabs.harvard.edu/abs/2023ApJ...951..106B}
}

@ARTICLE{2000A&A...360...57R,
       author = {{Romero}, G.~E. and {Chajet}, L. and {Abraham}, Z. and {Fan}, J.~H.},
        title = "{Beaming and precession in the inner jet of 3C 273 --- II. The central engine}",
      journal = {\aap},
         year = 2000,
        month = aug,
       volume = {360},
        pages = {57-64},
       adsurl = {https://ui.adsabs.harvard.edu/abs/2000A&A...360...57R}
}

@ARTICLE{2000A&A...355..915A,
       author = {{Abraham}, Z.},
        title = "{Precession, beaming and the periodic light curve of OJ287}",
      journal = {\aap},
         year = 2000,
        month = mar,
       volume = {355},
        pages = {915-921},
       adsurl = {https://ui.adsabs.harvard.edu/abs/2000A&A...355..915A}
}

@ARTICLE{1997ApJ...478..527K,
       author = {{Katz}, J.~I.},
        title = "{A Precessing Disk in OJ 287?}",
      journal = {\apj},
         year = 1997,
        month = mar,
       volume = {478},
       number = {2},
        pages = {527-529},
          doi = {10.1086/303811},
       adsurl = {https://ui.adsabs.harvard.edu/abs/1997ApJ...478..527K}
}

@ARTICLE{2007MNRAS.376.1740K,
       author = {{King}, A.~R. and {Pringle}, J.~E. and {Livio}, M.},
        title = "{Accretion disc viscosity: how big is alpha?}",
      journal = {\mnras},
         year = 2007,
        month = apr,
       volume = {376},
       number = {4},
        pages = {1740-1746},
          doi = {10.1111/j.1365-2966.2007.11556.x},
archivePrefix = {arXiv},
       eprint = {astro-ph/0701803},
 primaryClass = {astro-ph},
       adsurl = {https://ui.adsabs.harvard.edu/abs/2007MNRAS.376.1740K}
}

@ARTICLE{2021ARA&A..59..117R,
       author = {{Reynolds}, Christopher S.},
        title = "{Observational Constraints on Black Hole Spin}",
      journal = {\araa},
         year = 2021,
        month = sep,
       volume = {59},
        pages = {117-154},
          doi = {10.1146/annurev-astro-112420-035022},
archivePrefix = {arXiv},
       eprint = {2011.08948},
 primaryClass = {astro-ph.HE},
       adsurl = {https://ui.adsabs.harvard.edu/abs/2021ARA&A..59..117R}
}

@ARTICLE{2005ApJ...635L..17L,
       author = {{Lu}, Ju-Fu and {Zhou}, Bo-Yan},
        title = "{Observational Evidence of Jet Precession in Galactic Nuclei Caused by Accretion Disks}",
      journal = {\apjl},
         year = 2005,
        month = dec,
       volume = {635},
       number = {1},
        pages = {L17-L20},
          doi = {10.1086/499333},
archivePrefix = {arXiv},
       eprint = {astro-ph/0511212},
 primaryClass = {astro-ph},
       adsurl = {https://ui.adsabs.harvard.edu/abs/2005ApJ...635L..17L}
}

@ARTICLE{1980ApJ...238L.129S,
       author = {{Sarazin}, C.~L. and {Begelman}, M.~C. and {Hatchett}, S.~P.},
        title = "{Disk-driven precession in SS 433}",
      journal = {\apjl},
         year = 1980,
        month = jun,
       volume = {238},
        pages = {L129-L132},
          doi = {10.1086/183272},
       adsurl = {https://ui.adsabs.harvard.edu/abs/1980ApJ...238L.129S}
}

@ARTICLE{1990A&A...229..424L,
       author = {{Lu}, J.~F.},
        title = "{Accretion disk-driven jet precession in active galactic nuclei}",
      journal = {\aap},
         year = 1990,
        month = mar,
       volume = {229},
       number = {2},
        pages = {424-426},
       adsurl = {https://ui.adsabs.harvard.edu/abs/1990A&A...229..424L}
}

@ARTICLE{2006PASJ...58..797F,
       author = {{Fan}, Jun Hui and {Tao}, Jun and {Qian}, Bo Chen and {Gupta}, Alok C. and {Liu}, Yi and {Yuan}, Yu-Hai and {Yang}, Jiang-He and {Wang}, Hong Guang and {Huang}, Y.},
        title = "{Optical Photometrical Observations and Variability for Quasar 4C 29.45}",
      journal = {\pasj},
         year = 2006,
        month = oct,
       volume = {58},
       number = {5},
        pages = {797-808},
          doi = {10.1093/pasj/58.5.797},
archivePrefix = {arXiv},
       eprint = {astro-ph/0610227},
 primaryClass = {astro-ph},
       adsurl = {https://ui.adsabs.harvard.edu/abs/2006PASJ...58..797F}
}

@ARTICLE{2021A&A...648A..27V,
       author = {{Volvach}, A.~E. and {Volvach}, L.~N. and {Larionov}, M.~G.},
        title = "{Most massive double black hole 3C 454.3 and powerful gravitational wave radiation}",
      journal = {\aap},
         year = 2021,
        month = apr,
       volume = {648},
          eid = {A27},
        pages = {A27},
          doi = {10.1051/0004-6361/202039124},
       adsurl = {https://ui.adsabs.harvard.edu/abs/2021A&A...648A..27V}
}

@ARTICLE{2024A&A...691L...9V,
       author = {{Volvach}, A.~E. and {Volvach}, L.~N. and {Larionov}, M.~G.},
        title = "{Blazar S 0528+134 is possibly the most powerful emitter in the Universe, including in the range of gravitational waves}",
      journal = {\aap},
         year = 2024,
        month = nov,
       volume = {691},
          eid = {L9},
        pages = {L9},
          doi = {10.1051/0004-6361/202451911},
       adsurl = {https://ui.adsabs.harvard.edu/abs/2024A&A...691L...9V}
}

@ARTICLE{2024iSci...27j9427V,
       author = {{Volvach}, Alexandr and {Volvach}, Larisa and {Larionov}, Mikhail},
        title = "{Electromagnetic and gravitational radiation of blazar OJ 287}",
      journal = {iScience},
         year = 2024,
        month = apr,
       volume = {27},
       number = {4},
        pages = {109427},
          doi = {10.1016/j.isci.2024.109427},
       adsurl = {https://ui.adsabs.harvard.edu/abs/2024iSci...27j9427V}
}

@ARTICLE{2022ApJS..260...53A,
       author = {{Abdollahi}, S. and {Acero}, F. and {Baldini}, L. and {Ballet}, J. and {Bastieri}, D. and {Bellazzini}, R. and {Berenji}, B. and {Berretta}, A. and {Bissaldi}, E. and {Blandford}, R.~D. and {Bloom}, E. and {Bonino}, R. and {Brill}, A. and {Britto}, R.~J. and {Bruel}, P. and {Burnett}, T.~H. and {Buson}, S. and {Cameron}, R.~A. and {Caputo}, R. and {Caraveo}, P.~A. and {Castro}, D. and {Chaty}, S. and {Cheung}, C.~C. and {Chiaro}, G. and {Cibrario}, N. and {Ciprini}, S. and {Coronado-Bl{\'a}zquez}, J. and {Crnogorcevic}, M. and {Cutini}, S. and {D'Ammando}, F. and {De Gaetano}, S. and {Digel}, S.~W. and {Di Lalla}, N. and {Dirirsa}, F. and {Di Venere}, L. and {Dom{\'\i}nguez}, A. and {Fallah Ramazani}, V. and {Fegan}, S.~J. and {Ferrara}, E.~C. and {Fiori}, A. and {Fleischhack}, H. and {Franckowiak}, A. and {Fukazawa}, Y. and {Funk}, S. and {Fusco}, P. and {Galanti}, G. and {Gammaldi}, V. and {Gargano}, F. and {Garrappa}, S. and {Gasparrini}, D. and {Giacchino}, F. and {Giglietto}, N. and {Giordano}, F. and {Giroletti}, M. and {Glanzman}, T. and {Green}, D. and {Grenier}, I.~A. and {Grondin}, M. -H. and {Guillemot}, L. and {Guiriec}, S. and {Gustafsson}, M. and {Harding}, A.~K. and {Hays}, E. and {Hewitt}, J.~W. and {Horan}, D. and {Hou}, X. and {J{\'o}hannesson}, G. and {Karwin}, C. and {Kayanoki}, T. and {Kerr}, M. and {Kuss}, M. and {Landriu}, D. and {Larsson}, S. and {Latronico}, L. and {Lemoine-Goumard}, M. and {Li}, J. and {Liodakis}, I. and {Longo}, F. and {Loparco}, F. and {Lott}, B. and {Lubrano}, P. and {Maldera}, S. and {Malyshev}, D. and {Manfreda}, A. and {Mart{\'\i}-Devesa}, G. and {Mazziotta}, M.~N. and {Mereu}, I. and {Meyer}, M. and {Michelson}, P.~F. and {Mirabal}, N. and {Mitthumsiri}, W. and {Mizuno}, T. and {Moiseev}, A.~A. and {Monzani}, M.~E. and {Morselli}, A. and {Moskalenko}, I.~V. and {Negro}, M. and {Nuss}, E. and {Omodei}, N. and {Orienti}, M. and {Orlando}, E. and {Paneque}, D. and {Pei}, Z. and {Perkins}, J.~S. and {Persic}, M. and {Pesce-Rollins}, M. and {Petrosian}, V. and {Pillera}, R. and {Poon}, H. and {Porter}, T.~A. and {Principe}, G. and {Rain{\`o}}, S. and {Rando}, R. and {Rani}, B. and {Razzano}, M. and {Razzaque}, S. and {Reimer}, A. and {Reimer}, O. and {Reposeur}, T. and {S{\'a}nchez-Conde}, M. and {Saz Parkinson}, P.~M. and {Scotton}, L. and {Serini}, D. and {Sgr{\`o}}, C. and {Siskind}, E.~J. and {Smith}, D.~A. and {Spandre}, G. and {Spinelli}, P. and {Sueoka}, K. and {Suson}, D.~J. and {Tajima}, H. and {Tak}, D. and {Thayer}, J.~B. and {Thompson}, D.~J. and {Torres}, D.~F. and {Troja}, E. and {Valverde}, J. and {Wood}, K. and {Zaharijas}, G.},
        title = "{Incremental Fermi Large Area Telescope Fourth Source Catalog}",
      journal = {\apjs},
         year = 2022,
        month = jun,
       volume = {260},
       number = {2},
          eid = {53},
        pages = {53},
          doi = {10.3847/1538-4365/ac6751},
archivePrefix = {arXiv},
       eprint = {2201.11184},
 primaryClass = {astro-ph.HE},
       adsurl = {https://ui.adsabs.harvard.edu/abs/2022ApJS..260...53A}
}

@inproceedings{seabold2010statsmodels,
  author    = {Seabold, Skipper and Perktold, Josef},
  title     = {Statsmodels: Econometric and Statistical Modeling with Python},
  booktitle = {Proceedings of the 9th Python in Science Conference (SciPy 2010)},
  editor    = {van der Walt, Stéfan and Millman, Jarrod},
  year      = {2010},
  address   = {Austin, Texas, USA},
  publisher = {SciPy},
  pages     = {57--61}
}

@ARTICLE{2005A&A...431..831L,
       author = {{Lobanov}, A.~P. and {Roland}, J.},
        title = "{A supermassive binary black hole in the quasar 3C 345}",
      journal = {\aap},
         year = 2005,
        month = mar,
       volume = {431},
       number = {3},
        pages = {831-846},
          doi = {10.1051/0004-6361:20041831},
archivePrefix = {arXiv},
       eprint = {astro-ph/0411417},
 primaryClass = {astro-ph},
       adsurl = {https://ui.adsabs.harvard.edu/abs/2005A&A...431..831L}
}

@ARTICLE{2022arXiv220201915Q,
       author = {{Qian}, S.~J.},
        title = "{Possible evidence for double precessing nozzle structure in QSO 3C345}",
      journal = {arXiv e-prints},
         year = 2022,
        month = feb,
          eid = {arXiv:2202.01915},
        pages = {arXiv:2202.01915},
          doi = {10.48550/arXiv.2202.01915},
archivePrefix = {arXiv},
       eprint = {2202.01915},
 primaryClass = {astro-ph.GA},
       adsurl = {https://ui.adsabs.harvard.edu/abs/2022arXiv220201915Q}
}

@ARTICLE{1999ApJ...521..509L,
       author = {{Lobanov}, Andrew P. and {Zensus}, J. Anton},
        title = "{Spectral Evolution of the Parsec-Scale Jet in the Quasar 3C 345}",
      journal = {\apj},
         year = 1999,
        month = aug,
       volume = {521},
       number = {2},
        pages = {509-525},
          doi = {10.1086/307555},
archivePrefix = {arXiv},
       eprint = {astro-ph/9903318},
 primaryClass = {astro-ph},
       adsurl = {https://ui.adsabs.harvard.edu/abs/1999ApJ...521..509L}
}

@ARTICLE{2010CeMDA.106..235V,
       author = {{Valtonen}, M.~J. and {Mikkola}, S. and {Lehto}, H.~J. and {Hyv{\"o}nen}, T. and {Nilsson}, K. and {Merritt}, D. and {Gopakumar}, A. and {Rampadarath}, H. and {Hudec}, R. and {Basta}, M. and {Saunders}, R.},
        title = "{Measuring black hole spin in OJ287}",
      journal = {Celestial Mechanics and Dynamical Astronomy},
         year = 2010,
        month = mar,
       volume = {106},
       number = {3},
        pages = {235-243},
          doi = {10.1007/s10569-009-9252-z},
archivePrefix = {arXiv},
       eprint = {1001.1284},
 primaryClass = {astro-ph.CO},
       adsurl = {https://ui.adsabs.harvard.edu/abs/2010CeMDA.106..235V}
}

@ARTICLE{2012MNRAS.427...77V,
       author = {{Valtonen}, M.~J. and {Ciprini}, S. and {Lehto}, H.~J.},
        title = "{On the masses of OJ287 black holes}",
      journal = {\mnras},
         year = 2012,
        month = nov,
       volume = {427},
       number = {1},
        pages = {77-83},
          doi = {10.1111/j.1365-2966.2012.21861.x},
archivePrefix = {arXiv},
       eprint = {1208.0906},
 primaryClass = {astro-ph.HE},
       adsurl = {https://ui.adsabs.harvard.edu/abs/2012MNRAS.427...77V}
}

@ARTICLE{2020Univ....6..191B,
       author = {{Butuzova}, Marina S. and {Pushkarev}, Alexander B.},
        title = "{Is OJ 287 a Single Supermassive Black Hole?}",
      journal = {Universe},
         year = 2020,
        month = oct,
       volume = {6},
       number = {11},
          eid = {191},
        pages = {191},
          doi = {10.3390/universe6110191},
archivePrefix = {arXiv},
       eprint = {2103.13845},
 primaryClass = {astro-ph.HE},
       adsurl = {https://ui.adsabs.harvard.edu/abs/2020Univ....6..191B}
}

@ARTICLE{2024APh...16002965G,
       author = {{Gorbachev}, M.~A. and {Butuzova}, M.~S. and {Nazarov}, S.~V. and {Zhovtan}, A.~V.},
        title = "{Evidence of jet-caused 12-year optical periodicity of blazar OJ 287}",
      journal = {Astroparticle Physics},
         year = 2024,
        month = aug,
       volume = {160},
          eid = {102965},
        pages = {102965},
          doi = {10.1016/j.astropartphys.2024.102965},
       adsurl = {https://ui.adsabs.harvard.edu/abs/2024APh...16002965G}
}

@ARTICLE{2024A&A...683A.248C,
       author = {{Cho}, Ilje and {G{\'o}mez}, Jos{\'e} L. and {Lico}, Rocco and {Zhao}, Guang-Yao and {Traianou}, Efthalia and {Dahale}, Rohan and {Fuentes}, Antonio and {Toscano}, Teresa and {Foschi}, Marianna and {Kovalev}, Yuri Y. and {Lobanov}, Andrei and {Pushkarev}, Alexander B. and {Gurvits}, Leonid I. and {Kim}, Jae-Young and {Lisakov}, Mikhail and {Voitsik}, Petr and {Myserlis}, Ioannis and {P{\"o}tzl}, Felix and {Ros}, Eduardo},
        title = "{Unveiling the bent-jet structure and polarization of OJ 287 at 1.7 GHz with space VLBI}",
      journal = {\aap},
         year = 2024,
        month = mar,
       volume = {683},
          eid = {A248},
        pages = {A248},
          doi = {10.1051/0004-6361/202347157},
archivePrefix = {arXiv},
       eprint = {2312.08643},
 primaryClass = {astro-ph.HE},
       adsurl = {https://ui.adsabs.harvard.edu/abs/2024A&A...683A.248C}
}

@ARTICLE{2005ApJ...620..646H,
       author = {{Hardee}, P.~E. and {Walker}, R.~C. and {G{\'o}mez}, J.~L.},
        title = "{Modeling the 3C 120 Radio Jet from 1 to 30 Milliarcseconds}",
      journal = {\apj},
         year = 2005,
        month = feb,
       volume = {620},
       number = {2},
        pages = {646-664},
          doi = {10.1086/427083},
archivePrefix = {arXiv},
       eprint = {astro-ph/0410720},
 primaryClass = {astro-ph},
       adsurl = {https://ui.adsabs.harvard.edu/abs/2005ApJ...620..646H}
}

@ARTICLE{2020ApJ...901..149Z,
       author = {{Zhang}, Haocheng and {Li}, Xiaocan and {Giannios}, Dimitrios and {Guo}, Fan and {Liu}, Yi-Hsin and {Dong}, Lingyi},
        title = "{Radiation and Polarization Signatures from Magnetic Reconnection in Relativistic Jets. I. A Systematic Study}",
      journal = {\apj},
         year = 2020,
        month = oct,
       volume = {901},
       number = {2},
          eid = {149},
        pages = {149},
          doi = {10.3847/1538-4357/abb1b0},
archivePrefix = {arXiv},
       eprint = {2008.09444},
 primaryClass = {astro-ph.HE},
       adsurl = {https://ui.adsabs.harvard.edu/abs/2020ApJ...901..149Z}
}

@article{BOLLERSLEV1986307,
title = {Generalized autoregressive conditional heteroskedasticity},
journal = {Journal of Econometrics},
volume = {31},
number = {3},
pages = {307-327},
year = {1986},
issn = {0304-4076},
doi = {https://doi.org/10.1016/0304-4076(86)90063-1},
url = {https://www.sciencedirect.com/science/article/pii/0304407686900631},
author = {Tim Bollerslev}
}

@INPROCEEDINGS{2020gbar.conf...32S,
       author = {{Sotnikova}, Yu. V.},
        title = "{RATAN-600 Radio Telescope: Observing Programs and Outlook}",
    booktitle = {Ground-Based Astronomy in Russia. 21st Century},
         year = 2020,
       editor = {{Romanyuk}, I.~I. and {Yakunin}, I.~A. and {Valeev}, A.~F. and {Kudryavtsev}, D.~O.},
        month = dec,
        pages = {32-40},
          doi = {10.26119/978-5-6045062-0-2_2020_32},
       adsurl = {https://ui.adsabs.harvard.edu/abs/2020gbar.conf...32S}
}

@ARTICLE{2023Galax..11...96V,
       author = {{Volvach}, Alexandr and {Volvach}, Larisa and {Larionov}, Mikhail},
        title = "{A Close Binary Supermassive Black Hole Model for the Galaxy 3C 273}",
      journal = {Galaxies},
         year = 2023,
        month = sep,
       volume = {11},
       number = {5},
          eid = {96},
        pages = {96},
          doi = {10.3390/galaxies11050096},
       adsurl = {https://ui.adsabs.harvard.edu/abs/2023Galax..11...96V}
}

@ARTICLE{2009A&A...494..527H,
       author = {{Hovatta}, T. and {Valtaoja}, E. and {Tornikoski}, M. and {L{\"a}hteenm{\"a}ki}, A.},
        title = "{Doppler factors, Lorentz factors and viewing angles for quasars, BL Lacertae objects and radio galaxies}",
      journal = {\aap},
         year = 2009,
        month = feb,
       volume = {494},
       number = {2},
        pages = {527-537},
          doi = {10.1051/0004-6361:200811150},
archivePrefix = {arXiv},
       eprint = {0811.4278},
 primaryClass = {astro-ph},
       adsurl = {https://ui.adsabs.harvard.edu/abs/2009A&A...494..527H}
}

@ARTICLE{2018AstBu..73..494T,
       author = {{Tsybulev}, P.~G. and {Nizhelskii}, N.~A. and {Dugin}, M.~V. and {Borisov}, A.~N. and {Kratov}, D.~V. and {Udovitskii}, R. Yu.},
        title = "{C-Band Radiometer for Continuum Observations at RATAN-600 Radio Telescope}",
      journal = {Astrophysical Bulletin},
         year = 2018,
        month = oct,
       volume = {73},
       number = {4},
        pages = {494-500},
          doi = {10.1134/S1990341318040132},
       adsurl = {https://ui.adsabs.harvard.edu/abs/2018AstBu..73..494T}
}

@ARTICLE{2003MNRAS.345.1271V,
       author = {{Vaughan}, S. and {Edelson}, R. and {Warwick}, R.~S. and {Uttley}, P.},
        title = "{On characterizing the variability properties of X-ray light curves from active galaxies}",
      journal = {\mnras},
         year = 2003,
        month = nov,
       volume = {345},
       number = {4},
        pages = {1271-1284},
          doi = {10.1046/j.1365-2966.2003.07042.x},
archivePrefix = {arXiv},
       eprint = {astro-ph/0307420},
 primaryClass = {astro-ph},
       adsurl = {https://ui.adsabs.harvard.edu/abs/2003MNRAS.345.1271V}
}

@STRING(pasp="PASP")

@ARTICLE{2008A&A...488..897H,
   author = {{Hovatta}, T. and {Lehto}, H.~J. and {Tornikoski}, M.},
    title = "{Wavelet analysis of a large sample of AGN at high radio frequencies}",
  journal = {\aap},
   eprint = {0807.1796},
     year = 2008,
    month = sep,
   volume = 488,
    pages = {897-903},
      doi ={10.1051/0004-6361:200810200},
   adsurl = {http://adsabs.harvard.edu/abs/2008A%26A...488..897H}
}

@ARTICLE{2007A&A...469..899H,
   author = {{Hovatta}, T. and {Tornikoski}, M. and {Lainela}, M. and {Lehto}, H.~J. and
    {Valtaoja}, E. and {Torniainen}, I and {Aller}, M.~F. and {Aller}, H.~D.},
    title = "{Statistical analyses of long-term variability of AGN at high radio frequencies}",
  journal = {\aap},
   eprint = {0705.3293},
     year = 2007,
    month = jul,
   volume = 469,
    pages = {899-912},
      doi ={10.1051/0004-6361:20077529},
   adsurl = {http://adsabs.harvard.edu/abs/2007A%26A...469..899H}
}

@ARTICLE{1994A&A...284..331O,
   author = {{Ott}, M. and {Witzel}, A. and {Quirrenbach}, A. and {Krichbaum}, T.~P. and
	{Standke}, K.~J. and {Schalinski}, C.~J. and {Hummel}, C.~A.},
    title = "{An updated list of radio flux density calibrators}",
  journal = {\aap},
     year = 1994,
    month = apr,
   volume = 284,
    pages = {331-339},
   adsurl = {http://adsabs.harvard.edu/abs/1994A%26A...284..331O}
}

@ARTICLE{1980A&AS...39..379T,
   author = {{Tabara}, H. and {Inoue}, M.},
    title = "{A catalogue of linear polarization of radio sources}",
  journal = {\aaps},
     year = 1980,
    month = mar,
   volume = 39,
    pages = {379-393},
   adsurl = {http://adsabs.harvard.edu/abs/980A%26AS...39..379T}
}

@ARTICLE{1997ASPC..125...46V,
   author = {{Verkhodanov}, O.~V.},
    title = "{Multiwave Continuum Data Reduction at RATAN-600}",
  journal = {Astronomical Data Analysis Software and Systems VI, A.S.P. Conference Series},
     year = 1997,
    month = mar,
   volume = 125,
    pages = {46-49},
   adsurl = {http://adsabs.harvard.edu/abs/1997ASPC..125...46V}
}

@ARTICLE{2013MNRAS.434.3487A,
       author = {{An}, Tao and {Baan}, Willem A. and {Wang}, Jun-Yi and {Wang}, Yu and {Hong}, Xiao-Yu},
        title = "{Periodic radio variabilities in NRAO 530: a jet-disc connection?}",
      journal = {\mnras},
         year = 2013,
        month = oct,
       volume = {434},
       number = {4},
        pages = {3487-3496},
          doi = {10.1093/mnras/stt1265},
       adsurl = {https://ui.adsabs.harvard.edu/abs/2013MNRAS.434.3487A}
}

@ARTICLE{2014MNRAS.443...58W,
       author = {{Wang}, Jun-Yi and {An}, Tao and {Baan}, Willem A. and {Lu}, Xiang-Long},
        title = "{Periodic radio variabilities of the blazar 1156+295: harmonic oscillations}",
      journal = {\mnras},
         year = 2014,
        month = sep,
       volume = {443},
       number = {1},
        pages = {58-66},
          doi = {10.1093/mnras/stu1135},
       adsurl = {https://ui.adsabs.harvard.edu/abs/2014MNRAS.443...58W}
}

@ARTICLE{1993IAPM...35....7P,
   author = {{Parijskij}, Y.~N.},
    title = "{RATAN-600 - The world's biggest reflector at the 'cross roads'}",
  journal = {IEEE Antennas and Propagation Magazine},
     year = 1993,
    month = aug,
   volume = 35,
    pages = {7-12},
      doi = {10.1109/74.229840},
   adsurl = {http://adsabs.harvard.edu/abs/1993IAPM...35....7P}
}

@ARTICLE{2011AstBu..66..109T,
   author = {{Tsybulev}, P.~G.},
    title = "{New-generation data acquisition and control system for continuum radio-astronomic observations with RATAN-600 radio telescope: Development, observations, and measurements}",
  journal = {Astrophysical Bulletin},
     year = 2011,
    month = jan,
   volume = 66,
    pages = {109-122},
      doi = {10.1134/S199034131101010X},
   adsurl = {http://adsabs.harvard.edu/abs/2011AstBu..66..109T}
}

@ARTICLE{1977A&A....61...99B,
   author = {{Baars}, J.~W.~M. and {Genzel}, R. and {Pauliny-Toth}, I.~I.~K. and
    {Witzel}, A.},
    title = "{The absolute spectrum of CAS A - an accurate flux density scale and a set of secondary calibrators}",
  journal = {\aap},
     year = 1977,
    month = oct,
   volume = 61,
    pages = {99-106},
   adsurl = {http://adsabs.harvard.edu/abs/1977A%26A....61...99B}
}

@ARTICLE{2001A&A...377..396R,
   author = {{Raiteri}, C.~M. and {Villata}, M. and {Aller}, H.~D. and {Aller}, M.~F. and
    {Heidt}, J. and {Kurtanidze}, O.~M. and {Lanteri}, L. and {Maesano}, M. and
    {Massaro}, E. and {Montagni}, F. and {Nesci}, R. and {Nilsson}, K. and
    {Nikolashvili}, M.~G. and {Nurmi}, P. and {Ostorero}, L. and
    {Pursimo}, T. and {Rekola}, R. and {Sillanp{\"a}{\"a}}, A. and
    {Takalo}, L.~O. and {Ter{\"a}sranta}, H. and {Tosti}, G. and
    {Balonek}, T.~J. and {Feldt}, M. and {Heines}, A. and {Heisler}, C. and
    {Hu}, J. and {Kidger}, M. and {Mattox}, J.~R. and {McGrath}, E.~J. and
    {Pati}, A. and {Robb}, R. and {Sadun}, A.~C. and {Shastri}, P. and
    {Wagner}, S.~J. and {Wei}, J. and {Wu}, X.},
    title = "{Optical and radio variability of the BL Lacertae object <ASTROBJ>AO 0235+16</ASTROBJ>: A possible 5-6 year periodicity}",
  journal = {\aap},
   eprint = {astro-ph/0108165},
     year = 2001,
    month = oct,
   volume = 377,
    pages = {396-412},
      doi = {10.1051/0004-6361:20011112},
   adsurl = {http://adsabs.harvard.edu/abs/2001A%26A...377..396R}
}

@ARTICLE{2016AstBu..71..496U,
       author = {{Udovitskiy}, R. Yu. and {Sotnikova}, Yu. V. and {Mingaliev}, M.~G. and
         {Tsybulev}, P.~G. and {Zhekanis}, G.~V. and {Nizhelskij}, N.~A.},
        title = "{Automated system for reduction of observational data on RATAN-600 radio telescope}",
      journal = {Astrophysical Bulletin},
         year = 2016,
        month = oct,
       volume = {71},
       number = {4},
        pages = {496-505},
          doi = {10.1134/S1990341316040131},
       adsurl = {https://ui.adsabs.harvard.edu/abs/2016AstBu..71..496U}
}

@ARTICLE{2017ApJS..230....7P,
       author = {{Perley}, R.~A. and {Butler}, B.~J.},
        title = "{An Accurate Flux Density Scale from 50 MHz to 50 GHz}",
      journal = {\apjs},
         year = 2017,
        month = may,
       volume = {230},
       number = {1},
          eid = {7},
        pages = {7},
          doi = {10.3847/1538-4365/aa6df9},
archivePrefix = {arXiv},
       eprint = {1609.05940},
 primaryClass = {astro-ph.IM},
       adsurl = {https://ui.adsabs.harvard.edu/abs/2017ApJS..230....7P}
}

@ARTICLE{2013ApJS..204...19P,
       author = {{Perley}, R.~A. and {Butler}, B.~J.},
        title = "{An Accurate Flux Density Scale from 1 to 50 GHz}",
      journal = {\apjs},
         year = 2013,
        month = feb,
       volume = {204},
       number = {2},
          eid = {19},
        pages = {19},
          doi = {10.1088/0067-0049/204/2/19},
archivePrefix = {arXiv},
       eprint = {1211.1300},
 primaryClass = {astro-ph.IM},
       adsurl = {https://ui.adsabs.harvard.edu/abs/2013ApJS..204...19P}
}

@ARTICLE{2020arXiv200511208B,
       author = {{Ballet}, J. and {Burnett}, T.~H. and {Digel}, S.~W. and {Lott}, B.},
        title = "{Fermi Large Area Telescope Fourth Source Catalog Data Release 2}",
      journal = {arXiv e-prints},
         year = 2020,
        month = may,
          eid = {arXiv:2005.11208},
        pages = {arXiv:2005.11208},
archivePrefix = {arXiv},
       eprint = {2005.11208},
 primaryClass = {astro-ph.HE},
       adsurl = {https://ui.adsabs.harvard.edu/abs/2020arXiv200511208B}
}

@ARTICLE{2010MNRAS.405.2302H,
       author = {{Hewett}, Paul C. and {Wild}, Vivienne},
        title = "{Improved redshifts for SDSS quasar spectra}",
      journal = {\mnras},
         year = 2010,
        month = jul,
       volume = {405},
       number = {4},
        pages = {2302-2316},
          doi = {10.1111/j.1365-2966.2010.16648.x},
archivePrefix = {arXiv},
       eprint = {1003.3017},
 primaryClass = {astro-ph.CO},
       adsurl = {https://ui.adsabs.harvard.edu/abs/2010MNRAS.405.2302H}
}

@ARTICLE{1985ApJ...298..114M,
       author = {{Marscher}, A.~P. and {Gear}, W.~K.},
        title = "{Models for high-frequency radio outbursts in extragalactic sources, with application to the early 1983 millimeter-to-infrared flare of 3C 273.}",
      journal = {\apj},
         year = 1985,
        month = nov,
       volume = {298},
        pages = {114-127},
          doi = {10.1086/163592},
       adsurl = {https://ui.adsabs.harvard.edu/abs/1985ApJ...298..114M}
}

@ARTICLE{1999A&A...347...30V,
       author = {{Villata}, M. and {Raiteri}, C.~M.},
        title = "{Helical jets in blazars. I. The case of MKN 501}",
      journal = {\aap},
         year = 1999,
        month = jul,
       volume = {347},
        pages = {30-36},
       adsurl = {https://ui.adsabs.harvard.edu/abs/1999A&A...347...30V}
}

@article{kharinov2012,
  author = {M.~A. Kharinov and A.~E. Yablokova},
  issue = {24},
  journal = {Tr. IPA RAN},
  note = {russian},
  pages = {342--347},
  title = {Class Visual: modernization of the package for processing the single dish observations data},
  url = {http://iaaras.ru/library/paper/877/},
  year = {2012}
}

@ARTICLE{1999A&AS..139..545K,
       author = {{Kovalev}, Y.~Y. and {Nizhelsky}, N.~A. and {Kovalev}, Yu. A. and {Berlin}, A.~B. and {Zhekanis}, G.~V. and {Mingaliev}, M.~G. and {Bogdantsov}, A.~V.},
        title = "{Survey of instantaneous 1-22 GHz spectra of 550 compact extragalactic objects with declinations from -30$^{deg}$ to +43$^{deg}$}",
      journal = {\aaps},
         year = 1999,
        month = nov,
       volume = {139},
        pages = {545-554},
          doi = {10.1051/aas:1999406},
archivePrefix = {arXiv},
       eprint = {astro-ph/0408264},
 primaryClass = {astro-ph},
       adsurl = {https://ui.adsabs.harvard.edu/abs/1999A&AS..139..545K}
}

@ARTICLE{1993BSAO...36..107V, 
       author = {{Vlasyuk}, V.~V.},
        title = "{Software for reduction of spectral data obtained with panoramic detectors of the 6 m telescope}",
      journal = {Bulletin of Special Astrophysical Observatory},
         year = 1993,
        month = dec,
       volume = {36},
       number = {3},
        pages = {107-117},
       adsurl = {https://ui.adsabs.harvard.edu/abs/1993BSAO...36..107V}
}

@ARTICLE{1988ApJ...333..646E,
   author = {{Edelson}, R.~A. and {Krolik}, J.~H.},
    title = "{The Discrete Correlation Function: A New Method for Analyzing Unevenly Sampled Variability Data}",
  journal = {\apj},
     year = 1988,
    month = october,
   volume = 333,
    pages = {646},
      doi = {10.1086/166773},
   adsurl = {https://ui.adsabs.harvard.edu/abs/1988ApJ...333..646E/abstract}
}

@ARTICLE{2015MNRAS.453.3455R,
       author = {{Robertson}, D.~R.~S. and {Gallo}, L.~C. and {Zoghbi}, A. and
         {Fabian}, A.~C.},
        title = "{Searching for correlations in simultaneous X-ray and UV emission in the narrow-line Seyfert 1 galaxy 1H 0707-495}",
      journal = {\mnras},
         year = 2015,
        month = nov,
       volume = {453},
       number = {4},
        pages = {3455-3460},
          doi = {10.1093/mnras/stv1575},
archivePrefix = {arXiv},
       eprint = {1507.05201},
 primaryClass = {astro-ph.HE},
       adsurl = {https://ui.adsabs.harvard.edu/abs/2015MNRAS.453.3455R}
}

@ARTICLE{2013MNRAS.433..907E,
       author = {{Emmanoulopoulos}, D. and {McHardy}, I.~M. and {Papadakis}, I.~E.},
        title = "{Generating artificial light curves: revisited and updated}",
      journal = {\mnras},
         year = 2013,
        month = aug,
       volume = {433},
       number = {2},
        pages = {907-927},
          doi = {10.1093/mnras/stt764},
archivePrefix = {arXiv},
       eprint = {1305.0304},
 primaryClass = {astro-ph.IM},
       adsurl = {https://ui.adsabs.harvard.edu/abs/2013MNRAS.433..907E}
}

@ARTICLE{1990A&A...239..443S,
       author = {{Steffen}, M.},
        title = "{A simple method for monotonic interpolation in one dimension.}",
      journal = {\aap},
         year = 1990,
        month = nov,
       volume = {239},
        pages = {443-450},
       adsurl = {https://ui.adsabs.harvard.edu/abs/1990A&A...239..443S}
}

@ARTICLE{1990A&AS...83..183M,
       author = {{Mead}, A.~R.~G. and {Ballard}, K.~R. and {Brand}, P.~W.~J.~L. and {Hough}, J.~H. and {Brindle}, C. and {Bailey}, J.~A.},
        title = "{Optical and infrared polarimetry and photometry of blazars.}",
      journal = {\aaps},
         year = 1990,
        month = apr,
       volume = {83},
        pages = {183-204},
       adsurl = {https://ui.adsabs.harvard.edu/abs/1990A&AS...83..183M}
}

@ARTICLE{2023AstBu..78..105S,
       author = {{Sotnikova}, Yu. V. and {Kovalev}, Yu. A. and {Ermakov}, A.~N. and {Volvach}, L.~N. and {Volvach}, A.~E.},
        title = "{Study of Calibration Sources at 22 and 37 GHz Frequency Bands with RT-22 CrAO RAS}",
      journal = {Astrophysical Bulletin},
         year = 2023,
        month = mar,
       volume = {78},
       number = {1},
        pages = {105-115},
          doi = {10.1134/S1990341323010091},
       adsurl = {https://ui.adsabs.harvard.edu/abs/2023AstBu..78..105S}
}

@ARTICLE{1999ApJ...521..493L,
       author = {{L{\"a}hteenm{\"a}ki}, A. and {Valtaoja}, E.},
        title = "{Total Flux Density Variations in Extragalactic Radio Sources. III. Doppler Boosting Factors, Lorentz Factors, and Viewing Angles for Active Galactic Nuclei}",
      journal = {\apj},
         year = 1999,
        month = aug,
       volume = {521},
       number = {2},
        pages = {493-501},
          doi = {10.1086/307587},
       adsurl = {https://ui.adsabs.harvard.edu/abs/1999ApJ...521..493L}
}

@ARTICLE{2009ApJS..183...46A,
       author = {{Abdo}, A.~A. and {Ackermann}, M. and {Ajello}, M. and {Atwood}, W.~B. and {Axelsson}, M. and {Baldini}, L. and {Ballet}, J. and {Band}, D.~L. and {Barbiellini}, G. and {Bastieri}, D. and {Battelino}, M. and {Baughman}, B.~M. and {Bechtol}, K. and {Bellazzini}, R. and {Berenji}, B. and {Bignami}, G.~F. and {Blandford}, R.~D. and {Bloom}, E.~D. and {Bonamente}, E. and {Borgland}, A.~W. and {Bouvier}, A. and {Bregeon}, J. and {Brez}, A. and {Brigida}, M. and {Bruel}, P. and {Burnett}, T.~H. and {Caliandro}, G.~A. and {Cameron}, R.~A. and {Caraveo}, P.~A. and {Casandjian}, J.~M. and {Cavazzuti}, E. and {Cecchi}, C. and {Charles}, E. and {Chekhtman}, A. and {Cheung}, C.~C. and {Chiang}, J. and {Ciprini}, S. and {Claus}, R. and {Cohen-Tanugi}, J. and {Cominsky}, L.~R. and {Conrad}, J. and {Corbet}, R. and {Costamante}, L. and {Cutini}, S. and {Davis}, D.~S. and {Dermer}, C.~D. and {de Angelis}, A. and {de Luca}, A. and {de Palma}, F. and {Digel}, S.~W. and {Dormody}, M. and {do Couto e Silva}, E. and {Drell}, P.~S. and {Dubois}, R. and {Dumora}, D. and {Farnier}, C. and {Favuzzi}, C. and {Fegan}, S.~J. and {Ferrara}, E.~C. and {Focke}, W.~B. and {Frailis}, M. and {Fukazawa}, Y. and {Funk}, S. and {Fusco}, P. and {Gargano}, F. and {Gasparrini}, D. and {Gehrels}, N. and {Germani}, S. and {Giebels}, B. and {Giglietto}, N. and {Giommi}, P. and {Giordano}, F. and {Glanzman}, T. and {Godfrey}, G. and {Grenier}, I.~A. and {Grondin}, M. -H. and {Grove}, J.~E. and {Guillemot}, L. and {Guiriec}, S. and {Hanabata}, Y. and {Harding}, A.~K. and {Hartman}, R.~C. and {Hayashida}, M. and {Hays}, E. and {Healey}, S.~E. and {Horan}, D. and {Hughes}, R.~E. and {J{\'o}hannesson}, G. and {Johnson}, A.~S. and {Johnson}, R.~P. and {Johnson}, T.~J. and {Johnson}, W.~N. and {Kamae}, T. and {Katagiri}, H. and {Kataoka}, J. and {Kawai}, N. and {Kerr}, M. and {Kn{\"o}dlseder}, J. and {Kocevski}, D. and {Kocian}, M.~L. and {Komin}, N. and {Kuehn}, F. and {Kuss}, M. and {Lande}, J. and {Latronico}, L. and {Lee}, S. -H. and {Lemoine-Goumard}, M. and {Longo}, F. and {Loparco}, F. and {Lott}, B. and {Lovellette}, M.~N. and {Lubrano}, P. and {Madejski}, G.~M. and {Makeev}, A. and {Marelli}, M. and {Mazziotta}, M.~N. and {McConville}, W. and {McEnery}, J.~E. and {McGlynn}, S. and {Meurer}, C. and {Michelson}, P.~F. and {Mitthumsiri}, W. and {Mizuno}, T. and {Moiseev}, A.~A. and {Monte}, C. and {Monzani}, M.~E. and {Moretti}, E. and {Morselli}, A. and {Moskalenko}, I.~V. and {Murgia}, S. and {Nakamori}, T. and {Nolan}, P.~L. and {Norris}, J.~P. and {Nuss}, E. and {Ohno}, M. and {Ohsugi}, T. and {Omodei}, N. and {Orlando}, E. and {Ormes}, J.~F. and {Ozaki}, M. and {Paneque}, D. and {Panetta}, J.~H. and {Parent}, D. and {Pelassa}, V. and {Pepe}, M. and {Pesce-Rollins}, M. and {Piron}, F. and {Porter}, T.~A. and {Poupard}, L. and {Rain{\`o}}, S. and {Rando}, R. and {Ray}, P.~S. and {Razzano}, M. and {Rea}, N. and {Reimer}, A. and {Reimer}, O. and {Reposeur}, T. and {Ritz}, S. and {Rochester}, L.~S. and {Rodriguez}, A.~Y. and {Romani}, R.~W. and {Roth}, M. and {Ryde}, F. and {Sadrozinski}, H.~F. -W. and {Sanchez}, D. and {Sander}, A. and {Saz Parkinson}, P.~M. and {Scargle}, J.~D. and {Schalk}, T.~L. and {Sellerholm}, A. and {Sgr{\`o}}, C. and {Shaw}, M.~S. and {Shrader}, C. and {Sierpowska-Bartosik}, A. and {Siskind}, E.~J. and {Smith}, D.~A. and {Smith}, P.~D. and {Spandre}, G. and {Spinelli}, P. and {Starck}, J. -L. and {Stephens}, T.~E. and {Strickman}, M.~S. and {Strong}, A.~W. and {Suson}, D.~J. and {Tajima}, H. and {Takahashi}, H. and {Takahashi}, T. and {Tanaka}, T. and {Thayer}, J.~B. and {Thayer}, J.~G. and {Thompson}, D.~J. and {Tibaldo}, L. and {Tibolla}, O. and {Torres}, D.~F. and {Tosti}, G. and {Tramacere}, A. and {Uchiyama}, Y. and {Usher}, T.~L. and {Van Etten}, A. and {Vilchez}, N. and {Vitale}, V. and {Waite}, A.~P. and {Wallace}, E. and {Wang}, P. and {Watters}, K. and {Winer}, B.~L. and {Wood}, K.~S. and {Ylinen}, T. and {Ziegler}, M. and {Fermi/LAT Collaboration}},
        title = "{Fermi/Large Area Telescope Bright Gamma-Ray Source List}",
      journal = {\apjs},
         year = 2009,
        month = jul,
       volume = {183},
       number = {1},
        pages = {46-66},
          doi = {10.1088/0067-0049/183/1/46},
archivePrefix = {arXiv},
       eprint = {0902.1340},
 primaryClass = {astro-ph.HE},
       adsurl = {https://ui.adsabs.harvard.edu/abs/2009ApJS..183...46A}
}

@ARTICLE{2018ApJ...866...11D,
       author = {{Dey}, Lankeswar and {Valtonen}, M.~J. and {Gopakumar}, A. and {Zola}, S. and {Hudec}, R. and {Pihajoki}, P. and {Ciprini}, S. and {Matsumoto}, K. and {Sadakane}, K. and {Kidger}, M. and {Nilsson}, K. and {Mikkola}, S. and {Sillanp{\"a}{\"a}}, A. and {Takalo}, L.~O. and {Lehto}, H.~J. and {Berdyugin}, A. and {Piirola}, V. and {Jermak}, H. and {Baliyan}, K.~S. and {Pursimo}, T. and {Caton}, D.~B. and {Alicavus}, F. and {Baransky}, A. and {Blay}, P. and {Boumis}, P. and {Boyd}, D. and {Campas Torrent}, M. and {Campos}, F. and {Carrillo G{\'o}mez}, J. and {Chandra}, S. and {Chavushyan}, V. and {Dalessio}, J. and {Debski}, B. and {Drozdz}, M. and {Er}, H. and {Erdem}, A. and {Escartin P{\'e}rez}, A. and {Fallah Ramazani}, V. and {Filippenko}, A.~V. and {Gafton}, E. and {Ganesh}, S. and {Garcia}, F. and {Gazeas}, K. and {Godunova}, V. and {G{\'o}mez Pinilla}, F. and {Gopinathan}, M. and {Haislip}, J.~B. and {Harmanen}, J. and {Hurst}, G. and {Jan{\'\i}k}, J. and {Jelinek}, M. and {Joshi}, A. and {Kagitani}, M. and {Karjalainen}, R. and {Kaur}, N. and {Keel}, W.~C. and {Kouprianov}, V.~V. and {Kundera}, T. and {Kurowski}, S. and {Kvammen}, A. and {LaCluyze}, A.~P. and {Lee}, B.~C. and {Liakos}, A. and {Lindfors}, E. and {Lozano de Haro}, J. and {Mugrauer}, M. and {Naves Nogues}, R. and {Neely}, A.~W. and {Nelson}, R.~H. and {Ogloza}, W. and {Okano}, S. and {Pajdosz-{\'S}mierciak}, U. and {Pandey}, J.~C. and {Perri}, M. and {Poyner}, G. and {Provencal}, J. and {Raj}, A. and {Reichart}, D.~E. and {Reinthal}, R. and {Reynolds}, T. and {Saario}, J. and {Sadegi}, S. and {Sakanoi}, T. and {Salto Gonz{\'a}lez}, J. -L. and {Sameer} and {Schweyer}, T. and {Simon}, A. and {Siwak}, M. and {Sold{\'a}n Alfaro}, F.~C. and {Sonbas}, E. and {Steele}, I. and {Stocke}, J.~T. and {Strobl}, J. and {Tomov}, T. and {Tremosa Espasa}, L. and {Valdes}, J.~R. and {Valero P{\'e}rez}, J. and {Verrecchia}, F. and {Vasylenko}, V. and {Webb}, J.~R. and {Yoneda}, M. and {Zejmo}, M. and {Zheng}, W. and {Zielinski}, P.},
        title = "{Authenticating the Presence of a Relativistic Massive Black Hole Binary in OJ 287 Using Its General Relativity Centenary Flare: Improved Orbital Parameters}",
      journal = {\apj},
         year = 2018,
        month = oct,
       volume = {866},
       number = {1},
          eid = {11},
        pages = {11},
          doi = {10.3847/1538-4357/aadd95},
archivePrefix = {arXiv},
       eprint = {1808.09309},
 primaryClass = {astro-ph.HE},
       adsurl = {https://ui.adsabs.harvard.edu/abs/2018ApJ...866...11D}
}

@ARTICLE{2019A&A...621A..11Q,
       author = {{Qian}, S.~J. and {Britzen}, S. and {Krichbaum}, T.~P. and {Witzel}, A.},
        title = "{Possible evidence of a supermassive black hole binary with two radio jets in blazar 3C279}",
      journal = {\aap},
         year = 2019,
        month = jan,
       volume = {621},
          eid = {A11},
        pages = {A11},
          doi = {10.1051/0004-6361/201833508},
       adsurl = {https://ui.adsabs.harvard.edu/abs/2019A&A...621A..11Q}
}

@ARTICLE{1979ApJ...232...34B,
       author = {{Blandford}, R.~D. and {K{\"o}nigl}, A.},
        title = "{Relativistic jets as compact radio sources.}",
      journal = {\apj},
         year = 1979,
        month = aug,
       volume = {232},
        pages = {34-48},
          doi = {10.1086/157262},
       adsurl = {https://ui.adsabs.harvard.edu/abs/1979ApJ...232...34B}
}

@ARTICLE{2010A&A...512A..24S,
       author = {{Savolainen}, T. and {Homan}, D.~C. and {Hovatta}, T. and {Kadler}, M. and {Kovalev}, Y.~Y. and {Lister}, M.~L. and {Ros}, E. and {Zensus}, J.~A.},
        title = "{Relativistic beaming and gamma-ray brightness of blazars}",
      journal = {\aap},
         year = 2010,
        month = mar,
       volume = {512},
          eid = {A24},
        pages = {A24},
          doi = {10.1051/0004-6361/200913740},
archivePrefix = {arXiv},
       eprint = {0911.4924},
 primaryClass = {astro-ph.CO},
       adsurl = {https://ui.adsabs.harvard.edu/abs/2010A&A...512A..24S}
}

@ARTICLE{2023ApJS..265...31A,
       author = {{Abdollahi}, S. and {Ajello}, M. and {Baldini}, L. and {Ballet}, J. and {Bastieri}, D. and {Becerra Gonzalez}, J. and {Bellazzini}, R. and {Berretta}, A. and {Bissaldi}, E. and {Bonino}, R. and {Brill}, A. and {Bruel}, P. and {Burns}, E. and {Buson}, S. and {Cameron}, R.~A. and {Caputo}, R. and {Caraveo}, P.~A. and {Cibrario}, N. and {Ciprini}, S. and {Cristarella Orestano}, P. and {Crnogorcevic}, M. and {Cutini}, S. and {D'Ammando}, F. and {De Gaetano}, S. and {Digel}, S.~W. and {Di Lalla}, N. and {Di Venere}, L. and {Dom{\'\i}nguez}, A. and {Ramazani}, V. Fallah and {Fegan}, S.~J. and {Ferrara}, E.~C. and {Fiori}, A. and {Fleischhack}, H. and {Franckowiak}, A. and {Fukazawa}, Y. and {Fusco}, P. and {Gammaldi}, V. and {Gargano}, F. and {Garrappa}, S. and {Gasbarra}, C. and {Gasparrini}, D. and {Giglietto}, N. and {Giordano}, F. and {Giroletti}, M. and {Green}, D. and {Grenier}, I.~A. and {Guiriec}, S. and {Gustafsson}, M. and {Hays}, E. and {Horan}, D. and {Hou}, X. and {J{\'o}hannesson}, G. and {Kerr}, M. and {Kocevski}, D. and {Kuss}, M. and {Latronico}, L. and {Li}, J. and {Liodakis}, I. and {Longo}, F. and {Loparco}, F. and {Lorusso}, L. and {Lott}, B. and {Lovellette}, M.~N. and {Lubrano}, P. and {Maldera}, S. and {Manfreda}, A. and {Mart{\'\i}-Devesa}, G. and {Mazziotta}, M.~N. and {Mereu}, I. and {Meyer}, M. and {Michelson}, P.~F. and {Mizuno}, T. and {Monzani}, M.~E. and {Morselli}, A. and {Moskalenko}, I.~V. and {Negro}, M. and {Omodei}, N. and {Orlando}, E. and {Ormes}, J.~F. and {Paneque}, D. and {Panzarini}, G. and {Perkins}, J.~S. and {Persic}, M. and {Pesce-Rollins}, M. and {Pillera}, R. and {Porter}, T.~A. and {Principe}, G. and {Racusin}, J.~L. and {Rain{\`o}}, S. and {Rando}, R. and {Rani}, B. and {Razzano}, M. and {Razzaque}, S. and {Reimer}, A. and {Reimer}, O. and {S{\'a}nchez-Conde}, M. and {Parkinson}, P.~M. Saz and {Scargle}, Jeff and {Scotton}, L. and {Serini}, D. and {Sgr{\`o}}, C. and {Siskind}, E.~J. and {Spandre}, G. and {Spinelli}, P. and {Suson}, D.~J. and {Tajima}, H. and {Thompson}, D.~J. and {Torres}, D.~F. and {Valverde}, J. and {Venters}, T. and {Wadiasingh}, Z. and {Wagner}, S. and {Wood}, K.},
        title = "{The Fermi-LAT Lightcurve Repository}",
      journal = {\apjs},
         year = 2023,
        month = apr,
       volume = {265},
       number = {2},
          eid = {31},
        pages = {31},
          doi = {10.3847/1538-4365/acbb6a},
archivePrefix = {arXiv},
       eprint = {2301.01607},
 primaryClass = {astro-ph.HE},
       adsurl = {https://ui.adsabs.harvard.edu/abs/2023ApJS..265...31A}
}

@ARTICLE{2002A&A...390..407V,
       author = {{Villata}, M. and {Raiteri}, C.~M. and {Kurtanidze}, O.~M. and {Nikolashvili}, M.~G. and {Ibrahimov}, M.~A. and {Papadakis}, I.~E. and {Tsinganos}, K. and {Sadakane}, K. and {Okada}, N. and {Takalo}, L.~O. and {Sillanp{\"a}{\"a}}, A. and {Tosti}, G. and {Ciprini}, S. and {Frasca}, A. and {Marilli}, E. and {Robb}, R.~M. and {Noble}, J.~C. and {Jorstad}, S.~G. and {Hagen-Thorn}, V.~A. and {Larionov}, V.~M. and {Nesci}, R. and {Maesano}, M. and {Schwartz}, R.~D. and {Basler}, J. and {Gorham}, P.~W. and {Iwamatsu}, H. and {Kato}, T. and {Pullen}, C. and {Ben{\'\i}tez}, E. and {de Diego}, J.~A. and {Moilanen}, M. and {Oksanen}, A. and {Rodriguez}, D. and {Sadun}, A.~C. and {Kelly}, M. and {Carini}, M.~T. and {Miller}, H.~R. and {Catalano}, S. and {Dultzin-Hacyan}, D. and {Fan}, J.~H. and {Ishioka}, R. and {Karttunen}, H. and {Kein{\"a}nen}, P. and {Kudryavtseva}, N.~A. and {Lainela}, M. and {Lanteri}, L. and {Larionova}, E.~G. and {Matsumoto}, K. and {Mattox}, J.~R. and {Montagni}, F. and {Nucciarelli}, G. and {Ostorero}, L. and {Papamastorakis}, J. and {Pasanen}, M. and {Sobrito}, G. and {Uemura}, M.},
        title = "{The WEBT <ASTROBJ>BL Lacertae</ASTROBJ> Campaign 2000}",
      journal = {\aap},
         year = 2002,
        month = aug,
       volume = {390},
        pages = {407-421},
          doi = {10.1051/0004-6361:20020662},
archivePrefix = {arXiv},
       eprint = {astro-ph/0205479},
 primaryClass = {astro-ph},
       adsurl = {https://ui.adsabs.harvard.edu/abs/2002A&A...390..407V}
}

@ARTICLE{2014Ap&SS.351..281W,
       author = {{Wang}, Hongtao},
        title = "{The optical periodic analysis of BL Lac object AO 0235+164}",
      journal = {\apss},
         year = 2014,
        month = may,
       volume = {351},
       number = {1},
        pages = {281-287},
          doi = {10.1007/s10509-014-1840-z},
       adsurl = {https://ui.adsabs.harvard.edu/abs/2014Ap&SS.351..281W}
}

@ARTICLE{2022MNRAS.513.5238R,
       author = {{Roy}, Abhradeep and {Chitnis}, Varsha R. and {Gupta}, Alok C. and {Wiita}, Paul J. and {Romero}, Gustavo E. and {Cellone}, Sergio A. and {Chatterjee}, Anshu and {Combi}, Jorge A. and {Raiteri}, Claudia M. and {Sarkar}, Arkadipta and {Villata}, Massimo},
        title = "{Detection of a quasi-periodic oscillation in the optical light curve of the remarkable blazar AO 0235+164}",
      journal = {\mnras},
         year = 2022,
        month = jul,
       volume = {513},
       number = {4},
        pages = {5238-5244},
          doi = {10.1093/mnras/stac1287},
archivePrefix = {arXiv},
       eprint = {2205.03586},
 primaryClass = {astro-ph.HE},
       adsurl = {https://ui.adsabs.harvard.edu/abs/2022MNRAS.513.5238R}
}

@article{2019..VLBI..Quasar,
author = {Shuygina, N. and Ivanov, Dmitrii and Ipatov, A. and Gayazov, I. and Marshalov, D. and Melnikov, Alexey and Kurdubov, Sergei and Vasilyev, Mikhail and Ilin, G. and Skurikhina, E. and Surkis, I. and Mardyshkin, V. and Mikhailov, A. and Salnikov, A. and Vytnov, A. and Rakhimov, I. and Dyakov, A. and Olifirov, V.},
year = {2019},
month = {3},
pages = {150-156},
title = {Russian VLBI network “Quasar”: Current status and outlook},
volume = {10},
journal = {Geodesy and Geodynamics},
doi = {10.1016/j.geog.2018.09.008}
}

@ARTICLE{2023AstBu..78..464V,
       author = {{Vlasyuk}, V.~V. and {Sotnikova}, Yu. V. and {Volvach}, A.~E. and {Spiridonova}, O.~I. and {Stolyarov}, V.~A. and {Mikhailov}, A.~G. and {Kovalev}, Yu. A. and {Kovalev}, Y.~Y. and {Khabibullina}, M.~L. and {Kharinov}, M.~A. and {Yang}, L. and {Mingaliev}, M.~G. and {Semenova}, T.~A. and {Zhekanis}, P.~G. and {Mufakharov}, T.~V. and {Udovitskiy}, R. Yu. and {Kudryashova}, A.~A. and {Volvach}, L.~N. and {Erkenov}, A.~K. and {Moskvitin}, A.~S. and {Emelianov}, E.~V. and {Fatkhullin}, T.~A. and {Tsybulev}, P.~G. and {Nizhelsky}, N.~A. and {Zhekanis}, G.~V. and {Kravchenko}, E.~V.},
        title = "{Optical and Radio Variability of the Blazar S4 0954+658}",
      journal = {Astrophysical Bulletin},
         year = 2023,
        month = dec,
       volume = {78},
       number = {4},
        pages = {464-486},
          doi = {10.1134/S1990341323600229},
archivePrefix = {arXiv},
       eprint = {2401.03992},
 primaryClass = {astro-ph.HE},
       adsurl = {https://ui.adsabs.harvard.edu/abs/2023AstBu..78..464V}
}

@ARTICLE{2015arXiv150306676C,
       author = {{Connolly}, S D},
        title = "{A Python Code for the Emmanoulopoulos et al. [arXiv:1305.0304] Light Curve Simulation Algorithm}",
      journal = {arXiv e-prints},
         year = 2015,
        month = mar,
          eid = {arXiv:1503.06676},
        pages = {arXiv:1503.06676},
          doi = {10.48550/arXiv.1503.06676},
archivePrefix = {arXiv},
       eprint = {1503.06676},
 primaryClass = {astro-ph.IM},
       adsurl = {https://ui.adsabs.harvard.edu/abs/2015arXiv150306676C}
}

@ARTICLE{2019PASP..131a8002B,
       author = {{Bellm}, Eric C. and {Kulkarni}, Shrinivas R. and {Graham}, Matthew J. and {Dekany}, Richard and {Smith}, Roger M. and {Riddle}, Reed and {Masci}, Frank J. and {Helou}, George and {Prince}, Thomas A. and {Adams}, Scott M. and {Barbarino}, C. and {Barlow}, Tom and {Bauer}, James and {Beck}, Ron and {Belicki}, Justin and {Biswas}, Rahul and {Blagorodnova}, Nadejda and {Bodewits}, Dennis and {Bolin}, Bryce and {Brinnel}, Valery and {Brooke}, Tim and {Bue}, Brian and {Bulla}, Mattia and {Burruss}, Rick and {Cenko}, S. Bradley and {Chang}, Chan-Kao and {Connolly}, Andrew and {Coughlin}, Michael and {Cromer}, John and {Cunningham}, Virginia and {De}, Kishalay and {Delacroix}, Alex and {Desai}, Vandana and {Duev}, Dmitry A. and {Eadie}, Gwendolyn and {Farnham}, Tony L. and {Feeney}, Michael and {Feindt}, Ulrich and {Flynn}, David and {Franckowiak}, Anna and {Frederick}, S. and {Fremling}, C. and {Gal-Yam}, Avishay and {Gezari}, Suvi and {Giomi}, Matteo and {Goldstein}, Daniel A. and {Golkhou}, V. Zach and {Goobar}, Ariel and {Groom}, Steven and {Hacopians}, Eugean and {Hale}, David and {Henning}, John and {Ho}, Anna Y.~Q. and {Hover}, David and {Howell}, Justin and {Hung}, Tiara and {Huppenkothen}, Daniela and {Imel}, David and {Ip}, Wing-Huen and {Ivezi{\'c}}, {\v{Z}}eljko and {Jackson}, Edward and {Jones}, Lynne and {Juric}, Mario and {Kasliwal}, Mansi M. and {Kaspi}, S. and {Kaye}, Stephen and {Kelley}, Michael S.~P. and {Kowalski}, Marek and {Kramer}, Emily and {Kupfer}, Thomas and {Landry}, Walter and {Laher}, Russ R. and {Lee}, Chien-De and {Lin}, Hsing Wen and {Lin}, Zhong-Yi and {Lunnan}, Ragnhild and {Giomi}, Matteo and {Mahabal}, Ashish and {Mao}, Peter and {Miller}, Adam A. and {Monkewitz}, Serge and {Murphy}, Patrick and {Ngeow}, Chow-Choong and {Nordin}, Jakob and {Nugent}, Peter and {Ofek}, Eran and {Patterson}, Maria T. and {Penprase}, Bryan and {Porter}, Michael and {Rauch}, Ludwig and {Rebbapragada}, Umaa and {Reiley}, Dan and {Rigault}, Mickael and {Rodriguez}, Hector and {van Roestel}, Jan and {Rusholme}, Ben and {van Santen}, Jakob and {Schulze}, S. and {Shupe}, David L. and {Singer}, Leo P. and {Soumagnac}, Maayane T. and {Stein}, Robert and {Surace}, Jason and {Sollerman}, Jesper and {Szkody}, Paula and {Taddia}, F. and {Terek}, Scott and {Van Sistine}, Angela and {van Velzen}, Sjoert and {Vestrand}, W. Thomas and {Walters}, Richard and {Ward}, Charlotte and {Ye}, Quan-Zhi and {Yu}, Po-Chieh and {Yan}, Lin and {Zolkower}, Jeffry},
        title = "{The Zwicky Transient Facility: System Overview, Performance, and First Results}",
      journal = {\pasp},
         year = 2019,
        month = jan,
       volume = {131},
       number = {995},
        pages = {018002},
          doi = {10.1088/1538-3873/aaecbe},
archivePrefix = {arXiv},
       eprint = {1902.01932},
 primaryClass = {astro-ph.IM},
       adsurl = {https://ui.adsabs.harvard.edu/abs/2019PASP..131a8002B}
}

@ARTICLE{2017ATel11075....1M,
       author = {{Mukherjee}, Reshmi and {VERITAS Collaboration}},
        title = "{VERITAS Detection of VHE Emission from Ton 599}",
      journal = {The Astronomer's Telegram},
         year = 2017,
        month = dec,
       volume = {11075},
        pages = {1},
       adsurl = {https://ui.adsabs.harvard.edu/abs/2017ATel11075....1M}
}

@ARTICLE{2004ApJ...613..725S,
       author = {{Sokolov}, Andrei and {Marscher}, Alan P. and {McHardy}, Ian M.},
        title = "{Synchrotron Self-Compton Model for Rapid Nonthermal Flares in Blazars with Frequency-dependent Time Lags}",
      journal = {\apj},
         year = 2004,
        month = oct,
       volume = {613},
       number = {2},
        pages = {725-746},
          doi = {10.1086/423165},
archivePrefix = {arXiv},
       eprint = {astro-ph/0406235},
 primaryClass = {astro-ph},
       adsurl = {https://ui.adsabs.harvard.edu/abs/2004ApJ...613..725S}
}

@ARTICLE{2001MNRAS.325.1559S,
       author = {{Spada}, Maddalena and {Ghisellini}, Gabriele and {Lazzati}, Davide and {Celotti}, Annalisa},
        title = "{Internal shocks in the jets of radio-loud quasars}",
      journal = {\mnras},
         year = 2001,
        month = aug,
       volume = {325},
       number = {4},
        pages = {1559-1570},
          doi = {10.1046/j.1365-8711.2001.04557.x},
archivePrefix = {arXiv},
       eprint = {astro-ph/0103424},
 primaryClass = {astro-ph},
       adsurl = {https://ui.adsabs.harvard.edu/abs/2001MNRAS.325.1559S}
}

@ARTICLE{2013ApJ...768...40L,
       author = {{Larionov}, V.~M. and {Jorstad}, S.~G. and {Marscher}, A.~P. and {Morozova}, D.~A. and {Blinov}, D.~A. and {Hagen-Thorn}, V.~A. and {Konstantinova}, T.~S. and {Kopatskaya}, E.~N. and {Larionova}, L.~V. and {Larionova}, E.~G. and {Troitsky}, I.~S.},
        title = "{The Outburst of the Blazar S5 0716+71 in 2011 October: Shock in a Helical Jet}",
      journal = {\apj},
         year = 2013,
        month = may,
       volume = {768},
       number = {1},
          eid = {40},
        pages = {40},
          doi = {10.1088/0004-637X/768/1/40},
archivePrefix = {arXiv},
       eprint = {1303.2218},
 primaryClass = {astro-ph.HE},
       adsurl = {https://ui.adsabs.harvard.edu/abs/2013ApJ...768...40L}
}

@INPROCEEDINGS{2007ASPC..375..234G,
       author = {{Gurwell}, M.~A. and {Peck}, A.~B. and {Hostler}, S.~R. and {Darrah}, M.~R. and {Katz}, C.~A.},
        title = "{Monitoring Phase Calibrators at Submillimeter Wavelengths}",
    booktitle = {From Z-Machines to ALMA: (Sub)Millimeter Spectroscopy of Galaxies},
         year = 2007,
       editor = {{Baker}, A.~J. and {Glenn}, J. and {Harris}, A.~I. and {Mangum}, J.~G. and {Yun}, M.~S.},
       series = {Astronomical Society of the Pacific Conference Series},
       volume = {375},
        month = oct,
        pages = {234},
       adsurl = {https://ui.adsabs.harvard.edu/abs/2007ASPC..375..234G}
}

@ARTICLE{2022ApJ...926..180H,
       author = {{Hallum}, Melissa K. and {Jorstad}, Svetlana G. and {Larionov}, Valeri M. and {Marscher}, Alan P. and {Joshi}, Manasvita and {Weaver}, Zachary R. and {Williamson}, Karen E. and {Agudo}, Iv{\'a}n and {Borman}, George A. and {Casadio}, Carolina and {Fuentes}, Antonio and {Grishina}, Tatiana S. and {Kopatskaya}, Evgenia N. and {Larionova}, Elena G. and {Larionova}, Liyudmila V. and {Morozova}, Daria A. and {Nikiforova}, Anna A. and {Savchenko}, Sergey S. and {Troitsky}, Ivan S. and {Troitskaya}, Yulia V. and {Vasilyev}, Andrey A.},
        title = "{Emission-line Variability during a Nonthermal Outburst in the Gamma-Ray Bright Quasar 1156+295}",
      journal = {\apj},
         year = 2022,
        month = feb,
       volume = {926},
       number = {2},
          eid = {180},
        pages = {180},
          doi = {10.3847/1538-4357/ac4710},
archivePrefix = {arXiv},
       eprint = {2202.00061},
 primaryClass = {astro-ph.HE},
       adsurl = {https://ui.adsabs.harvard.edu/abs/2022ApJ...926..180H}
}

@ARTICLE{2009A&A...507L..33P,
       author = {{Pushkarev}, A.~B. and {Kovalev}, Y.~Y. and {Lister}, M.~L. and {Savolainen}, T.},
        title = "{Jet opening angles and gamma-ray brightness of AGN}",
      journal = {\aap},
         year = 2009,
        month = nov,
       volume = {507},
       number = {2},
        pages = {L33-L36},
          doi = {10.1051/0004-6361/200913422},
archivePrefix = {arXiv},
       eprint = {0910.1813},
 primaryClass = {astro-ph.CO},
       adsurl = {https://ui.adsabs.harvard.edu/abs/2009A&A...507L..33P}
}

@ARTICLE{2024MNRAS.52711900R,
       author = {{Rajput}, Bhoomika and {Mandal}, Amit Kumar and {Pandey}, Ashwani and {Stalin}, C.~S. and {Max-Moerbeck}, Walter and {Mathew}, Blesson},
        title = "{Investigation of the correlation between optical and {\ensuremath{\gamma}}-ray flux variations in the blazar Ton 599}",
      journal = {\mnras},
         year = 2024,
        month = feb,
       volume = {527},
       number = {4},
        pages = {11900-11914},
          doi = {10.1093/mnras/stad4003},
archivePrefix = {arXiv},
       eprint = {2312.14857},
 primaryClass = {astro-ph.HE},
       adsurl = {https://ui.adsabs.harvard.edu/abs/2024MNRAS.52711900R}
}

@ARTICLE{2011A&A...529A.113Z,
       author = {{Zhao}, W. and {Hong}, X. -Y. and {An}, T. and {Jiang}, D. -R. and {Zhao}, J. -H. and {Gurvits}, L.~I. and {Yang}, J.},
        title = "{Radio structure of the blazar 1156 + 295 with sub-pc resolution}",
      journal = {\aap},
         year = 2011,
        month = may,
       volume = {529},
          eid = {A113},
        pages = {A113},
          doi = {10.1051/0004-6361/201016192},
archivePrefix = {arXiv},
       eprint = {1102.3046},
 primaryClass = {astro-ph.CO},
       adsurl = {https://ui.adsabs.harvard.edu/abs/2011A&A...529A.113Z}
}

@ARTICLE{2014MNRAS.445.1636R,
       author = {{Ramakrishnan}, Venkatessh and {Le{\'o}n-Tavares}, Jonathan and {Rastorgueva-Foi}, Elizaveta A. and {Wiik}, Kaj and {Jorstad}, Svetlana G. and {Marscher}, Alan P. and {Tornikoski}, Merja and {Agudo}, Iv{\'a}n and {L{\"a}hteenm{\"a}ki}, Anne and {Valtaoja}, Esko and {Aller}, Margo F. and {Blinov}, Dmitry A. and {Casadio}, Carolina and {Efimova}, Natalia V. and {Gurwell}, Mark A. and {G{\'o}mez}, Jos{\'e} L. and {Hagen-Thorn}, Vladimir A. and {Joshi}, Manasvita and {J{\"a}rvel{\"a}}, Emilia and {Konstantinova}, Tatiana S. and {Kopatskaya}, Evgenia N. and {Larionov}, Valeri M. and {Larionova}, Elena G. and {Larionova}, Liudmilla V. and {Lavonen}, Niko and {MacDonald}, Nicholas R. and {McHardy}, Ian M. and {Molina}, Sol N. and {Morozova}, Daria A. and {Nieppola}, Elina and {Tammi}, Joni and {Taylor}, Brian W. and {Troitsky}, Ivan S.},
        title = "{The connection between the parsec-scale radio jet and {\ensuremath{\gamma}}-ray flares in the blazar 1156+295}",
      journal = {\mnras},
         year = 2014,
        month = dec,
       volume = {445},
       number = {2},
        pages = {1636-1646},
          doi = {10.1093/mnras/stu1873},
archivePrefix = {arXiv},
       eprint = {1409.2542},
 primaryClass = {astro-ph.HE},
       adsurl = {https://ui.adsabs.harvard.edu/abs/2014MNRAS.445.1636R}
}

@ARTICLE{2014Ap&SS.352..215L,
       author = {{Liu}, Baorong and {Liu}, Xiang},
        title = "{Periodic oscillation in the long-term radio light curves of Quasar 1156+295}",
      journal = {\apss},
         year = 2014,
        month = jul,
       volume = {352},
       number = {1},
        pages = {215-220},
          doi = {10.1007/s10509-014-1894-y},
       adsurl = {https://ui.adsabs.harvard.edu/abs/2014Ap&SS.352..215L}
}

@ARTICLE{1976Ap&SS..39..447L,
       author = {{Lomb}, N.~R.},
        title = "{Least-Squares Frequency Analysis of Unequally Spaced Data}",
      journal = {\apss},
         year = 1976,
        month = feb,
       volume = {39},
       number = {2},
        pages = {447-462},
          doi = {10.1007/BF00648343},
       adsurl = {https://ui.adsabs.harvard.edu/abs/1976Ap&SS..39..447L}
}

@ARTICLE{1982ApJ...263..835S,
       author = {{Scargle}, J.~D.},
        title = "{Studies in astronomical time series analysis. II. Statistical aspects of spectral analysis of unevenly spaced data.}",
      journal = {\apj},
         year = 1982,
        month = dec,
       volume = {263},
        pages = {835-853},
          doi = {10.1086/160554},
       adsurl = {https://ui.adsabs.harvard.edu/abs/1982ApJ...263..835S}
}

@BOOK{2014sdmm.book.....I,
       author = {{Ivezi{\'c}}, {\v{Z}}eljko and {Connolly}, Andrew J. and {VanderPlas}, Jacob T. and {Gray}, Alexander},
        title = "{Statistics, Data Mining, and Machine Learning in Astronomy: A Practical Python Guide for the Analysis of Survey Data}",
          year = 2014,
          doi = {10.1515/9781400848911},
       adsurl = {https://ui.adsabs.harvard.edu/abs/2014sdmm.book.....I}
}

@ARTICLE{2006ApJ...637..669L,
       author = {{Liu}, Yi and {Jiang}, Dong Rong and {Gu}, Min Feng},
        title = "{The Jet Power, Radio Loudness, and Black Hole Mass in Radio-loud Active Galactic Nuclei}",
      journal = {\apj},
         year = 2006,
        month = feb,
       volume = {637},
       number = {2},
        pages = {669-681},
          doi = {10.1086/498639},
archivePrefix = {arXiv},
       eprint = {astro-ph/0510241},
 primaryClass = {astro-ph},
       adsurl = {https://ui.adsabs.harvard.edu/abs/2006ApJ...637..669L}
}

@ARTICLE{2002PASJ...54..159Z,
       author = {{Zhang}, Li Zhang and {Fan}, Jui-Hui and {Cheng}, Kwong-Sang},
        title = "{The Multiwavelength Doppler Factors for a Sample of Gamma-Ray Loud Blazars}",
      journal = {\pasj},
         year = 2002,
        month = apr,
       volume = {54},
       number = {2},
        pages = {159-169},
          doi = {10.1093/pasj/54.2.159},
       adsurl = {https://ui.adsabs.harvard.edu/abs/2002PASJ...54..159Z}
}

@ARTICLE{1996AJ....112.1709F,
       author = {{Foster}, Grant},
        title = "{Wavelets for period analysis of unevenly sampled time series}",
      journal = {\aj},
         year = 1996,
        month = oct,
       volume = {112},
        pages = {1709-1729},
          doi = {10.1086/118137},
       adsurl = {https://ui.adsabs.harvard.edu/abs/1996AJ....112.1709F}
}

@ARTICLE{1999A&A...344...61A,
       author = {{Abraham}, Z. and {Romero}, G.~E.},
        title = "{Beaming and precession in the inner jet of 3C 273}",
      journal = {\aap},
         year = 1999,
        month = apr,
       volume = {344},
        pages = {61-67},
       adsurl = {https://ui.adsabs.harvard.edu/abs/1999A&A...344...61A}
}

@ARTICLE{1992A&A...255...59C,
       author = {{Camenzind}, M. and {Krockenberger}, M.},
        title = "{The lighthouse effect of relativistic jets in blazars. A geometric originof intraday variability.}",
      journal = {\aap},
         year = 1992,
        month = feb,
       volume = {255},
        pages = {59-62},
       adsurl = {https://ui.adsabs.harvard.edu/abs/1992A&A...255...59C}
}

@ARTICLE{2021ARep...65.1233H,
       author = {{Hagen-Thorn}, V.~A. and {Morozova}, D.~A. and {Savchenko}, S.~S. and {Hagen-Thorn}, E.~I. and {Milanova}, Yu. V. and {Shalyapina}, L.~V. and {Vasil'ev}, A.~A.},
        title = "{Variability of the Blazar 1156+295 in 2005-2020}",
      journal = {Astronomy Reports},
         year = 2021,
        month = dec,
       volume = {65},
       number = {12},
        pages = {1233-1245},
          doi = {10.1134/S1063772921120039},
       adsurl = {https://ui.adsabs.harvard.edu/abs/2021ARep...65.1233H}
}

@ARTICLE{2022ApJ...926L..35O,
       author = {{O'Neill}, S. and {Kiehlmann}, S. and {Readhead}, A.~C.~S. and {Aller}, M.~F. and {Blandford}, R.~D. and {Liodakis}, I. and {Lister}, M.~L. and {Mr{\'o}z}, P. and {O'Dea}, C.~P. and {Pearson}, T.~J. and {Ravi}, V. and {Vallisneri}, M. and {Cleary}, K.~A. and {Graham}, M.~J. and {Grainge}, K.~J.~B. and {Hodges}, M.~W. and {Hovatta}, T. and {L{\"a}hteenm{\"a}ki}, A. and {Lamb}, J.~W. and {Lazio}, T.~J.~W. and {Max-Moerbeck}, W. and {Pavlidou}, V. and {Prince}, T.~A. and {Reeves}, R.~A. and {Tornikoski}, M. and {Vergara de la Parra}, P. and {Zensus}, J.~A.},
        title = "{The Unanticipated Phenomenology of the Blazar PKS 2131-021: A Unique Supermassive Black Hole Binary Candidate}",
      journal = {\apjl},
         year = 2022,
        month = feb,
       volume = {926},
       number = {2},
          eid = {L35},
        pages = {L35},
          doi = {10.3847/2041-8213/ac504b},
archivePrefix = {arXiv},
       eprint = {2111.02436},
 primaryClass = {astro-ph.HE},
       adsurl = {https://ui.adsabs.harvard.edu/abs/2022ApJ...926L..35O}
}

@ARTICLE{2005A&A...431..391V,
       author = {{Vaughan}, S.},
        title = "{A simple test for periodic signals in red noise}",
      journal = {\aap},
         year = 2005,
        month = feb,
       volume = {431},
        pages = {391-403},
          doi = {10.1051/0004-6361:20041453},
archivePrefix = {arXiv},
       eprint = {astro-ph/0412697},
 primaryClass = {astro-ph},
       adsurl = {https://ui.adsabs.harvard.edu/abs/2005A&A...431..391V}
}

@ARTICLE{2010ApJ...719L.153R,
       author = {{Rani}, Bindu and {Gupta}, Alok C. and {Joshi}, U.~C. and {Ganesh}, S. and {Wiita}, Paul J.},
        title = "{Quasi-periodic Oscillations of \raisebox{-0.5ex}\textasciitilde15 Minutes in the Optical Light Curve of the BL Lac S5 0716+714}",
      journal = {\apjl},
         year = 2010,
        month = aug,
       volume = {719},
       number = {2},
        pages = {L153-L157},
          doi = {10.1088/2041-8205/719/2/L153},
archivePrefix = {arXiv},
       eprint = {1007.2973},
 primaryClass = {astro-ph.CO},
       adsurl = {https://ui.adsabs.harvard.edu/abs/2010ApJ...719L.153R}
}

@ARTICLE{2007ApJ...664L..71A,
       author = {{Aharonian}, F. and {Akhperjanian}, A.~G. and {Bazer-Bachi}, A.~R. and {Behera}, B. and {Beilicke}, M. and {Benbow}, W. and {Berge}, D. and {Bernl{\"o}hr}, K. and {Boisson}, C. and {Bolz}, O. and {Borrel}, V. and {Boutelier}, T. and {Braun}, I. and {Brion}, E. and {Brown}, A.~M. and {B{\"u}hler}, R. and {B{\"u}sching}, I. and {Bulik}, T. and {Carrigan}, S. and {Chadwick}, P.~M. and {Clapson}, A.~C. and {Chounet}, L. -M. and {Coignet}, G. and {Cornils}, R. and {Costamante}, L. and {Degrange}, B. and {Dickinson}, H.~J. and {Djannati-Ata{\"\i}}, A. and {Domainko}, W. and {Drury}, L. O'C. and {Dubus}, G. and {Dyks}, J. and {Egberts}, K. and {Emmanoulopoulos}, D. and {Espigat}, P. and {Farnier}, C. and {Feinstein}, F. and {Fiasson}, A. and {F{\"o}rster}, A. and {Fontaine}, G. and {Funk}, Seb. and {Funk}, S. and {F{\"u}{\ss}ling}, M. and {Gallant}, Y.~A. and {Giebels}, B. and {Glicenstein}, J.~F. and {Gl{\"u}ck}, B. and {Goret}, P. and {Hadjichristidis}, C. and {Hauser}, D. and {Hauser}, M. and {Heinzelmann}, G. and {Henri}, G. and {Hermann}, G. and {Hinton}, J.~A. and {Hoffmann}, A. and {Hofmann}, W. and {Holleran}, M. and {Hoppe}, S. and {Horns}, D. and {Jacholkowska}, A. and {de Jager}, O.~C. and {Kendziorra}, E. and {Kerschhaggl}, M. and {Kh{\'e}lifi}, B. and {Komin}, Nu. and {Kosack}, K. and {Lamanna}, G. and {Latham}, I.~J. and {Le Gallou}, R. and {Lemi{\`e}re}, A. and {Lemoine-Goumard}, M. and {Lenain}, J. -P. and {Lohse}, T. and {Martin}, J.~M. and {Martineau-Huynh}, O. and {Marcowith}, A. and {Masterson}, C. and {Maurin}, G. and {McComb}, T.~J.~L. and {Moderski}, R. and {Moulin}, E. and {de Naurois}, M. and {Nedbal}, D. and {Nolan}, S.~J. and {Olive}, J. -P. and {Orford}, K.~J. and {Osborne}, J.~L. and {Ostrowski}, M. and {Panter}, M. and {Pedaletti}, G. and {Pelletier}, G. and {Petrucci}, P. -O. and {Pita}, S. and {P{\"u}hlhofer}, G. and {Punch}, M. and {Ranchon}, S. and {Raubenheimer}, B.~C. and {Raue}, M. and {Rayner}, S.~M. and {Renaud}, M. and {Ripken}, J. and {Rob}, L. and {Rolland}, L. and {Rosier-Lees}, S. and {Rowell}, G. and {Rudak}, B. and {Ruppel}, J. and {Sahakian}, V. and {Santangelo}, A. and {Saug{\'e}}, L. and {Schlenker}, S. and {Schlickeiser}, R. and {Schr{\"o}der}, R. and {Schwanke}, U. and {Schwarzburg}, S. and {Schwemmer}, S. and {Shalchi}, A. and {Sol}, H. and {Spangler}, D. and {Stawarz}, {\L}. and {Steenkamp}, R. and {Stegmann}, C. and {Superina}, G. and {Tam}, P.~H. and {Tavernet}, J. -P. and {Terrier}, R. and {van Eldik}, C. and {Vasileiadis}, G. and {Venter}, C. and {Vialle}, J.~P. and {Vincent}, P. and {Vivier}, M. and {V{\"o}lk}, H.~J. and {Volpe}, F. and {Wagner}, S.~J. and {Ward}, M. and {Zdziarski}, A.~A.},
        title = "{An Exceptional Very High Energy Gamma-Ray Flare of PKS 2155-304}",
      journal = {\apjl},
         year = 2007,
        month = aug,
       volume = {664},
       number = {2},
        pages = {L71-L74},
          doi = {10.1086/520635},
archivePrefix = {arXiv},
       eprint = {0706.0797},
 primaryClass = {astro-ph},
       adsurl = {https://ui.adsabs.harvard.edu/abs/2007ApJ...664L..71A}
}

@ARTICLE{2009A&A...506L..17L,
       author = {{Lachowicz}, P. and {Gupta}, A.~C. and {Gaur}, H. and {Wiita}, P.~J.},
        title = "{A \raisebox{-0.5ex}\textasciitilde4.6 h quasi-periodic oscillation in the BL Lacertae PKS 2155-304?}",
      journal = {\aap},
         year = 2009,
        month = nov,
       volume = {506},
       number = {2},
        pages = {L17-L20},
          doi = {10.1051/0004-6361/200913161},
archivePrefix = {arXiv},
       eprint = {0909.2113},
 primaryClass = {astro-ph.HE},
       adsurl = {https://ui.adsabs.harvard.edu/abs/2009A&A...506L..17L}
}

@ARTICLE{2019ARA&A..57..467B,
       author = {{Blandford}, Roger and {Meier}, David and {Readhead}, Anthony},
        title = "{Relativistic Jets from Active Galactic Nuclei}",
      journal = {\araa},
         year = 2019,
        month = aug,
       volume = {57},
        pages = {467-509},
          doi = {10.1146/annurev-astro-081817-051948},
archivePrefix = {arXiv},
       eprint = {1812.06025},
 primaryClass = {astro-ph.HE},
       adsurl = {https://ui.adsabs.harvard.edu/abs/2019ARA&A..57..467B}
}

@ARTICLE{2018Galax...6....1G,
       author = {{Gupta}, Alok},
        title = "{Multi-Wavelength Intra-Day Variability and Quasi-Periodic Oscillation in Blazars}",
      journal = {Galaxies},
         year = 2018,
        month = jan,
       volume = {6},
       number = {1},
          eid = {1},
        pages = {1},
          doi = {10.3390/galaxies6010001},
archivePrefix = {arXiv},
       eprint = {1712.02516},
 primaryClass = {astro-ph.HE},
       adsurl = {https://ui.adsabs.harvard.edu/abs/2018Galax...6....1G}
}

@ARTICLE{2016AJ....151...54S,
       author = {{Sandrinelli}, A. and {Covino}, S. and {Dotti}, M. and {Treves}, A.},
        title = "{Quasi-periodicities at Year-like Timescales in Blazars}",
      journal = {\aj},
         year = 2016,
        month = mar,
       volume = {151},
       number = {3},
          eid = {54},
        pages = {54},
          doi = {10.3847/0004-6256/151/3/54},
archivePrefix = {arXiv},
       eprint = {1512.04561},
 primaryClass = {astro-ph.GA},
       adsurl = {https://ui.adsabs.harvard.edu/abs/2016AJ....151...54S}
}

@ARTICLE{2013MNRAS.428..280C,
       author = {{Caproni}, A. and {Abraham}, Z. and {Monteiro}, H.},
        title = "{Parsec-scale jet precession in BL Lacertae (2200+420)}",
      journal = {\mnras},
         year = 2013,
        month = jan,
       volume = {428},
       number = {1},
        pages = {280-290},
          doi = {10.1093/mnras/sts014},
archivePrefix = {arXiv},
       eprint = {1210.2286},
 primaryClass = {astro-ph.HE},
       adsurl = {https://ui.adsabs.harvard.edu/abs/2013MNRAS.428..280C}
}

@ARTICLE{2004ApJ...615L...5R,
       author = {{Rieger}, Frank M.},
        title = "{On the Geometrical Origin of Periodicity in Blazar-type Sources}",
      journal = {\apjl},
         year = 2004,
        month = nov,
       volume = {615},
       number = {1},
        pages = {L5-L8},
          doi = {10.1086/426018},
archivePrefix = {arXiv},
       eprint = {astro-ph/0410188},
 primaryClass = {astro-ph},
       adsurl = {https://ui.adsabs.harvard.edu/abs/2004ApJ...615L...5R}
}

@ARTICLE{2017MNRAS.471.3036P,
       author = {{Prokhorov}, D.~A. and {Moraghan}, A.},
        title = "{A search for cyclical sources of {\ensuremath{\gamma}}-ray emission on the period range from days to years in the Fermi-LAT sky}",
      journal = {\mnras},
         year = 2017,
        month = nov,
       volume = {471},
       number = {3},
        pages = {3036-3042},
          doi = {10.1093/mnras/stx1742},
archivePrefix = {arXiv},
       eprint = {1707.05829},
 primaryClass = {astro-ph.HE},
       adsurl = {https://ui.adsabs.harvard.edu/abs/2017MNRAS.471.3036P}
}

@ARTICLE{2016ApJ...819L..37V,
       author = {{Valtonen}, M.~J. and {Zola}, S. and {Ciprini}, S. and {Gopakumar}, A. and {Matsumoto}, K. and {Sadakane}, K. and {Kidger}, M. and {Gazeas}, K. and {Nilsson}, K. and {Berdyugin}, A. and {Piirola}, V. and {Jermak}, H. and {Baliyan}, K.~S. and {Alicavus}, F. and {Boyd}, D. and {Campas Torrent}, M. and {Campos}, F. and {Carrillo G{\'o}mez}, J. and {Caton}, D.~B. and {Chavushyan}, V. and {Dalessio}, J. and {Debski}, B. and {Dimitrov}, D. and {Drozdz}, M. and {Er}, H. and {Erdem}, A. and {Escartin P{\'e}rez}, A. and {Fallah Ramazani}, V. and {Filippenko}, A.~V. and {Ganesh}, S. and {Garcia}, F. and {G{\'o}mez Pinilla}, F. and {Gopinathan}, M. and {Haislip}, J.~B. and {Hudec}, R. and {Hurst}, G. and {Ivarsen}, K.~M. and {Jelinek}, M. and {Joshi}, A. and {Kagitani}, M. and {Kaur}, N. and {Keel}, W.~C. and {LaCluyze}, A.~P. and {Lee}, B.~C. and {Lindfors}, E. and {Lozano de Haro}, J. and {Moore}, J.~P. and {Mugrauer}, M. and {Naves Nogues}, R. and {Neely}, A.~W. and {Nelson}, R.~H. and {Ogloza}, W. and {Okano}, S. and {Pandey}, J.~C. and {Perri}, M. and {Pihajoki}, P. and {Poyner}, G. and {Provencal}, J. and {Pursimo}, T. and {Raj}, A. and {Reichart}, D.~E. and {Reinthal}, R. and {Sadegi}, S. and {Sakanoi}, T. and {Salto Gonz{\'a}lez}, J. -L. and {Sameer} and {Schweyer}, T. and {Siwak}, M. and {Sold{\'a}n Alfaro}, F.~C. and {Sonbas}, E. and {Steele}, I. and {Stocke}, J.~T. and {Strobl}, J. and {Takalo}, L.~O. and {Tomov}, T. and {Tremosa Espasa}, L. and {Valdes}, J.~R. and {Valero P{\'e}rez}, J. and {Verrecchia}, F. and {Webb}, J.~R. and {Yoneda}, M. and {Zejmo}, M. and {Zheng}, W. and {Telting}, J. and {Saario}, J. and {Reynolds}, T. and {Kvammen}, A. and {Gafton}, E. and {Karjalainen}, R. and {Harmanen}, J. and {Blay}, P.},
        title = "{Primary Black Hole Spin in OJ 287 as Determined by the General Relativity Centenary Flare}",
      journal = {\apjl},
         year = 2016,
        month = mar,
       volume = {819},
       number = {2},
          eid = {L37},
        pages = {L37},
          doi = {10.3847/2041-8205/819/2/L37},
archivePrefix = {arXiv},
       eprint = {1603.04171},
 primaryClass = {astro-ph.HE},
       adsurl = {https://ui.adsabs.harvard.edu/abs/2016ApJ...819L..37V}
}

@ARTICLE{2015ApJ...813L..41A,
       author = {{Ackermann}, M. and {Ajello}, M. and {Albert}, A. and {Atwood}, W.~B. and {Baldini}, L. and {Ballet}, J. and {Barbiellini}, G. and {Bastieri}, D. and {Becerra Gonzalez}, J. and {Bellazzini}, R. and {Bissaldi}, E. and {Blandford}, R.~D. and {Bloom}, E.~D. and {Bonino}, R. and {Bottacini}, E. and {Bregeon}, J. and {Bruel}, P. and {Buehler}, R. and {Buson}, S. and {Caliandro}, G.~A. and {Cameron}, R.~A. and {Caputo}, R. and {Caragiulo}, M. and {Caraveo}, P.~A. and {Cavazzuti}, E. and {Cecchi}, C. and {Chekhtman}, A. and {Chiang}, J. and {Chiaro}, G. and {Ciprini}, S. and {Cohen-Tanugi}, J. and {Conrad}, J. and {Cutini}, S. and {D'Ammando}, F. and {de Angelis}, A. and {de Palma}, F. and {Desiante}, R. and {Di Venere}, L. and {Dom{\'\i}nguez}, A. and {Drell}, P.~S. and {Favuzzi}, C. and {Fegan}, S.~J. and {Ferrara}, E.~C. and {Focke}, W.~B. and {Fuhrmann}, L. and {Fukazawa}, Y. and {Fusco}, P. and {Gargano}, F. and {Gasparrini}, D. and {Giglietto}, N. and {Giommi}, P. and {Giordano}, F. and {Giroletti}, M. and {Godfrey}, G. and {Green}, D. and {Grenier}, I.~A. and {Grove}, J.~E. and {Guiriec}, S. and {Harding}, A.~K. and {Hays}, E. and {Hewitt}, J.~W. and {Hill}, A.~B. and {Horan}, D. and {Jogler}, T. and {J{\'o}hannesson}, G. and {Johnson}, A.~S. and {Kamae}, T. and {Kuss}, M. and {Larsson}, S. and {Latronico}, L. and {Li}, J. and {Li}, L. and {Longo}, F. and {Loparco}, F. and {Lott}, B. and {Lovellette}, M.~N. and {Lubrano}, P. and {Magill}, J. and {Maldera}, S. and {Manfreda}, A. and {Max-Moerbeck}, W. and {Mayer}, M. and {Mazziotta}, M.~N. and {McEnery}, J.~E. and {Michelson}, P.~F. and {Mizuno}, T. and {Monzani}, M.~E. and {Morselli}, A. and {Moskalenko}, I.~V. and {Murgia}, S. and {Nuss}, E. and {Ohno}, M. and {Ohsugi}, T. and {Ojha}, R. and {Omodei}, N. and {Orlando}, E. and {Ormes}, J.~F. and {Paneque}, D. and {Pearson}, T.~J. and {Perkins}, J.~S. and {Perri}, M. and {Pesce-Rollins}, M. and {Petrosian}, V. and {Piron}, F. and {Pivato}, G. and {Porter}, T.~A. and {Rain{\`o}}, S. and {Rando}, R. and {Razzano}, M. and {Readhead}, A. and {Reimer}, A. and {Reimer}, O. and {Schulz}, A. and {Sgr{\`o}}, C. and {Siskind}, E.~J. and {Spada}, F. and {Spandre}, G. and {Spinelli}, P. and {Suson}, D.~J. and {Takahashi}, H. and {Thayer}, J.~B. and {Thompson}, D.~J. and {Tibaldo}, L. and {Torres}, D.~F. and {Tosti}, G. and {Troja}, E. and {Uchiyama}, Y. and {Vianello}, G. and {Wood}, K.~S. and {Wood}, M. and {Zimmer}, S. and {Berdyugin}, A. and {Corbet}, R.~H.~D. and {Hovatta}, T. and {Lindfors}, E. and {Nilsson}, K. and {Reinthal}, R. and {Sillanp{\"a}{\"a}}, A. and {Stamerra}, A. and {Takalo}, L.~O. and {Valtonen}, M.~J.},
        title = "{Multiwavelength Evidence for Quasi-periodic Modulation in the Gamma-Ray Blazar PG 1553+113}",
      journal = {\apjl},
         year = 2015,
        month = nov,
       volume = {813},
       number = {2},
          eid = {L41},
        pages = {L41},
          doi = {10.1088/2041-8205/813/2/L41},
archivePrefix = {arXiv},
       eprint = {1509.02063},
 primaryClass = {astro-ph.HE},
       adsurl = {https://ui.adsabs.harvard.edu/abs/2015ApJ...813L..41A}
}

@ARTICLE{2024MNRAS.535.2775V,
       author = {{Vlasyuk}, V.~V. and {Sotnikova}, Y.~V. and {Volvach}, A.~E. and {Mufakharov}, T.~V. and {Kovalev}, Y.~A. and {Spiridonova}, O.~I. and {Khabibullina}, M.~L. and {Kovalev}, Y.~Y. and {Mikhailov}, A.~G. and {Stolyarov}, V.~A. and {Kudryavtsev}, D.~O. and {Mingaliev}, M.~G. and {Razzaque}, S. and {Semenova}, T.~A. and {Kudryashova}, A.~K. and {Bursov}, N.~N. and {Trushkin}, S.~A. and {Popkov}, A.~V. and {Erkenov}, A.~K. and {Rakhimov}, I.~A. and {Kharinov}, M.~A. and {Gurwell}, M.~A. and {Tsybulev}, P.~G. and {Moskvitin}, A.~S. and {Fatkhullin}, T.~A. and {Emelianov}, E.~V. and {Arshinova}, A. and {Iuzhanina}, K.~V. and {Andreeva}, T.~S. and {Volvach}, L.~N. and {Ghosh}, A.},
        title = "{Multiwavelength variability of the blazar AO 0235+164}",
      journal = {\mnras},
         year = 2024,
        month = dec,
       volume = {535},
       number = {3},
        pages = {2775-2799},
          doi = {10.1093/mnras/stae2491},
archivePrefix = {arXiv},
       eprint = {2411.01497},
 primaryClass = {astro-ph.HE},
       adsurl = {https://ui.adsabs.harvard.edu/abs/2024MNRAS.535.2775V}
}

@ARTICLE{2021PASP..133g4101Y,
       author = {{Yuan}, Y.~H. and {Fan}, J.~H.},
        title = "{Optical Monitoring and Intraday Variabilities of the BL Lac Object PKS 0735+178}",
      journal = {\pasp},
         year = 2021,
        month = jul,
       volume = {133},
       number = {1025},
          eid = {074101},
        pages = {074101},
          doi = {10.1088/1538-3873/ac015f},
       adsurl = {https://ui.adsabs.harvard.edu/abs/2021PASP..133g4101Y}
}

@ARTICLE{2025MNRAS.537.2380Z,
       author = {{Zhang}, Haiyun and {Yan}, Dahai and {Zhang}, Li and {Tang}, Niansheng},
        title = "{Evidence for magneto-gravitational processes in supermassive black hole binary PG 1553+113}",
      journal = {\mnras},
         year = 2025,
        month = mar,
       volume = {537},
       number = {3},
        pages = {2380-2386},
          doi = {10.1093/mnras/staf129},
archivePrefix = {arXiv},
       eprint = {2501.15169},
 primaryClass = {astro-ph.HE},
       adsurl = {https://ui.adsabs.harvard.edu/abs/2025MNRAS.537.2380Z}
}

@ARTICLE{2022ApJ...941L..25B,
       author = {{Becker Tjus}, Julia and {Jaroschewski}, Ilja and {Ghorbanietemad}, Armin and {Bartos}, Imre and {Kun}, Emma and {Biermann}, Peter L.},
        title = "{Neutrino Cadence of TXS 0506+056 Consistent with Supermassive Binary Origin}",
      journal = {\apjl},
         year = 2022,
        month = dec,
       volume = {941},
       number = {2},
          eid = {L25},
        pages = {L25},
          doi = {10.3847/2041-8213/aca65d},
archivePrefix = {arXiv},
       eprint = {2210.00202},
 primaryClass = {astro-ph.HE},
       adsurl = {https://ui.adsabs.harvard.edu/abs/2022ApJ...941L..25B}
}

@ARTICLE{2024MNRAS.527.6970K,
       author = {{Krishna Mohana}, A. and {Gupta}, Alok C. and {Marscher}, Alan P. and {Sotnikova}, Yulia V. and {Jorstad}, S.~G. and {Wiita}, Paul J. and {Cui}, Lang and {Aller}, Margo F. and {Aller}, Hugh D. and {Kovalev}, Yu A. and {Kovalev}, Y.~Y. and {Liu}, Xiang and {Mufakharov}, T.~V. and {Popkov}, A.~V. and {Mingaliev}, M.~G. and {Erkenov}, A.~K. and {Nizhelsky}, N.~A. and {Tsybulev}, P.~G. and {Zhao}, Wei and {Weaver}, Z.~R. and {Morozova}, D.~A.},
        title = "{Multiband cross-correlated radio variability of the blazar 3C 279}",
      journal = {\mnras},
         year = 2024,
        month = jan,
       volume = {527},
       number = {3},
        pages = {6970-6980},
          doi = {10.1093/mnras/stad3583},
archivePrefix = {arXiv},
       eprint = {2311.02395},
 primaryClass = {astro-ph.HE},
       adsurl = {https://ui.adsabs.harvard.edu/abs/2024MNRAS.527.6970K}
}

@ARTICLE{2017A&A...602A..29B,
       author = {{Britzen}, S. and {Qian}, S. -J. and {Steffen}, W. and {Kun}, E. and {Karouzos}, M. and {Gergely}, L. and {Schmidt}, J. and {Aller}, M. and {Aller}, H. and {Krause}, M. and {Fendt}, C. and {B{\"o}ttcher}, M. and {Witzel}, A. and {Eckart}, A. and {Moser}, L.},
        title = "{A swirling jet in the quasar 1308+326}",
      journal = {\aap},
         year = 2017,
        month = jun,
       volume = {602},
          eid = {A29},
        pages = {A29},
          doi = {10.1051/0004-6361/201629999},
       adsurl = {https://ui.adsabs.harvard.edu/abs/2017A&A...602A..29B}
}

@ARTICLE{2025arXiv250703967P,
       author = {{Penil}, P. and {Otero-Santos}, J. and {Banerjee}, A. and {Buson}, S. and {Rico}, A. and {Ajello}, M. and {Adhikari}, S.},
        title = "{Transient QPOs of Fermi-LAT blazars under the Curved Jet Model}",
      journal = {arXiv e-prints},
         year = 2025,
        month = jul,
          eid = {arXiv:2507.03967},
        pages = {arXiv:2507.03967},
          doi = {10.48550/arXiv.2507.03967},
archivePrefix = {arXiv},
       eprint = {2507.03967},
 primaryClass = {astro-ph.HE},
       adsurl = {https://ui.adsabs.harvard.edu/abs/2025arXiv250703967P}
}

@ARTICLE{2023MNRAS.523L..52B,
       author = {{Banerjee}, Anuvab and {Sharma}, Ajay and {Mandal}, Avijit and {Das}, Avik Kumar and {Bhatta}, Gopal and {Bose}, Debanjan},
        title = "{Detection of periodicity in the gamma-ray light curve of the BL Lac 4FGL J2202.7+4216}",
      journal = {\mnras},
         year = 2023,
        month = jul,
       volume = {523},
       number = {1},
        pages = {L52-L57},
          doi = {10.1093/mnrasl/slad057},
       adsurl = {https://ui.adsabs.harvard.edu/abs/2023MNRAS.523L..52B}
}

@ARTICLE{2021MNRAS.501.5478A,
       author = {{Ashton}, Dominic I. and {Middleton}, Matthew J.},
        title = "{Searching for energy-resolved quasi-periodic oscillations in AGN}",
      journal = {\mnras},
         year = 2021,
        month = mar,
       volume = {501},
       number = {4},
        pages = {5478-5499},
          doi = {10.1093/mnras/staa4024},
archivePrefix = {arXiv},
       eprint = {2101.01194},
 primaryClass = {astro-ph.HE},
       adsurl = {https://ui.adsabs.harvard.edu/abs/2021MNRAS.501.5478A}
}

@ARTICLE{2023NatAs...7.1368E,
       author = {{Evans}, P.~A. and {Nixon}, C.~J. and {Campana}, S. and {Charalampopoulos}, P. and {Perley}, D.~A. and {Breeveld}, A.~A. and {Page}, K.~L. and {Oates}, S.~R. and {Eyles-Ferris}, R.~A.~J. and {Malesani}, D.~B. and {Izzo}, L. and {Goad}, M.~R. and {O'Brien}, P.~T. and {Osborne}, J.~P. and {Sbarufatti}, B.},
        title = "{Monthly quasi-periodic eruptions from repeated stellar disruption by a massive black hole}",
      journal = {Nature Astronomy},
         year = 2023,
        month = nov,
       volume = {7},
        pages = {1368-1375},
          doi = {10.1038/s41550-023-02073-y},
archivePrefix = {arXiv},
       eprint = {2309.02500},
 primaryClass = {astro-ph.HE},
       adsurl = {https://ui.adsabs.harvard.edu/abs/2023NatAs...7.1368E}
}

@ARTICLE{2023A&A...672A..86R,
       author = {{Ren}, Helena X. and {Cerruti}, Matteo and {Sahakyan}, Narek},
        title = "{Quasi-periodic oscillations in the {\ensuremath{\gamma}}-ray light curves of bright active galactic nuclei}",
      journal = {\aap},
         year = 2023,
        month = apr,
       volume = {672},
          eid = {A86},
        pages = {A86},
          doi = {10.1051/0004-6361/202244754},
archivePrefix = {arXiv},
       eprint = {2204.13051},
 primaryClass = {astro-ph.HE},
       adsurl = {https://ui.adsabs.harvard.edu/abs/2023A&A...672A..86R}
}

\appendix
\clearpage
\newpage

\section{Statistical parameters of the observed light curves}
\label{app:stat}

To estimate the significance levels in the DCF and QPO calculations, Monte Carlo simulations are made, where a large number of synthetic light curves are generated to calculate the probability of observing a particular situation purely by chance. These synthetic light curves should mimic the observed light curves at each frequency as realistically as possible. We used the method of \cite{2013MNRAS.433..907E}, where synthetic light curves are generated based on the best fitting of the probability density function (PDF) and power spectral density (PSD) in the observed light curves. Figs.~\mbox{\ref{fig:sgamma_fit}--\ref{fig:s2_fit}} demonstrate the fitting results as they are provided by the 
{\tt DELightcurveSimulation} software \citep{2015arXiv150306676C}. The best-fitting PDF and PSD models underlying the observed PDF and periodogram estimates are shown by the red lines. To fit the PDF parameters, we have additionally modified the original code by using the nested sampling Monte Carlo algorithm MLFriends \citep{2019PASP..131j8005B} implemented in the UltraNest\footnote{\url{https://johannesbuchner.github.io/UltraNest/}} package \citep{2021JOSS....6.3001B}. The upper panels show the observed light curve interpolated evenly for each MJD by the Steffen spline \citep{1990A&A...239..443S} with added Gaussian noise whose standard deviation is equal to the median measurements uncertainty.

\cite{2013MNRAS.433..907E} estimated the PSD underlying a periodogram $P_{\rm obs}(\nu)$ by fitting a smoothly bending power-law model:
$$
\mathcal{P}(\nu) = \frac{A\nu^{-\alpha_{\rm low}}}{1+(\nu/\nu_{\rm bend})^{\alpha_{\rm high}-\alpha_{\rm low}}} + c,
$$
where $A$, $\nu_{\rm bend}$, $\alpha_{\rm low}$, and $\alpha_{\rm high}$ are the normalization, bend frequency, and low- and high-frequency slopes, respectively. 
The constant $c$ represents the Poisson noise level.

The PDF $f(x)$ is fitted as a mixture of the gamma distribution $\Gamma(\kappa,\theta)$, where $\kappa$ and $\theta$ are the shape and scale
parameters, and the log-normal distribution $\ln\mathcal{N}(\mu,\sigma^2)$
with the mean $\mu$ and variance $\sigma^2$:
$$
f_{\rm mix}(x) = 
w_\Gamma\frac{\theta^{-\kappa} e^{-x/\theta} x^{\kappa-1}}{\Gamma(\kappa)} +
w_{\ln\mathcal{N}}\frac{e^{-(\ln x-\mu)^2/(2\sigma^2)}}{\sqrt{2\pi}\,x\sigma},
$$
where $w_\Gamma$ and $w_{\ln\mathcal{N}}=1-w_\Gamma$ are the weights for the gamma and log-normal distributions.

The corresponding best-fitting parameters for the light curves are listed in Table~\ref{tab:lc_params}.

\begin{figure}
\centerline{\includegraphics[width=\columnwidth]{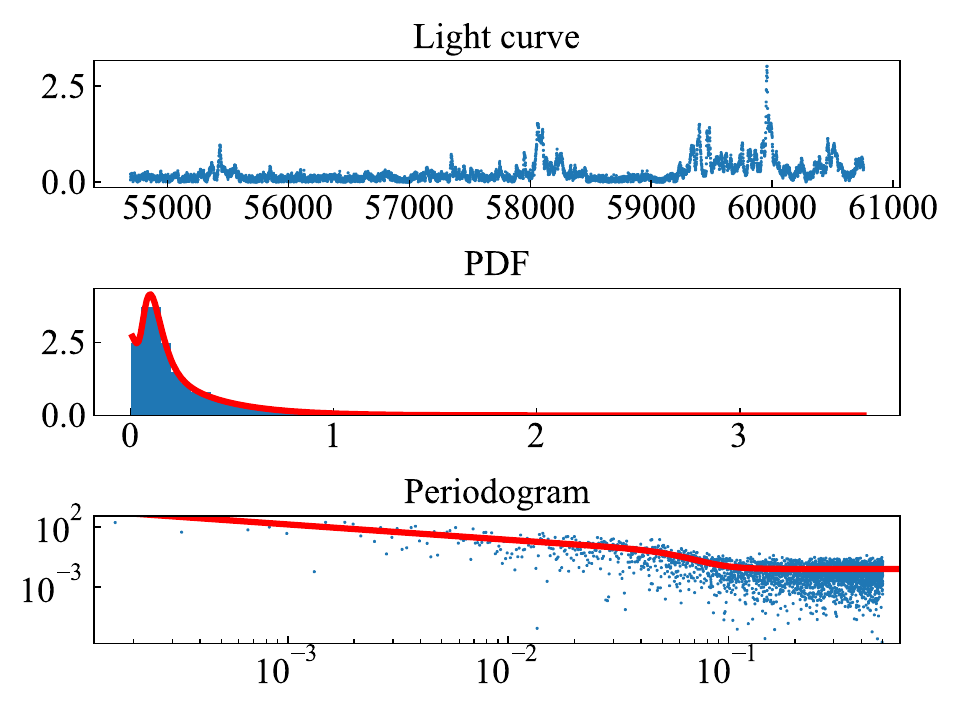}}
\caption{The interpolated $\gamma$-ray light curve (MJD--count\,cm$^{-2}$\,s$^{-1}$) and its PDF and PSD fitting.} 
\label{fig:sgamma_fit}
\end{figure}

\begin{figure}
\centerline{\includegraphics[width=\columnwidth]{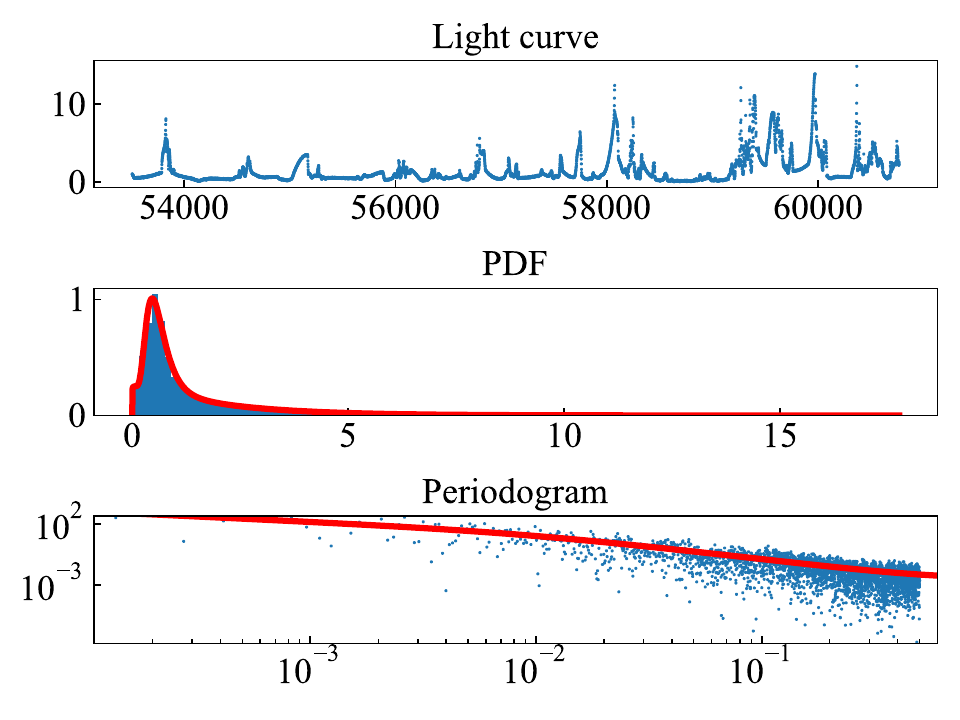}}
\caption{The interpolated $R$-band light curve (MJD--mJy) and its PDF and PSD fitting.} 
\label{fig:sR_fit}
\end{figure}

\begin{figure}
\centerline{\includegraphics[width=\columnwidth]{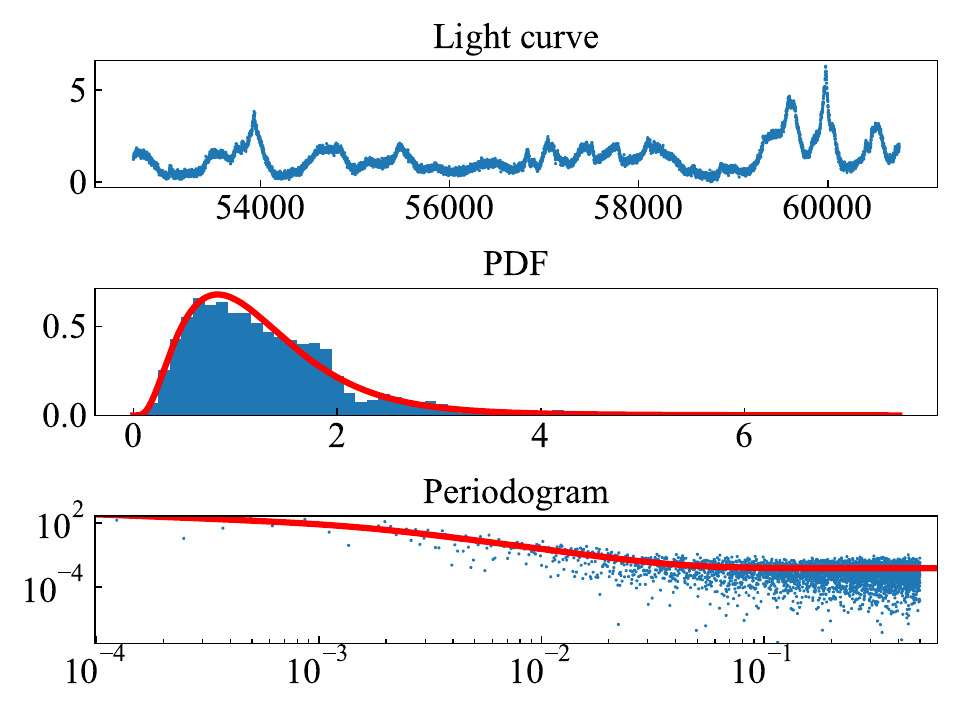}}
\caption{The interpolated 230 GHz light curve (MJD--Jy) and its PDF and PSD fitting.} 
\label{fig:s230_fit}
\end{figure}

\begin{figure}
\centerline{\includegraphics[width=\columnwidth]{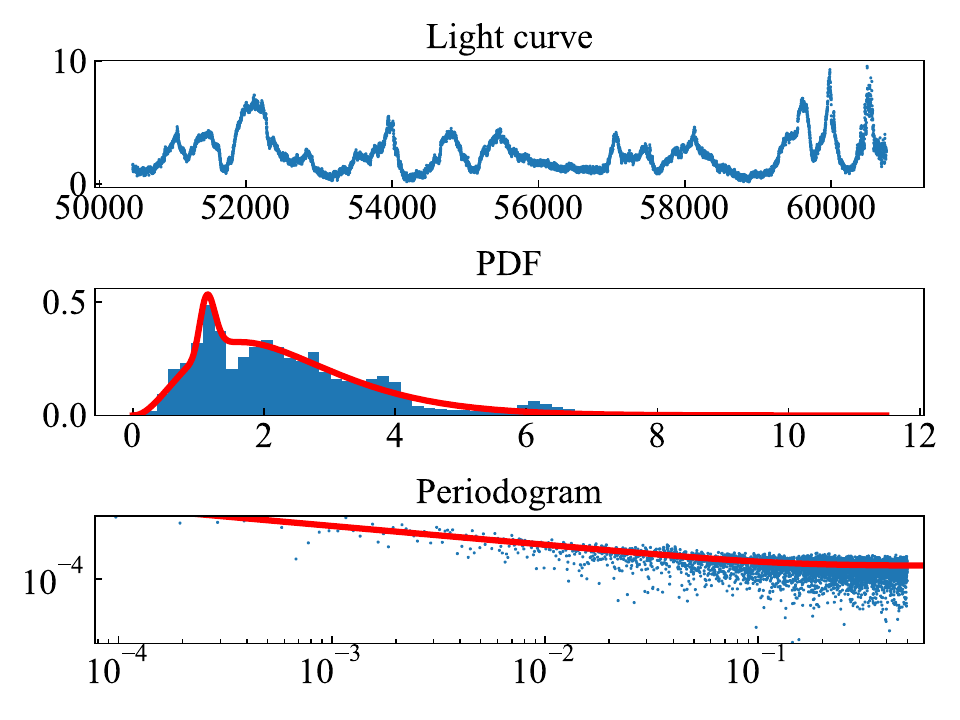}}
\caption{The interpolated 37 GHz light curve (MJD--Jy) and its PDF and PSD fitting.} 
\label{fig:s37_fit}
\end{figure}

\begin{figure}
\centerline{\includegraphics[width=\columnwidth]{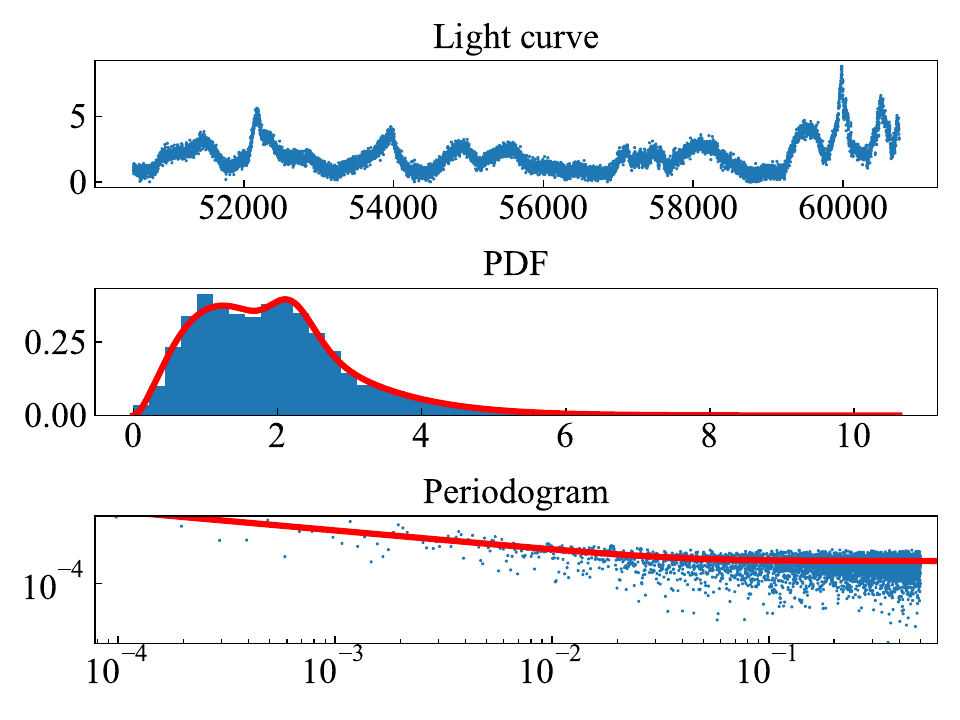}}
\caption{The interpolated 22 GHz light curve (MJD--Jy) and its PDF and PSD fitting.} 
\label{fig:s22_fit}
\end{figure}

\begin{figure}
\centerline{\includegraphics[width=\columnwidth]{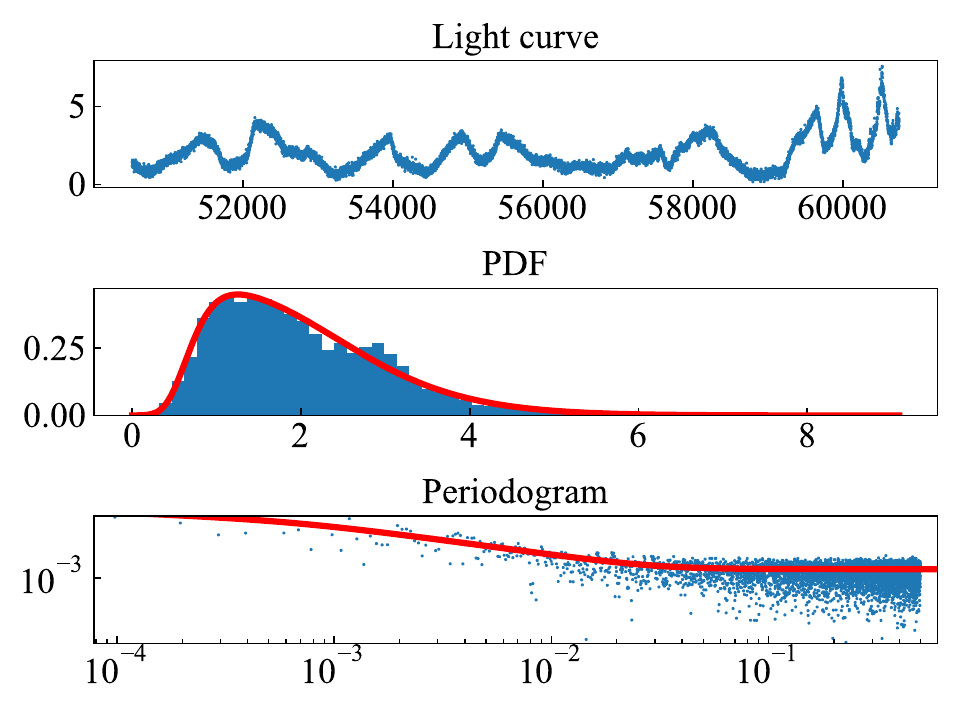}}
\caption{The interpolated 11 GHz light curve (MJD--Jy) and its PDF and PSD fitting.} 
\label{fig:s11_fit}
\end{figure}

\begin{figure}
\centerline{\includegraphics[width=\columnwidth]{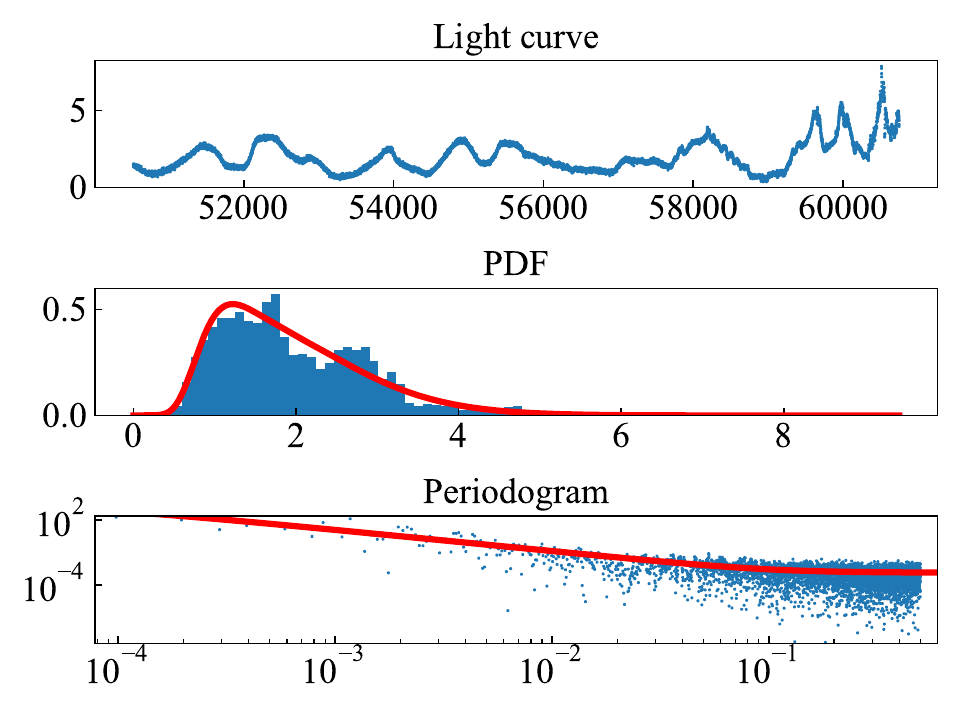}}
\caption{The interpolated 8 GHz light curve (MJD--Jy) and its PDF and PSD fitting.} 
\label{fig:s8_fit}
\end{figure}

\begin{figure}
\centerline{\includegraphics[width=\columnwidth]{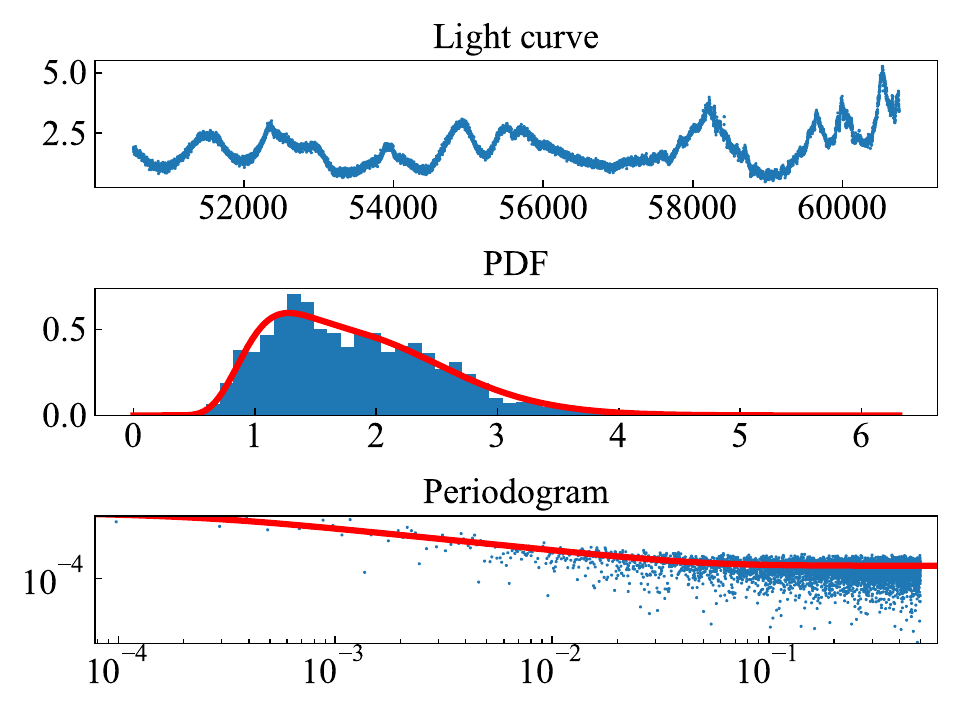}}
\caption{The interpolated 5 GHz light curve (MJD--Jy) and its PDF and PSD fitting.} 
\label{fig:s5_fit}
\end{figure}

\begin{figure}
\centerline{\includegraphics[width=\columnwidth]{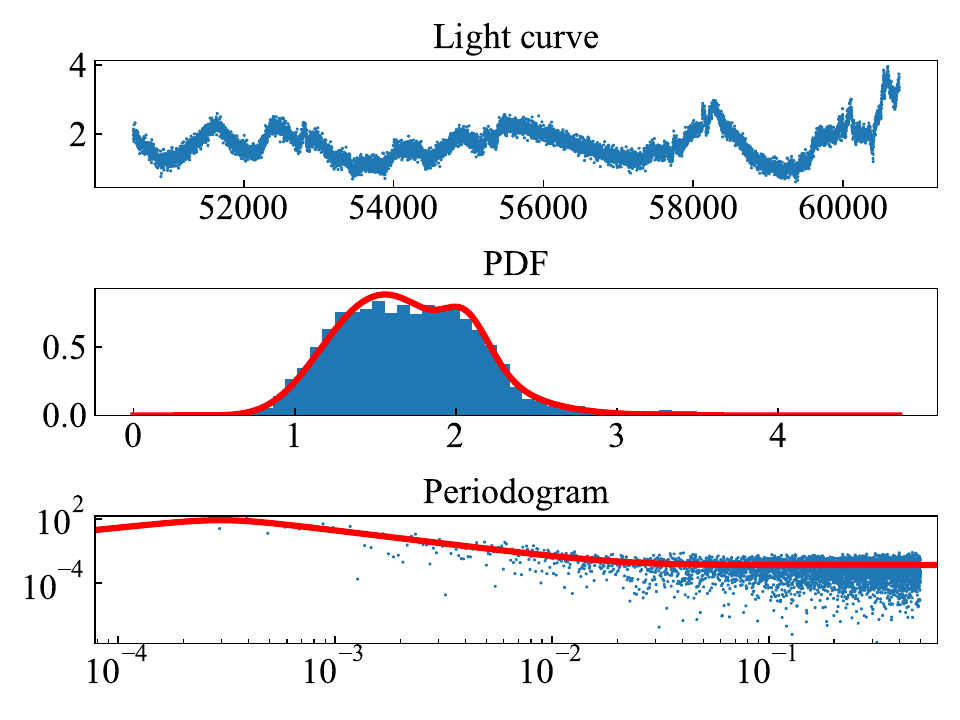}}
\caption{The interpolated 2 GHz light curve (MJD--Jy) and its PDF and PSD fitting.} 
\label{fig:s2_fit}
\end{figure}

\begin{table*}
\centering
\caption{Best-fitting PSD and PDF parameters for the light curves at different radio frequencies.}
\begin{tabular}{ccccccccccc}
\hline
Freq. & $A$ & $\nu_{\rm bend}$ & $\alpha_{\rm low}$ & $\alpha_{\rm high}$ & $c$ & $\kappa$ & $\theta$ & $\sigma$ & $\mu$ & $w_{\ln\mathcal{N}}$\\
\hline
$\gamma$ & 0.022 & 0.055 & 1.281 & 6.280 & 0.030 & 0.971 & 0.280 & 0.423 & 0.122 & 0.046 \\
$R$ & 0.388 & 0.010 & 0.864 & 2.174 & 0.003 & 1.041 & 1.857 & 0.445 & 0.572 & 0.483 \\
230~GHz & 0.445 & 0.002 & 0.790 & 2.837 & 0.006 & 4.029 & 0.314 & 0.680 & 1.045 & 0.860 \\
37~GHz & 4e-5 & 3.485 & 1.953 & 6.271 & 0.002 & 3.228 & 0.740 & 0.099 & 1.144 & 0.088 \\
22~GHz & 0.096 & 4e-12 & 1.625 & 1.927 & 0.018 & 2.809 & 0.686 & 0.134 & 2.259 & 0.241 \\
11~GHz & 0.358 & 0.001 & 0.821 & 2.641 & 0.006 & 4.622 & 0.478 & 0.382 & 1.105 & 0.273\\
8~GHz & 1.456 & 1e-4 &	0.749 & 1.976 & 0.002 & 6.260 & 0.362 & 0.334 & 1.198 & 0.594 \\
5~GHz & 8.884 &	3e-4 & 0.391 &	2.228 &	0.002 & 10.878 & 0.200 & 0.277 & 1.237 & 0.748 \\
2~GHz & 2e9 & 3e-4 & -2.049 & 2.437 & 0.005 & 15.982 & 0.104 & 0.067 & 2.084 & 0.244 \\
\hline
\end{tabular}
\label{tab:lc_params}
\end{table*}

\clearpage
\newpage

\section{DCF calculations}

Figs.~\ref{fig:dcf_ep1}--\ref{fig:dcf_ep4} present the DCF calculations for frequency pairs in epochs~1--4. The figures correspond to the measured lags in Table~\ref{tab:ccf-lags}. The upper panels present the light curves in the intervals corresponding to the epochs. In the lower panels the DCF values as a function of lags are shown by the black line with uncertainties indicated by the blue area. The 1, 2, and 3$\sigma$ significance levels, calculated by Monte simulations with synthetic light curves, are show by the blue, red, and green lines, respectively. The measured lag between the light curves is indicated by the orange vertical line, and the lag uncertainty is shown by the orange area. The lag value is derived by the method of flux redistribution and random subset selection (FR/RSS), introduced by \cite{1998PASP..110..660P}. Although in the figures we present the DCF values for a certain bin size, the measured best-fitting lags are independent of it, as we varied the bin size in the DCF calculation, in addition to FR/RSS. The uncertainties of the lags are measured as the 16th/84th percentiles in the FR/RSS lag distribution.


\begin{figure*}
\centerline{
\includegraphics[width=0.7\columnwidth]{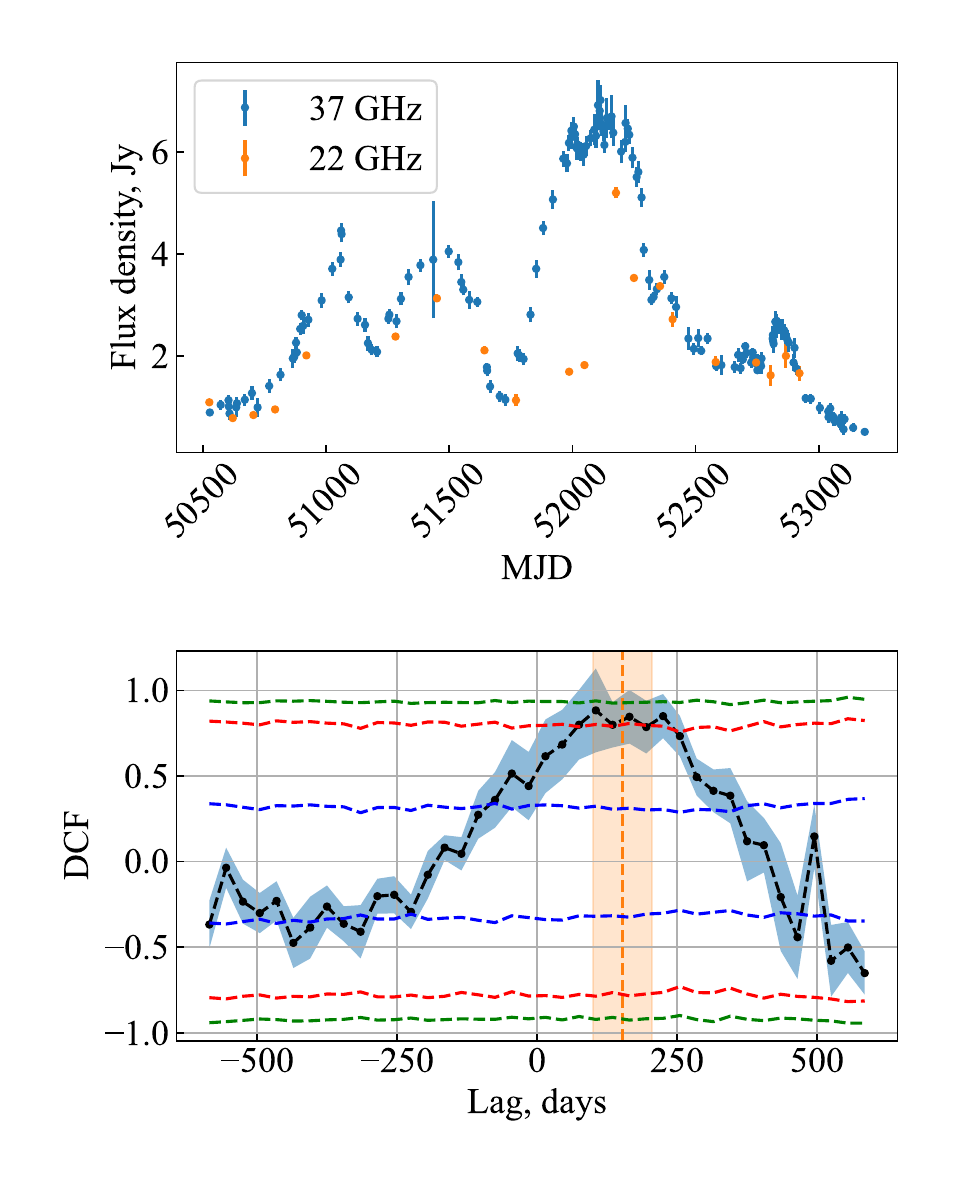}
\includegraphics[width=0.7\columnwidth]{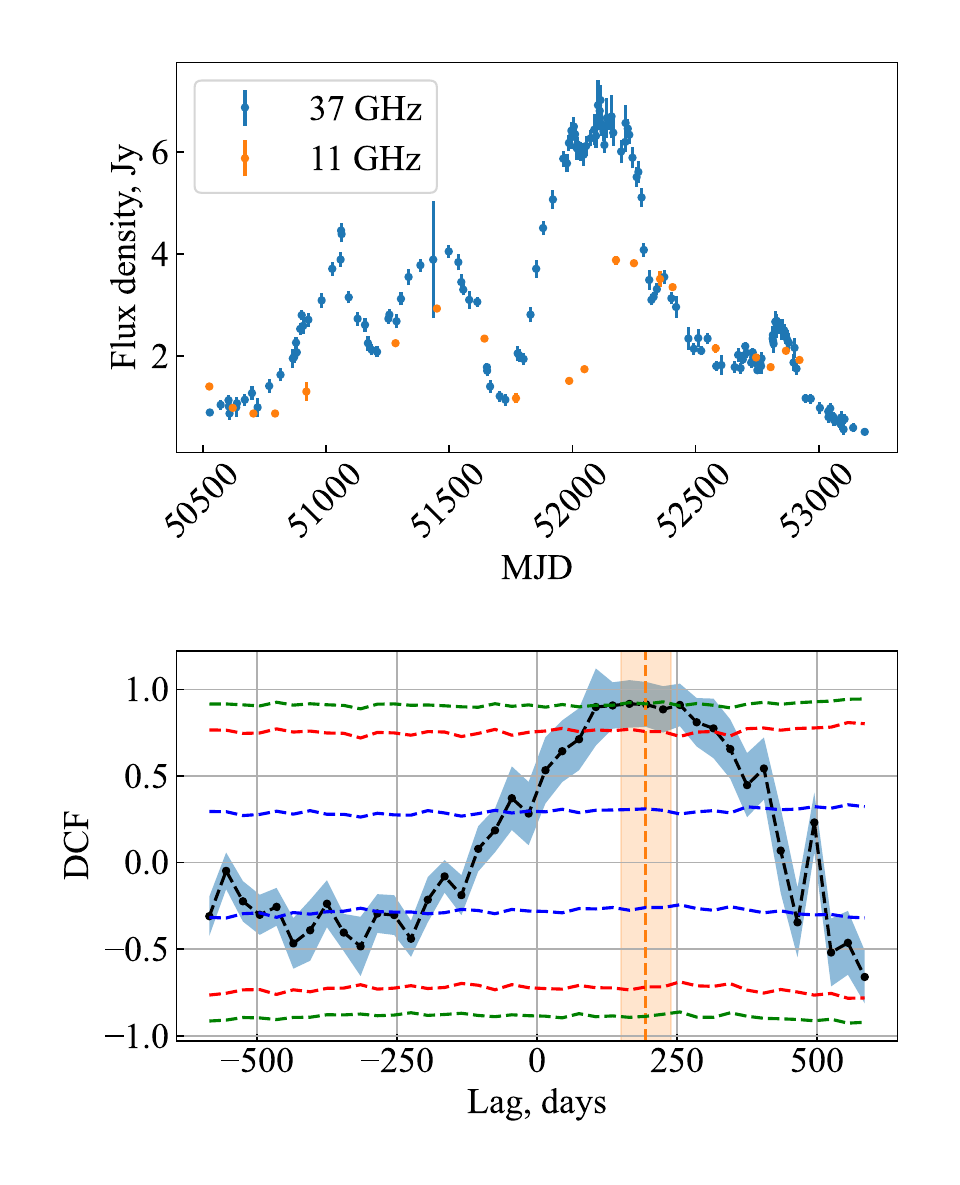}
\includegraphics[width=0.7\columnwidth]{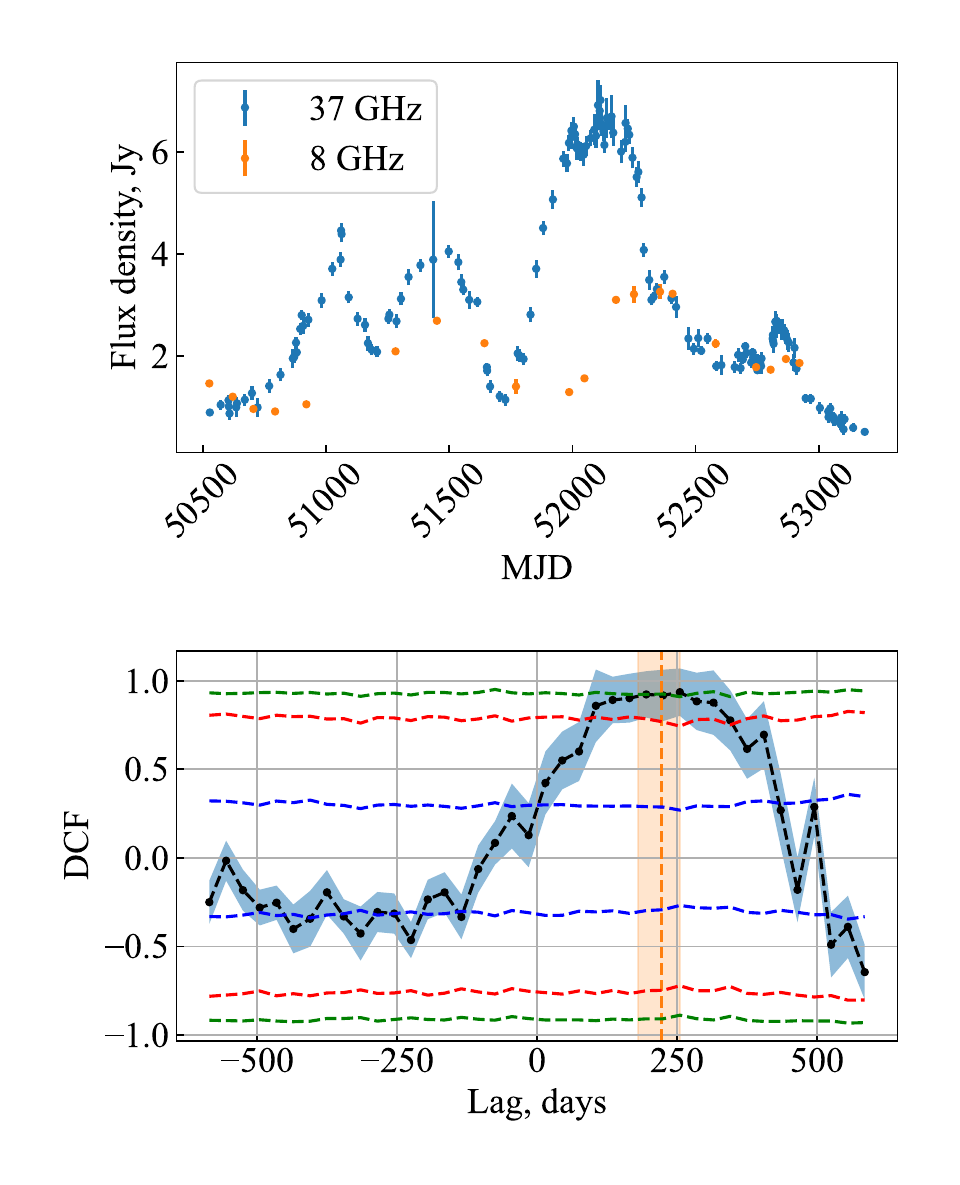}
}
\centerline{
\includegraphics[width=0.7\columnwidth]{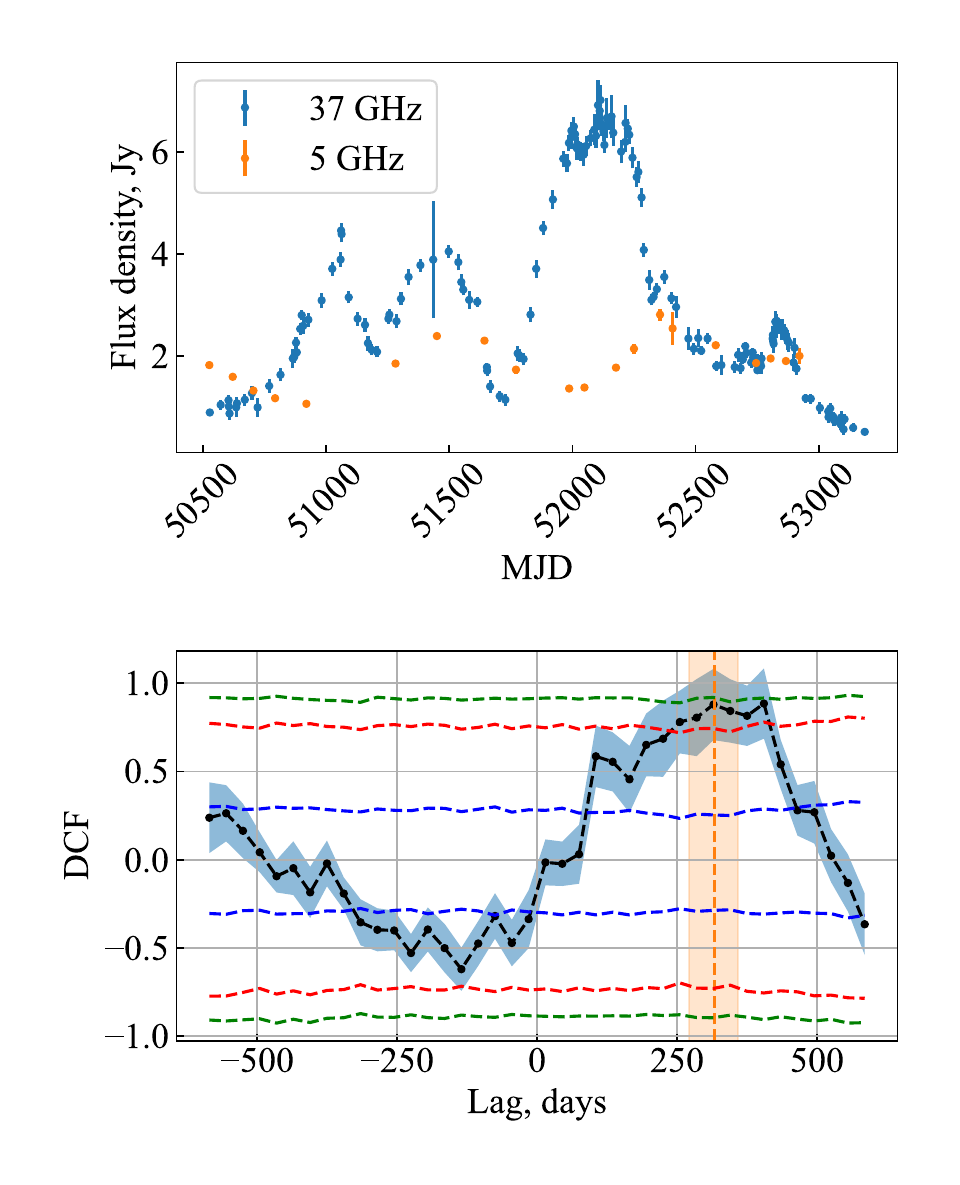}
\includegraphics[width=0.7\columnwidth]{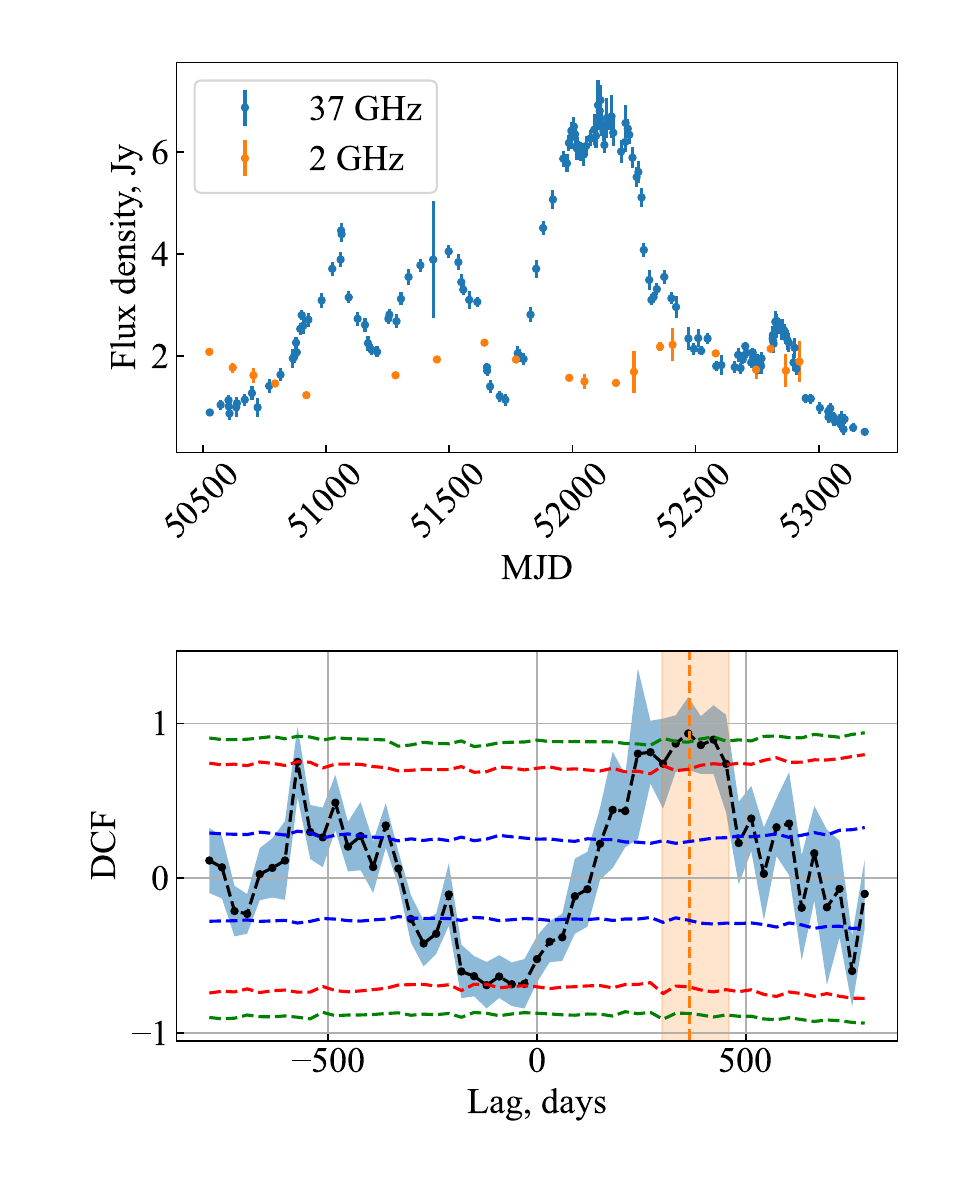}
\includegraphics[width=0.7\columnwidth]{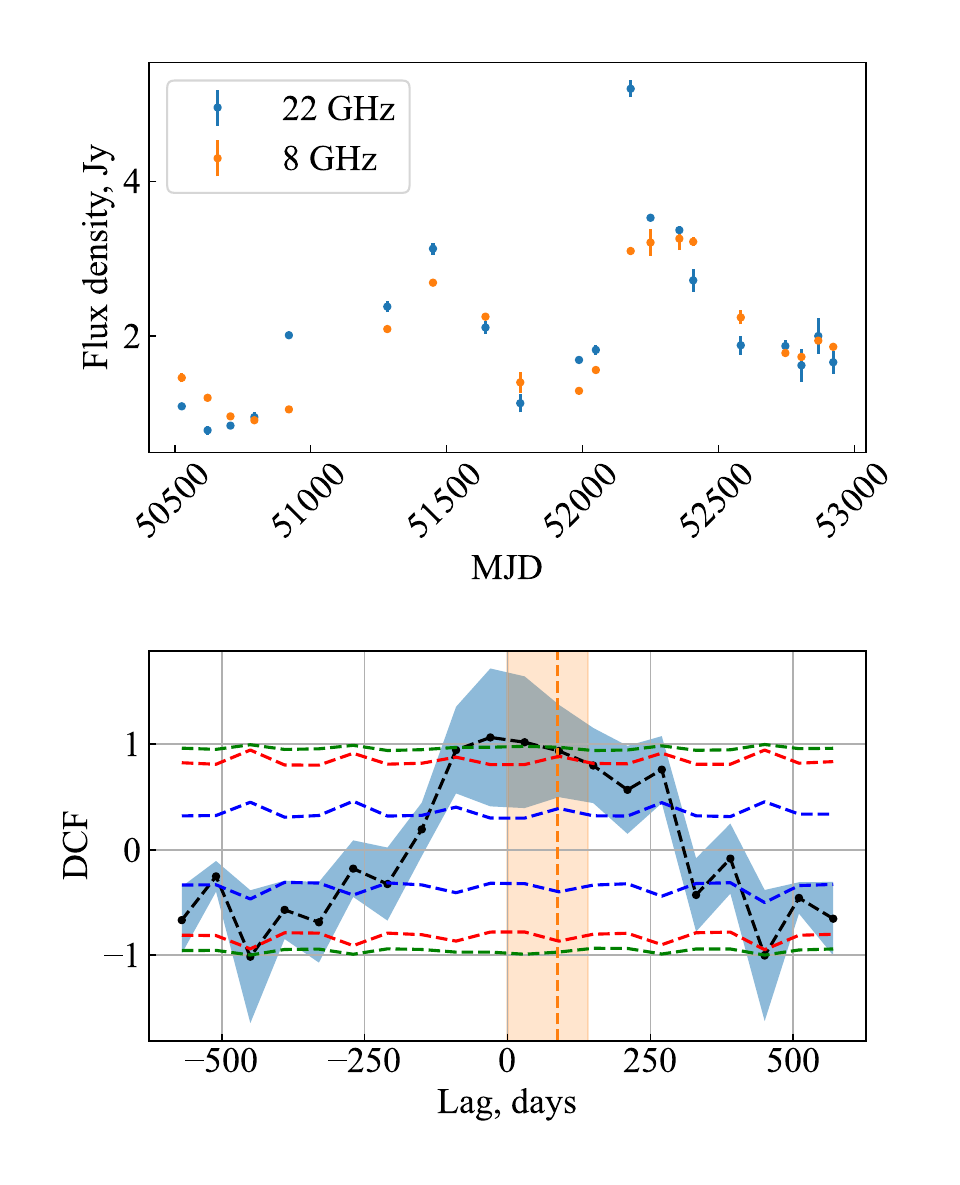}
}
\centerline{
\includegraphics[width=0.7\columnwidth]{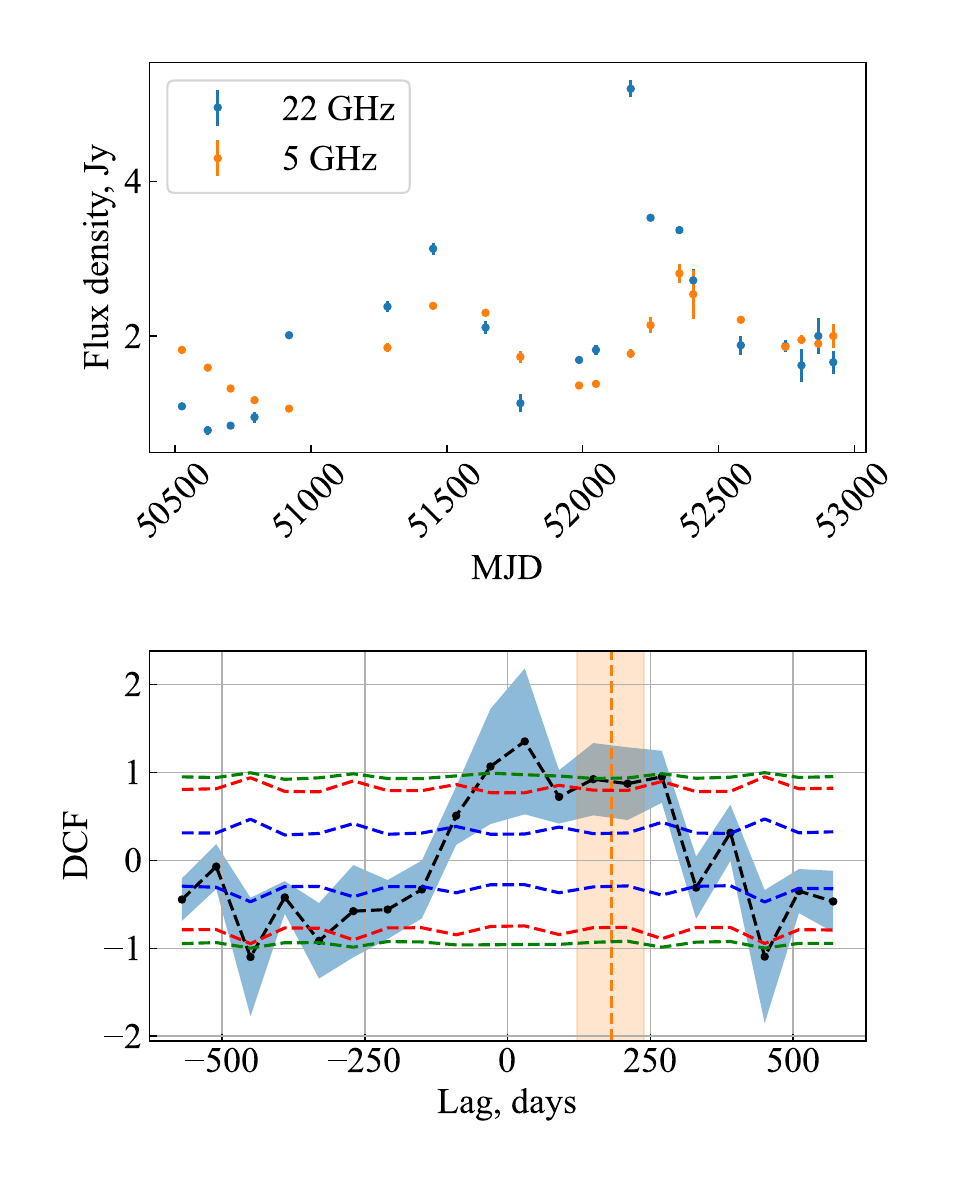}
\includegraphics[width=0.7\columnwidth]{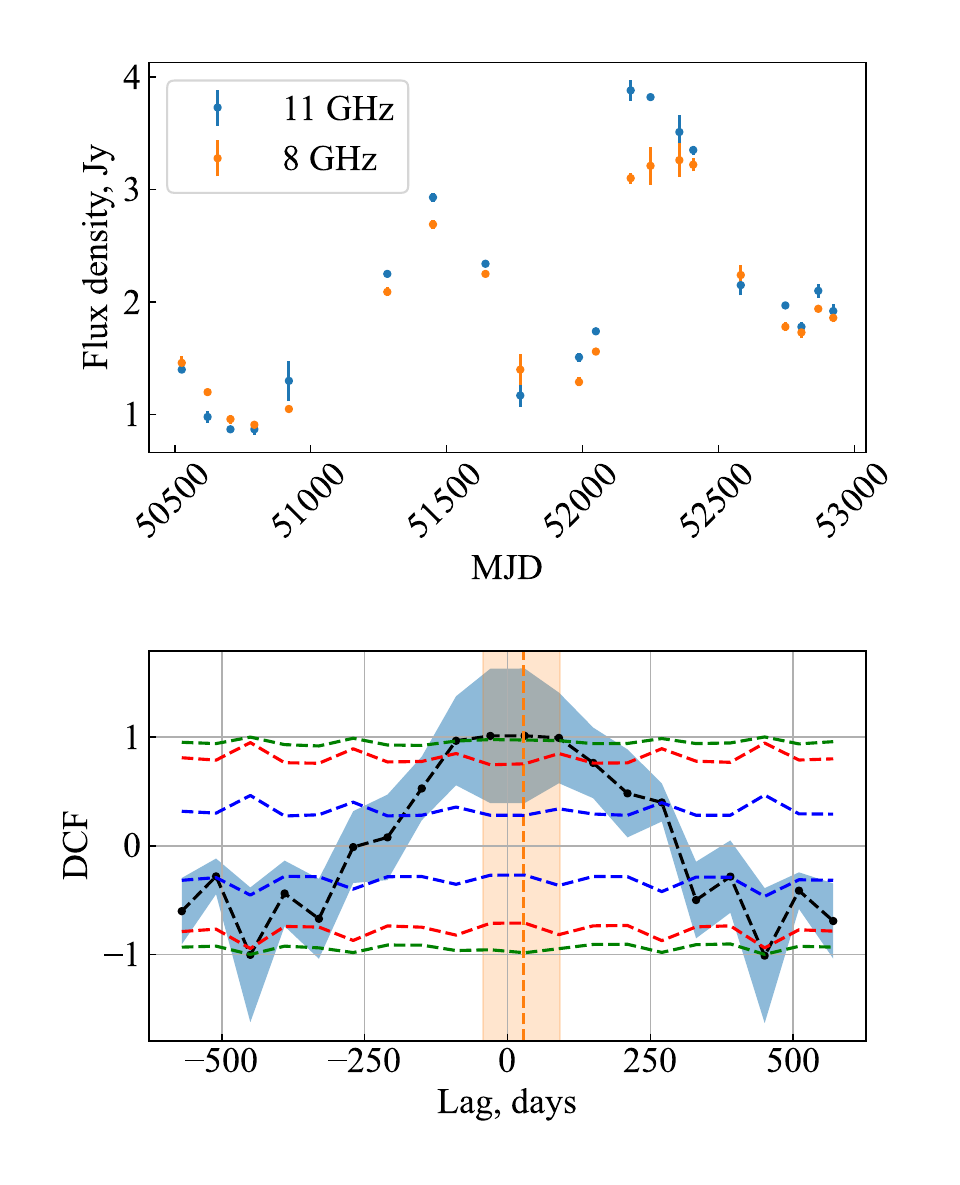}
\includegraphics[width=0.7\columnwidth]{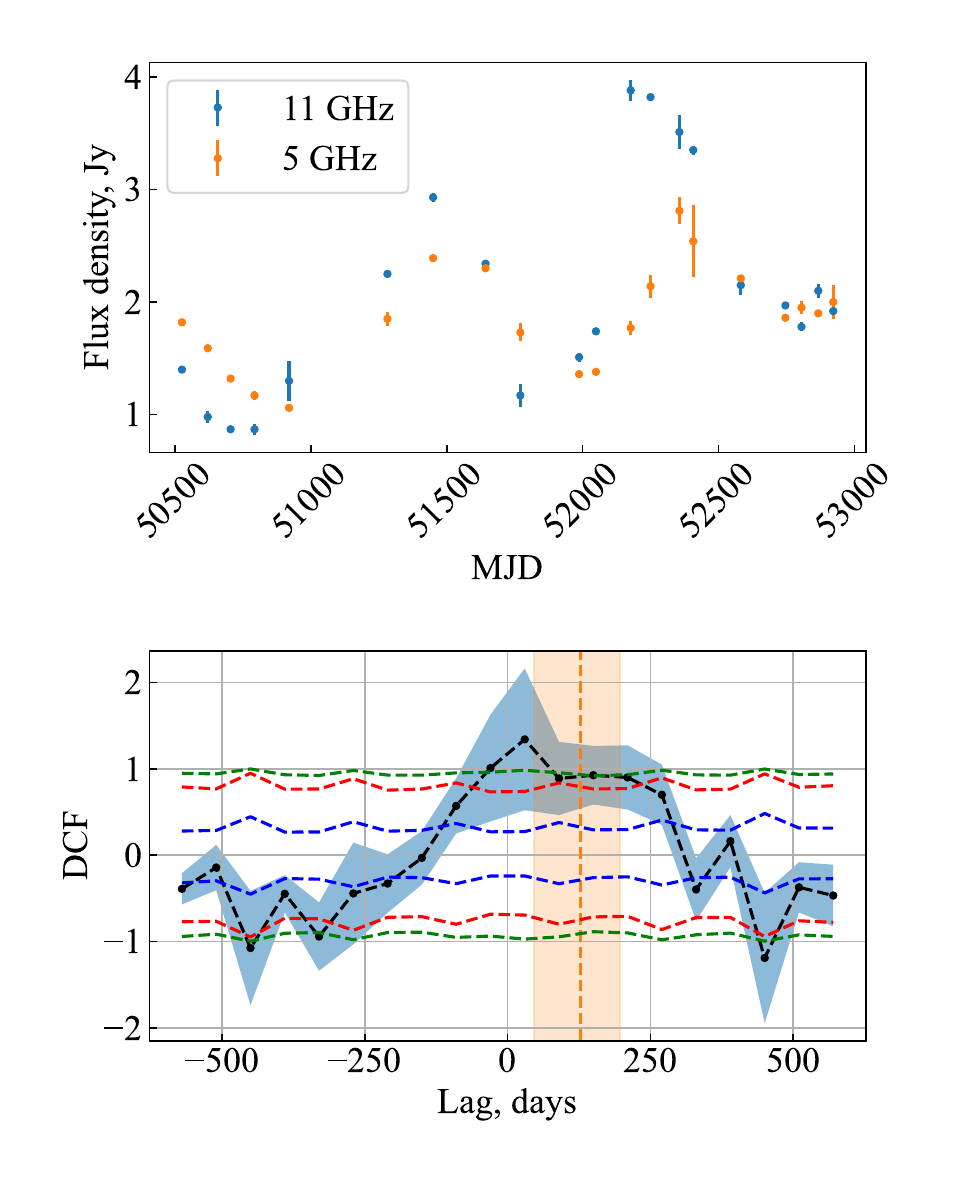}
}
\caption{The light curves (upper panels) and DCFs (lower panels) in epoch~1. The pairs of frequencies are designated in the legends.
The black lines with the blue areas are the DCF values with their uncertainties. The dashed blue, red, and green horizontal lines are the 1, 2, and 3$\sigma$ significance levels, respectively. The orange vertical lines are the lags measured by the FR/RSS method. The orange areas are the lag uncertainties (16th and 84th percentiles).} 
\label{fig:dcf_ep1}
\end{figure*}

\begin{figure*}
\centerline{
\includegraphics[width=0.7\columnwidth]{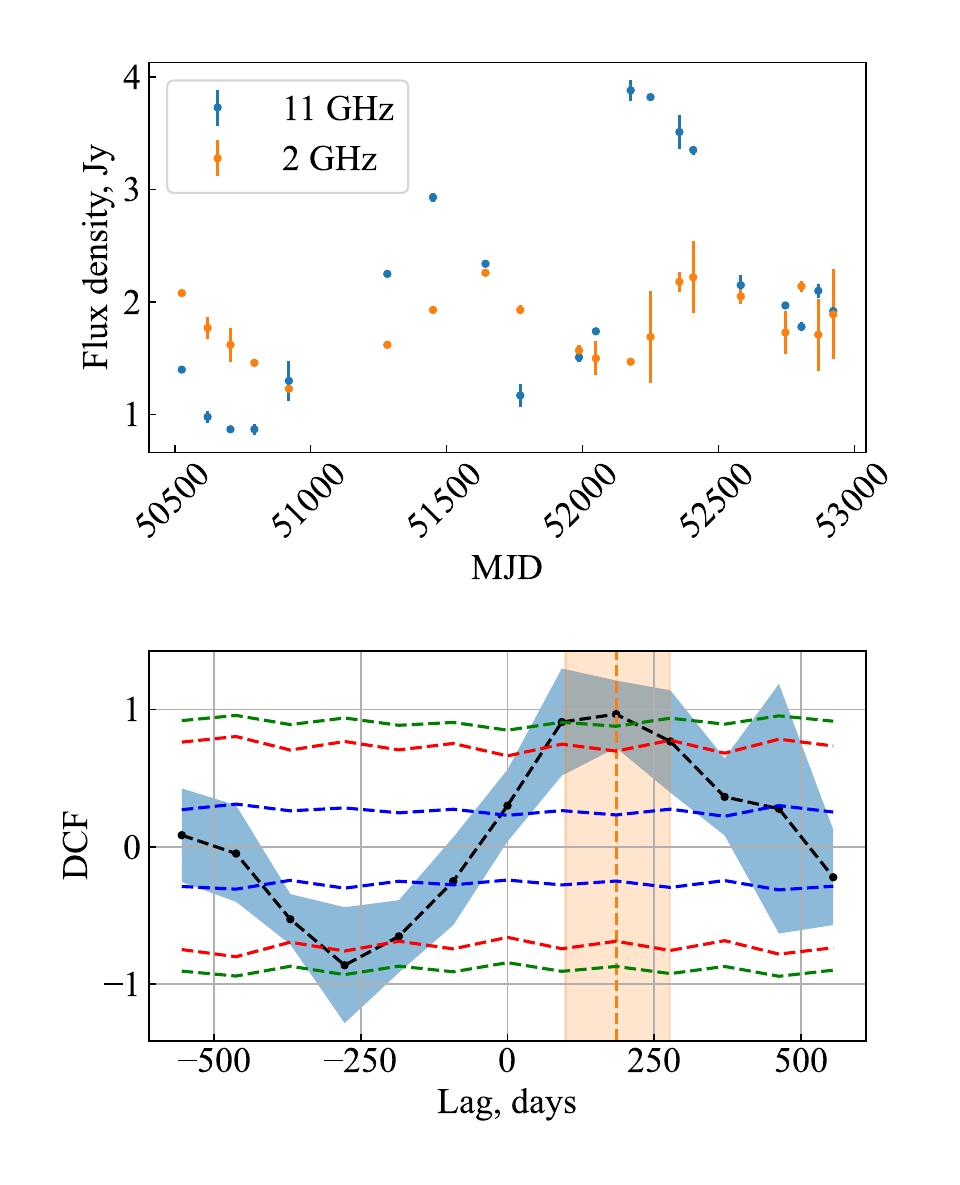}
\includegraphics[width=0.7\columnwidth]{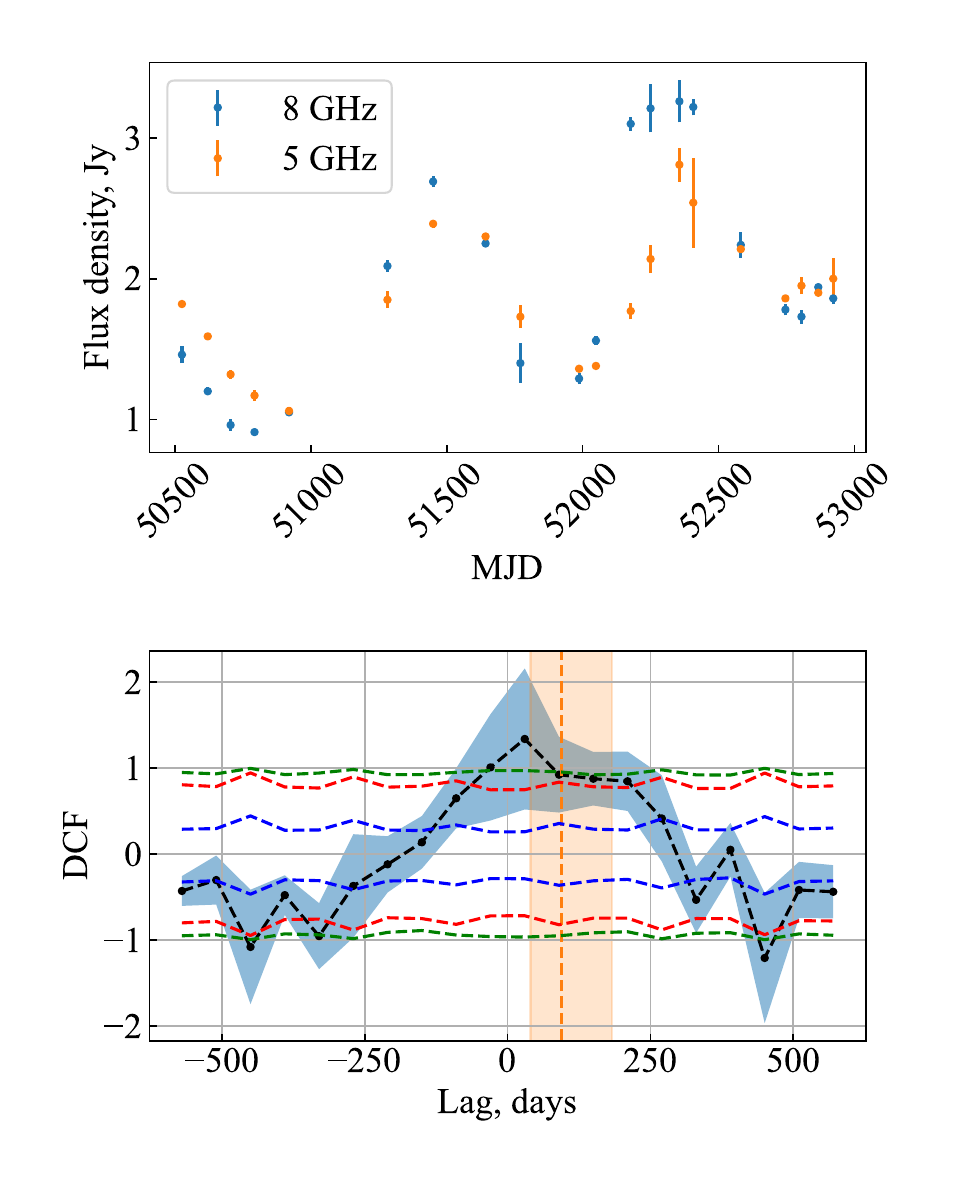}
\includegraphics[width=0.7\columnwidth]{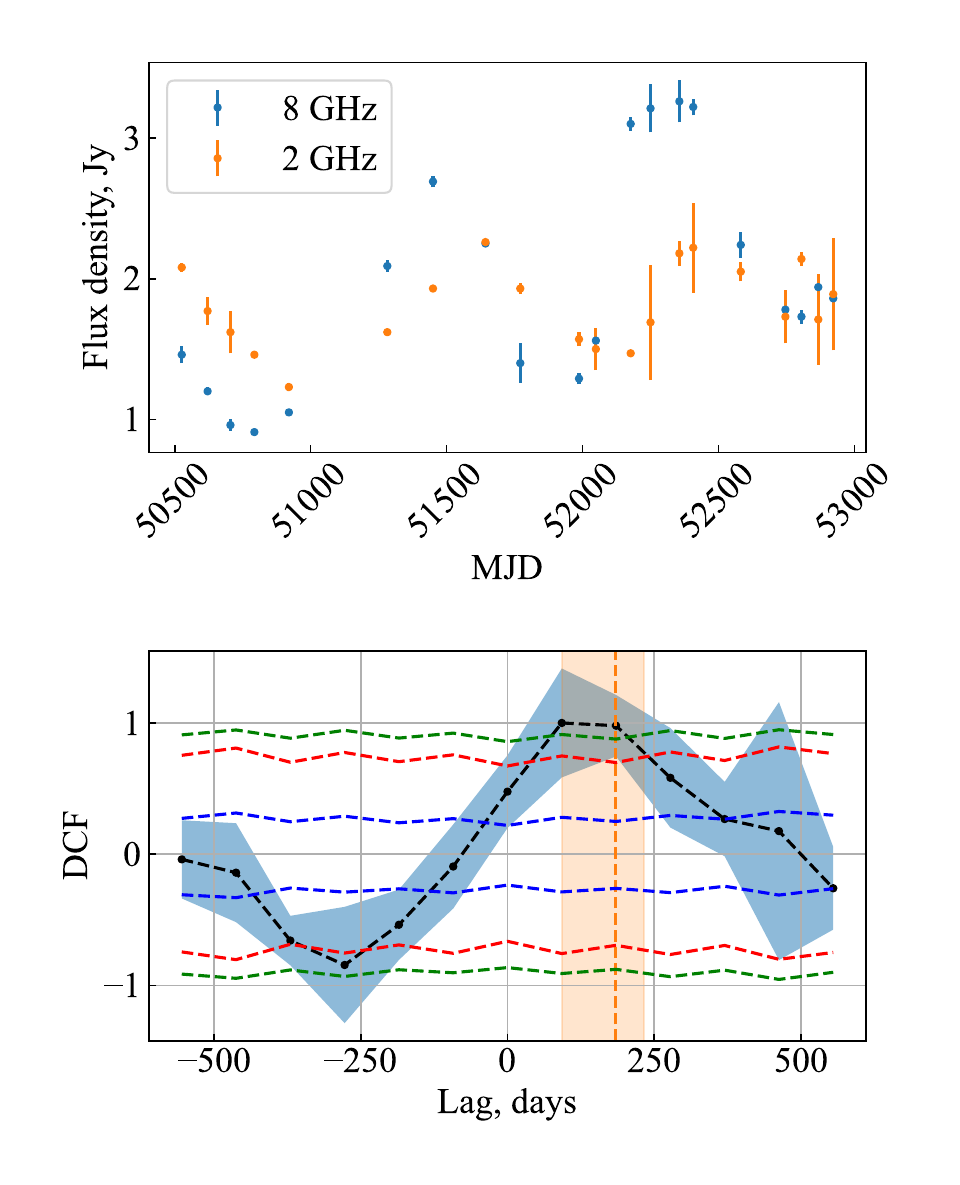}
}
\centerline{
\includegraphics[width=0.7\columnwidth]{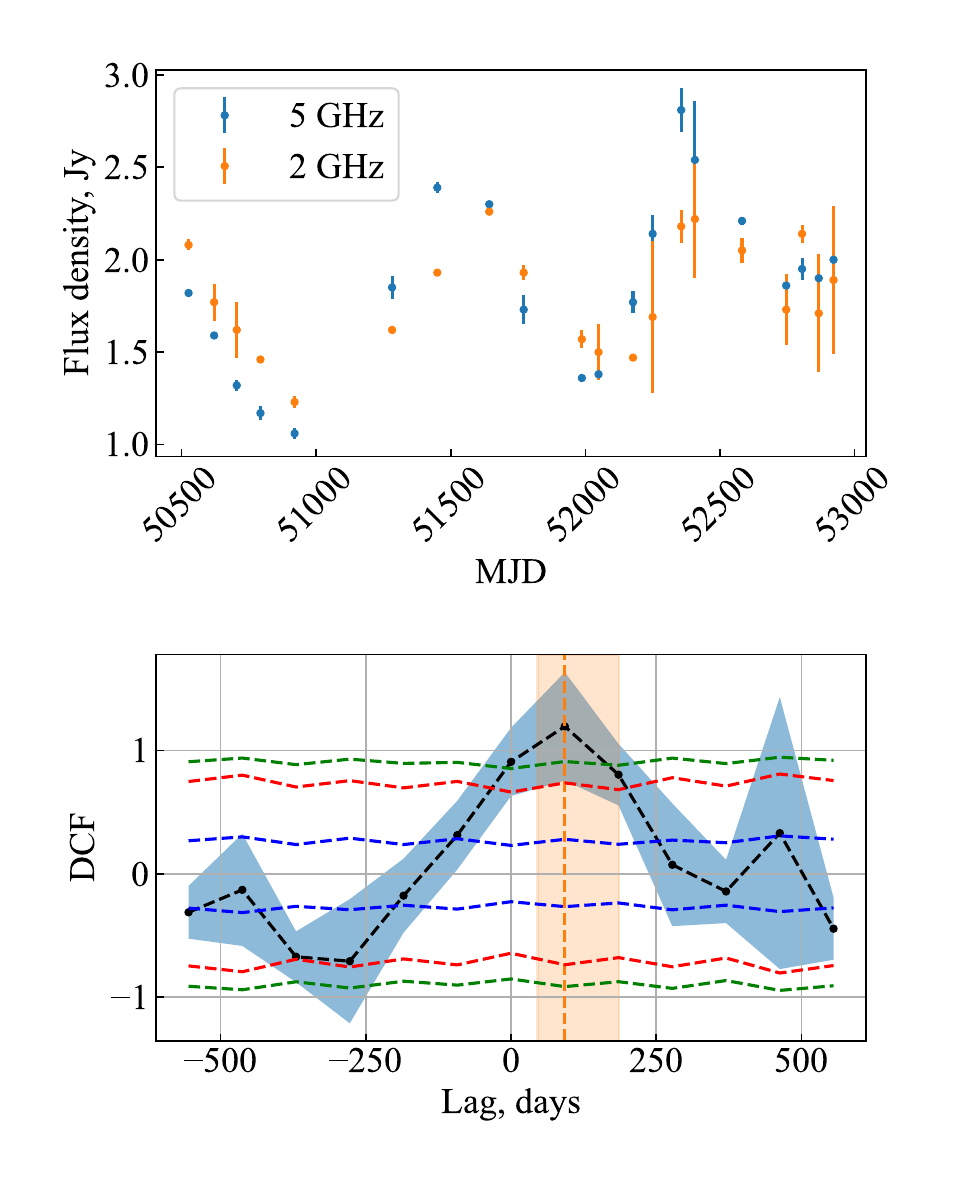}
}
\contcaption{The light curves and DCFs in epoch~1.} 
\end{figure*}


\begin{figure*}
\centerline{
\includegraphics[width=0.7\columnwidth]{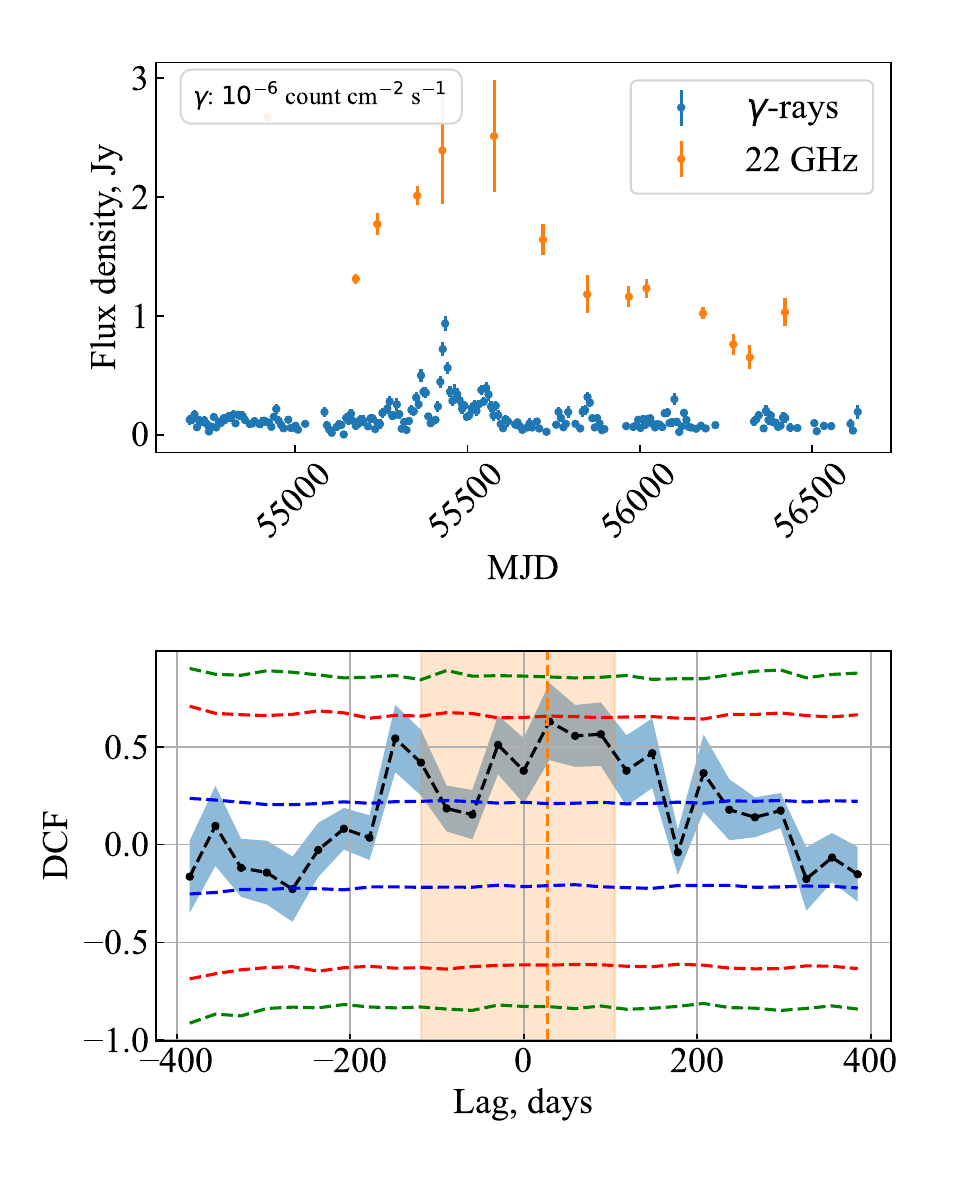}
\includegraphics[width=0.7\columnwidth]{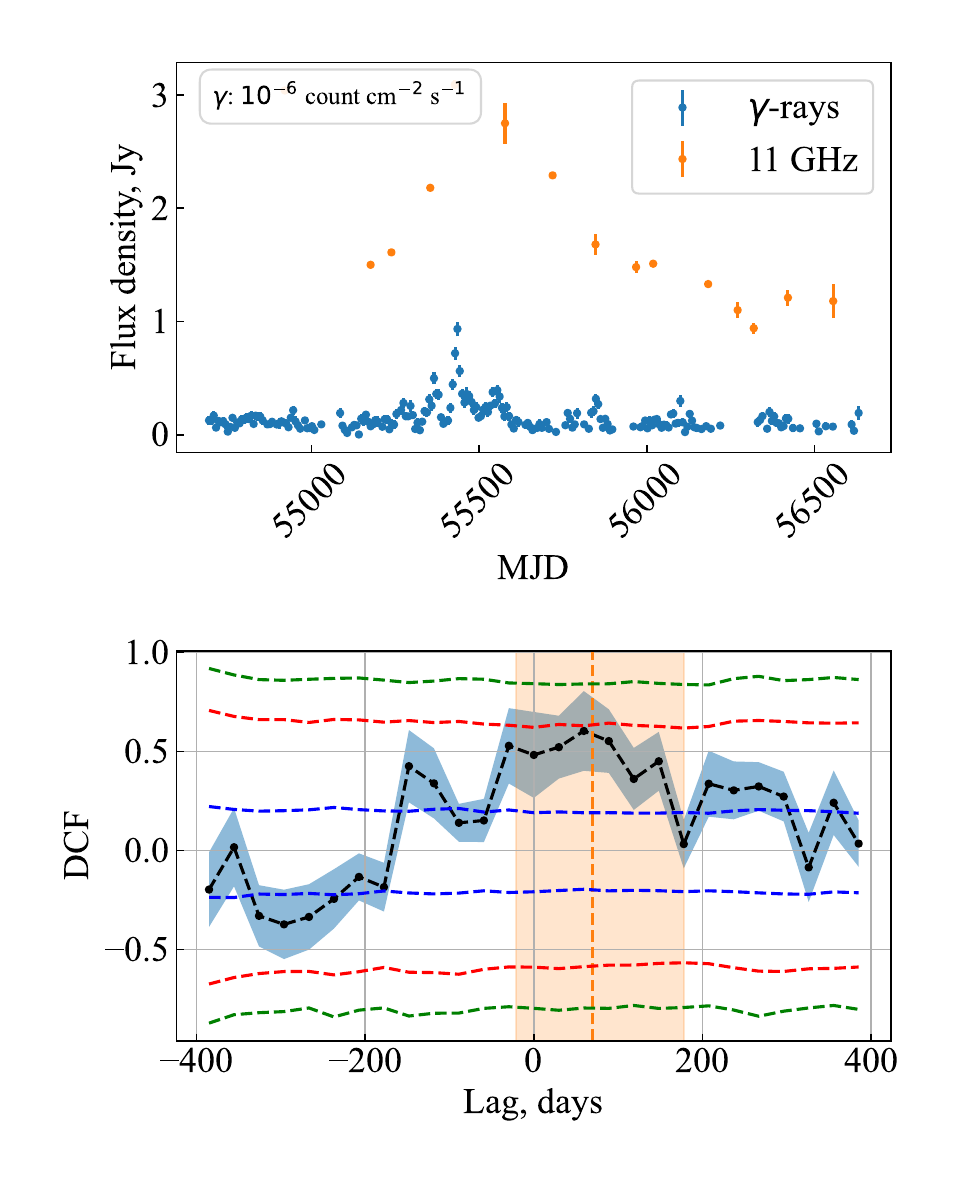}
\includegraphics[width=0.7\columnwidth]{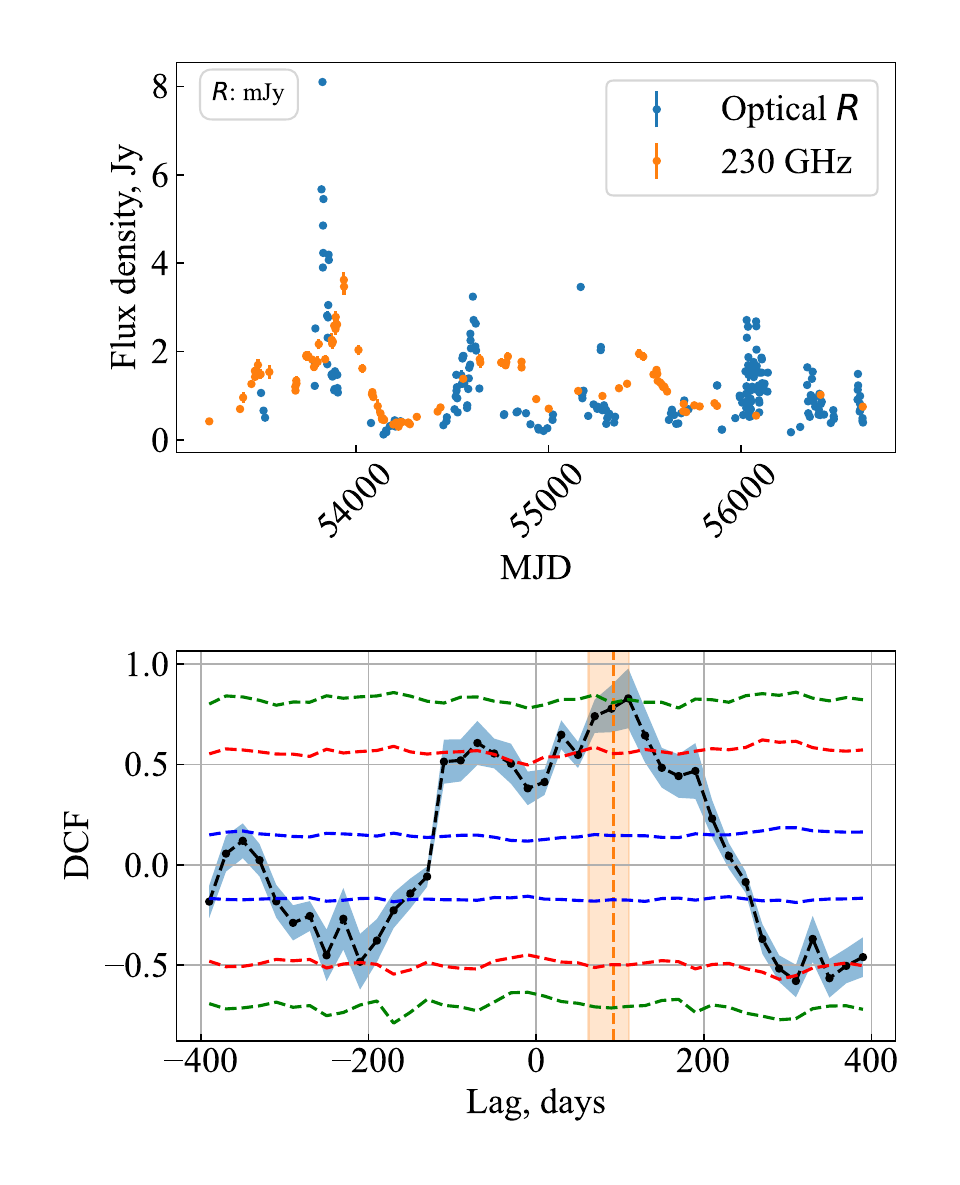}
}
\centerline{
\includegraphics[width=0.7\columnwidth]{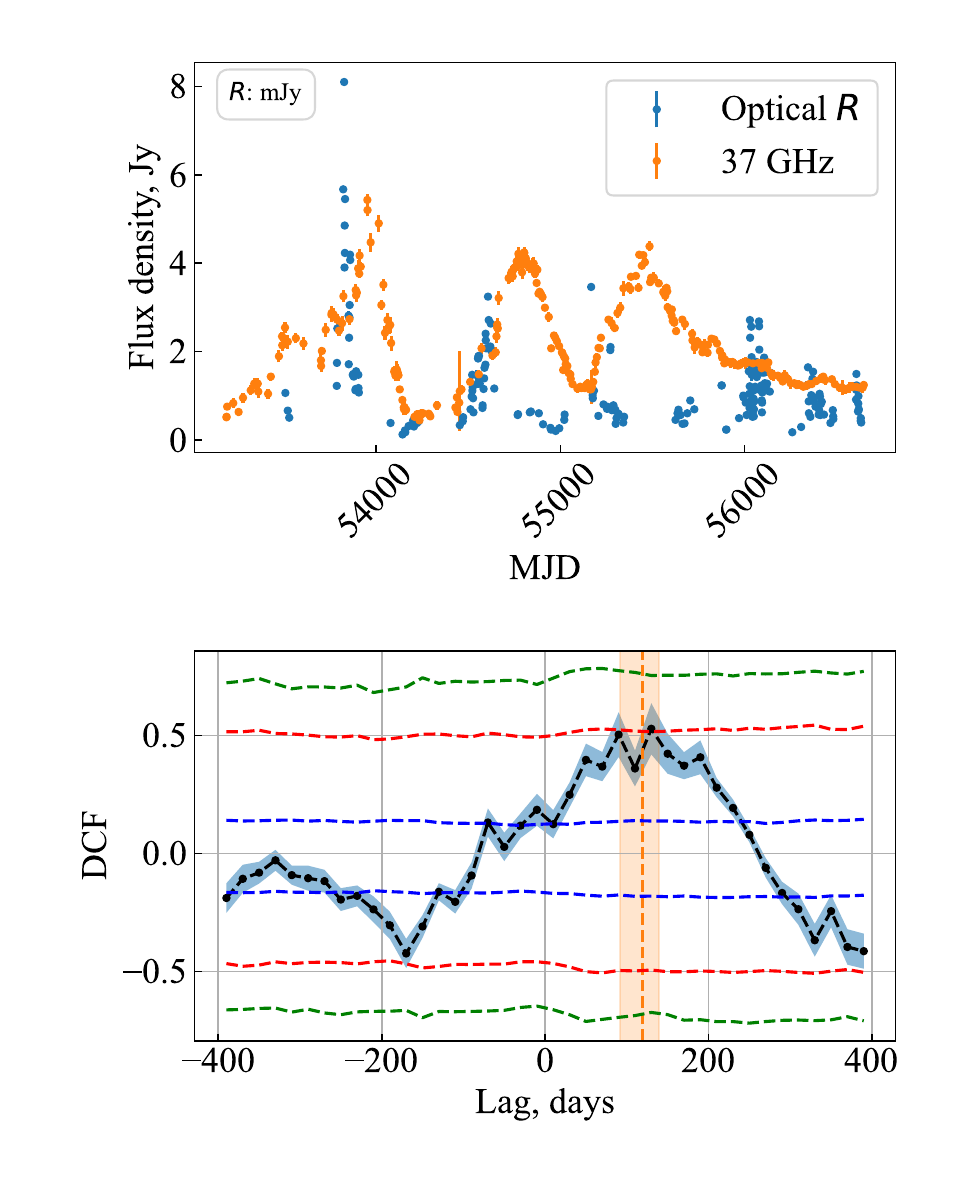}
\includegraphics[width=0.7\columnwidth]{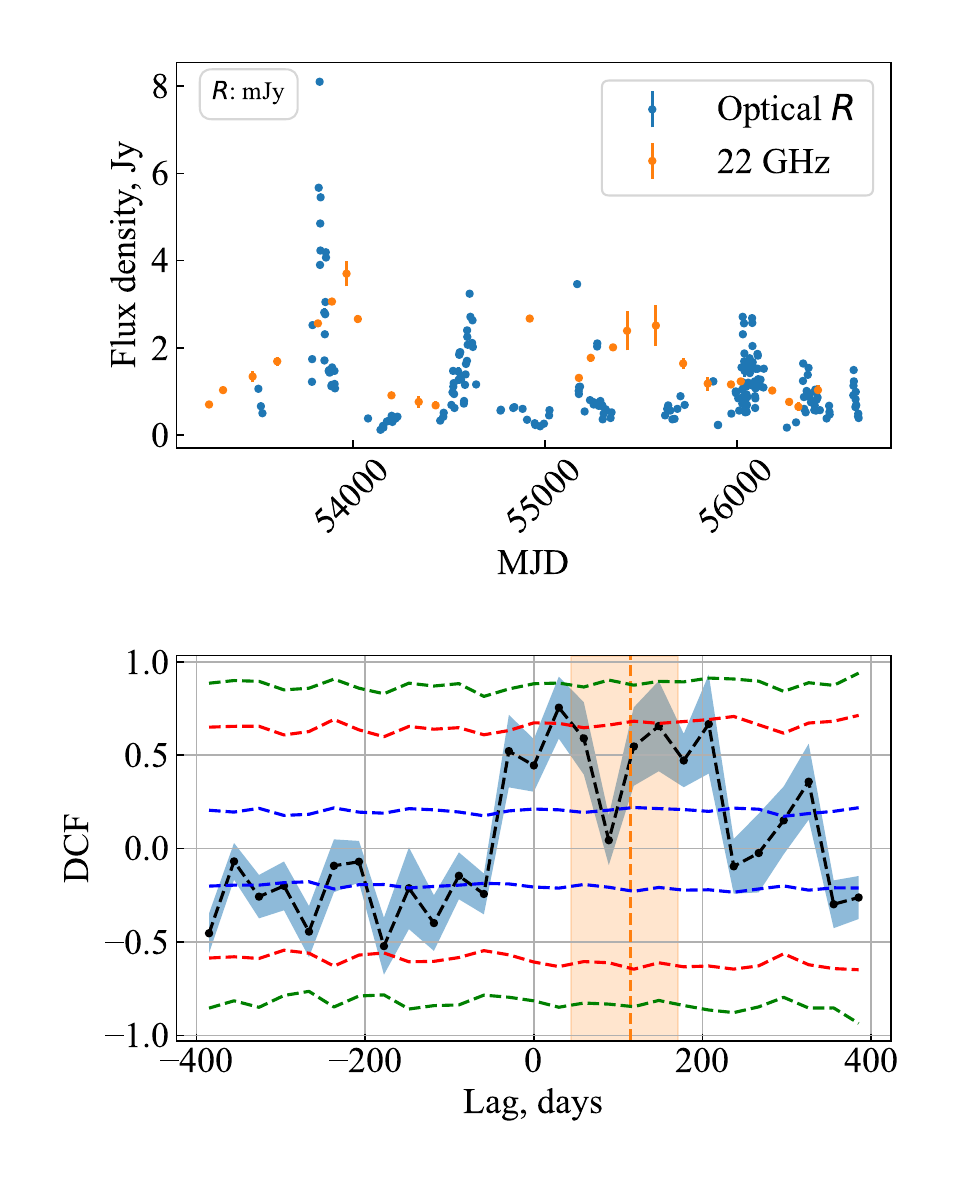}
\includegraphics[width=0.7\columnwidth]{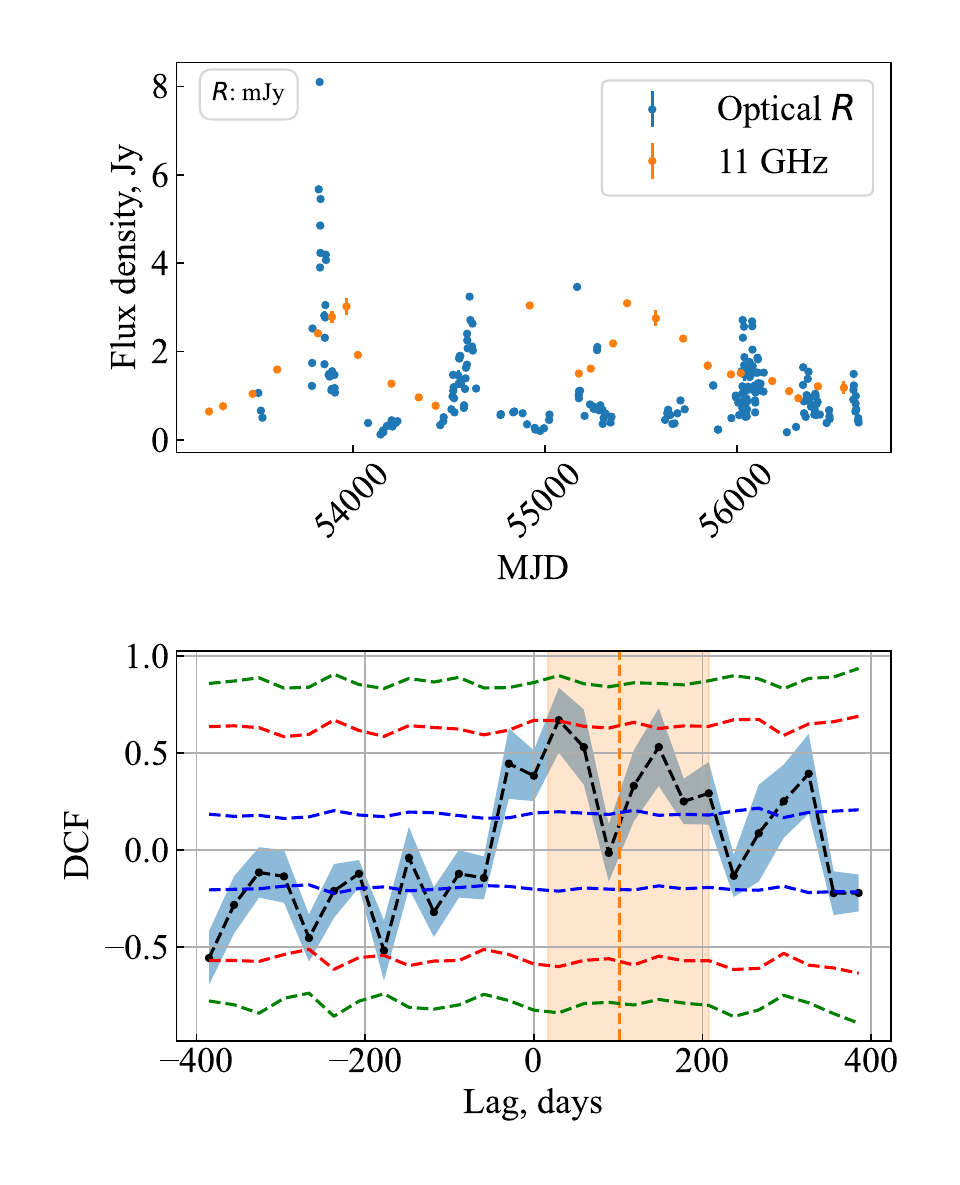}
}
\centerline{
\includegraphics[width=0.7\columnwidth]{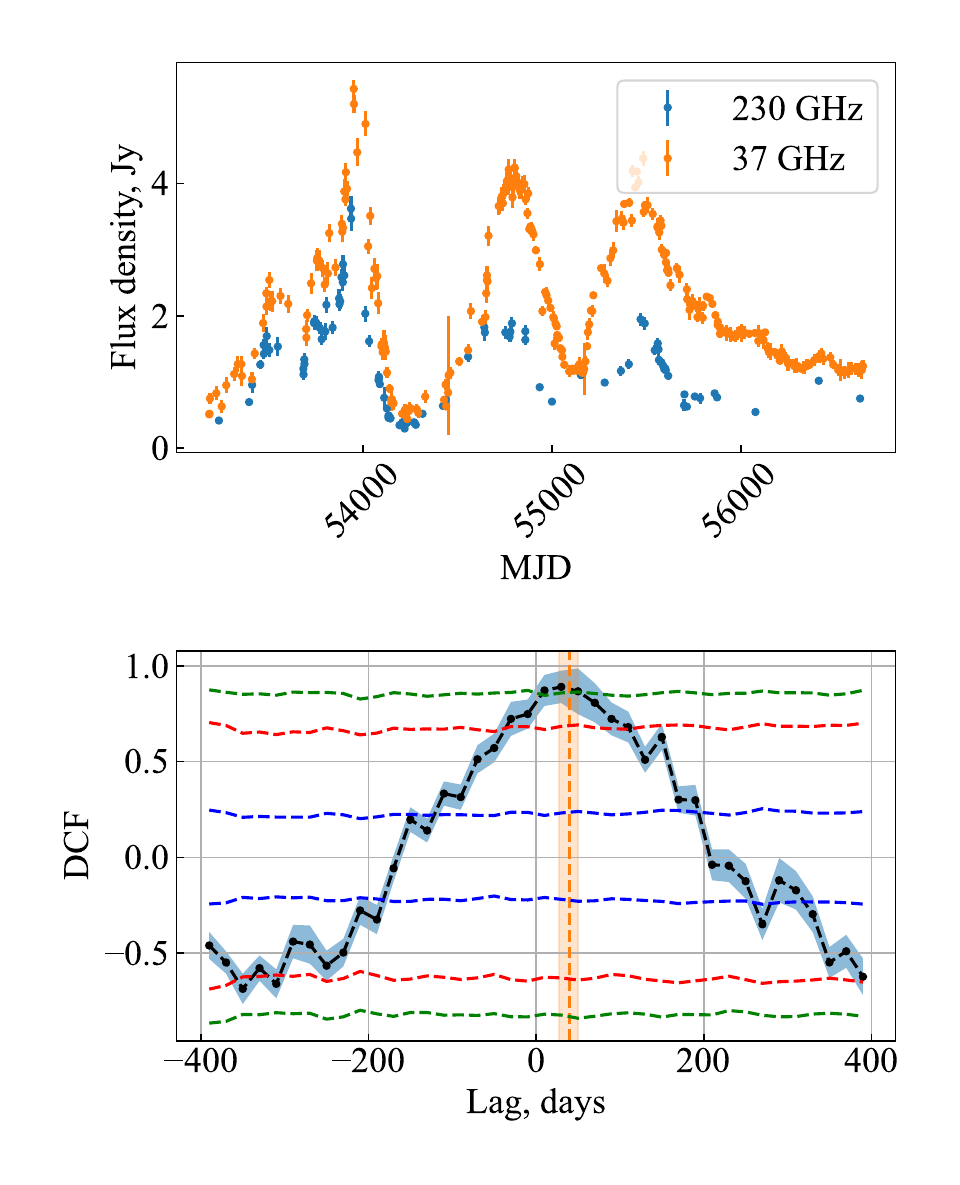}
\includegraphics[width=0.7\columnwidth]{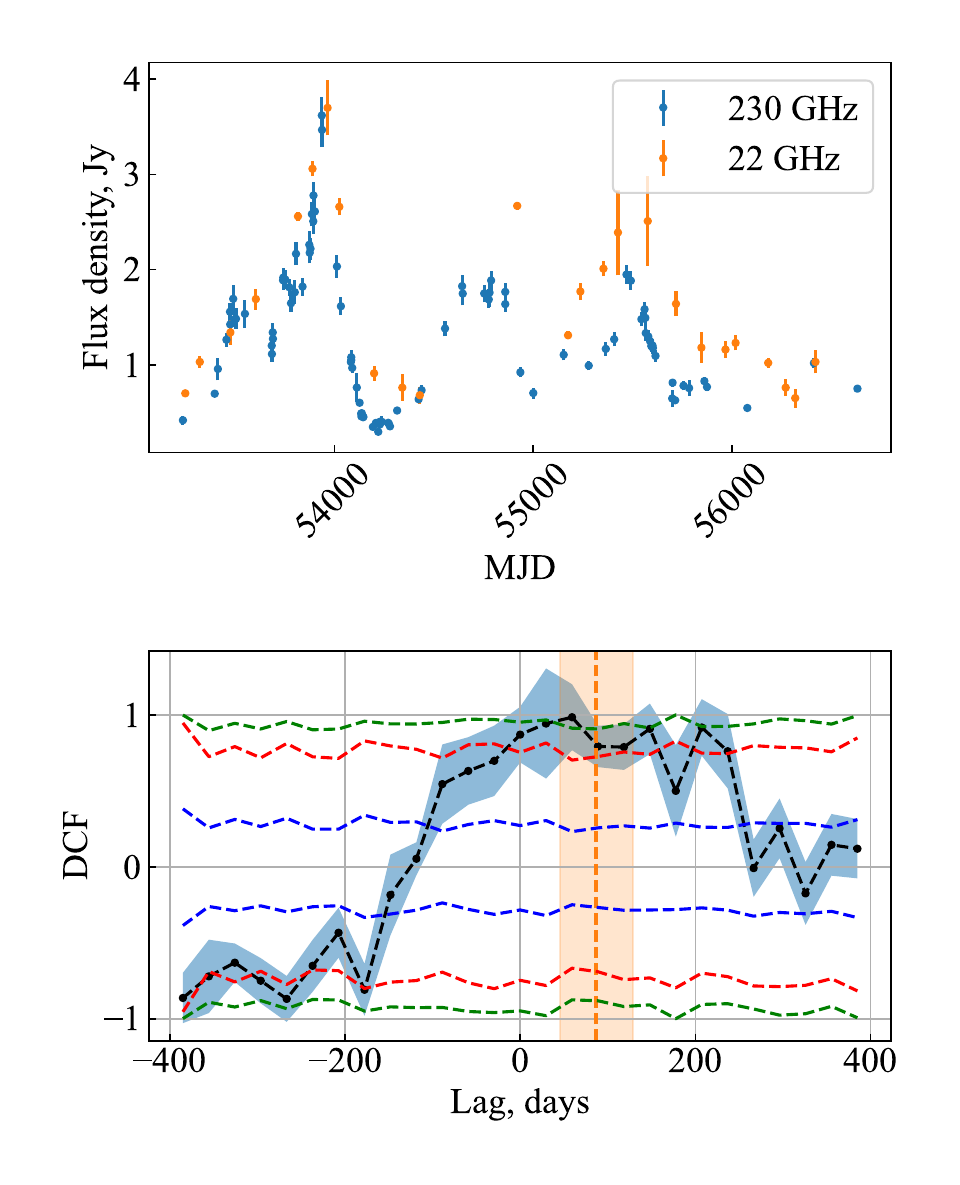}
\includegraphics[width=0.7\columnwidth]{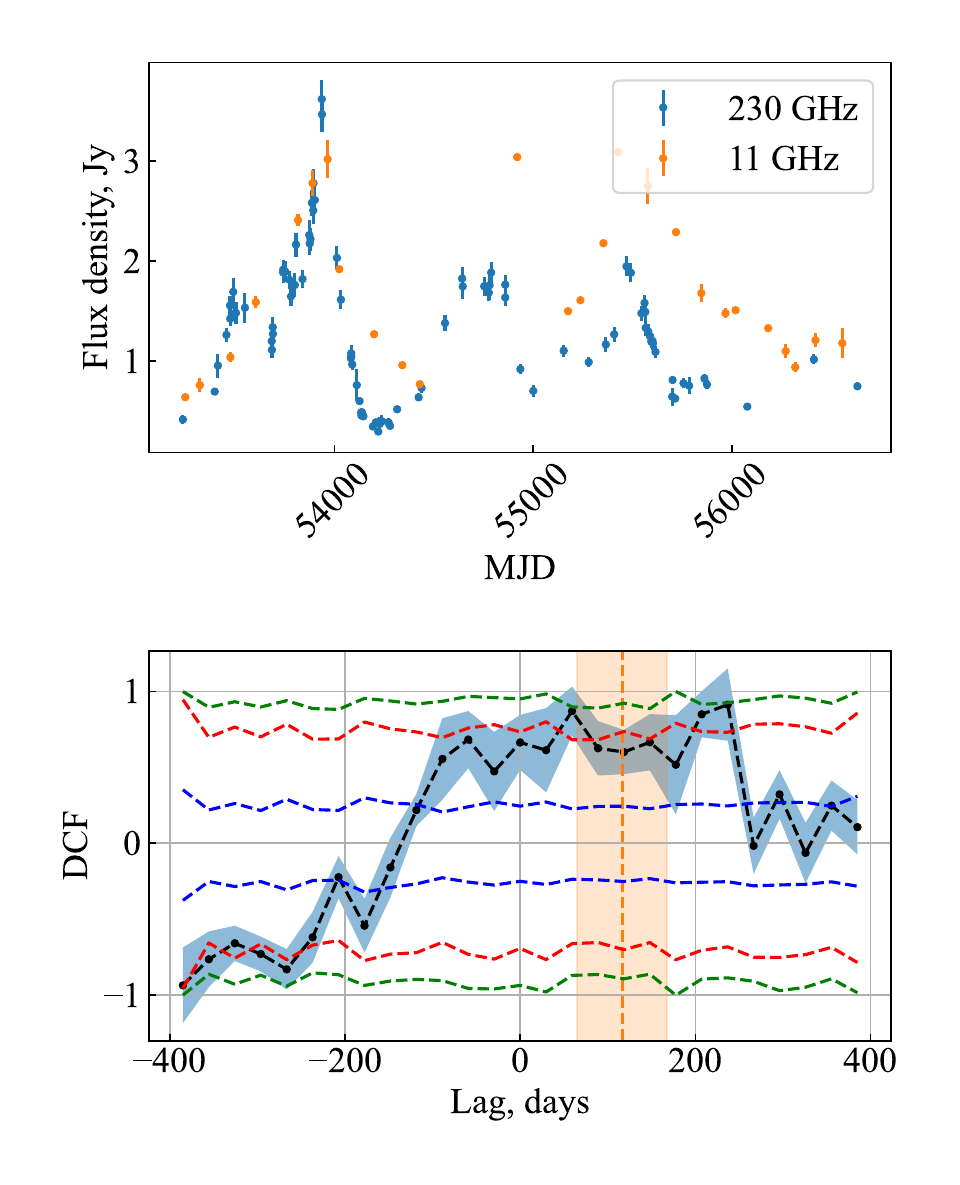}
}
\caption{The light curves and DCFs in epoch~2. Designations are as in Fig.~\ref{fig:dcf_ep1}} 
\label{fig:dcf_ep2}
\end{figure*}

\begin{figure*}
\centerline{
\includegraphics[width=0.7\columnwidth]{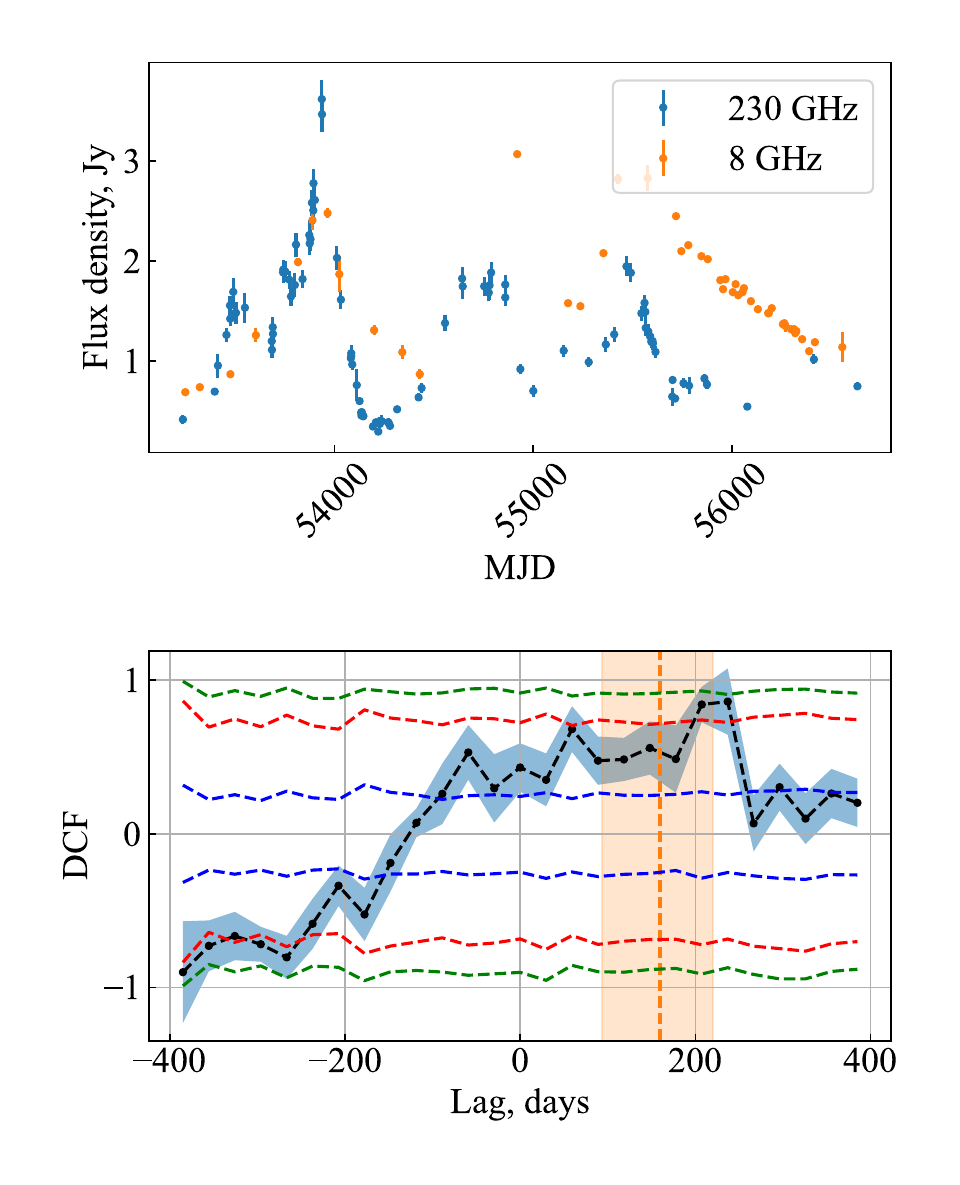}
\includegraphics[width=0.7\columnwidth]{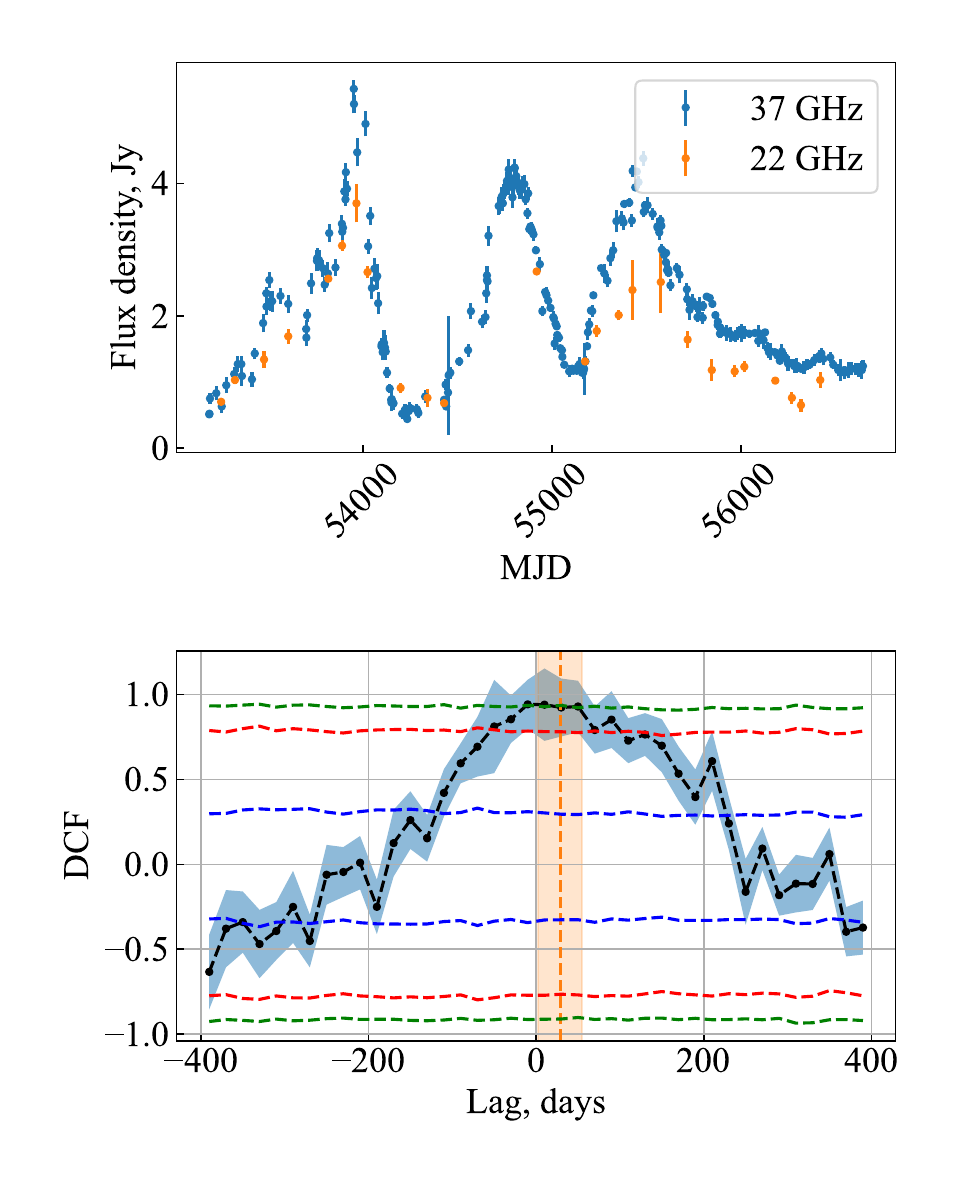}
\includegraphics[width=0.7\columnwidth]{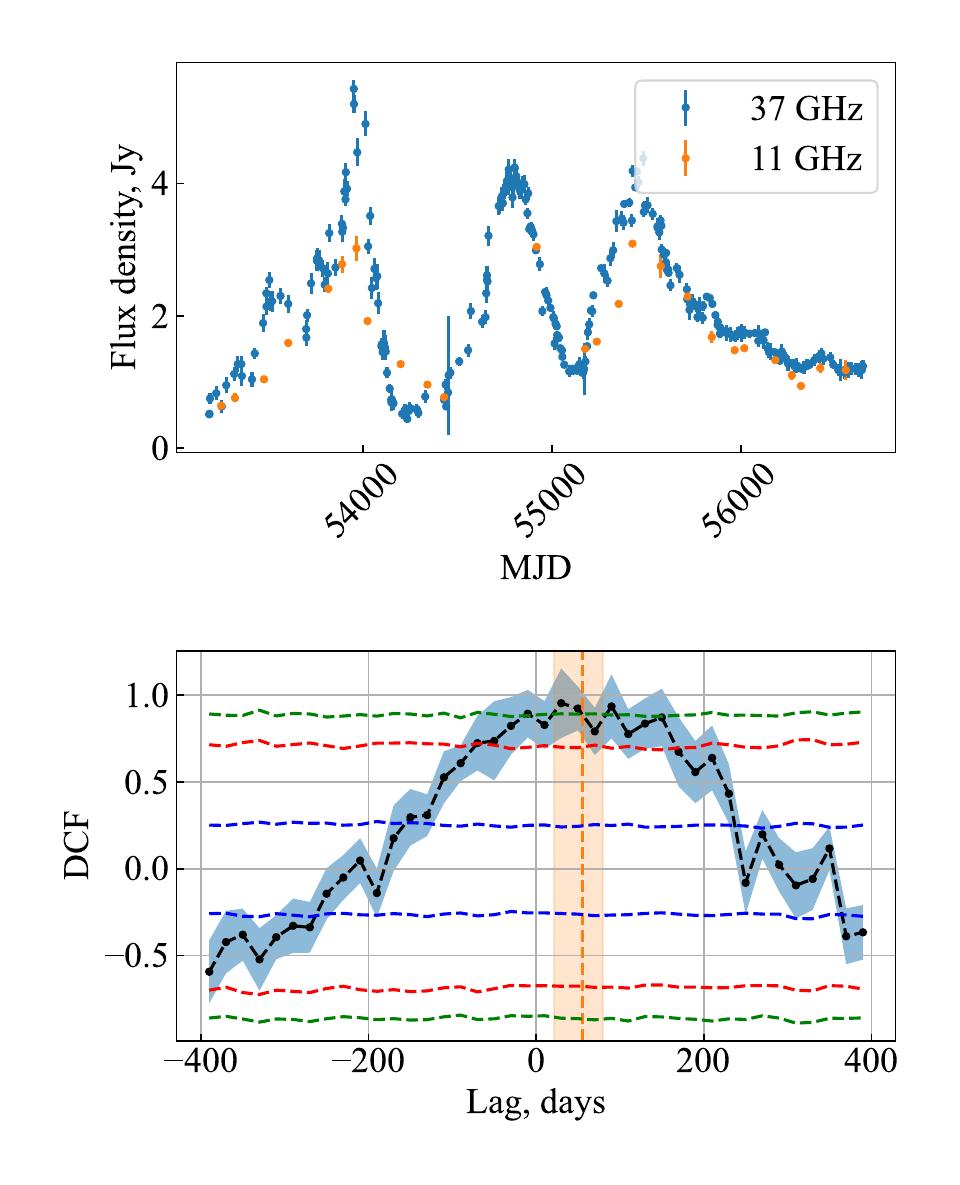}
}
\centerline{
\includegraphics[width=0.7\columnwidth]{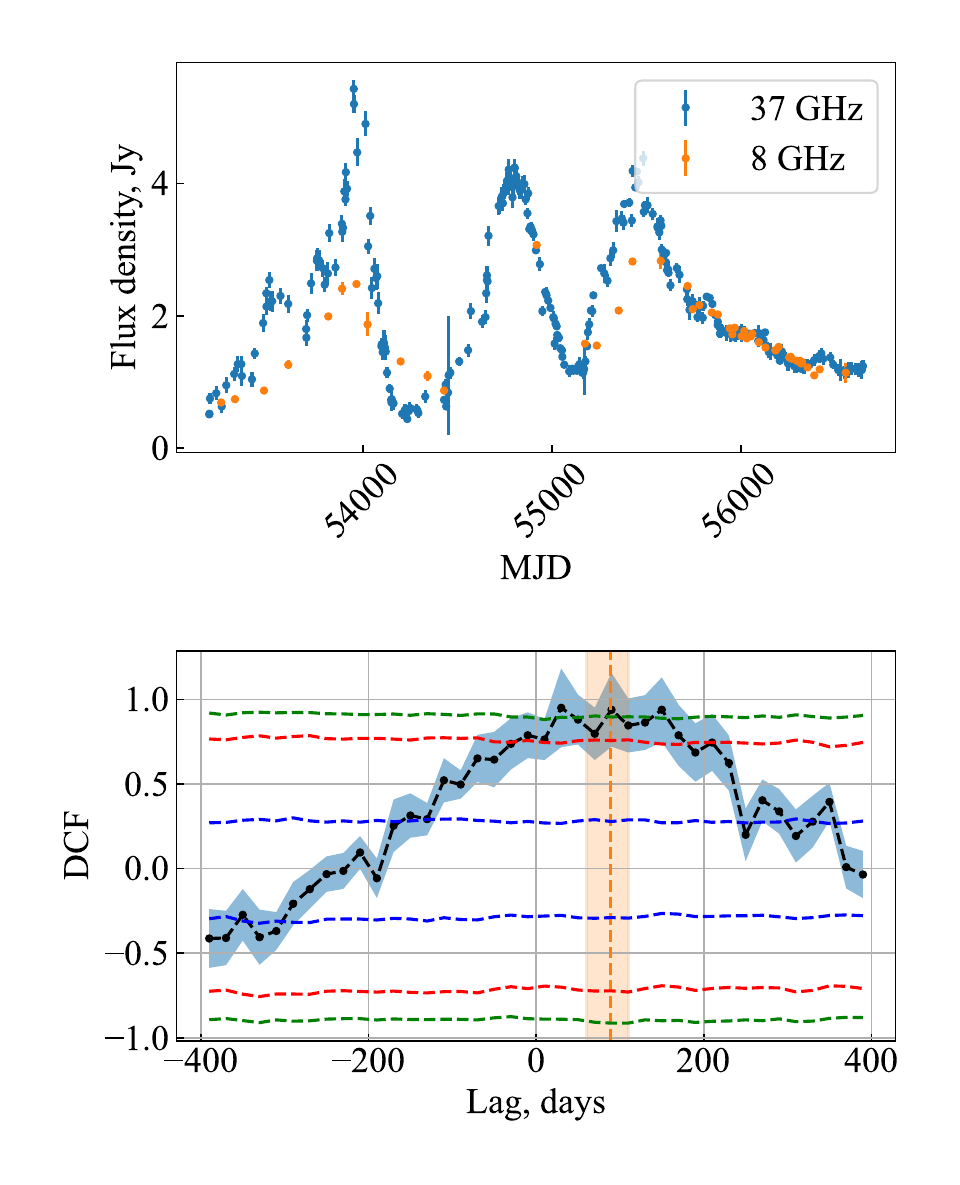}
\includegraphics[width=0.7\columnwidth]{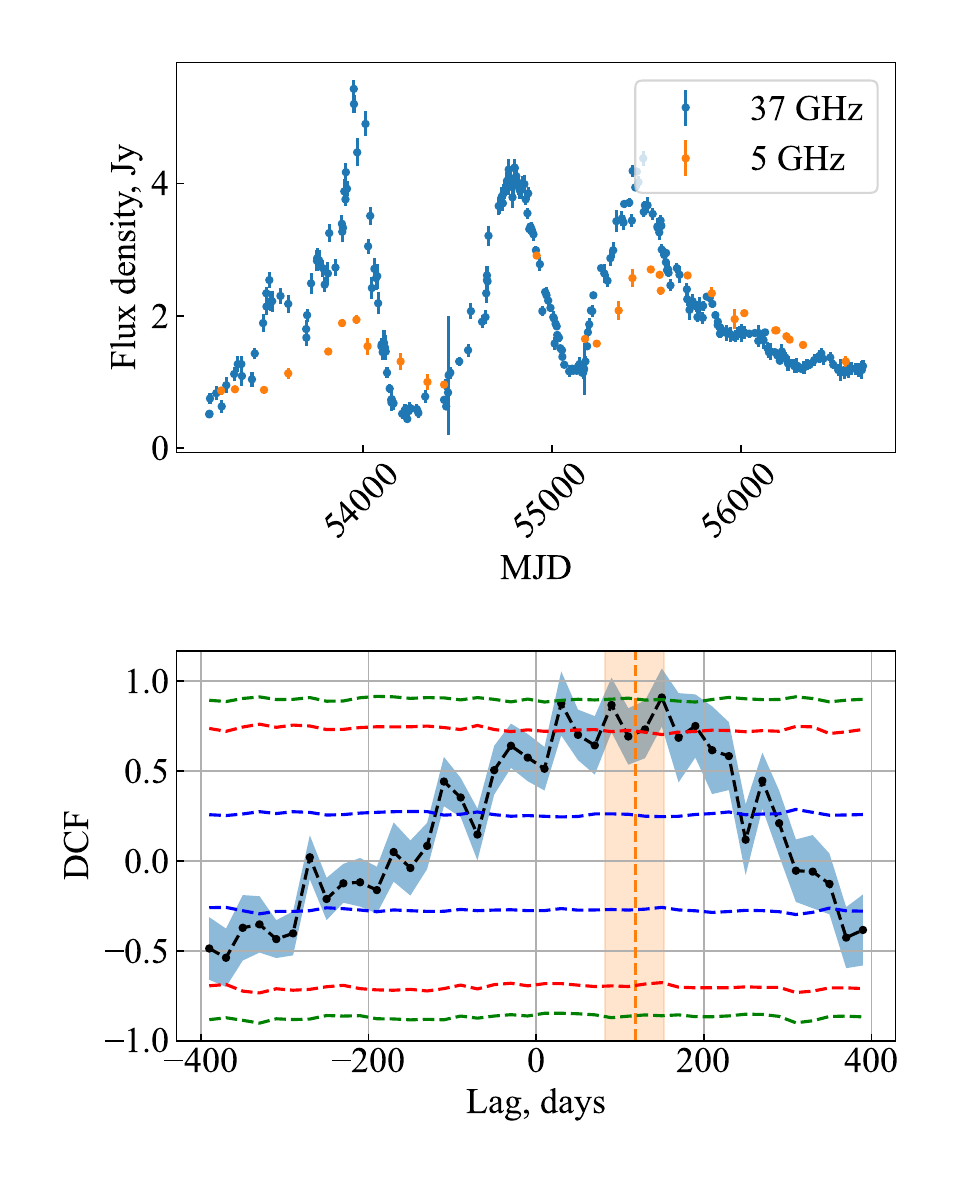}
\includegraphics[width=0.7\columnwidth]{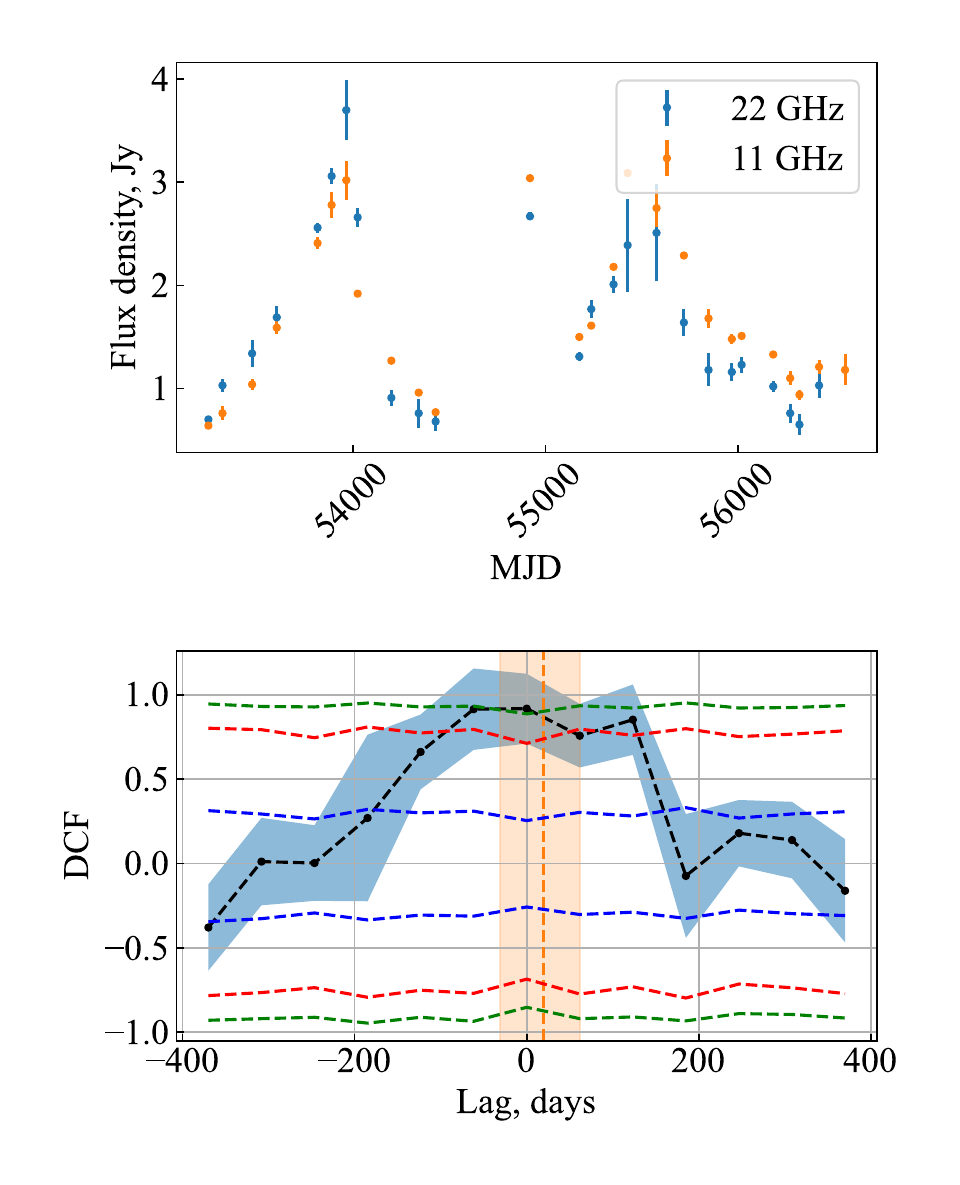}
}
\centerline{
\includegraphics[width=0.7\columnwidth]{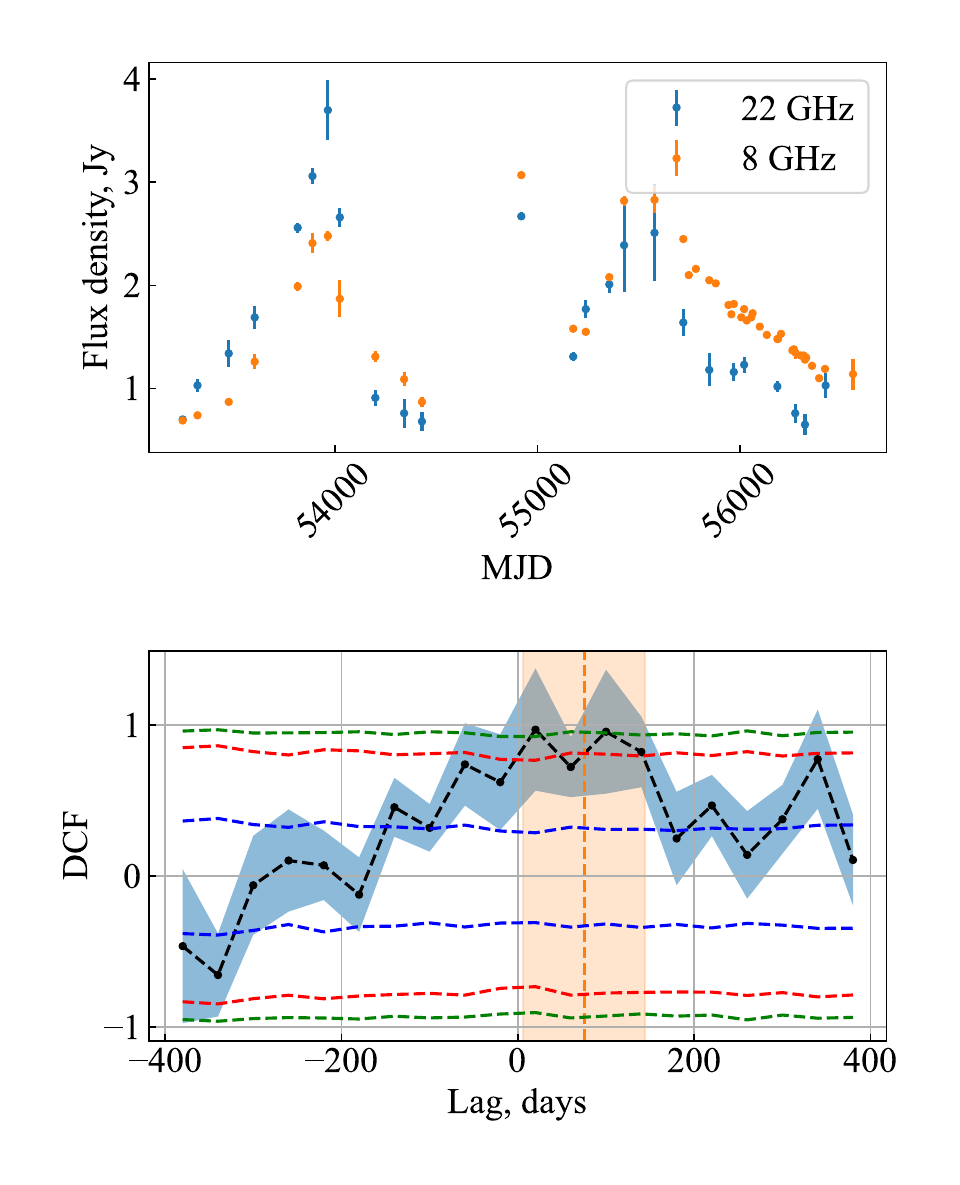}
\includegraphics[width=0.7\columnwidth]{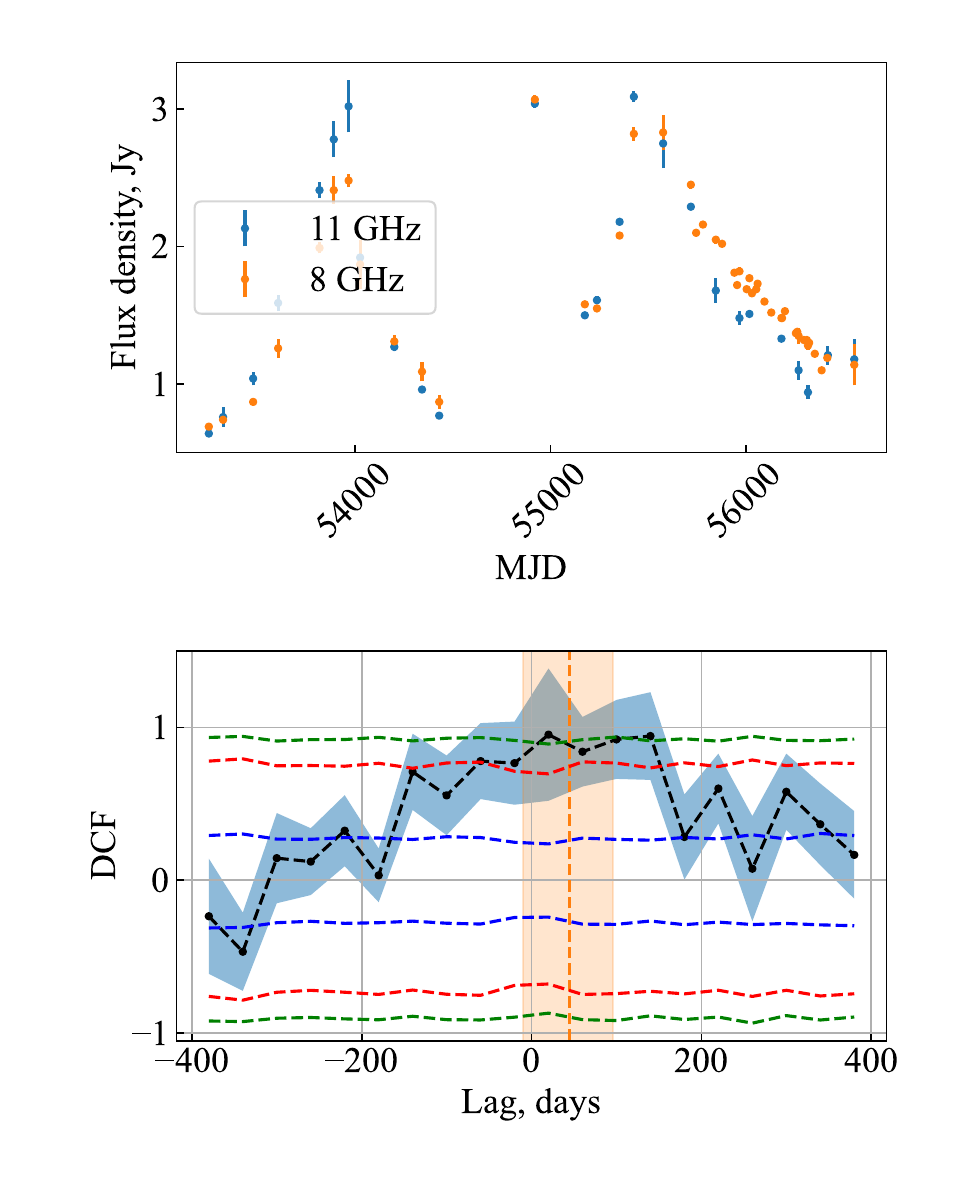}
\includegraphics[width=0.7\columnwidth]{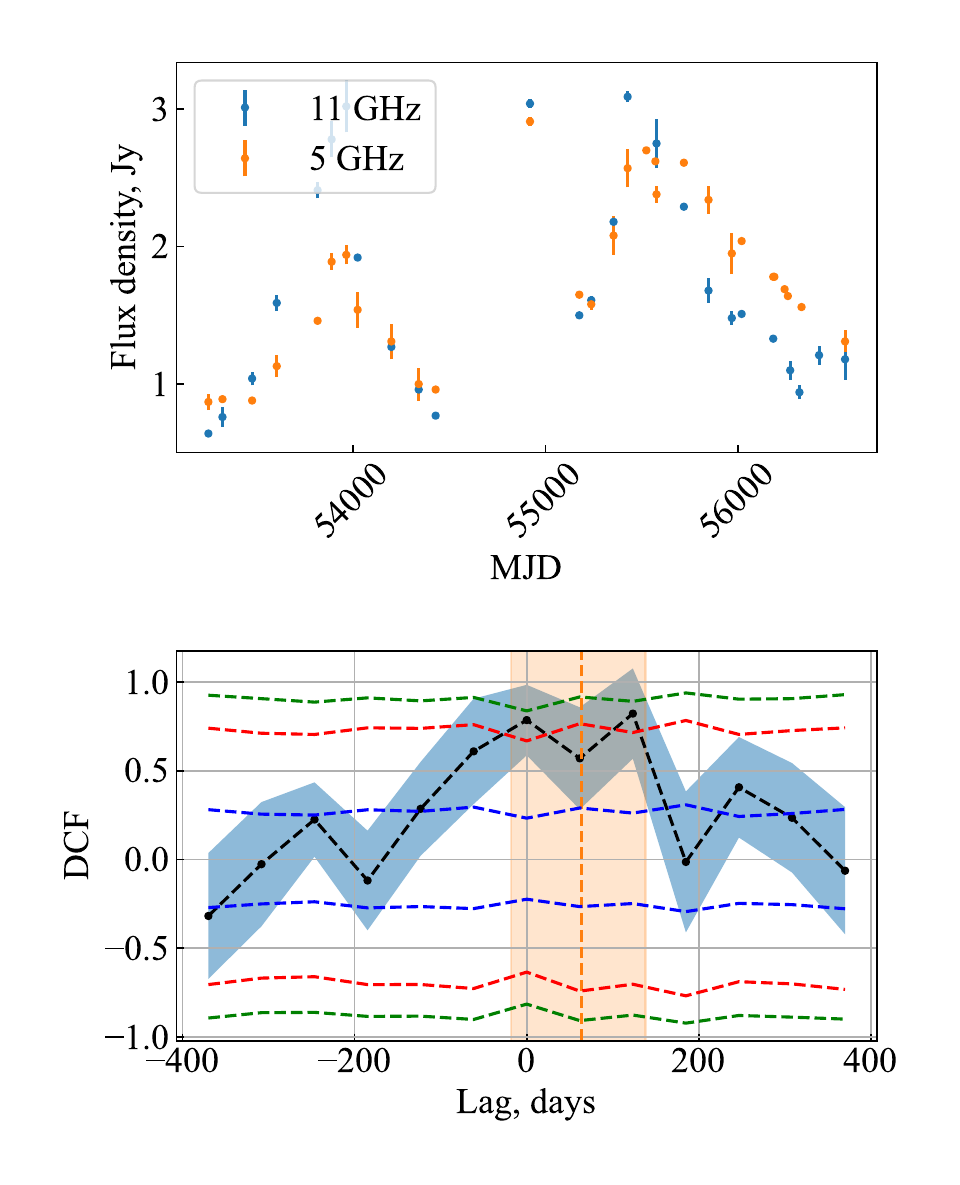}
}
\contcaption{The light curves and DCFs in epoch~2.} 
\end{figure*}

\begin{figure*}
\centerline{
\includegraphics[width=0.7\columnwidth]{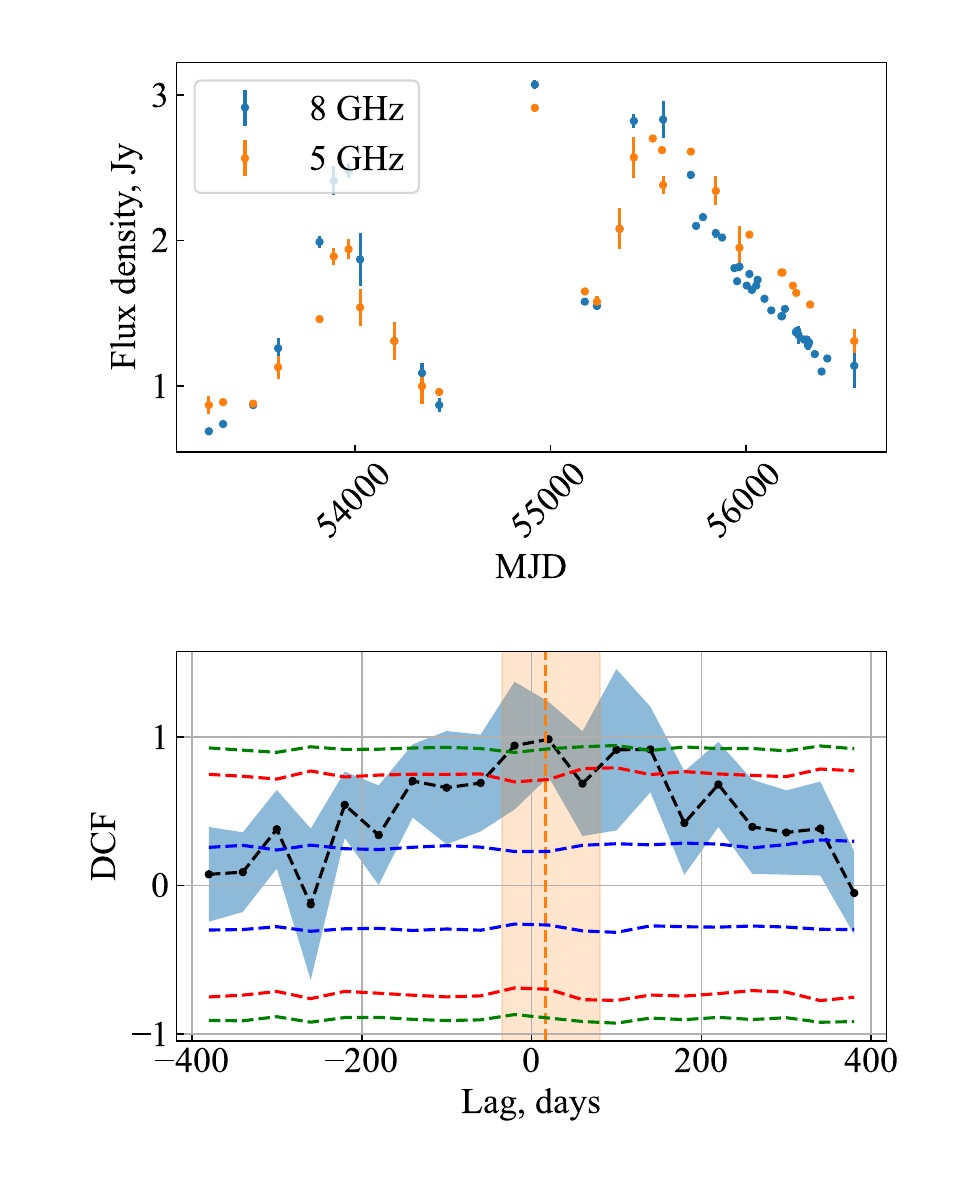}
}
\contcaption{The light curves and DCFs in epoch~2.} 
\end{figure*}


\begin{figure*}
\centerline{
\includegraphics[width=0.7\columnwidth]{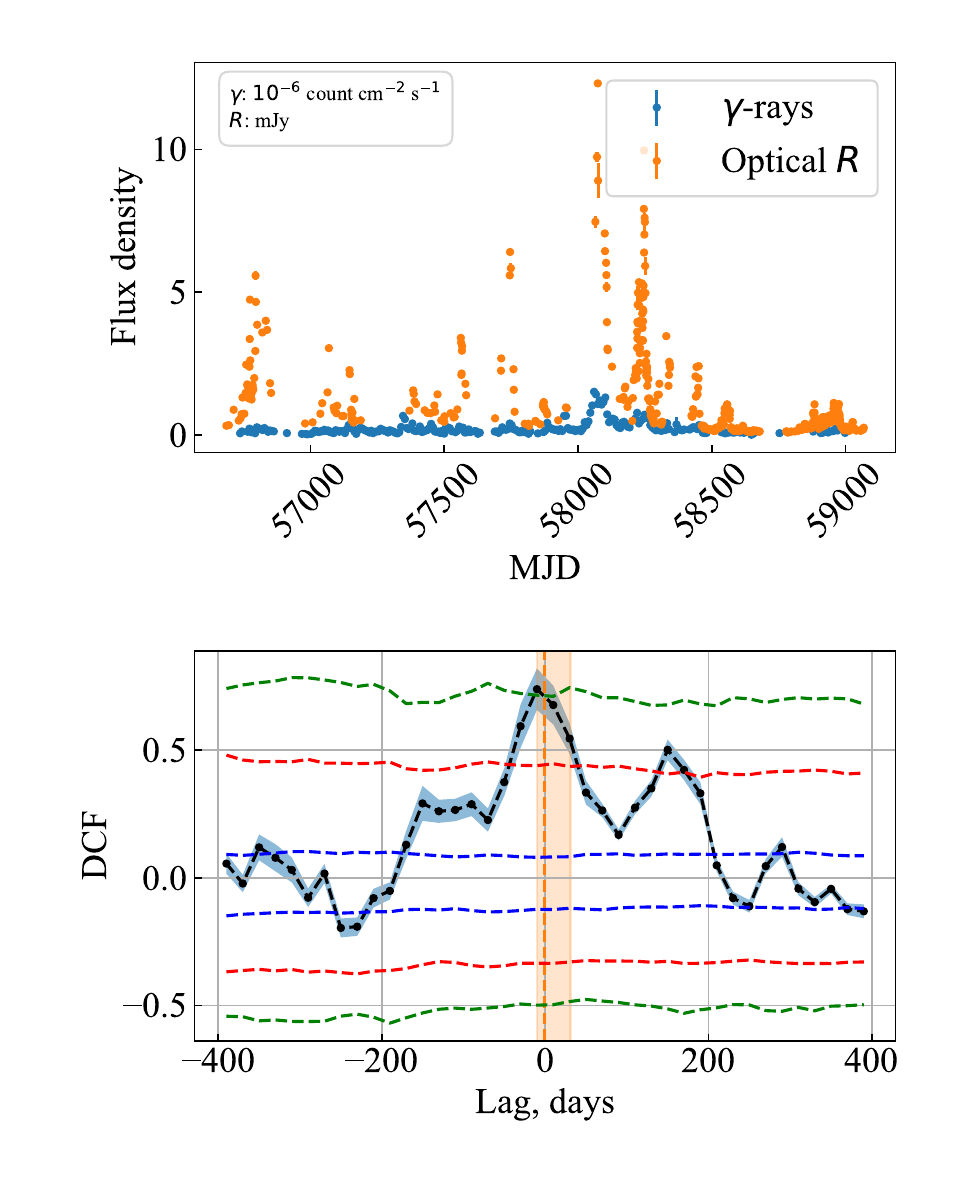}
\includegraphics[width=0.7\columnwidth]{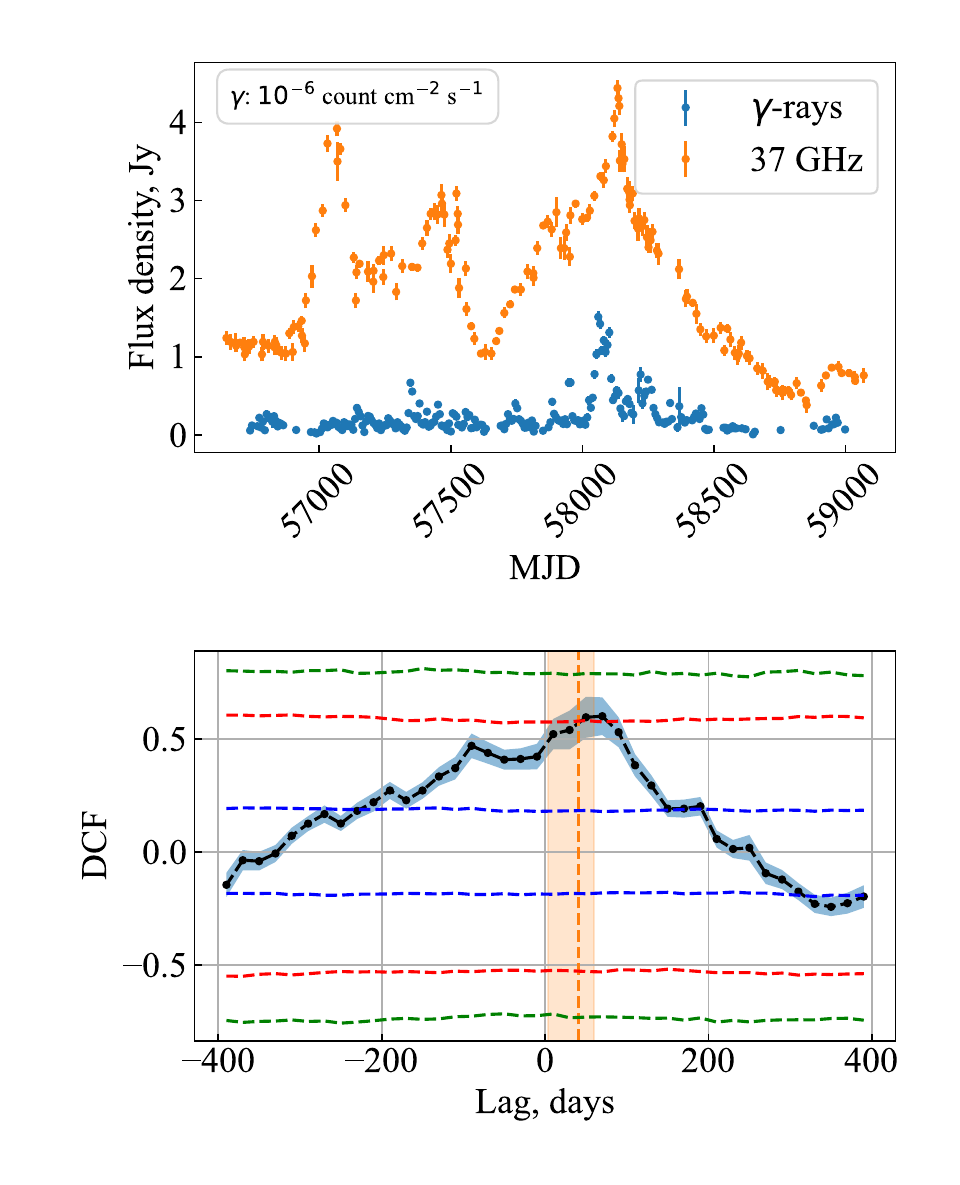}
\includegraphics[width=0.7\columnwidth]{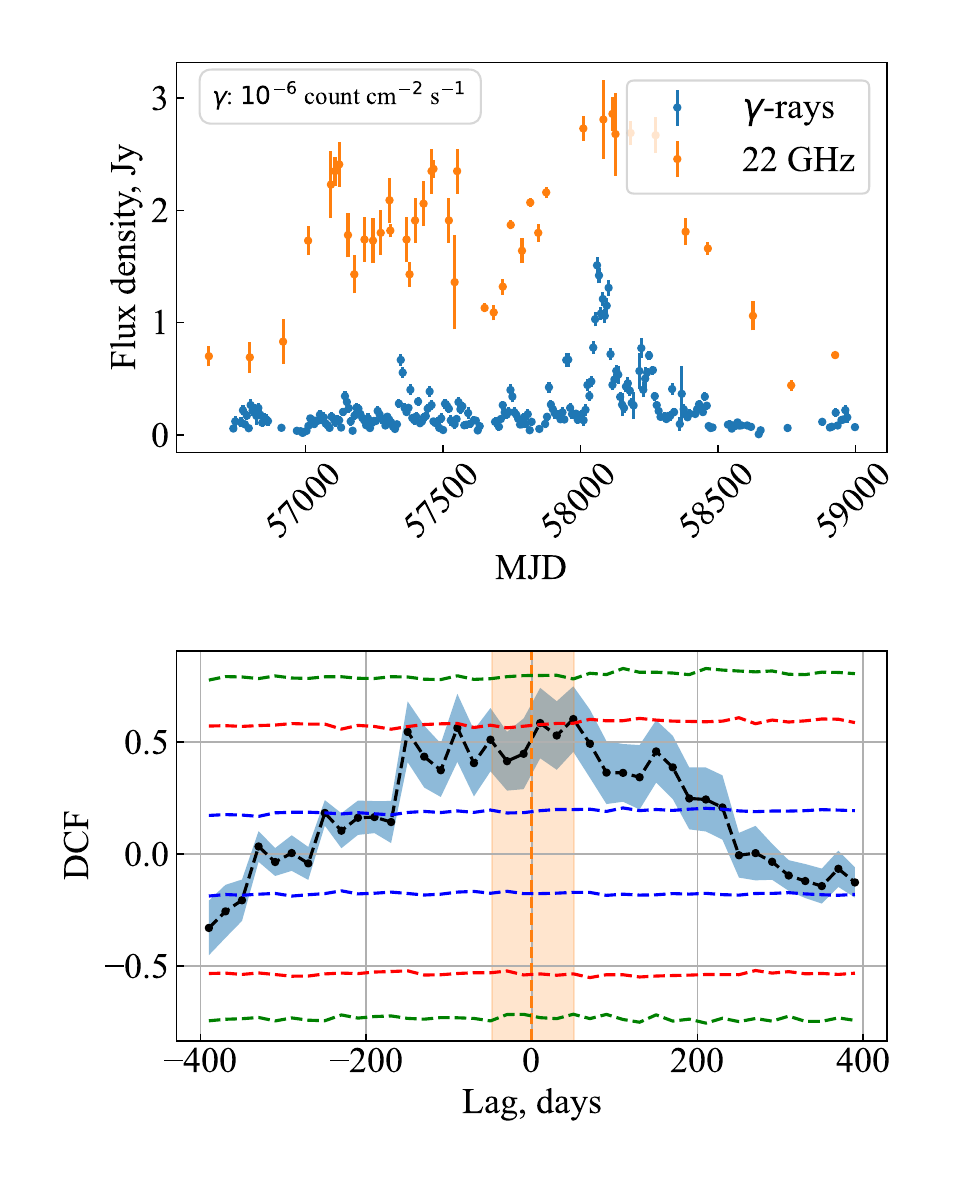}
}
\centerline{
\includegraphics[width=0.7\columnwidth]{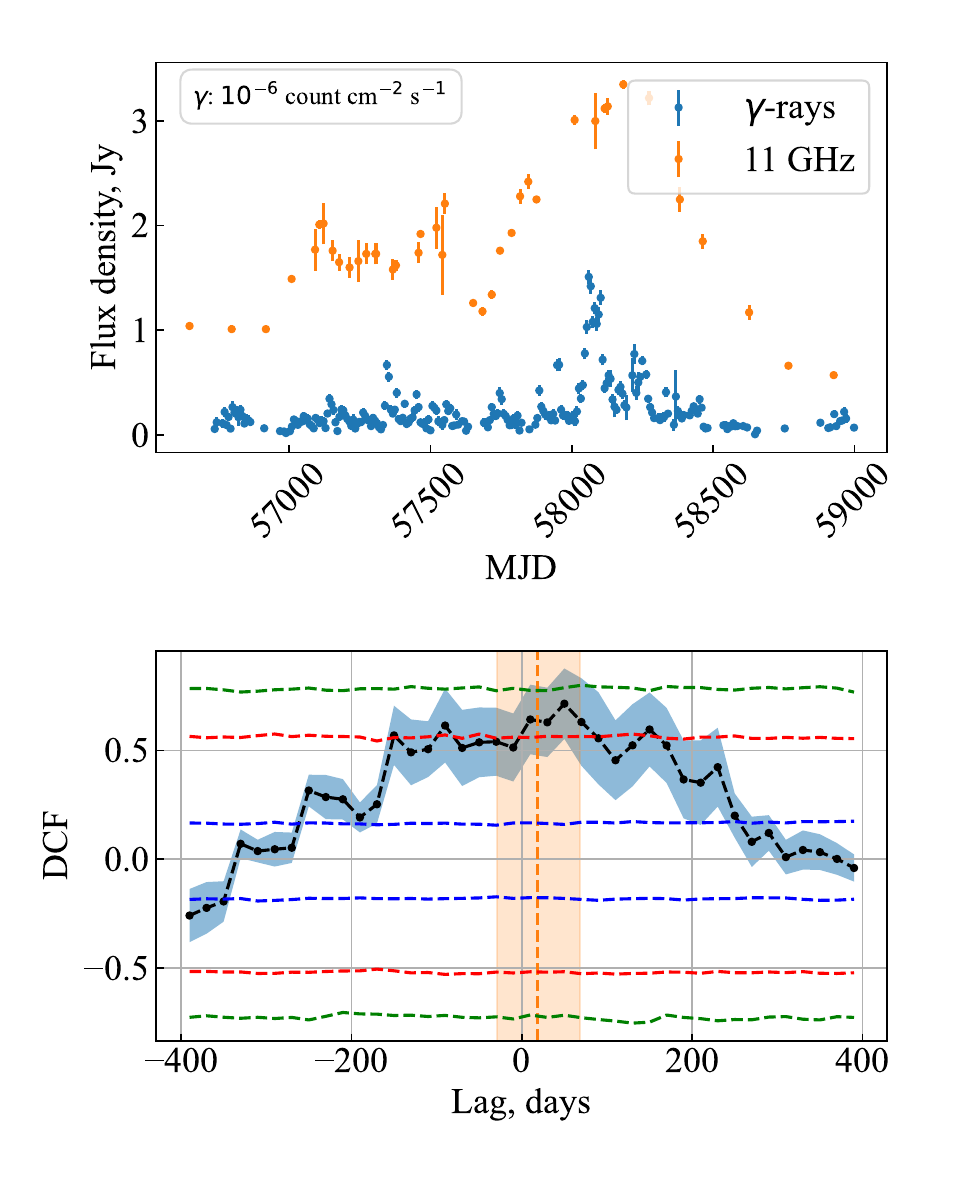}
\includegraphics[width=0.7\columnwidth]{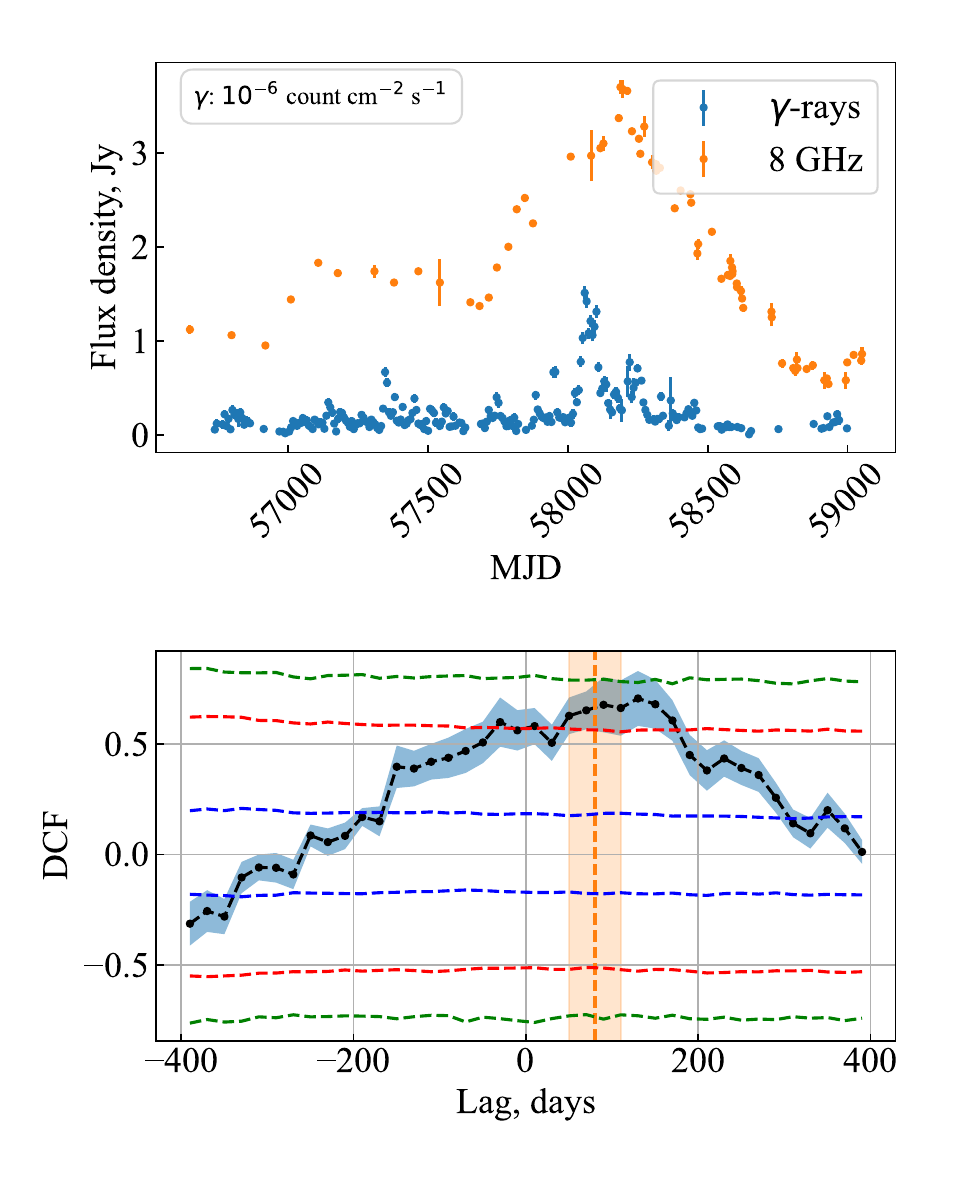}
\includegraphics[width=0.7\columnwidth]{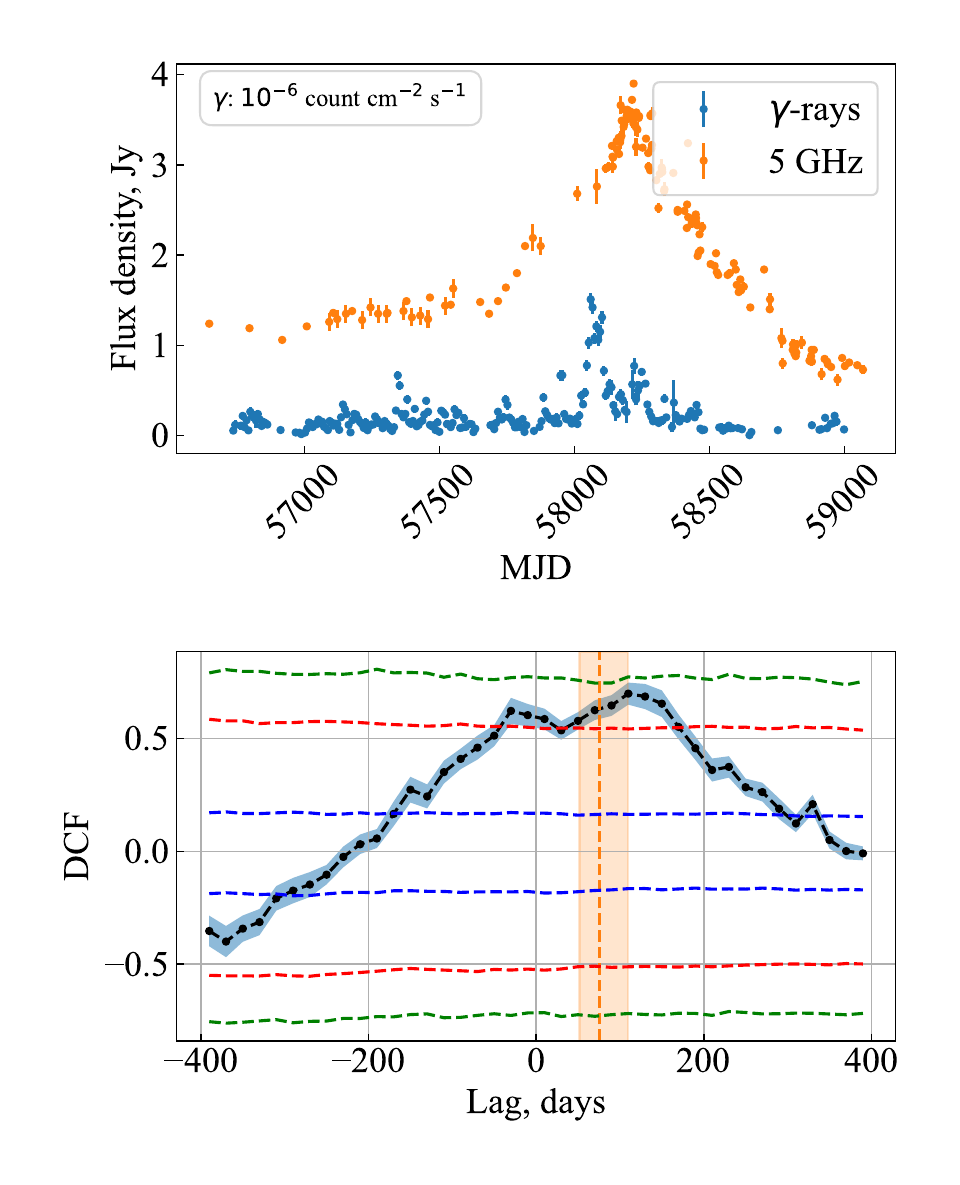}
}
\centerline{
\includegraphics[width=0.7\columnwidth]{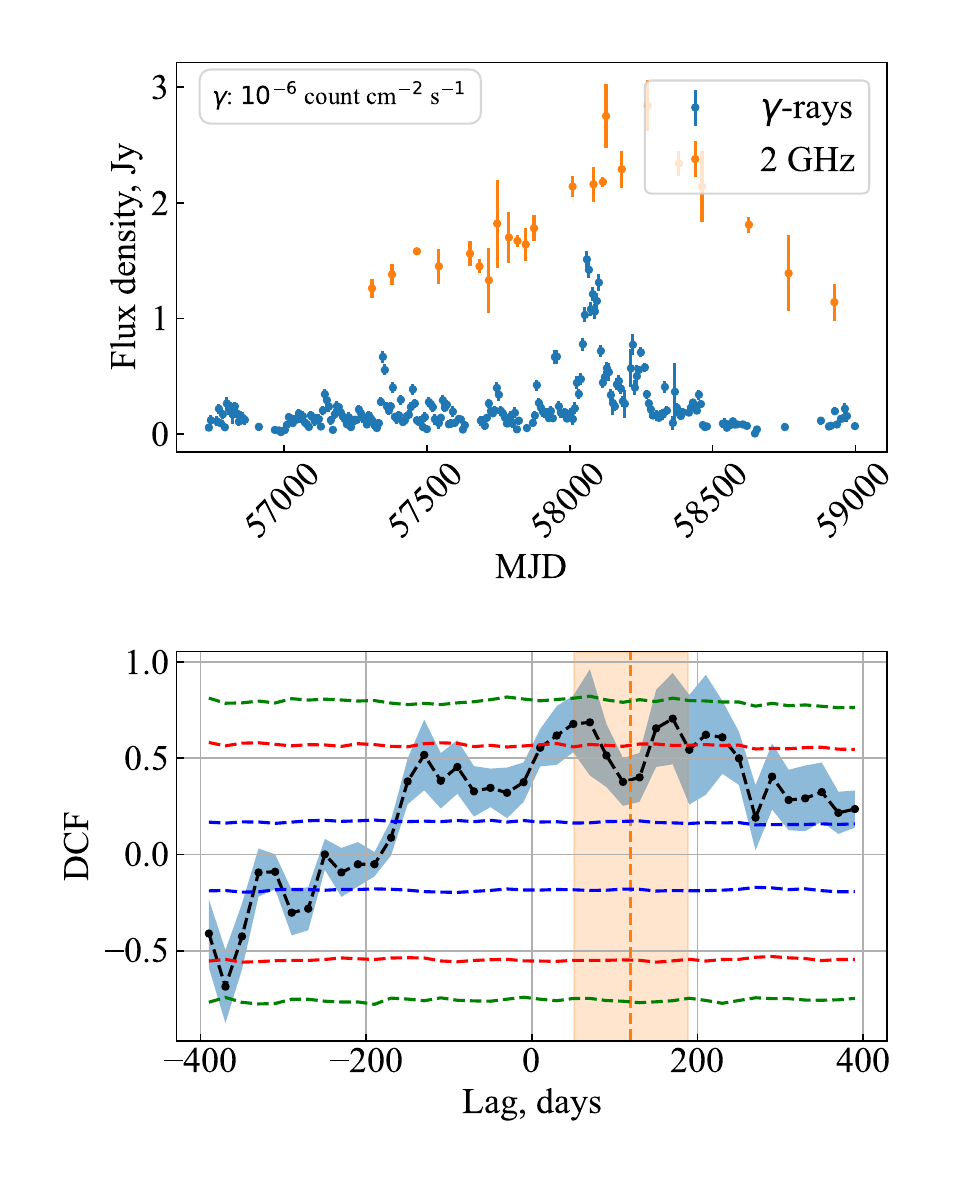}
\includegraphics[width=0.7\columnwidth]{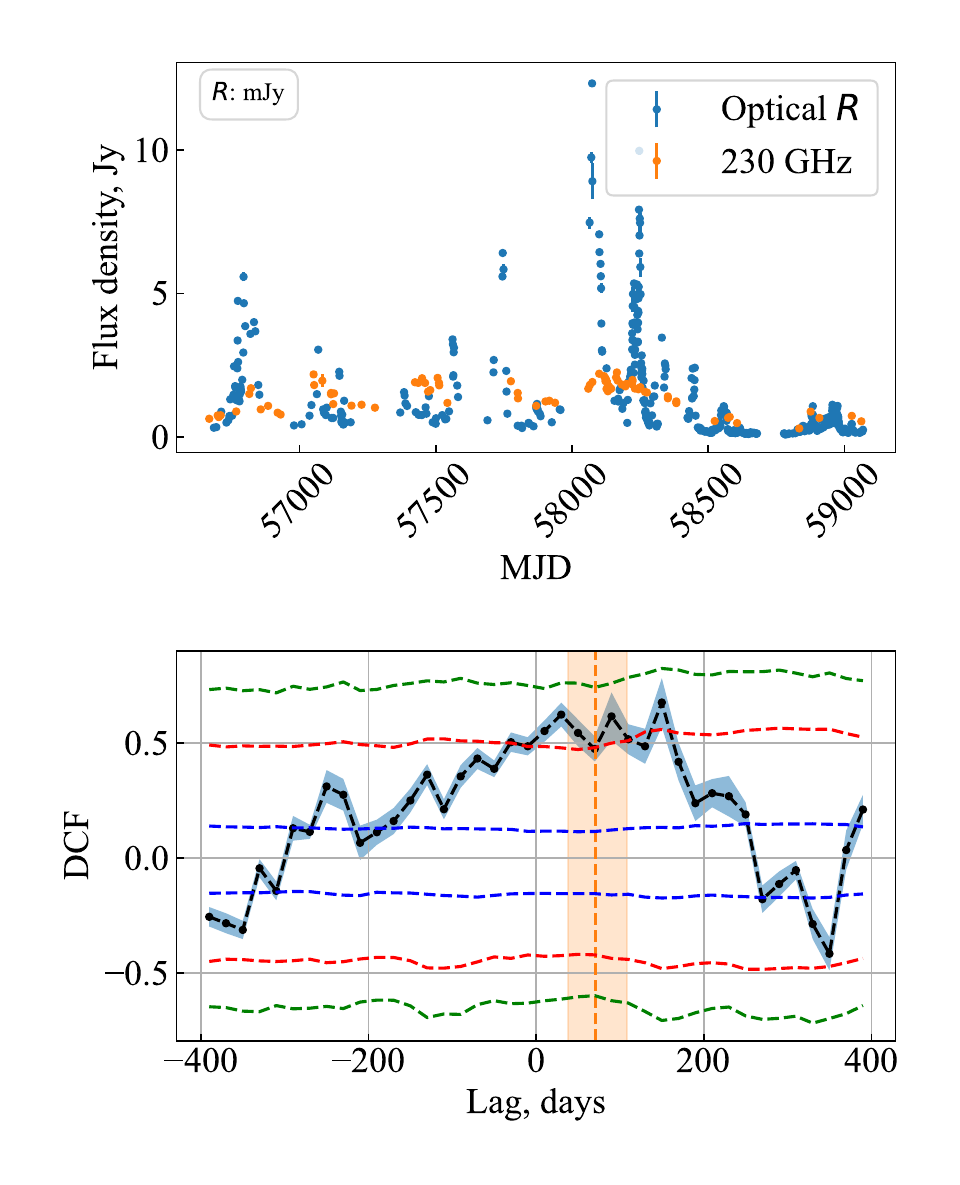}
\includegraphics[width=0.7\columnwidth]{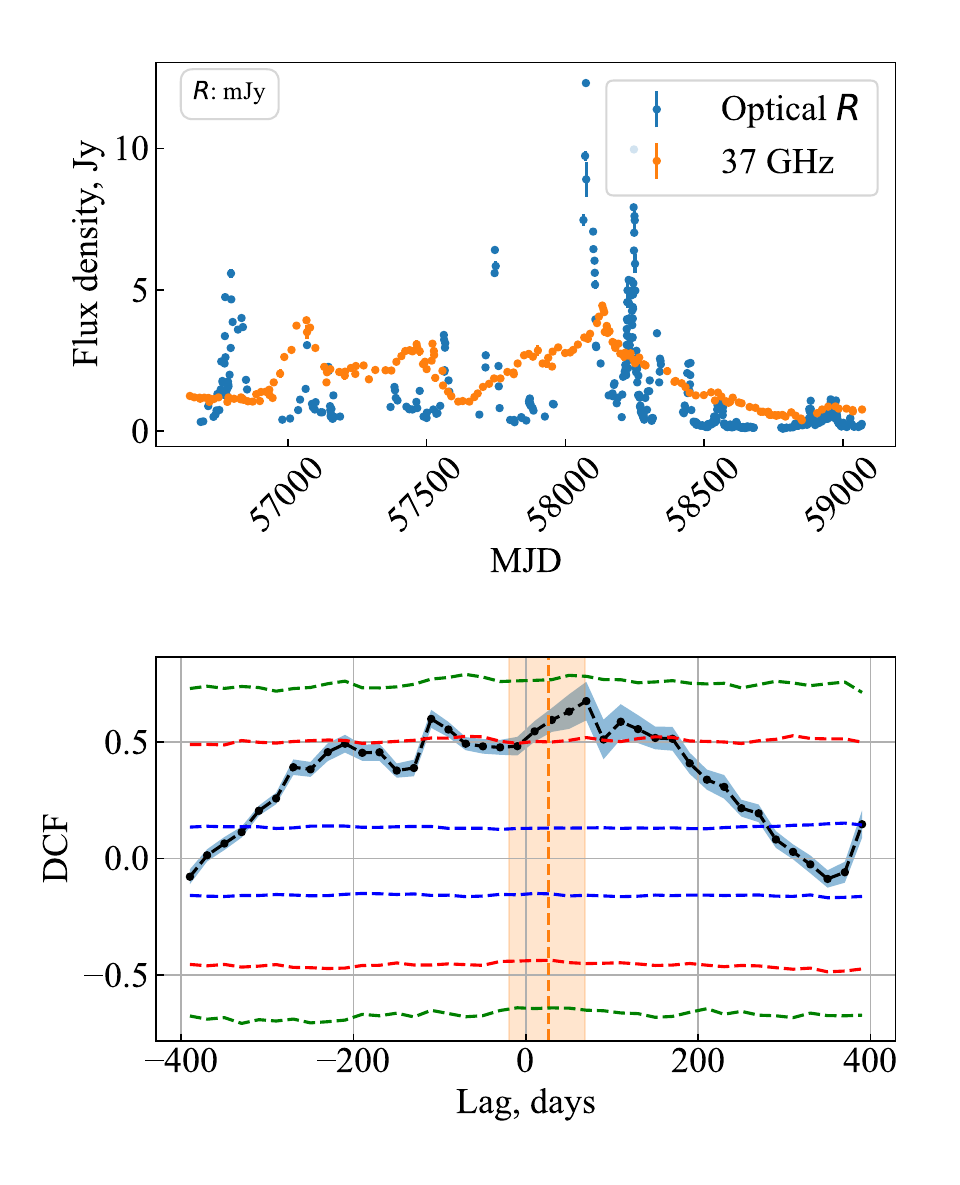}
}
\caption{The light curves and DCFs in epoch~3. Designations are as in Fig.~\ref{fig:dcf_ep1}} 
\label{fig:dcf_ep3}
\end{figure*}

\begin{figure*}
\centerline{
\includegraphics[width=0.7\columnwidth]{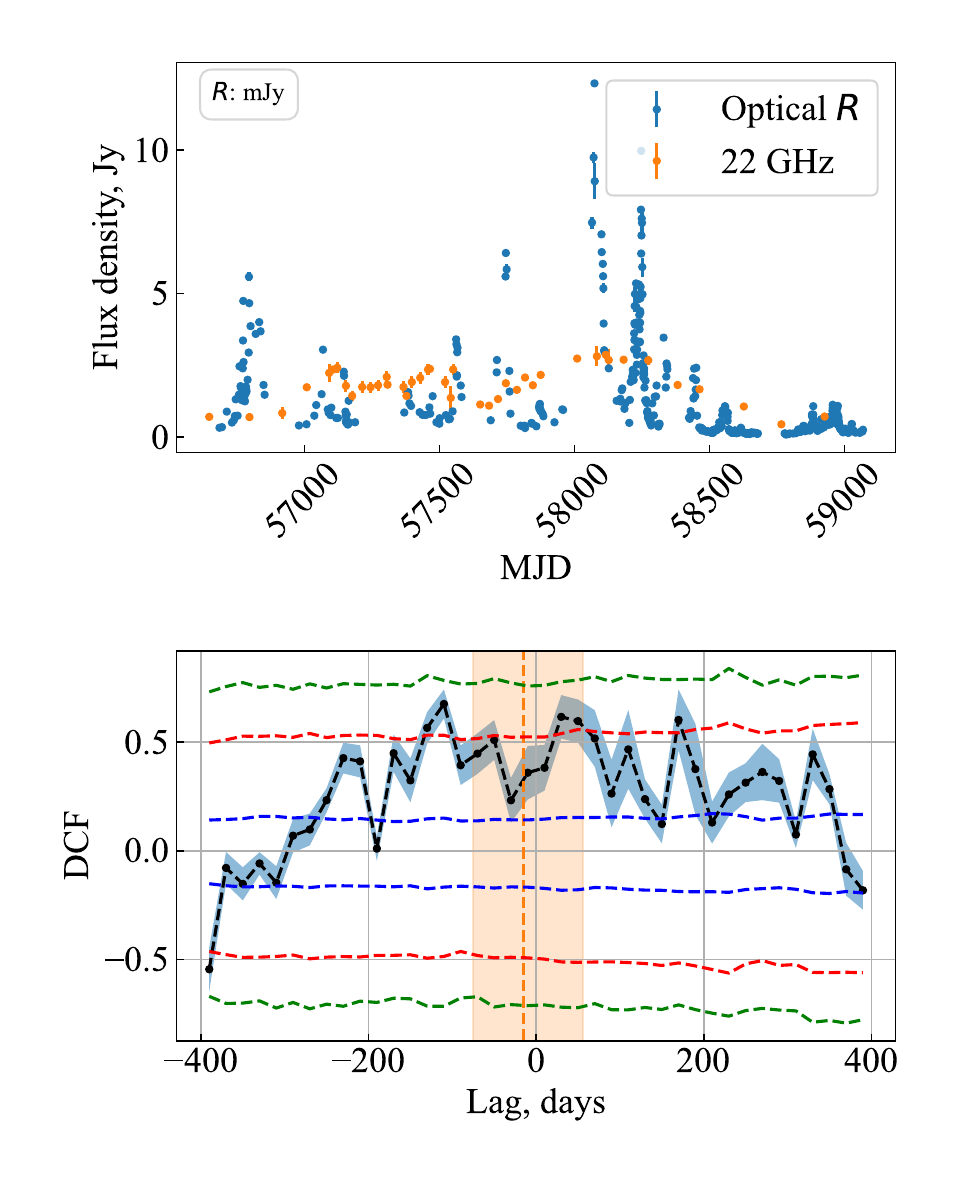}
\includegraphics[width=0.7\columnwidth]{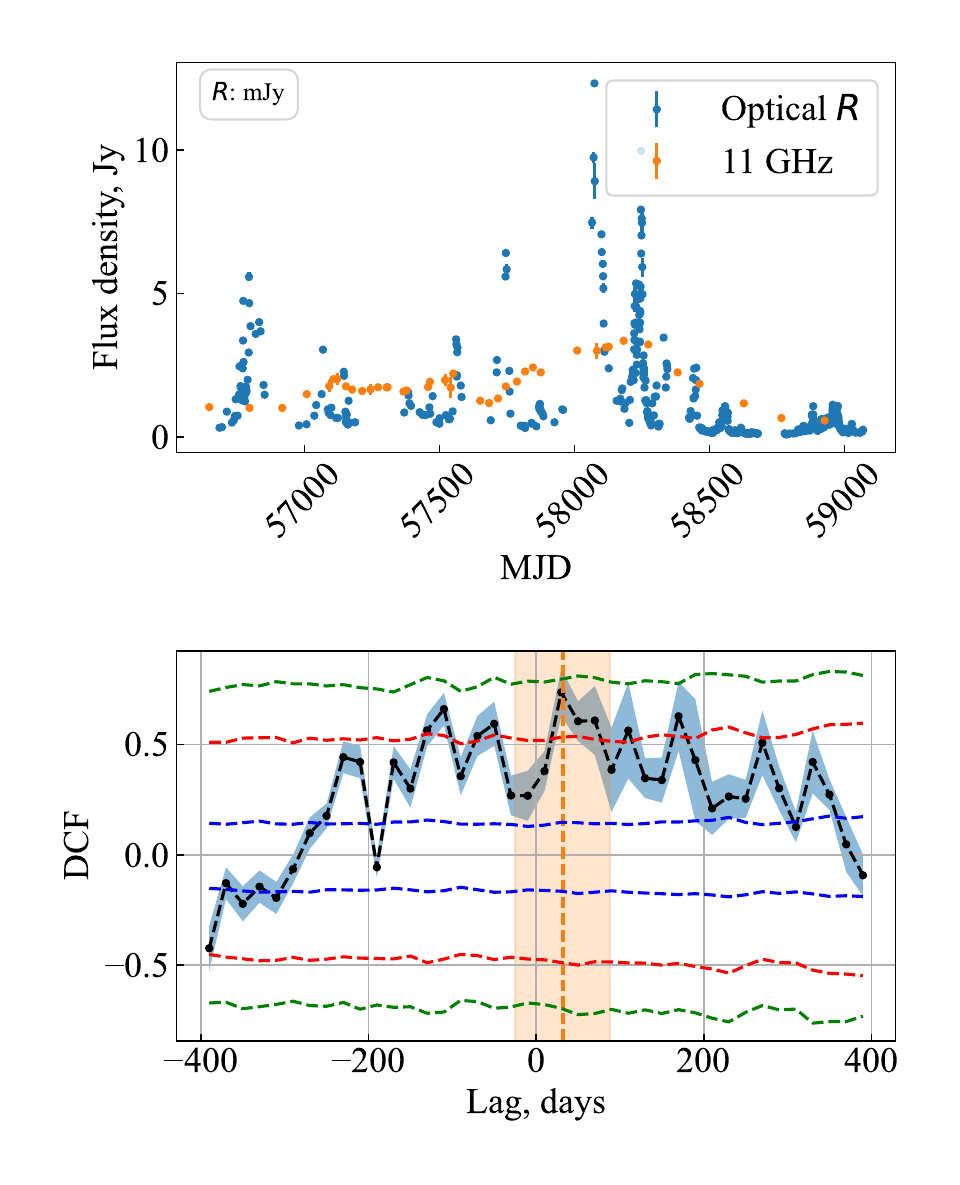}
\includegraphics[width=0.7\columnwidth]{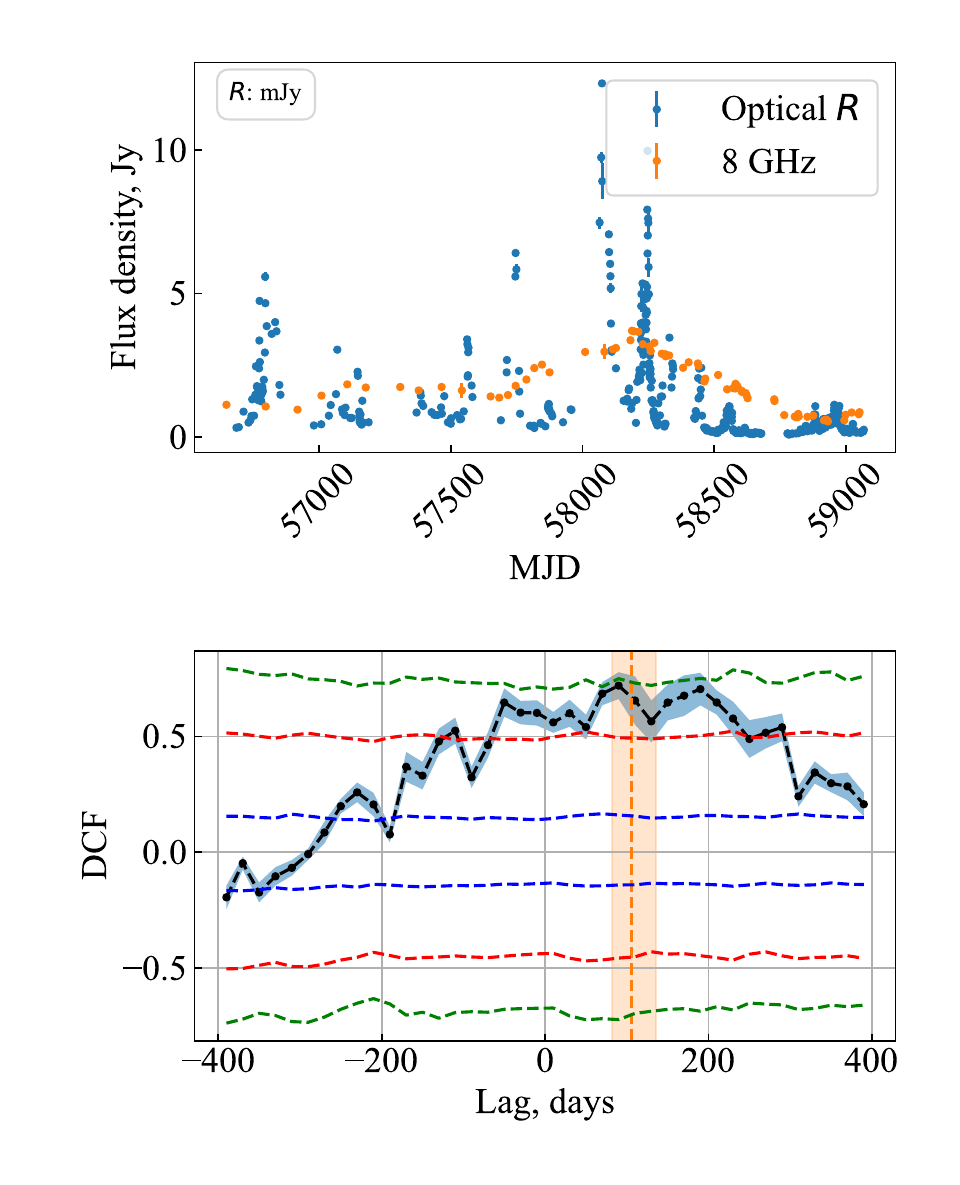}
}
\centerline{
\includegraphics[width=0.7\columnwidth]{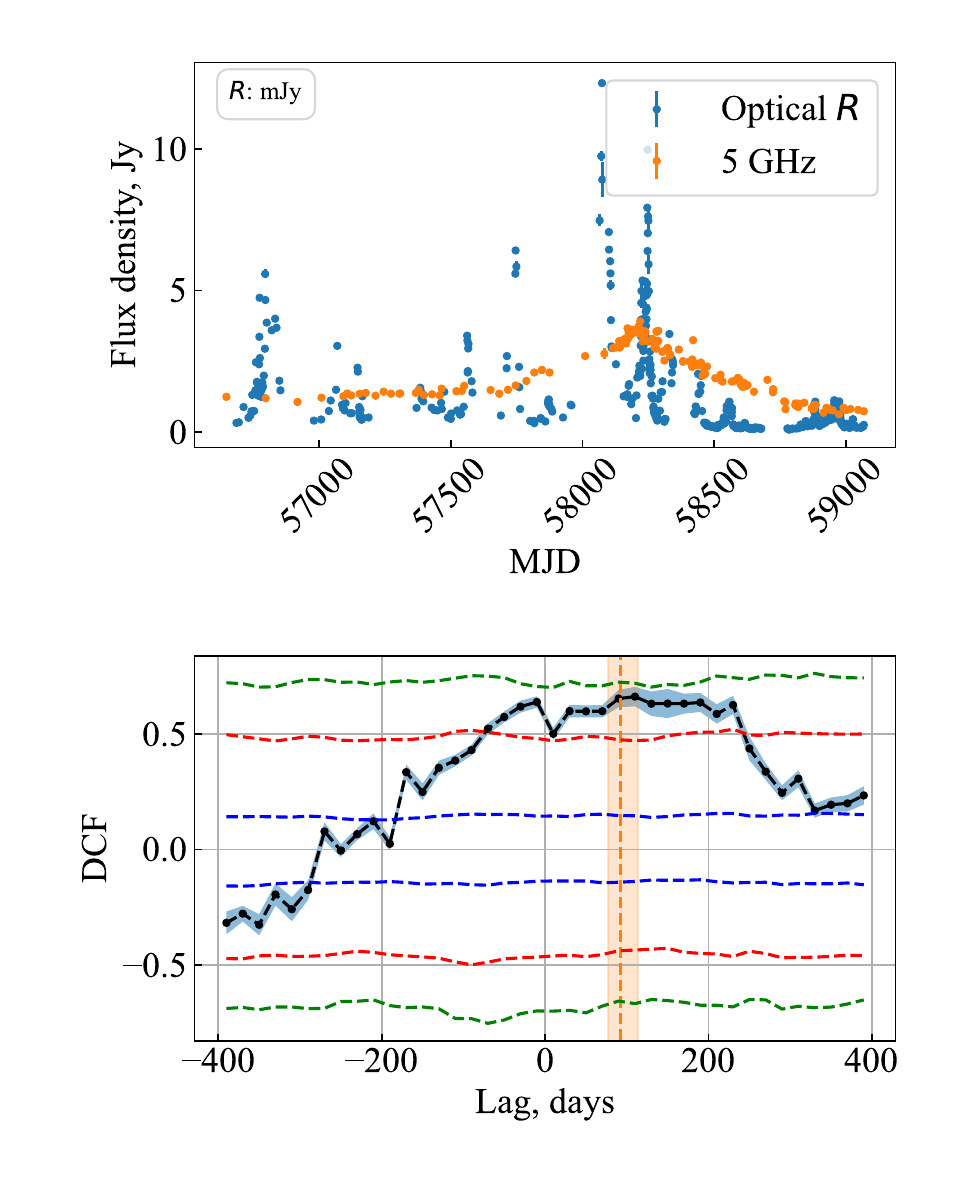}
\includegraphics[width=0.7\columnwidth]{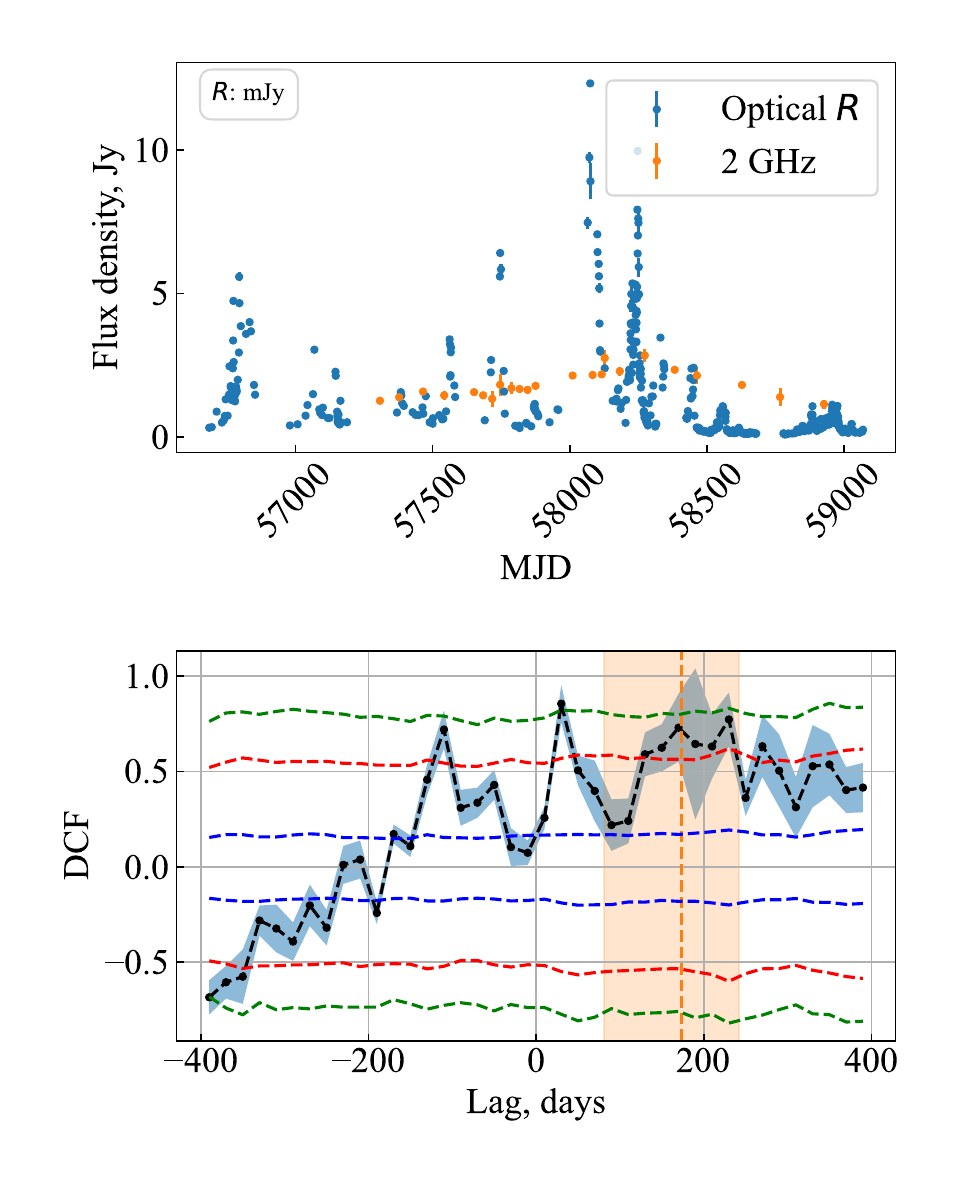}
\includegraphics[width=0.7\columnwidth]{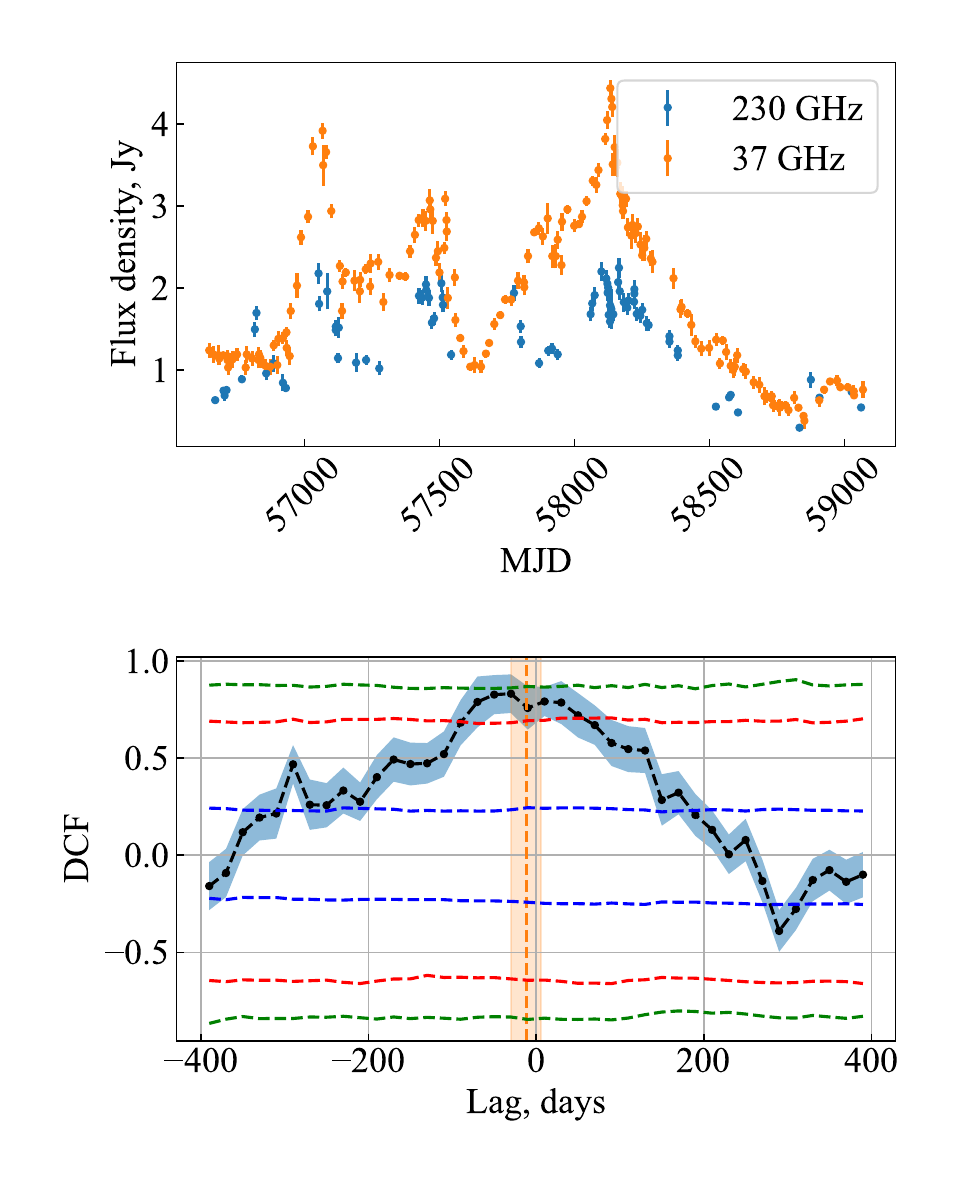}
}
\centerline{
\includegraphics[width=0.7\columnwidth]{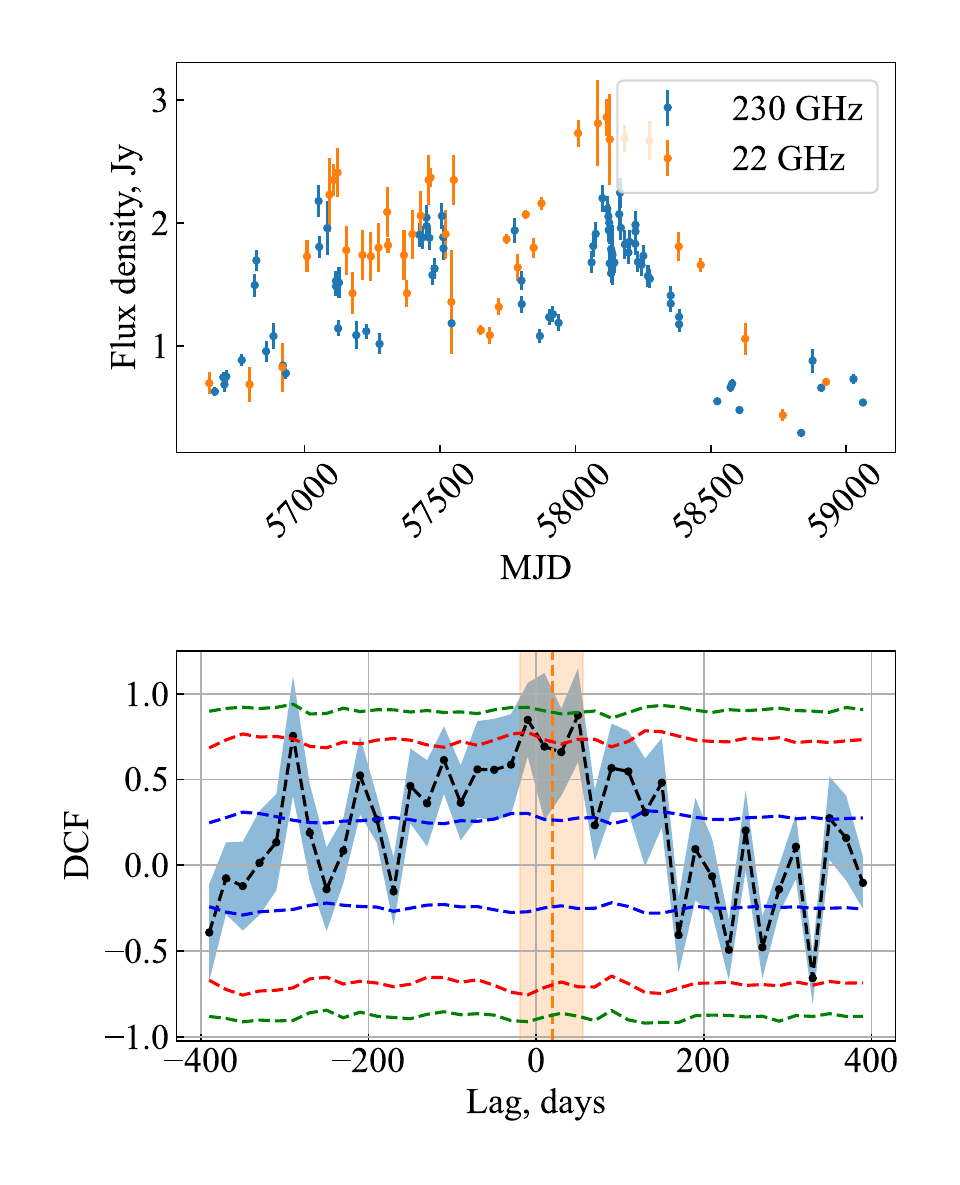}
\includegraphics[width=0.7\columnwidth]{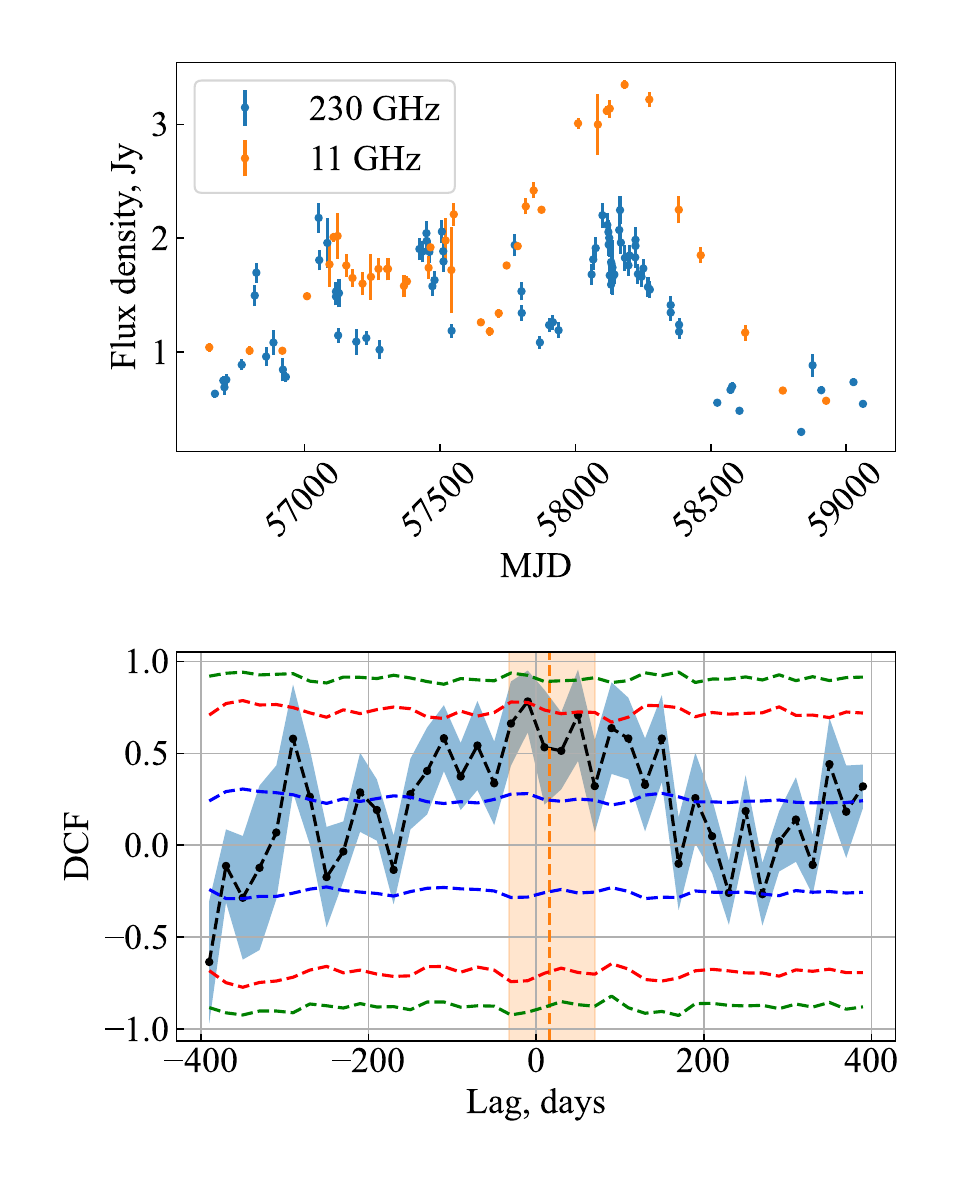}
\includegraphics[width=0.7\columnwidth]{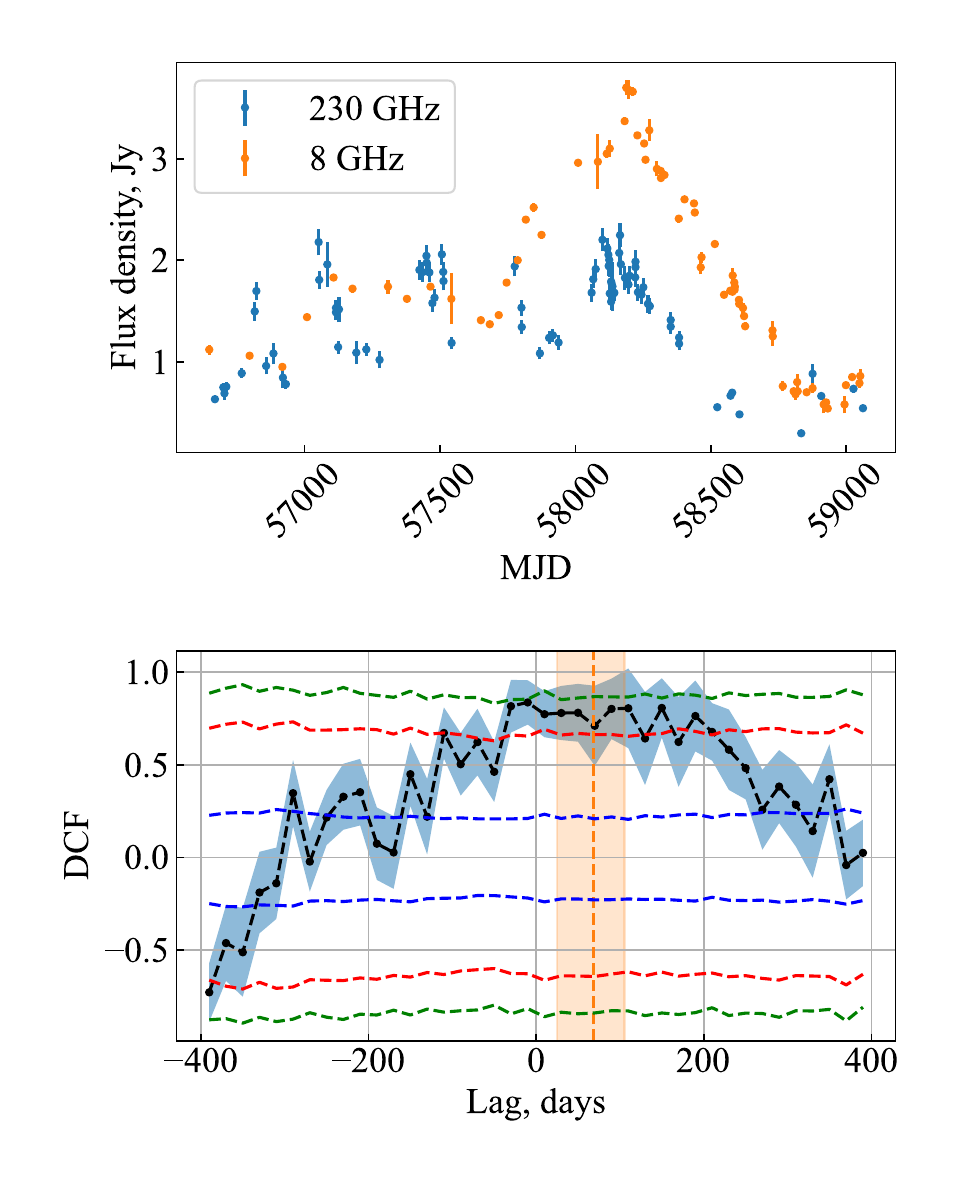}
}
\contcaption{The light curves and DCFs in epoch~3.} 
\end{figure*}

\begin{figure*}
\centerline{
\includegraphics[width=0.7\columnwidth]{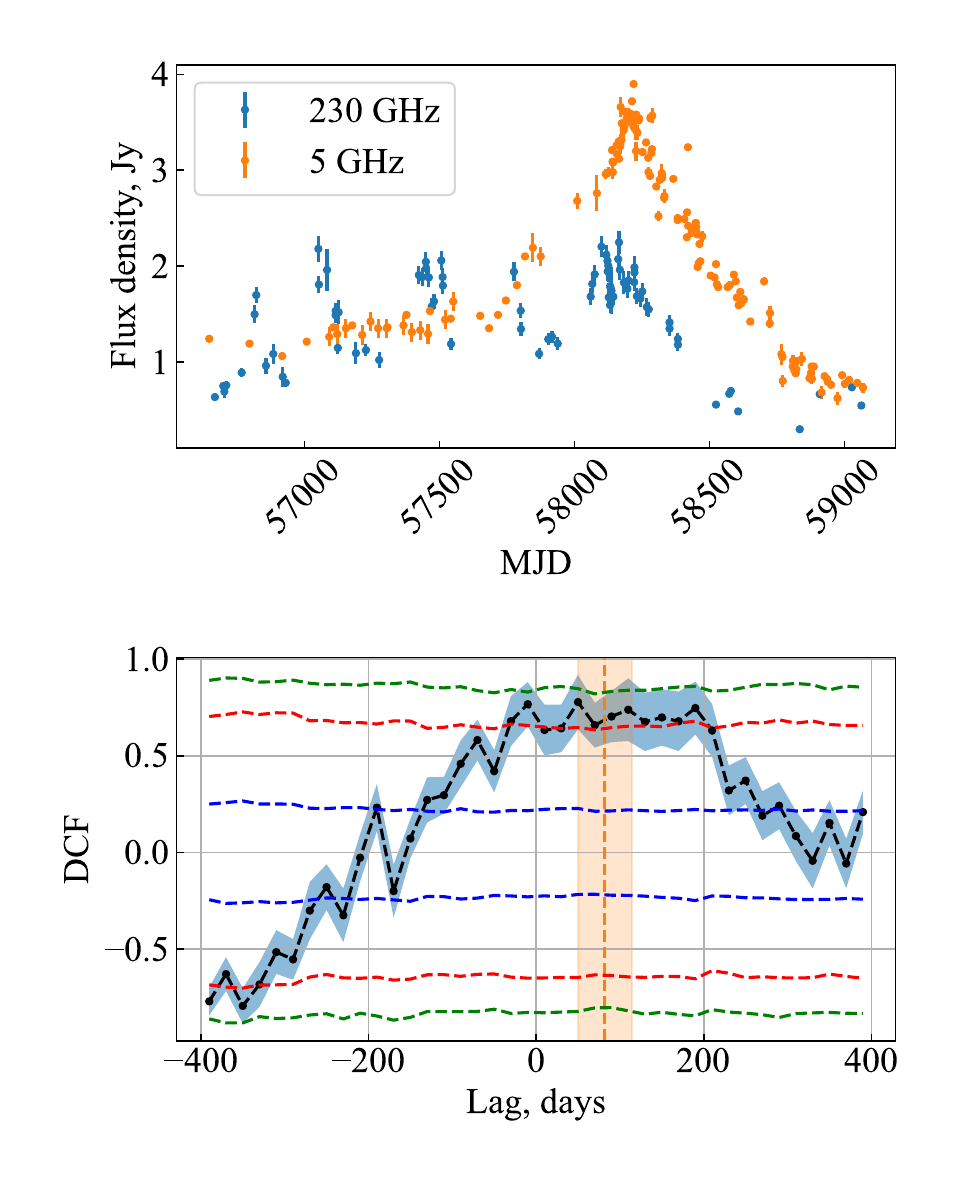}
\includegraphics[width=0.7\columnwidth]{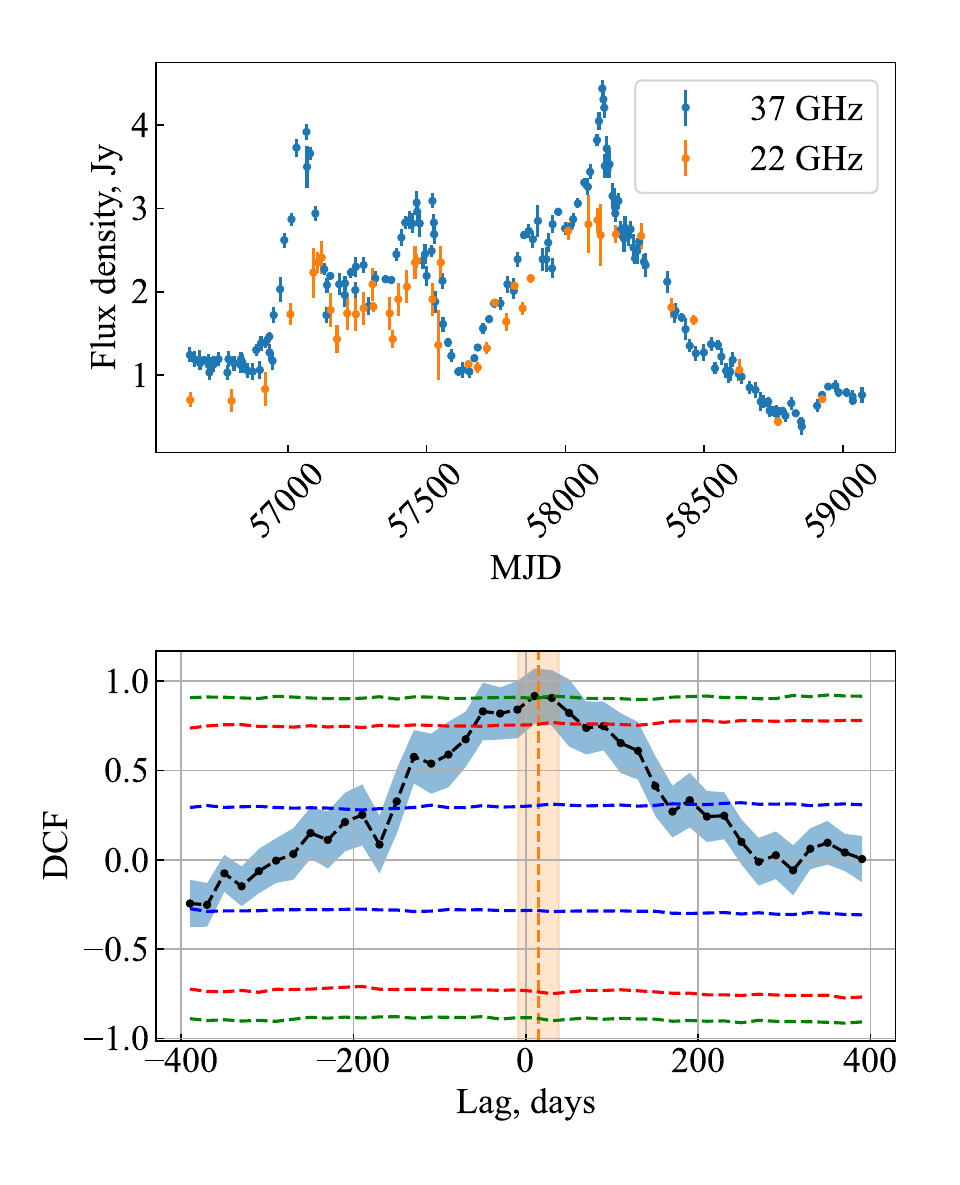}
\includegraphics[width=0.7\columnwidth]{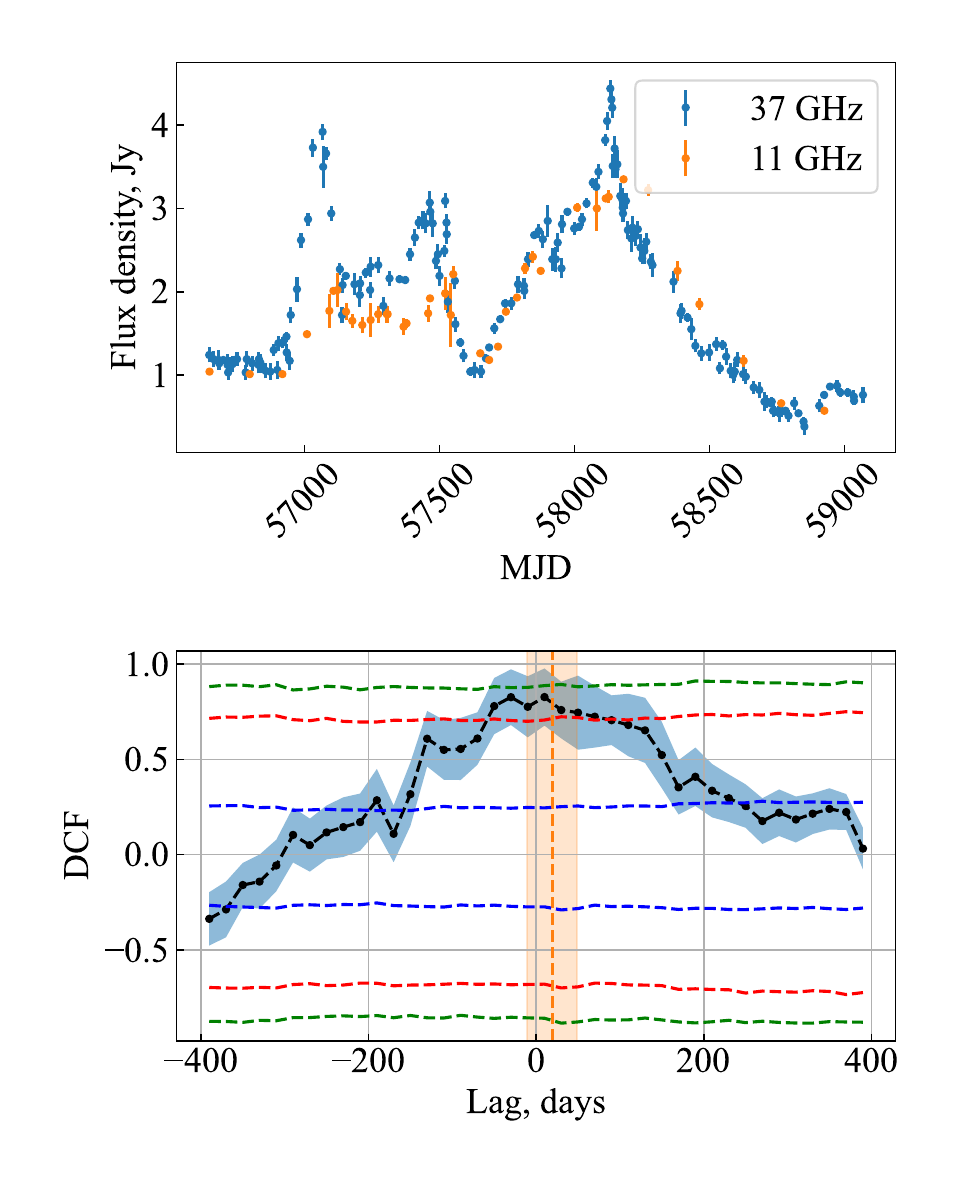}
}
\centerline{
\includegraphics[width=0.7\columnwidth]{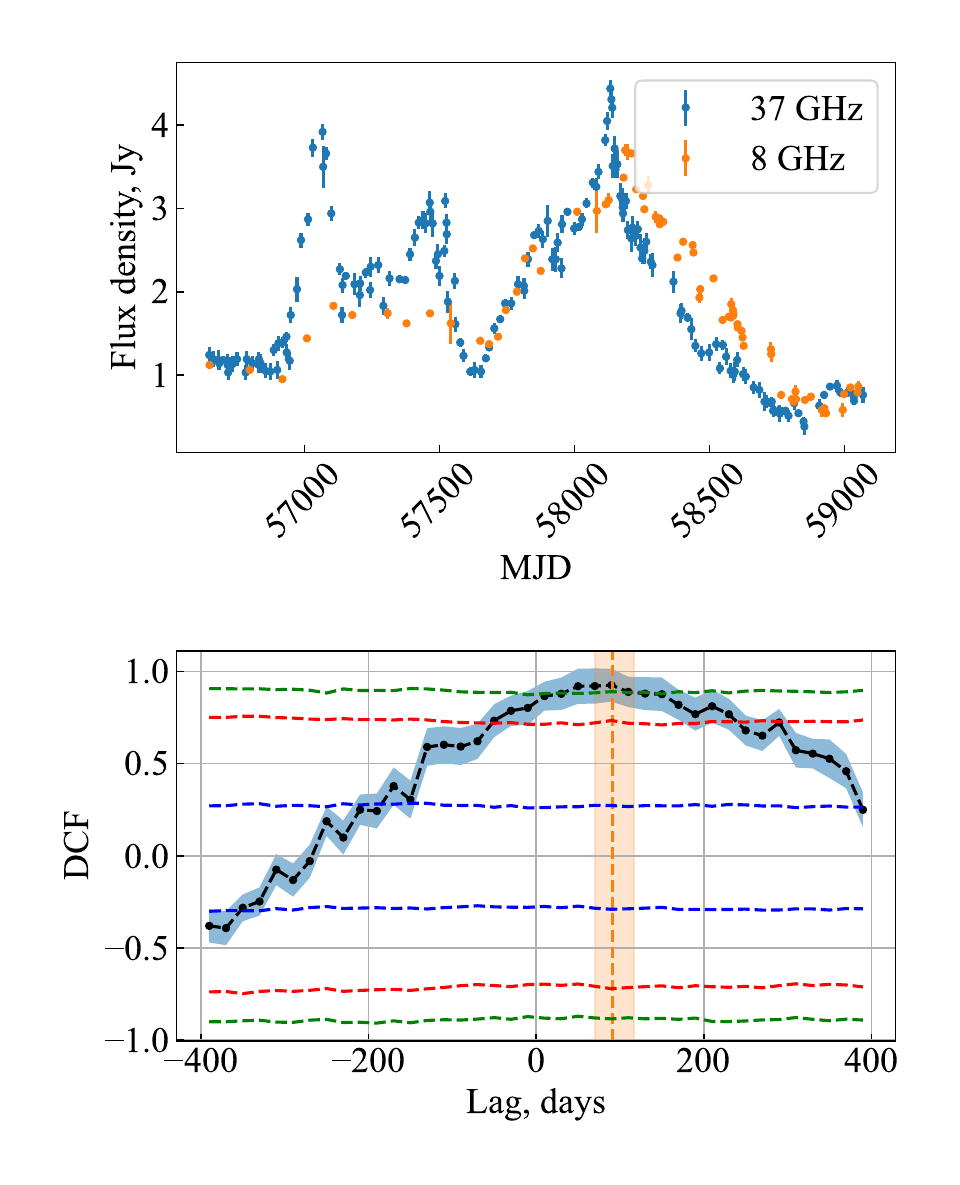}
\includegraphics[width=0.7\columnwidth]{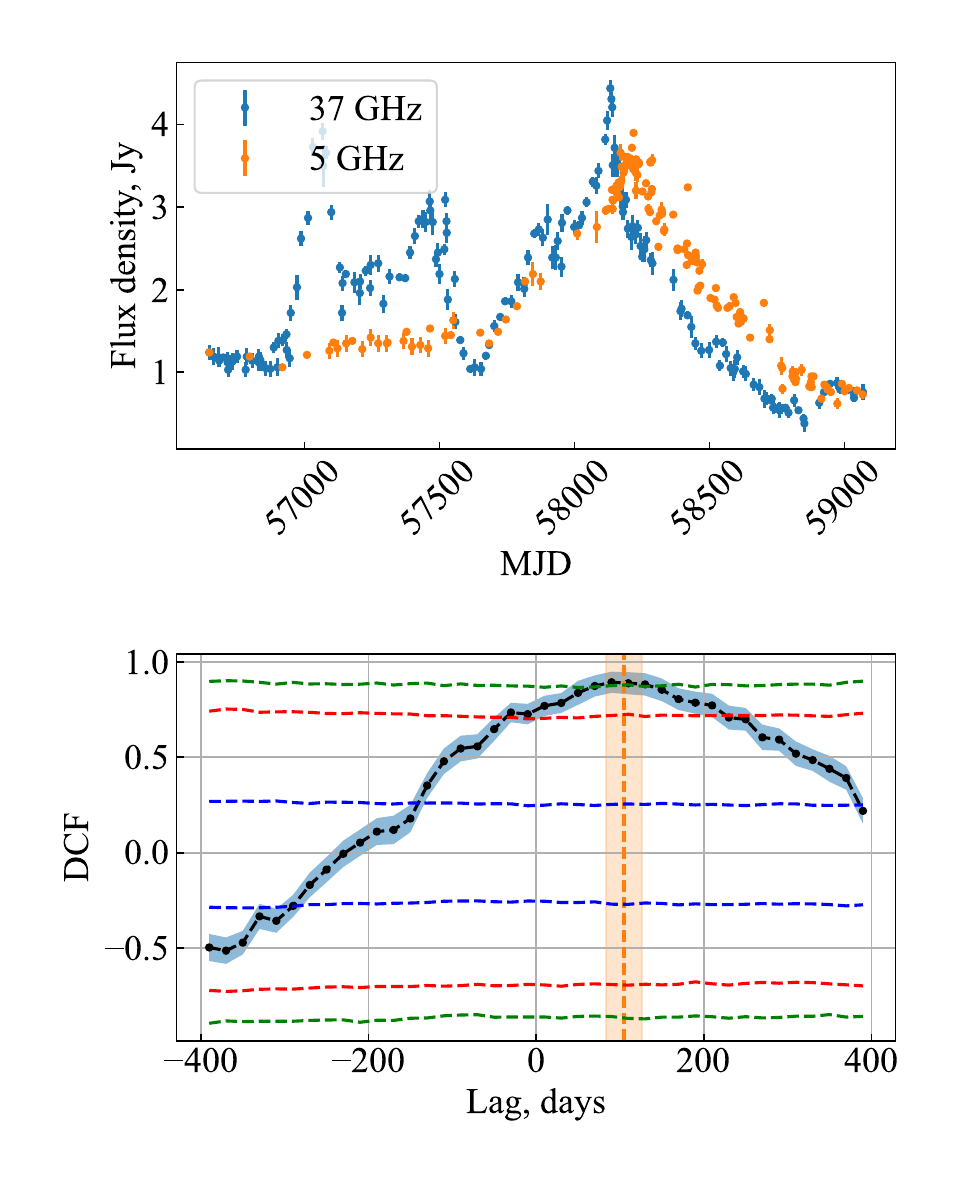}
\includegraphics[width=0.7\columnwidth]{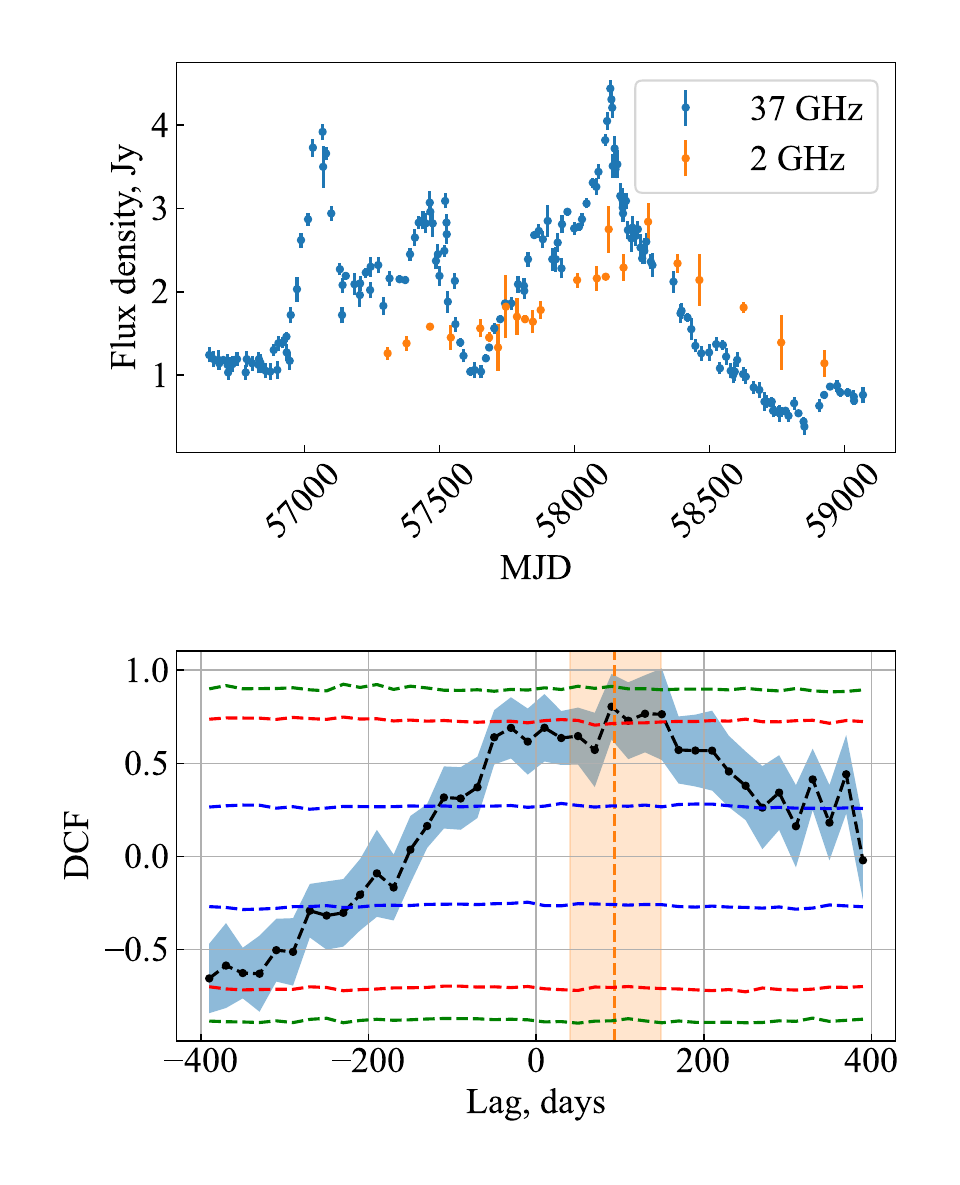}
}
\centerline{
\includegraphics[width=0.7\columnwidth]{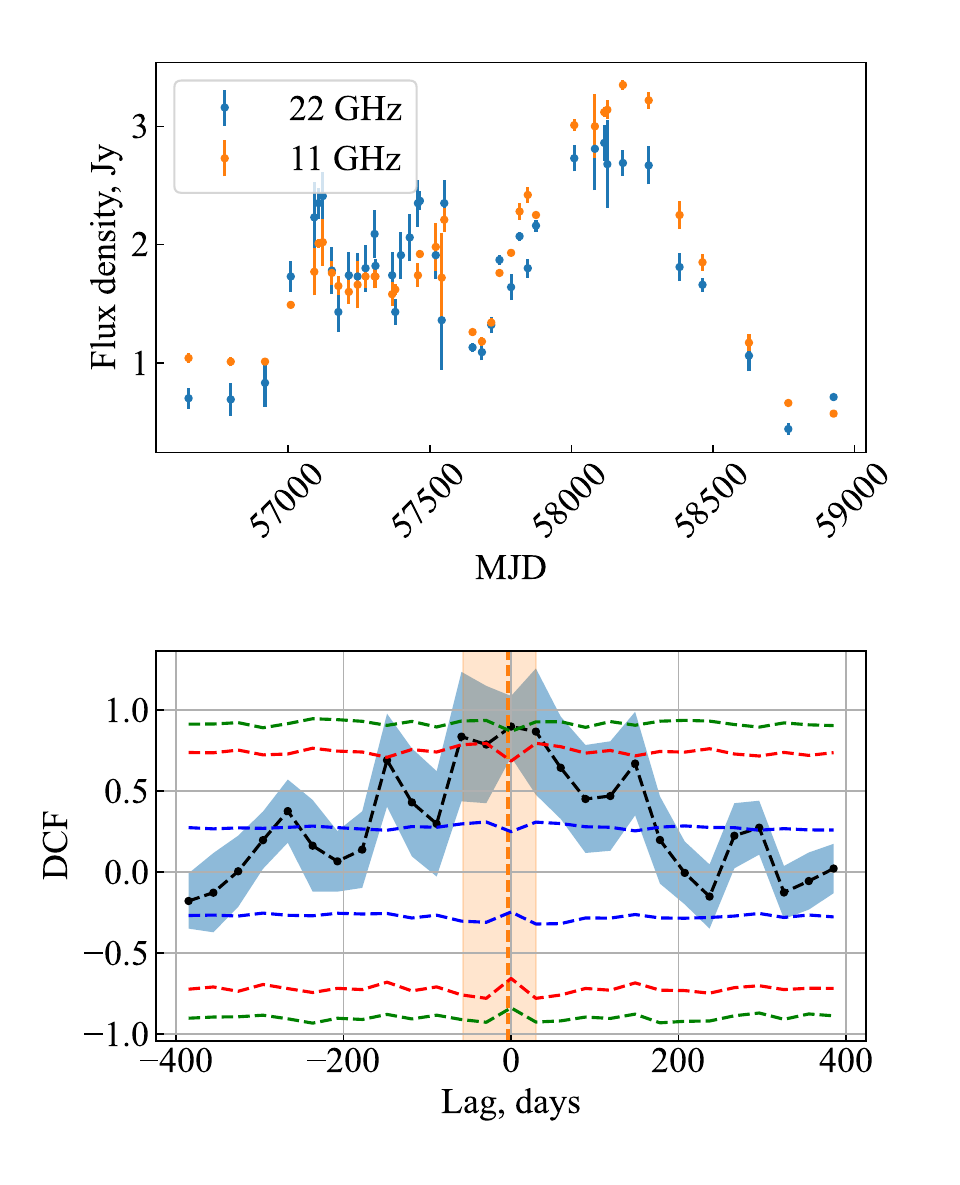}
\includegraphics[width=0.7\columnwidth]{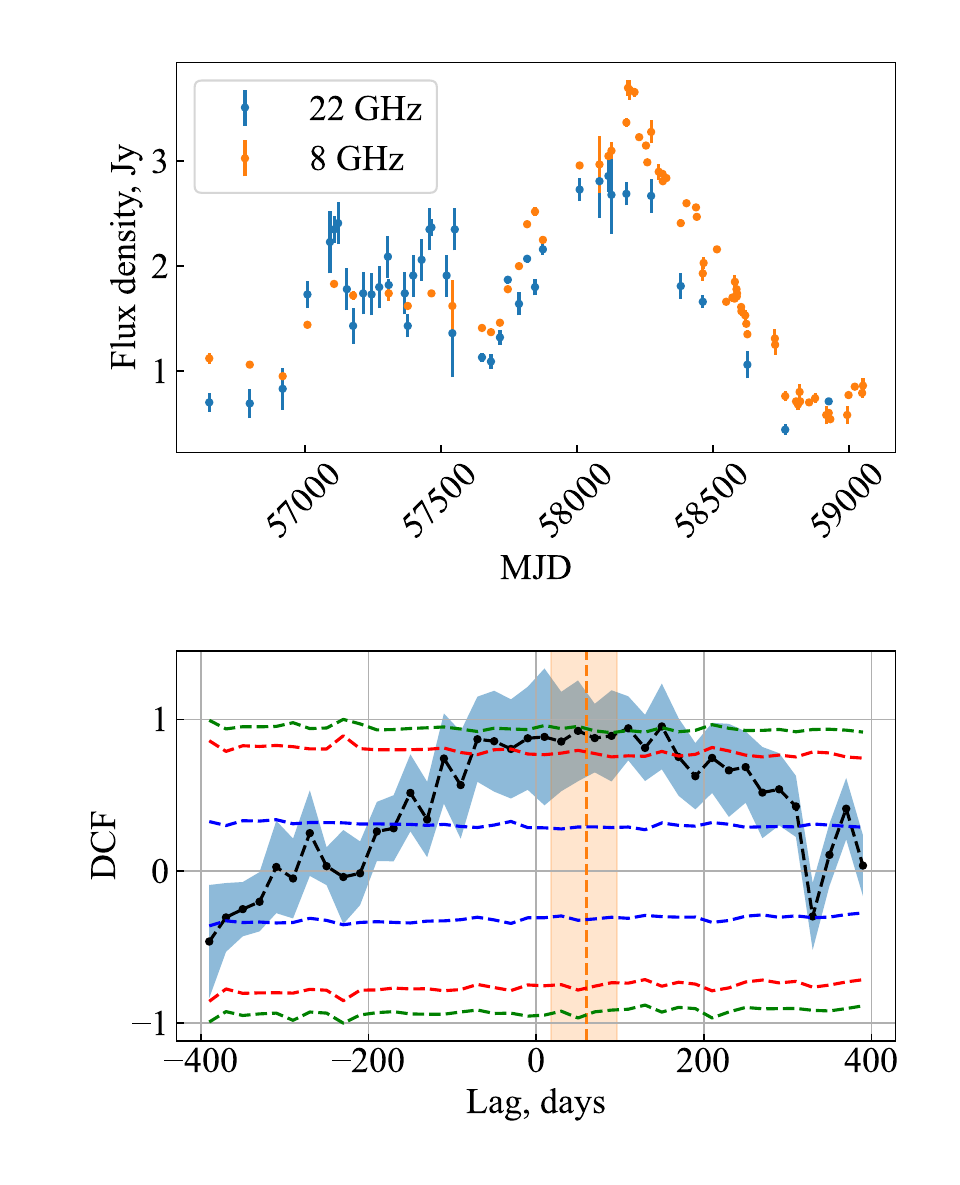}
\includegraphics[width=0.7\columnwidth]{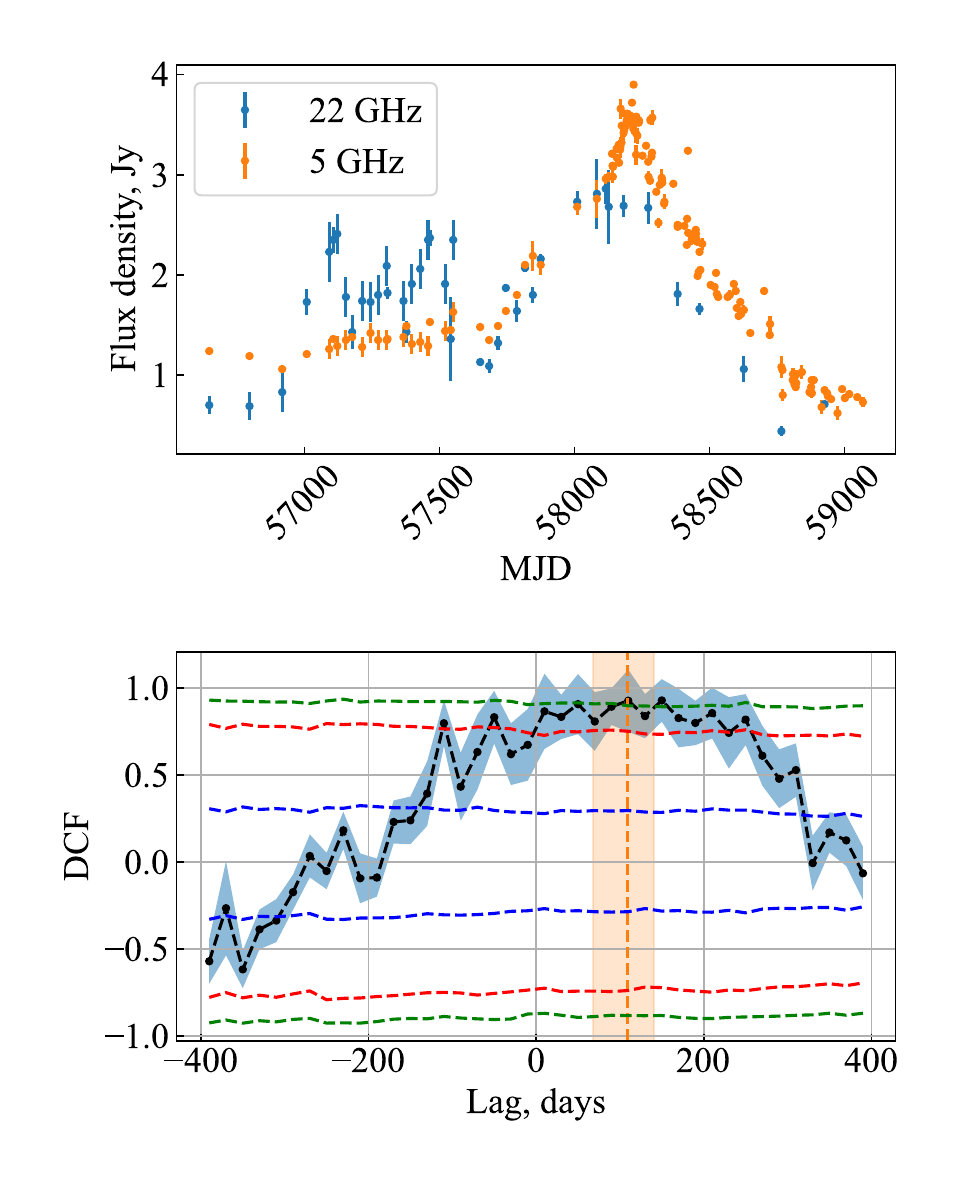}
}
\contcaption{The light curves and DCFs in epoch~3.} 
\end{figure*}

\begin{figure*}
\centerline{
\includegraphics[width=0.7\columnwidth]{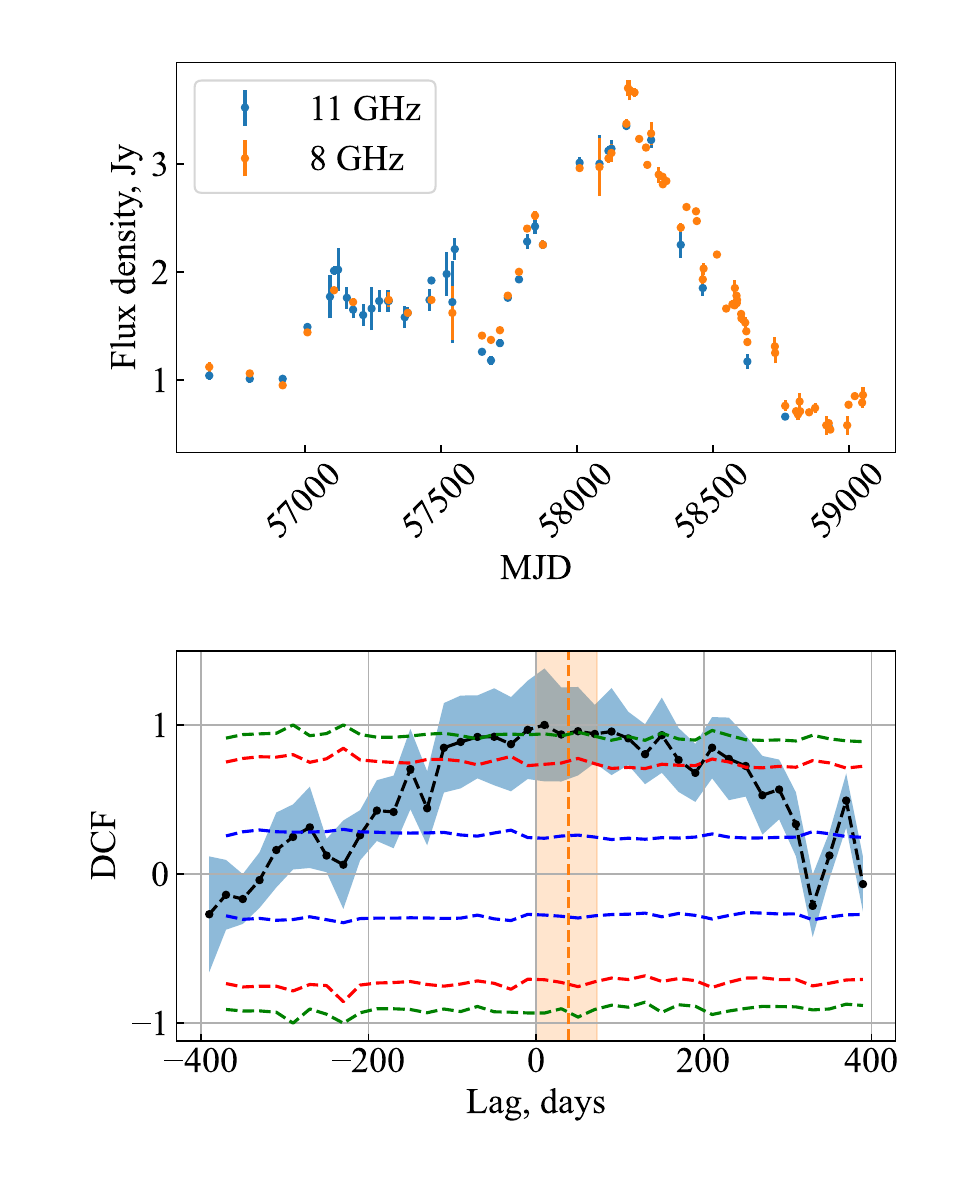}
\includegraphics[width=0.7\columnwidth]{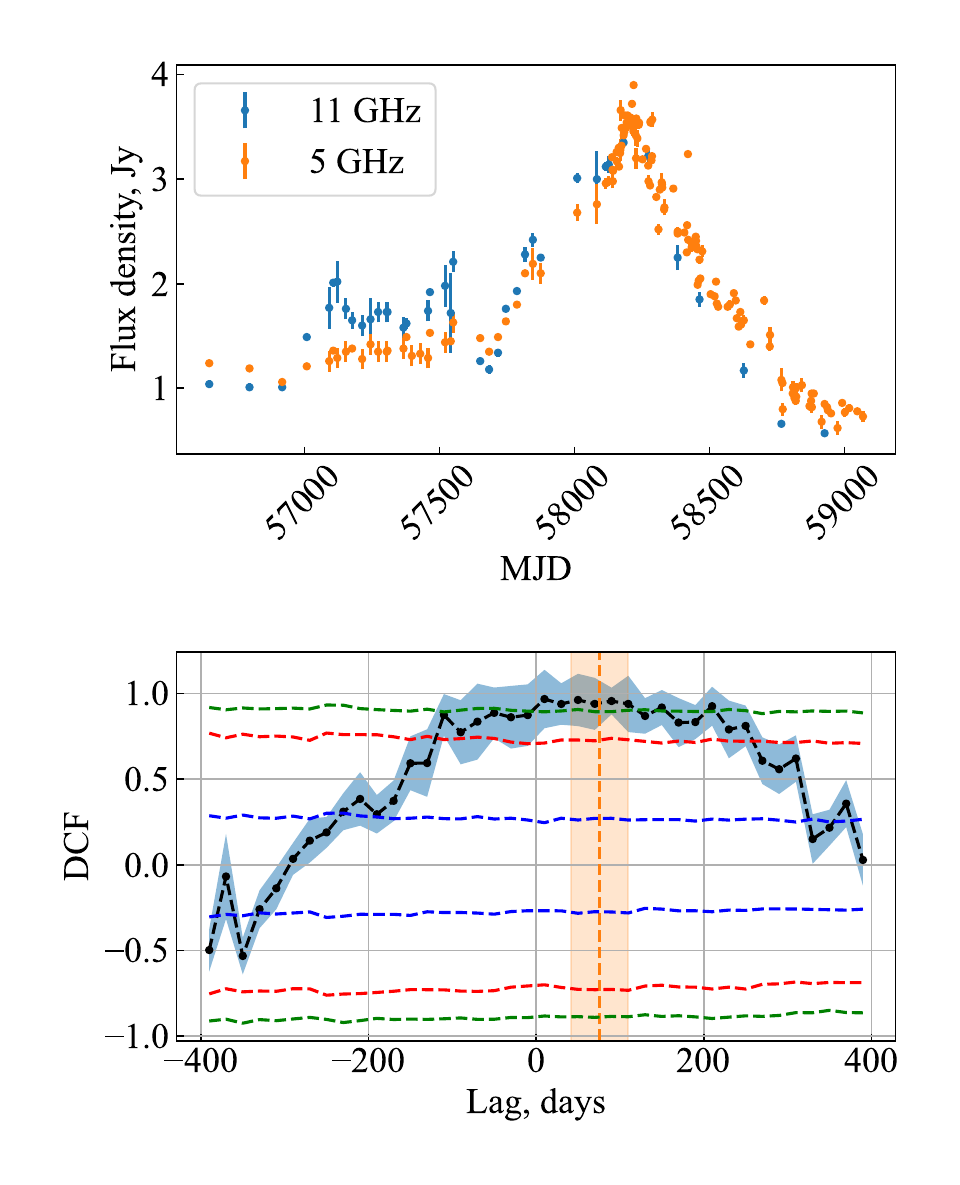}
\includegraphics[width=0.7\columnwidth]{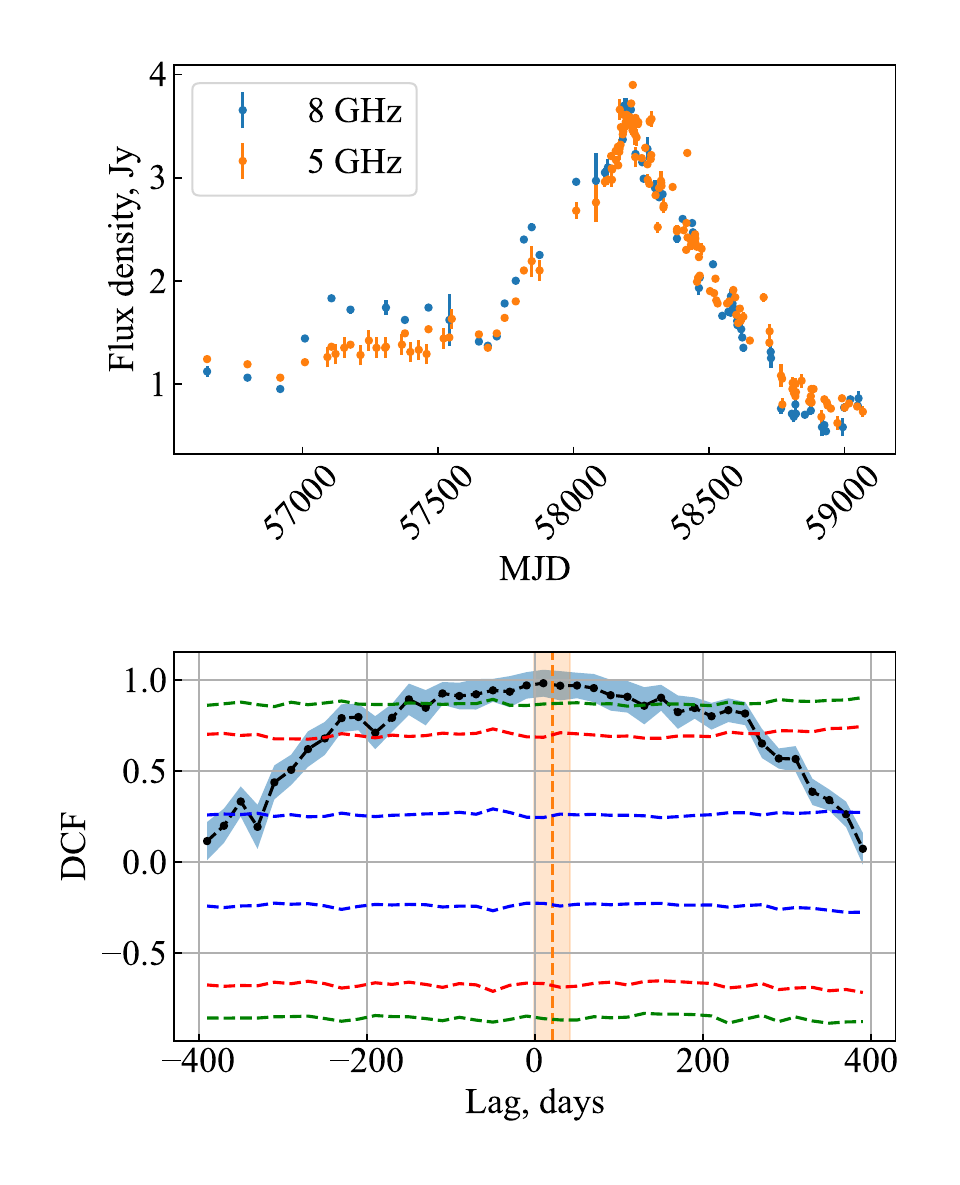}
}
\centerline{
\includegraphics[width=0.7\columnwidth]{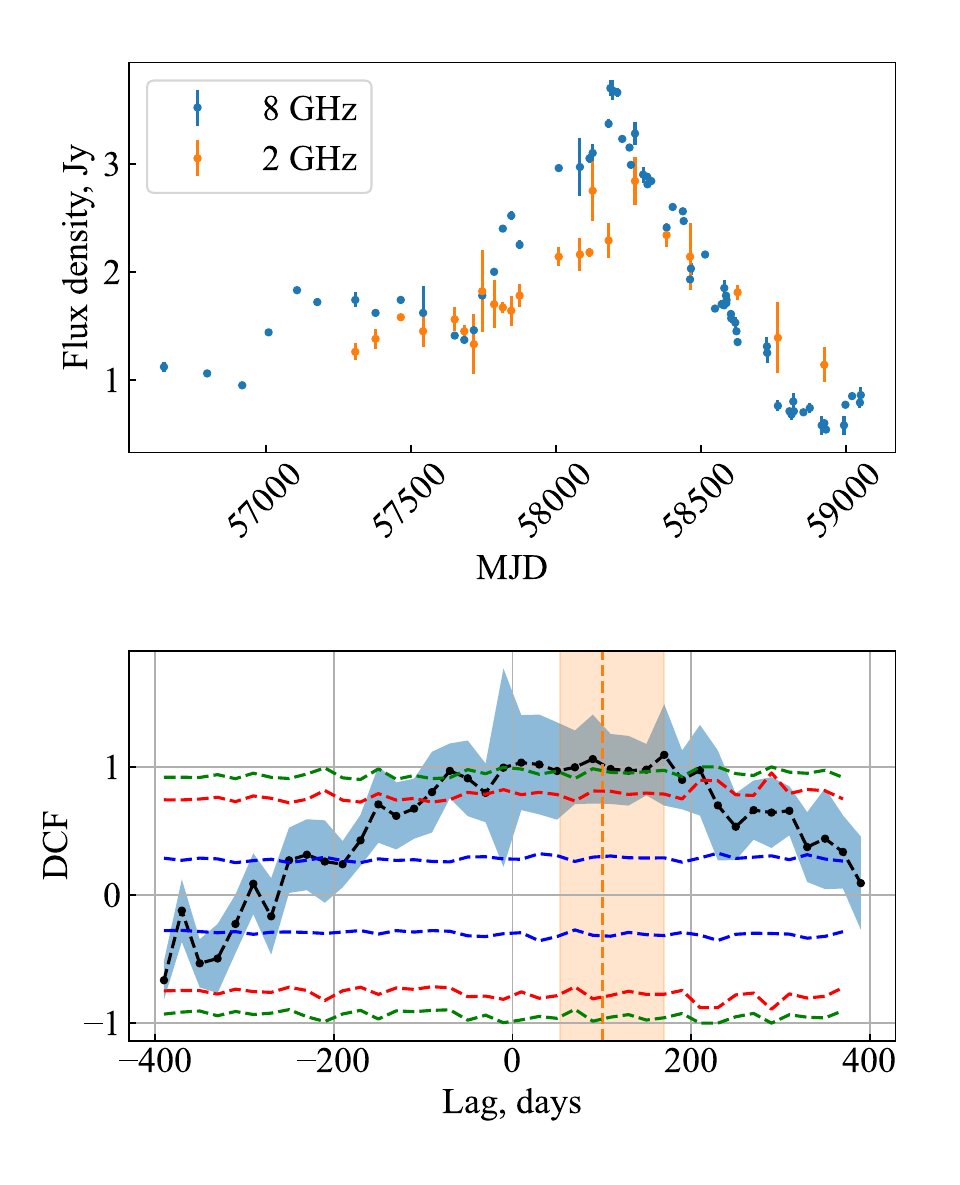}
\includegraphics[width=0.7\columnwidth]{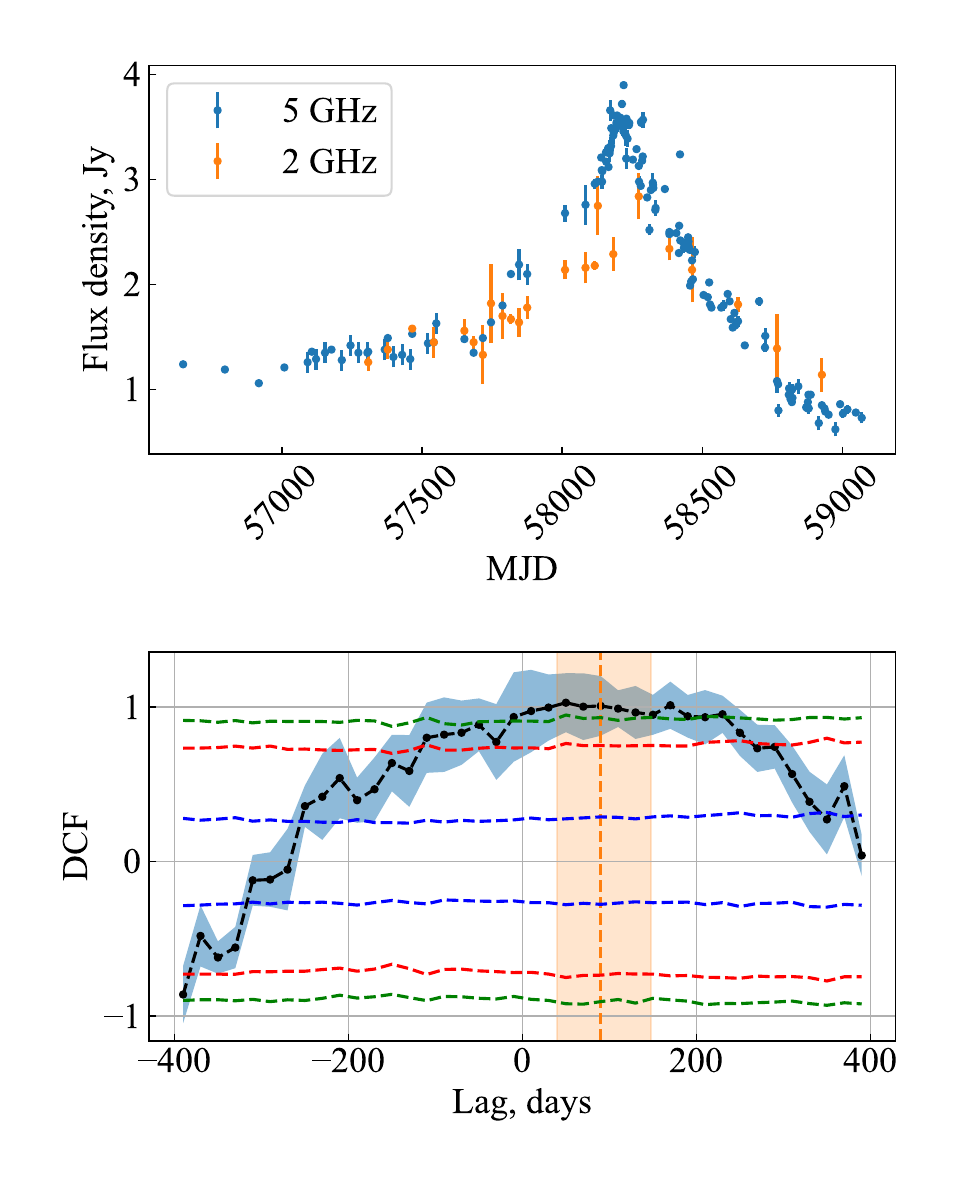}
}
\contcaption{The light curves and DCFs in epoch~3.} 
\end{figure*}


\begin{figure*}
\centerline{
\includegraphics[width=0.7\columnwidth]{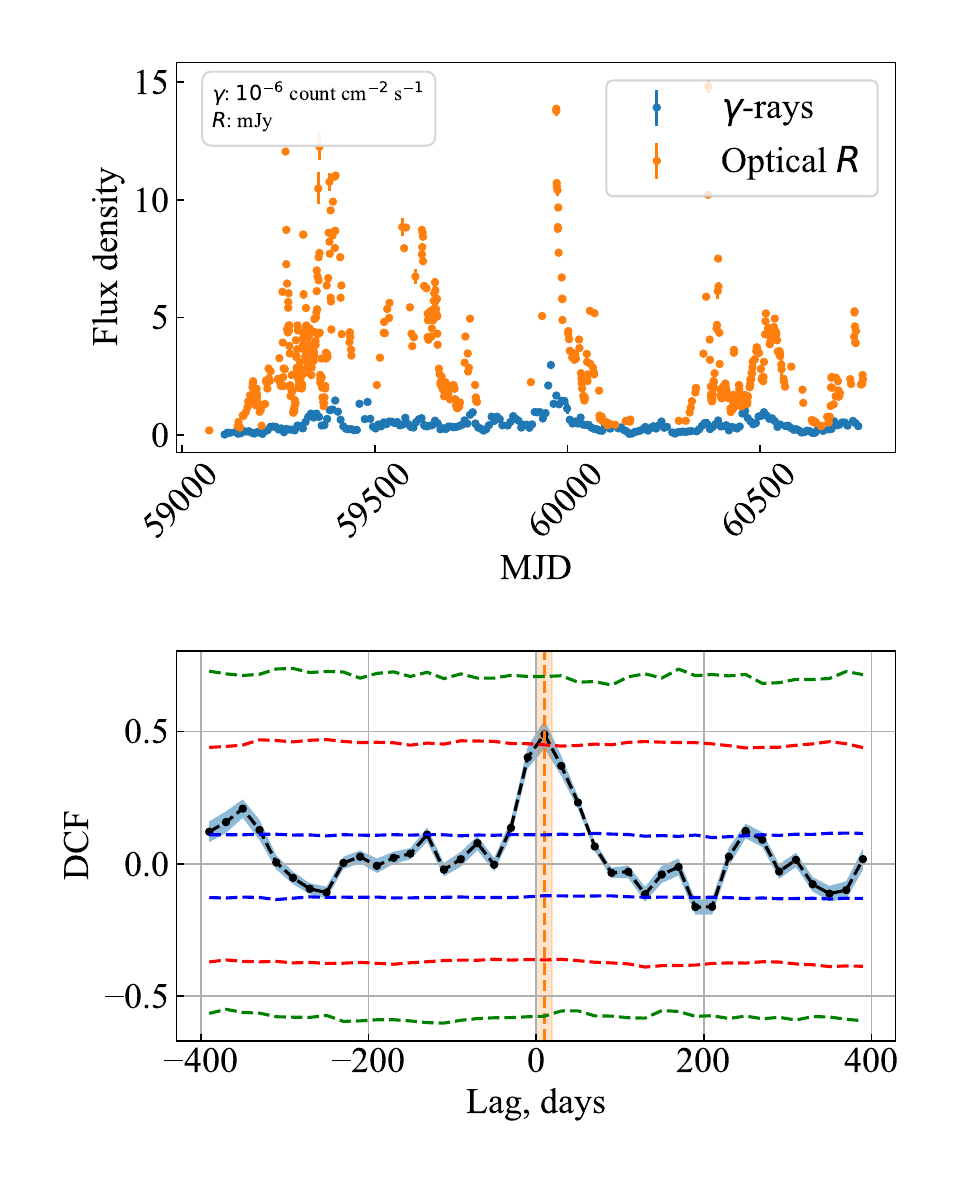}
\includegraphics[width=0.7\columnwidth]{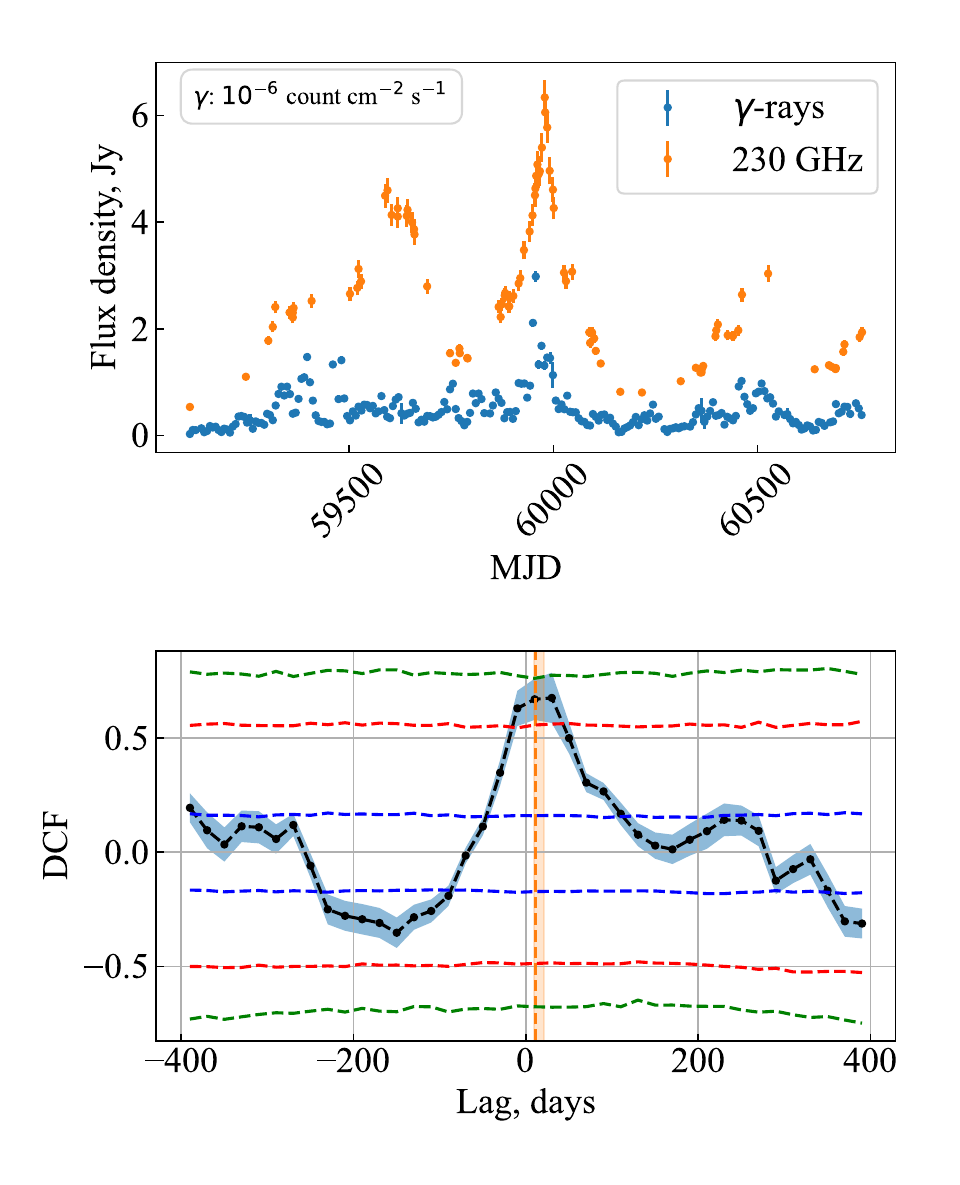}
\includegraphics[width=0.7\columnwidth]{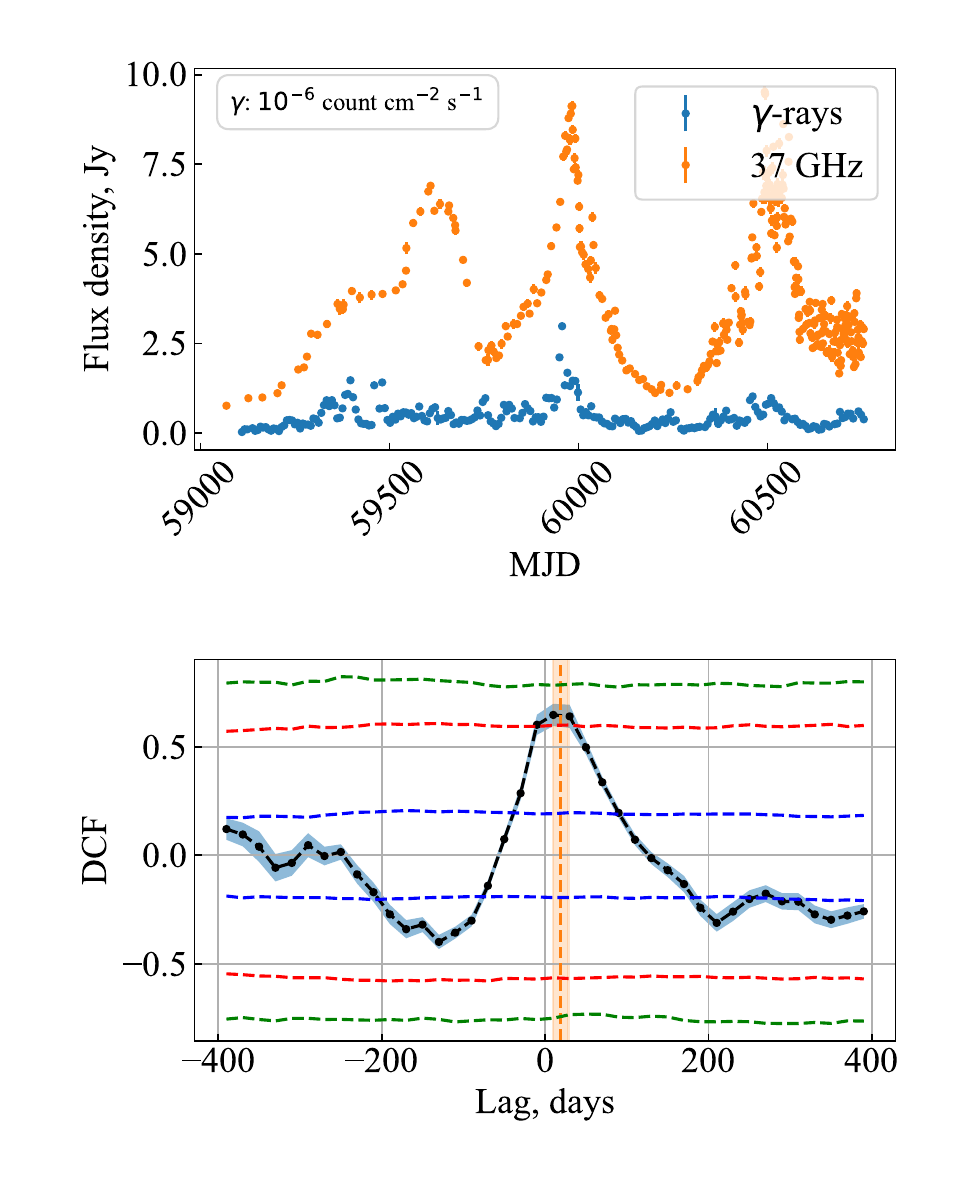}
}
\centerline{
\includegraphics[width=0.7\columnwidth]{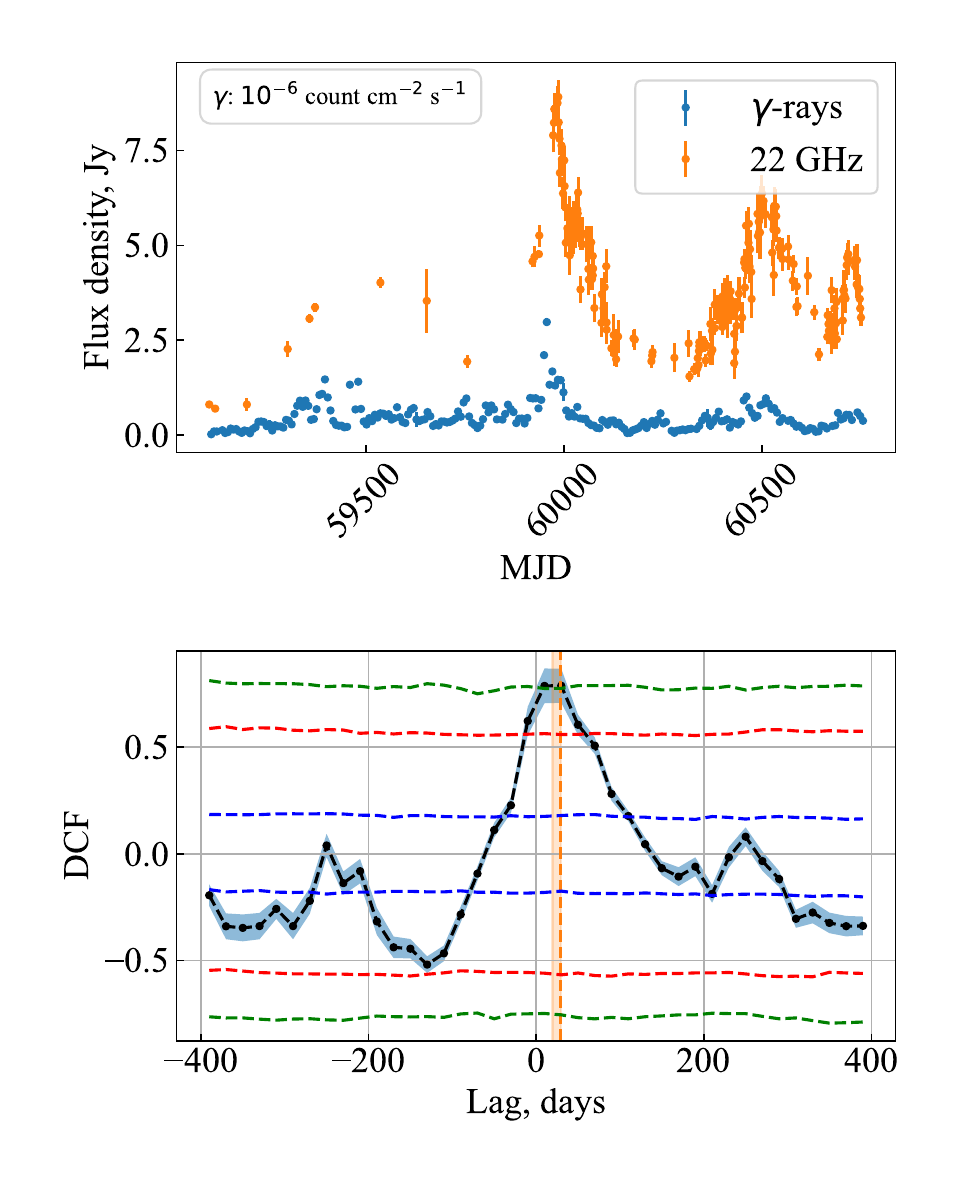}
\includegraphics[width=0.7\columnwidth]{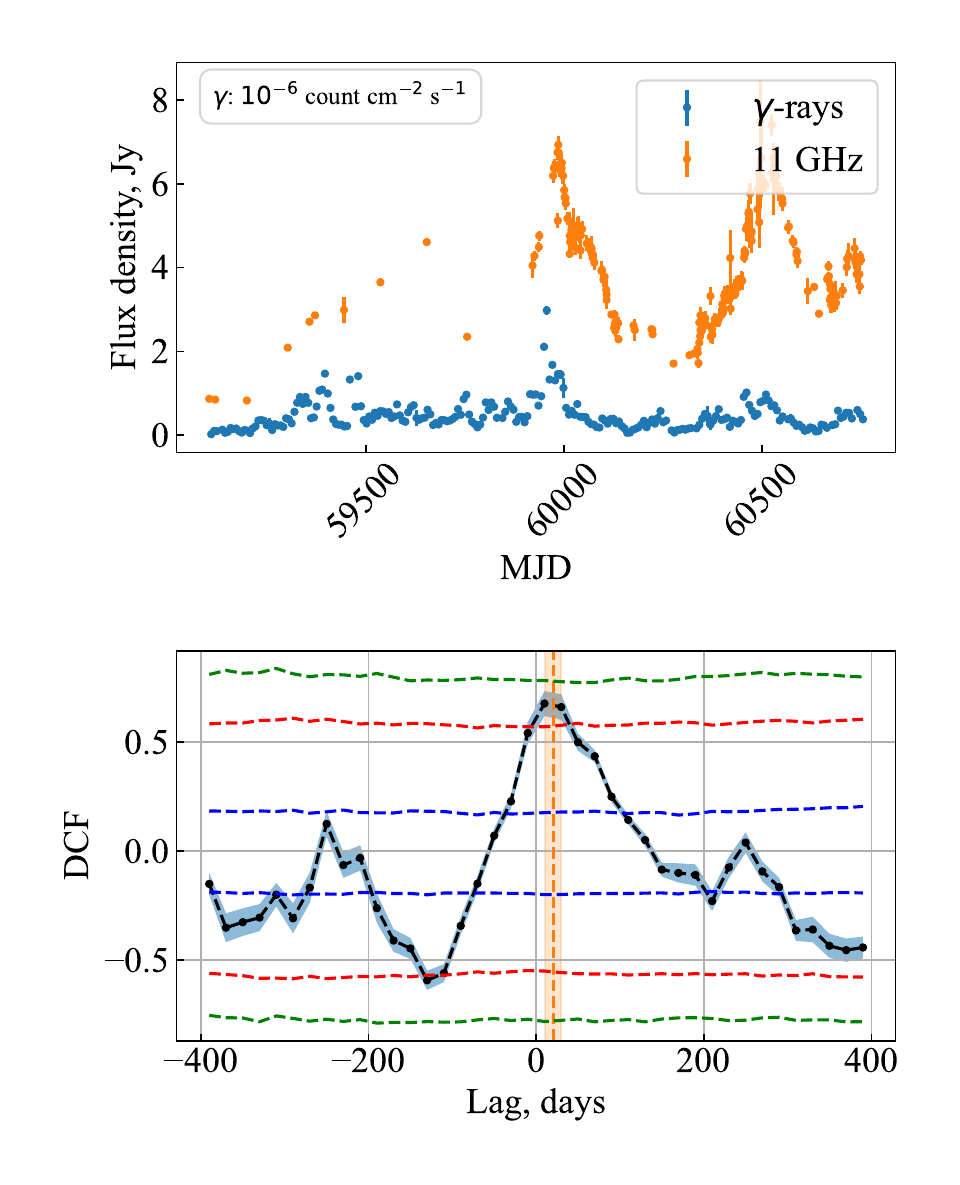}
\includegraphics[width=0.7\columnwidth]{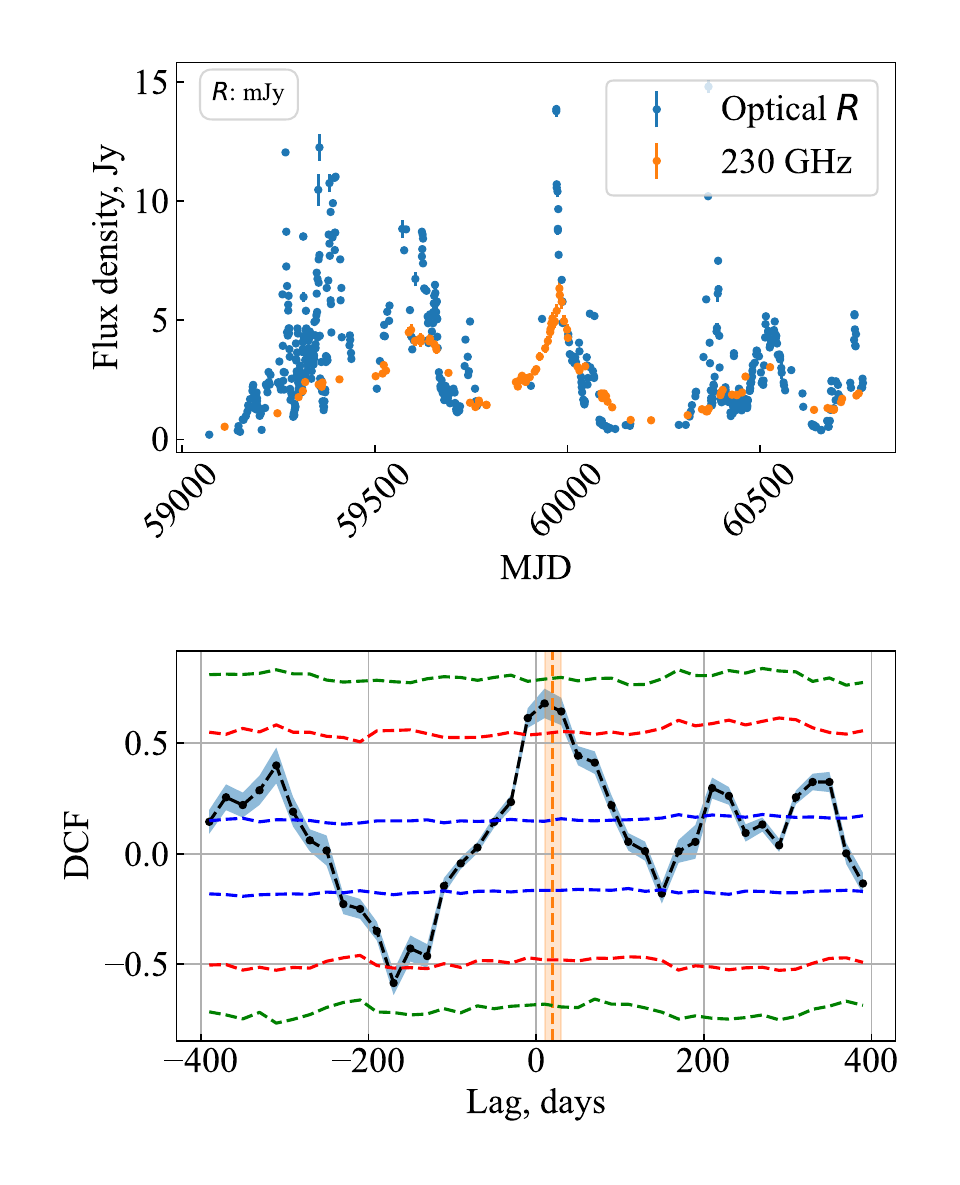}
}
\centerline{
\includegraphics[width=0.7\columnwidth]{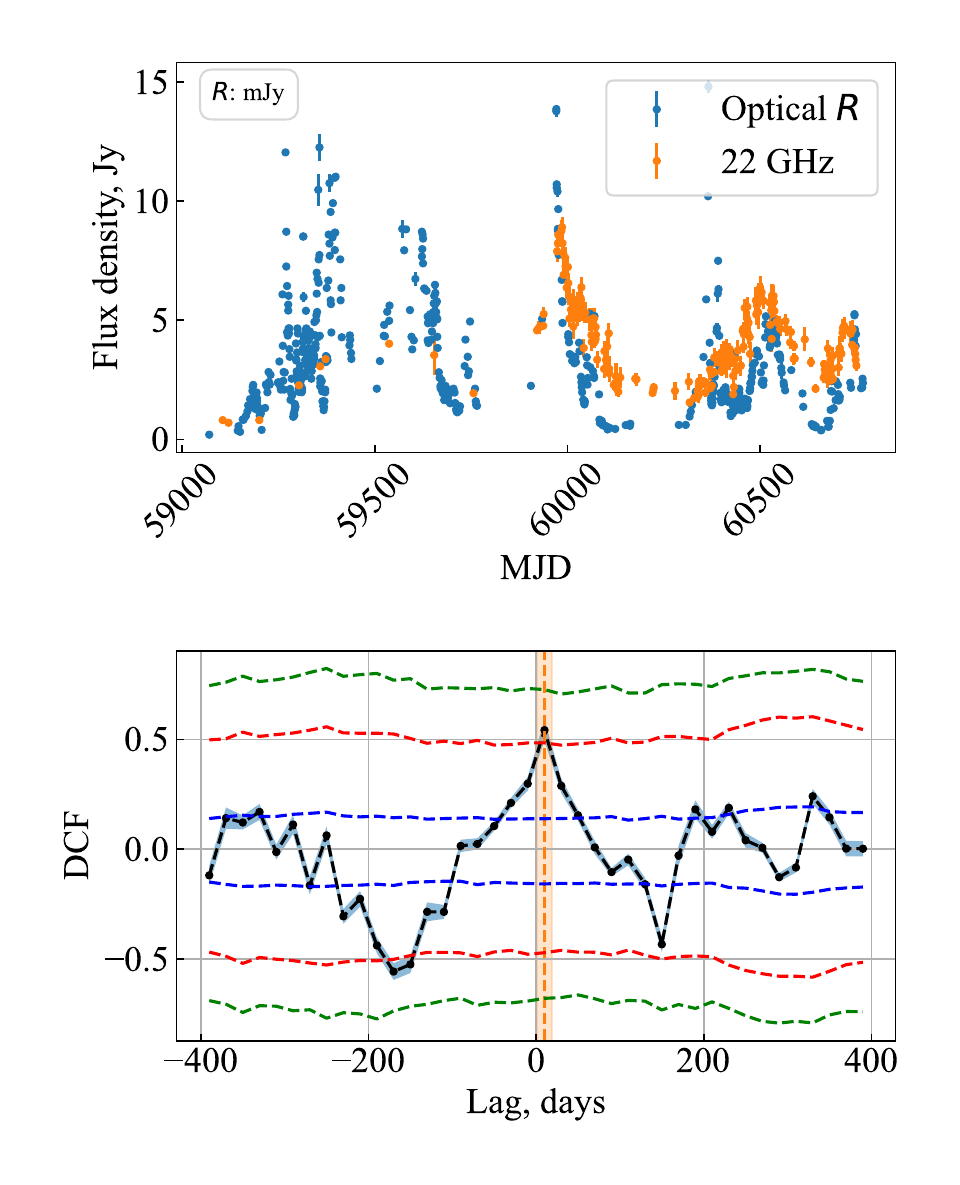}
\includegraphics[width=0.7\columnwidth]{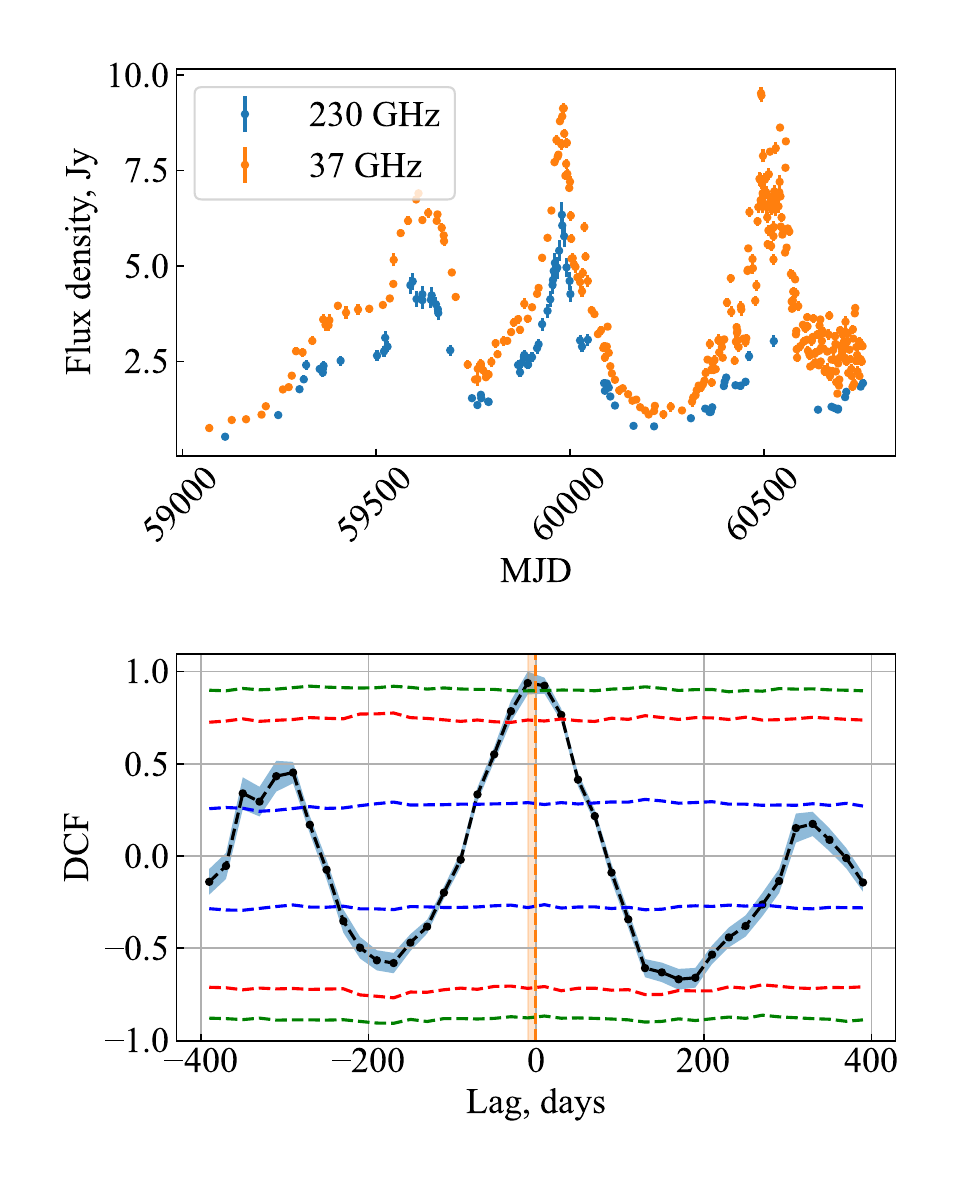}
\includegraphics[width=0.7\columnwidth]{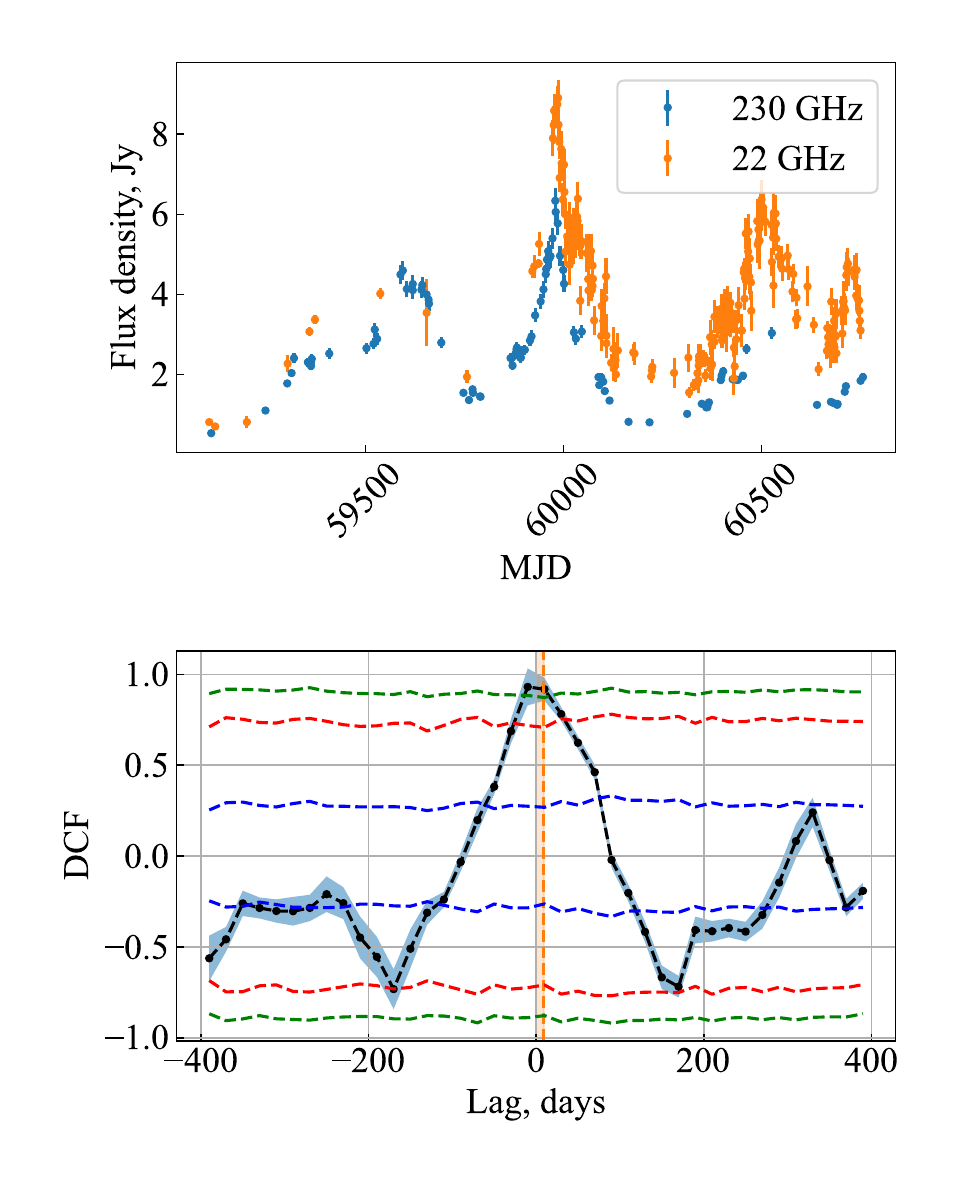}
}
\caption{The light curves and DCFs in epoch~4. Designations are as in Fig.~\ref{fig:dcf_ep1}} 
\label{fig:dcf_ep4}
\end{figure*}

\begin{figure*}
\centerline{
\includegraphics[width=0.7\columnwidth]{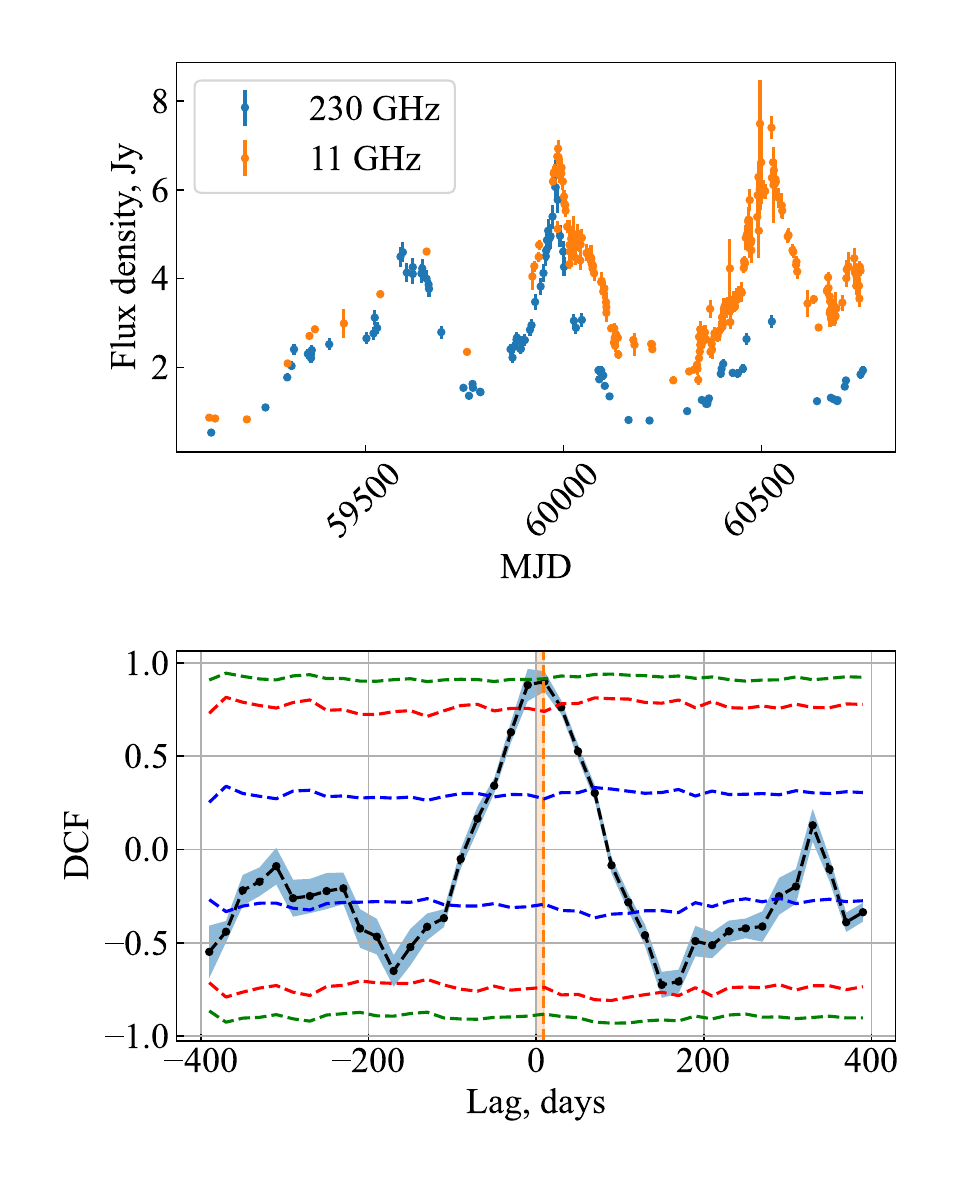}
\includegraphics[width=0.7\columnwidth]{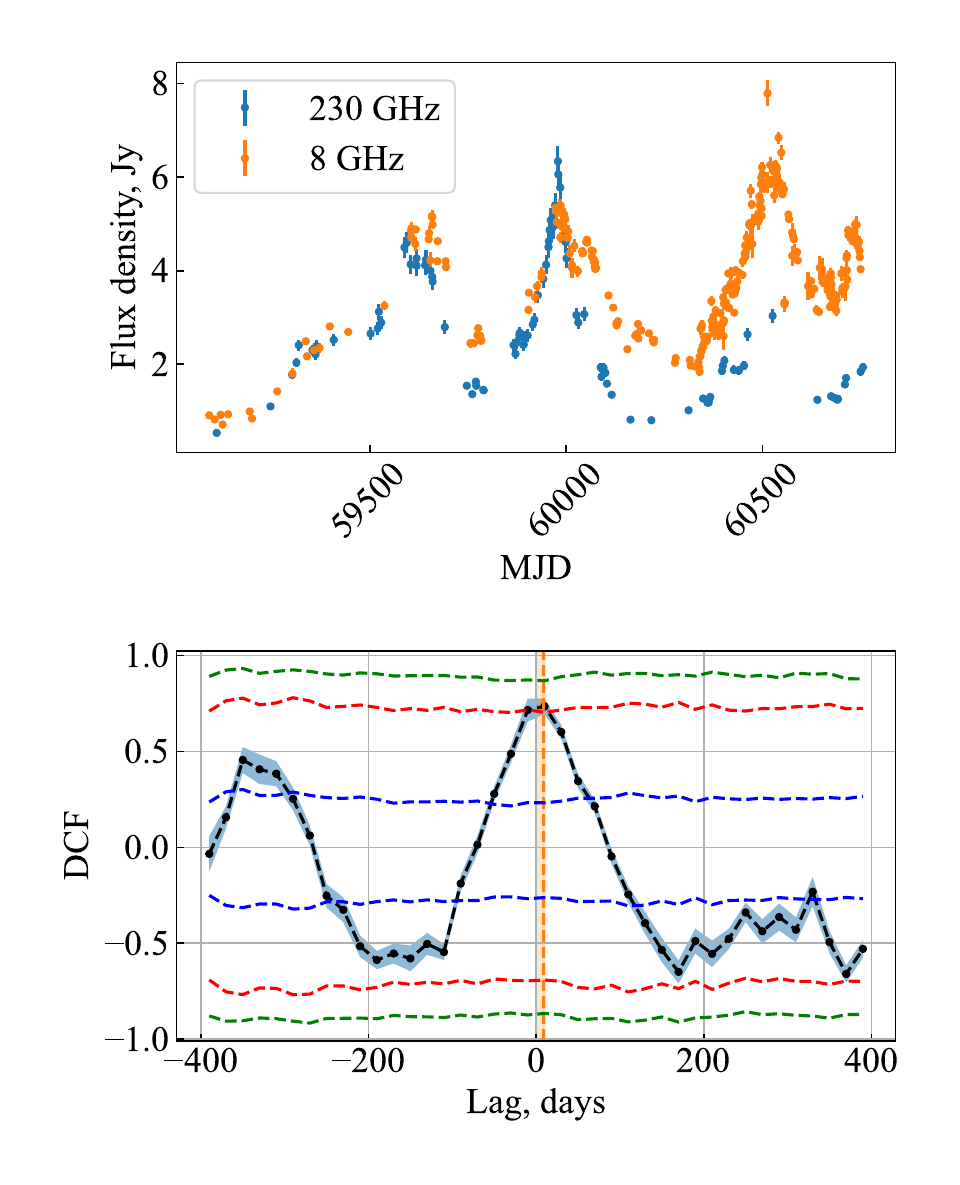}
\includegraphics[width=0.7\columnwidth]{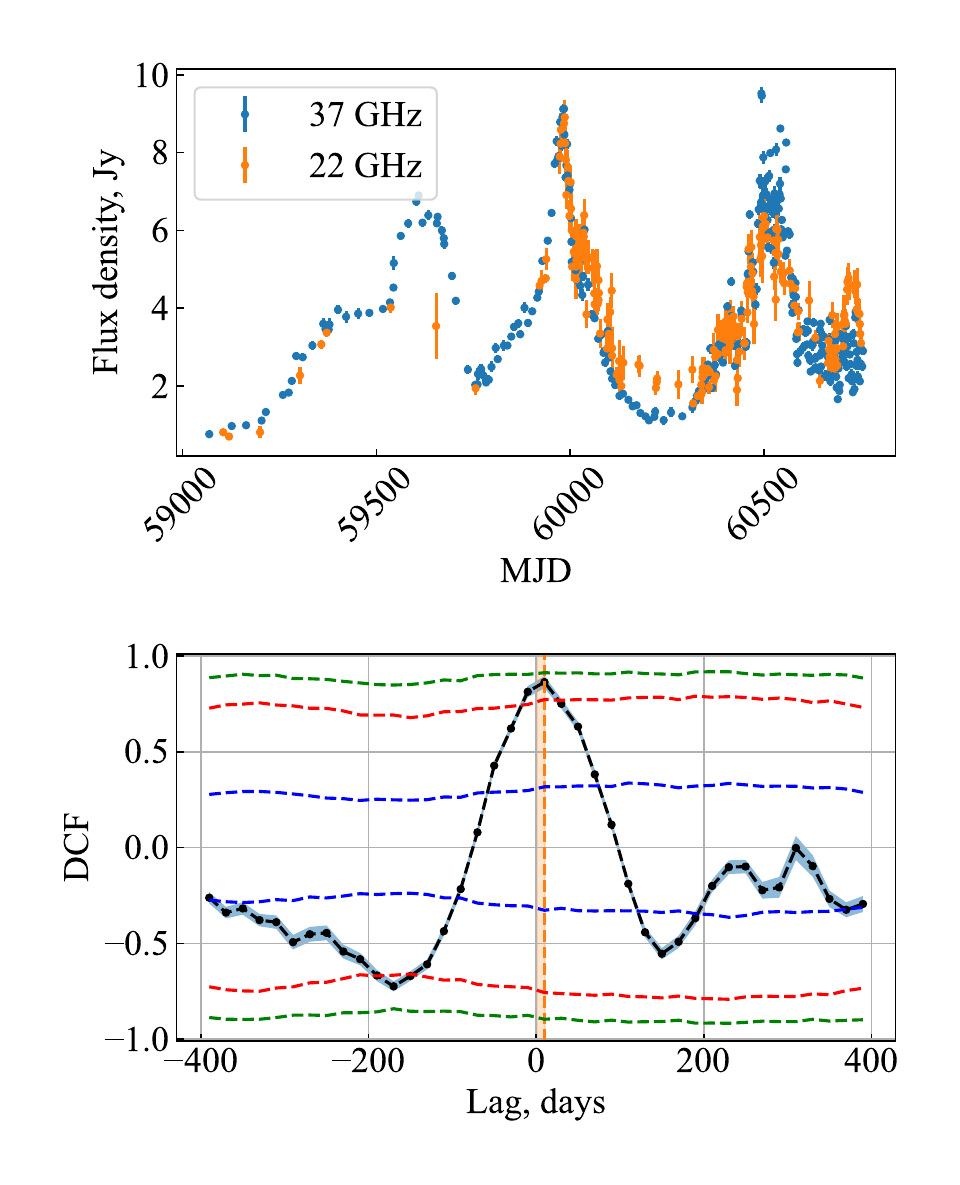}
}
\centerline{
\includegraphics[width=0.7\columnwidth]{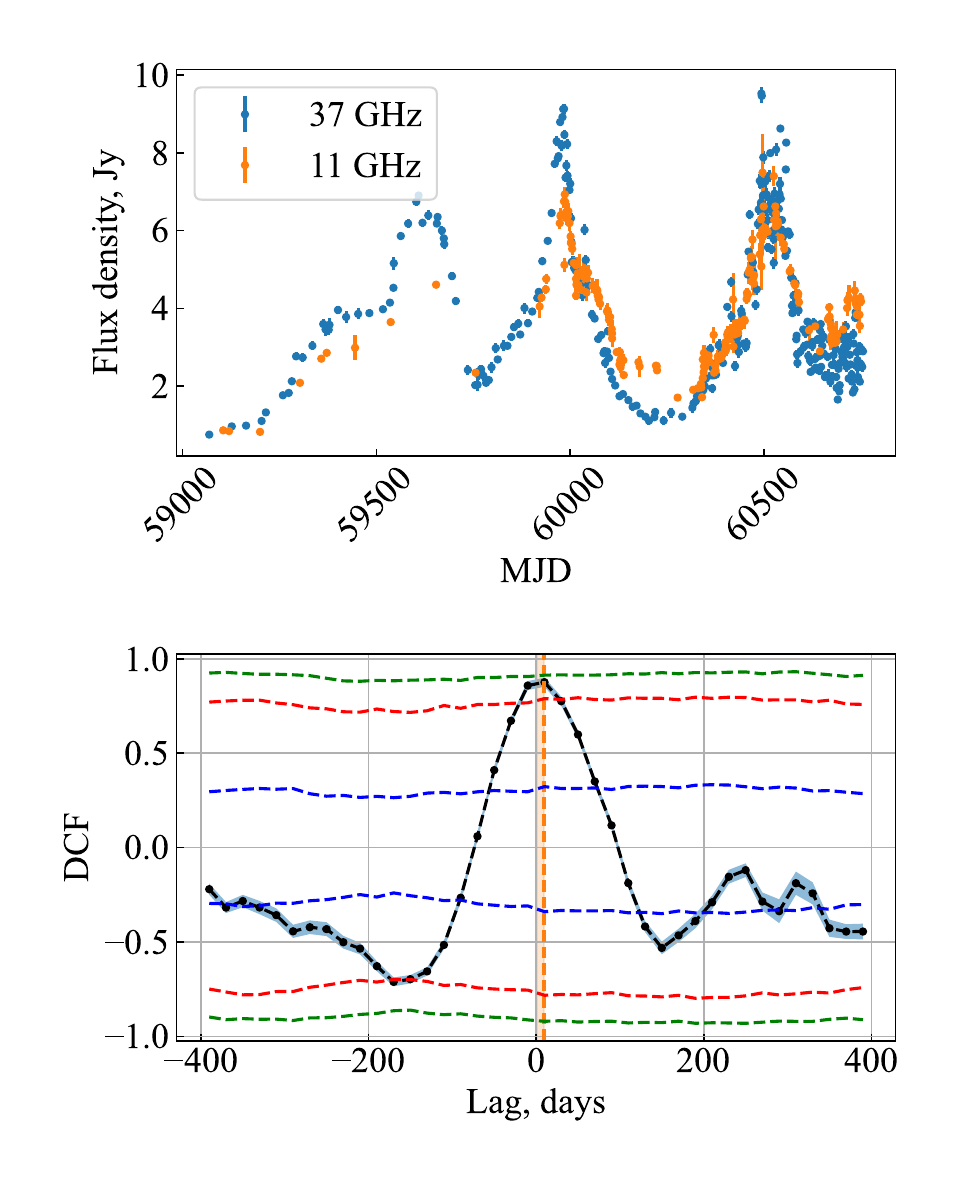}
\includegraphics[width=0.7\columnwidth]{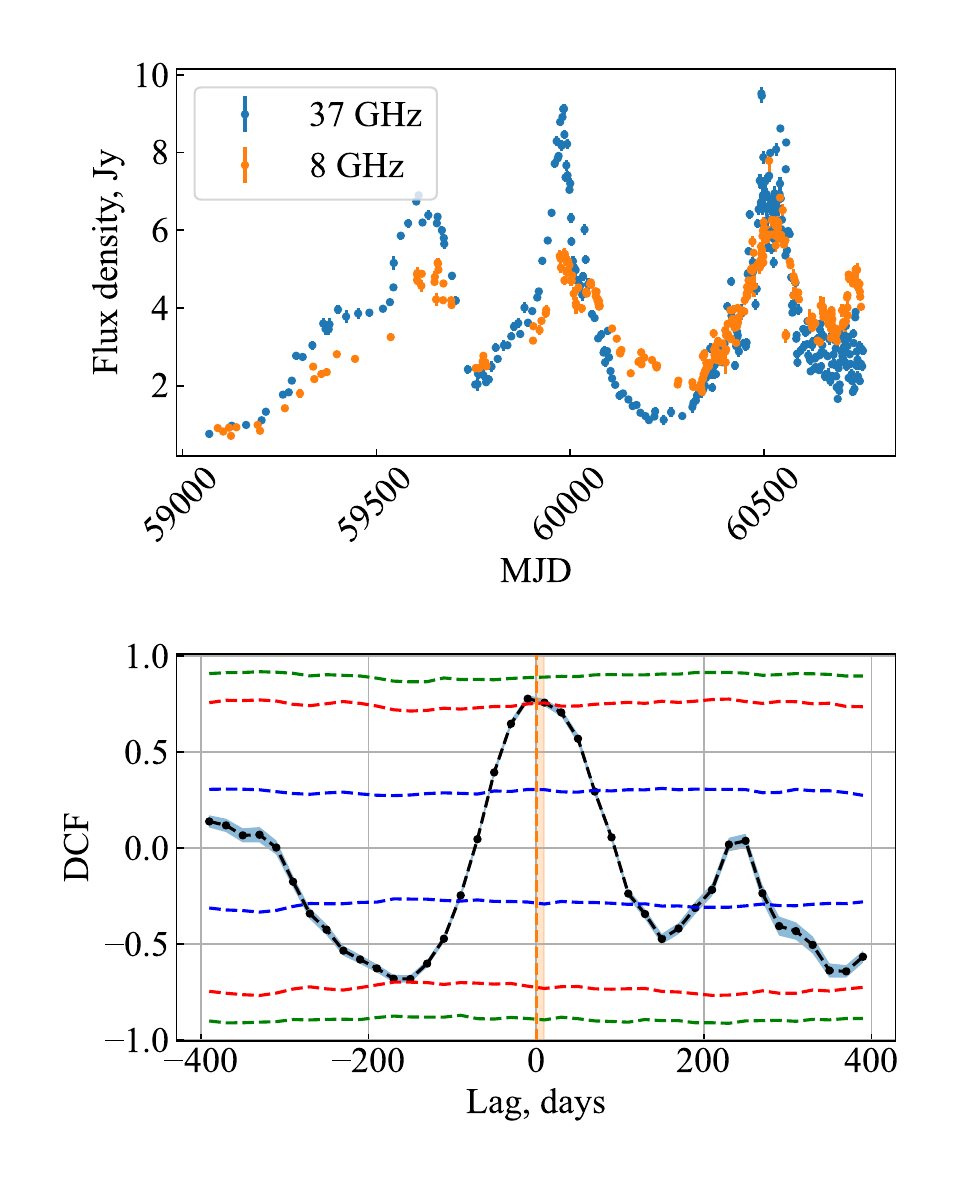}
\includegraphics[width=0.7\columnwidth]{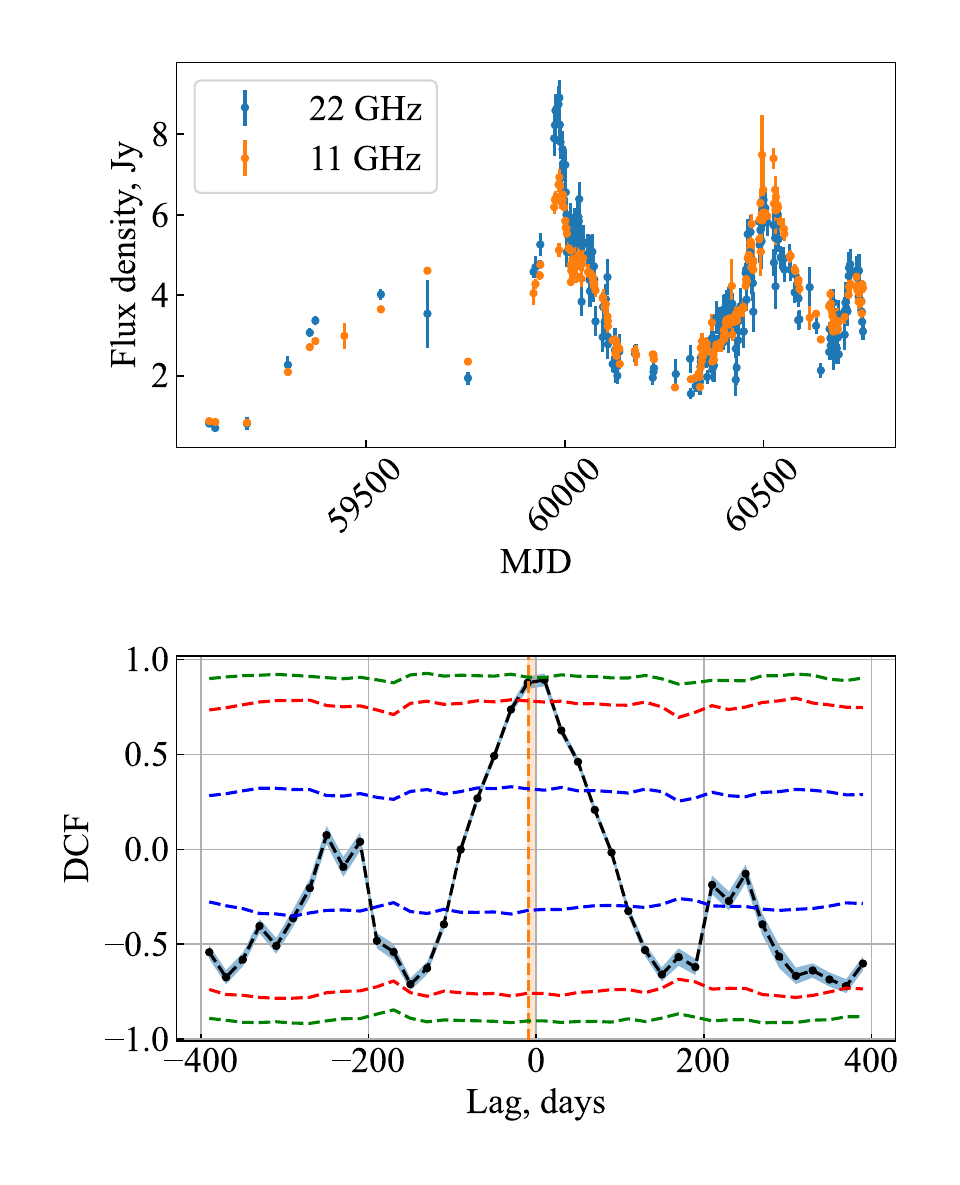}
}
\centerline{
\includegraphics[width=0.7\columnwidth]{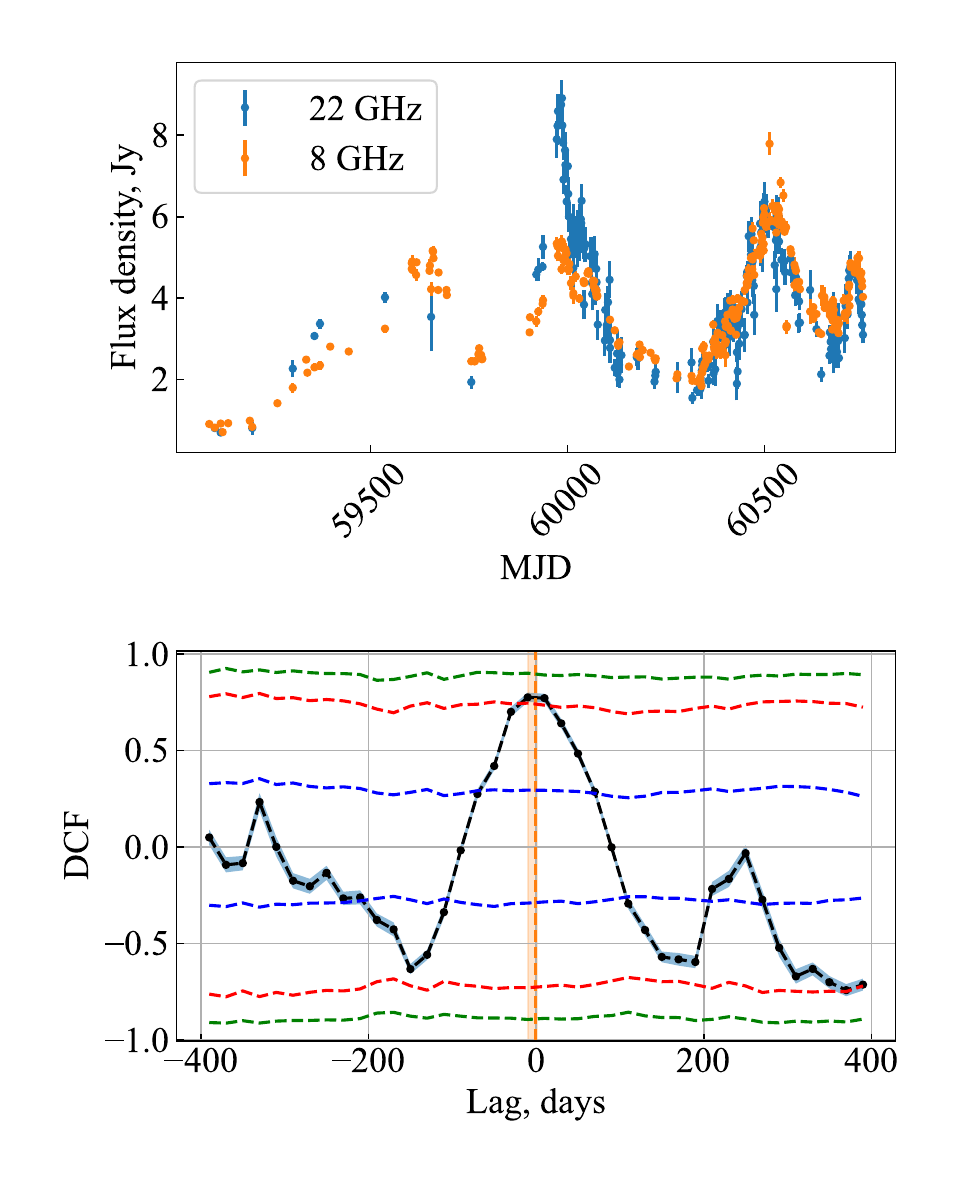}
\includegraphics[width=0.7\columnwidth]{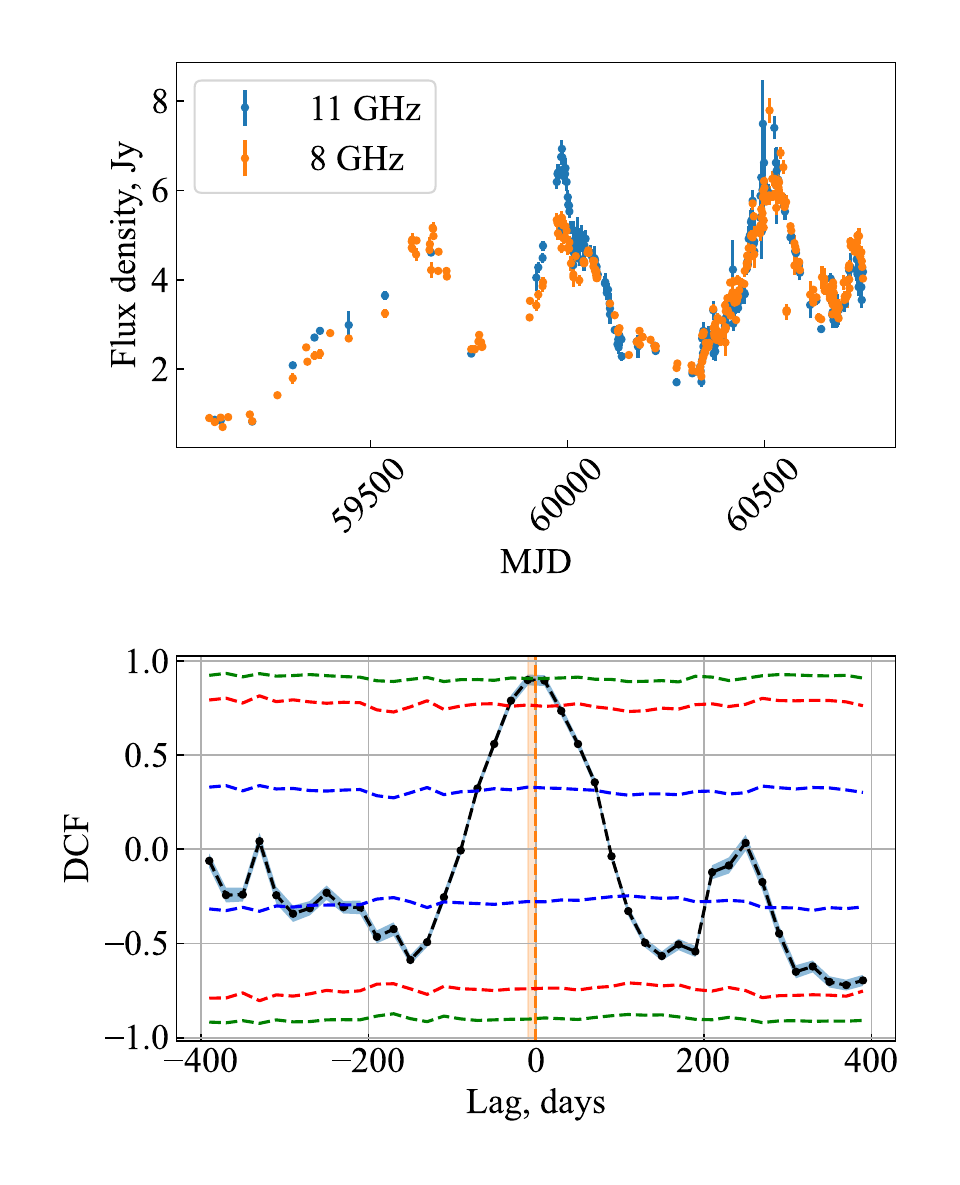}
\includegraphics[width=0.7\columnwidth]{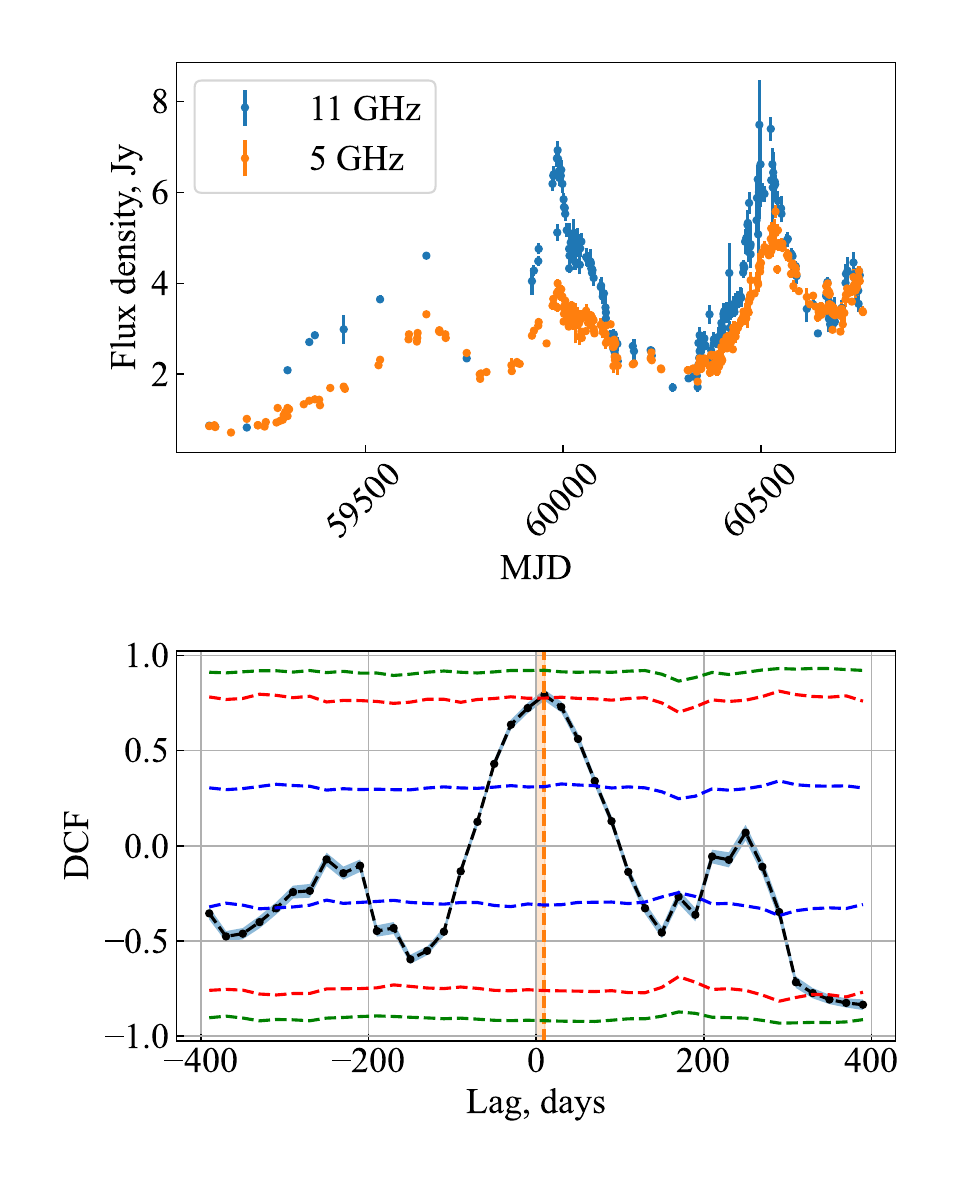}
}
\contcaption{The light curves and DCFs in epoch~4.} 
\end{figure*}

\begin{figure*}
\centerline{
\includegraphics[width=0.7\columnwidth]{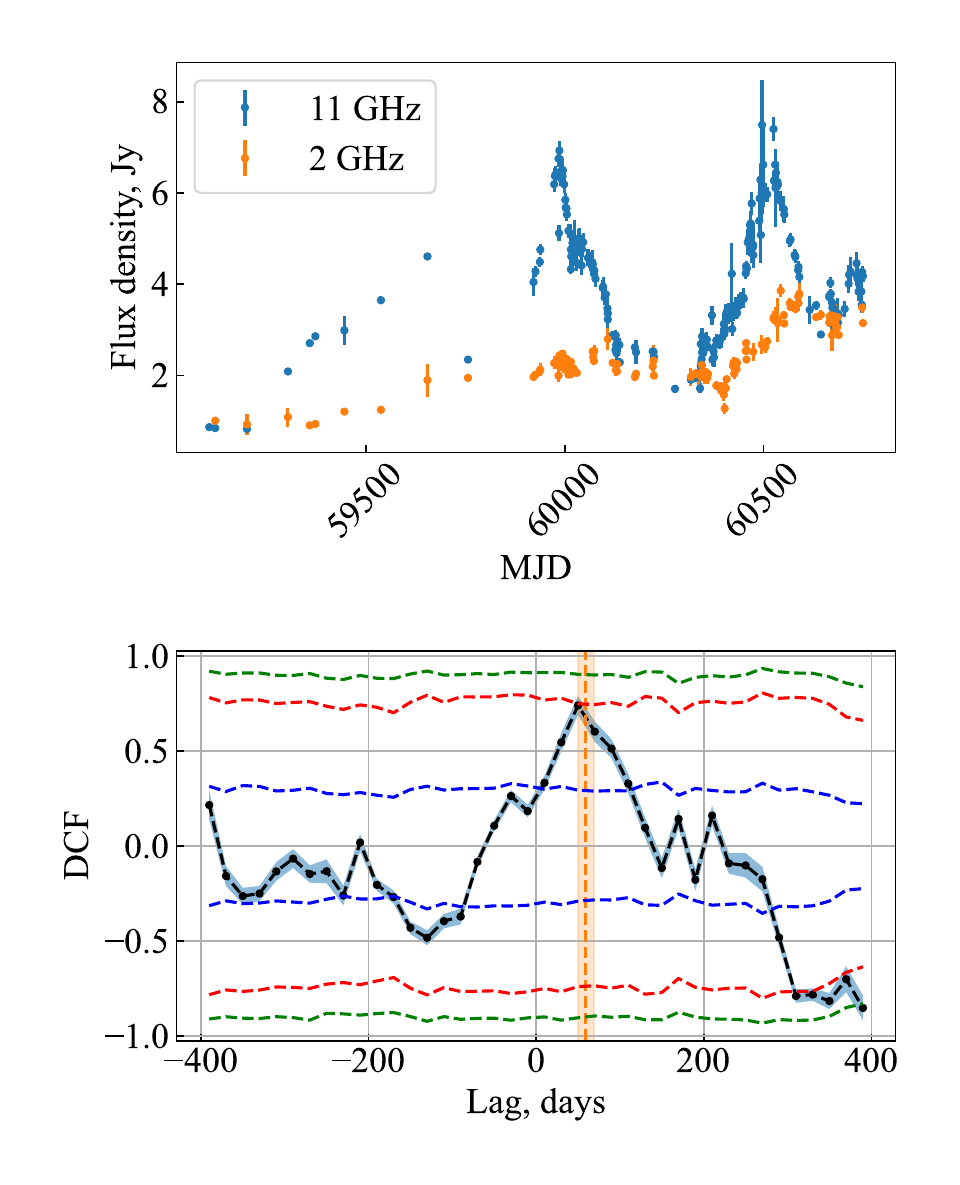}
\includegraphics[width=0.7\columnwidth]{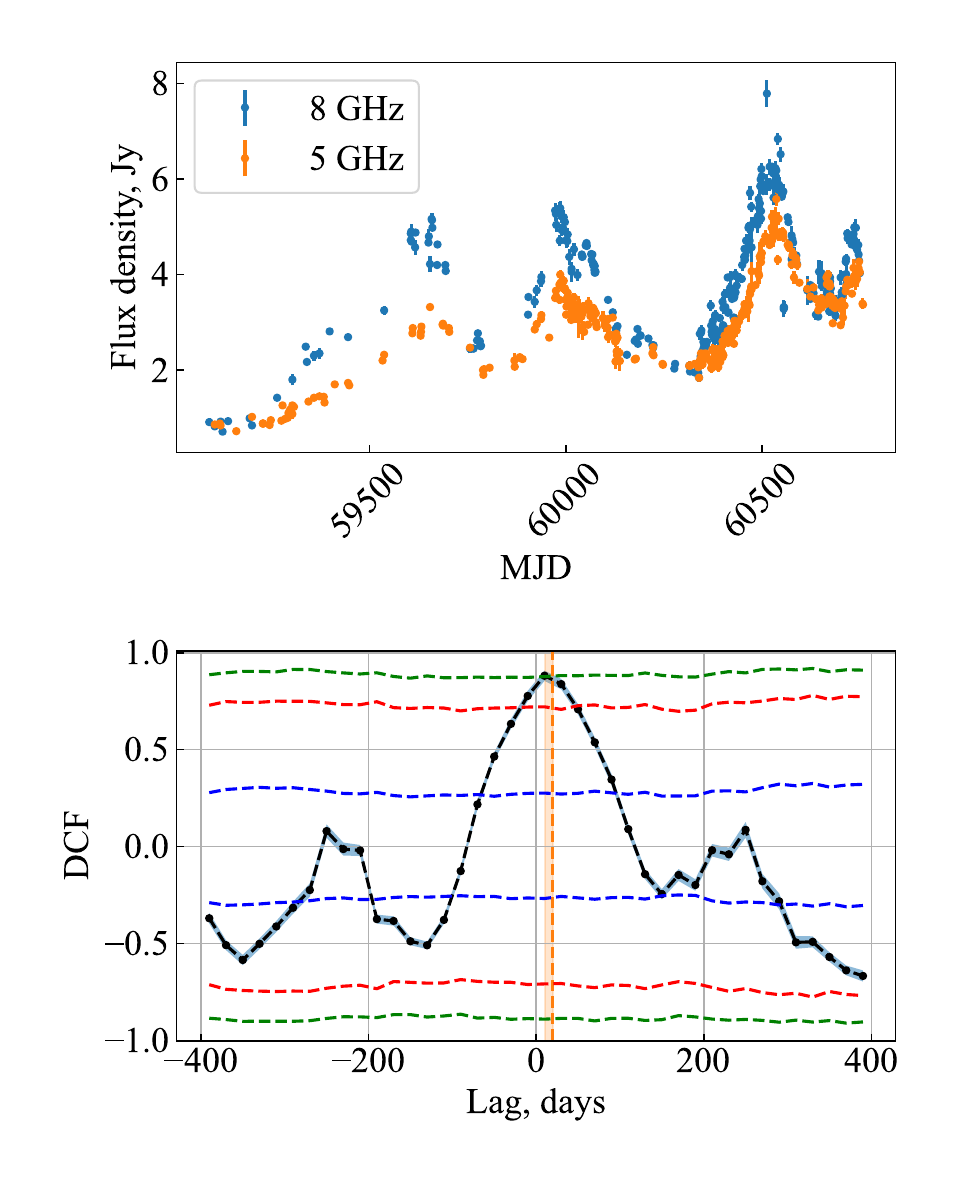}
\includegraphics[width=0.7\columnwidth]{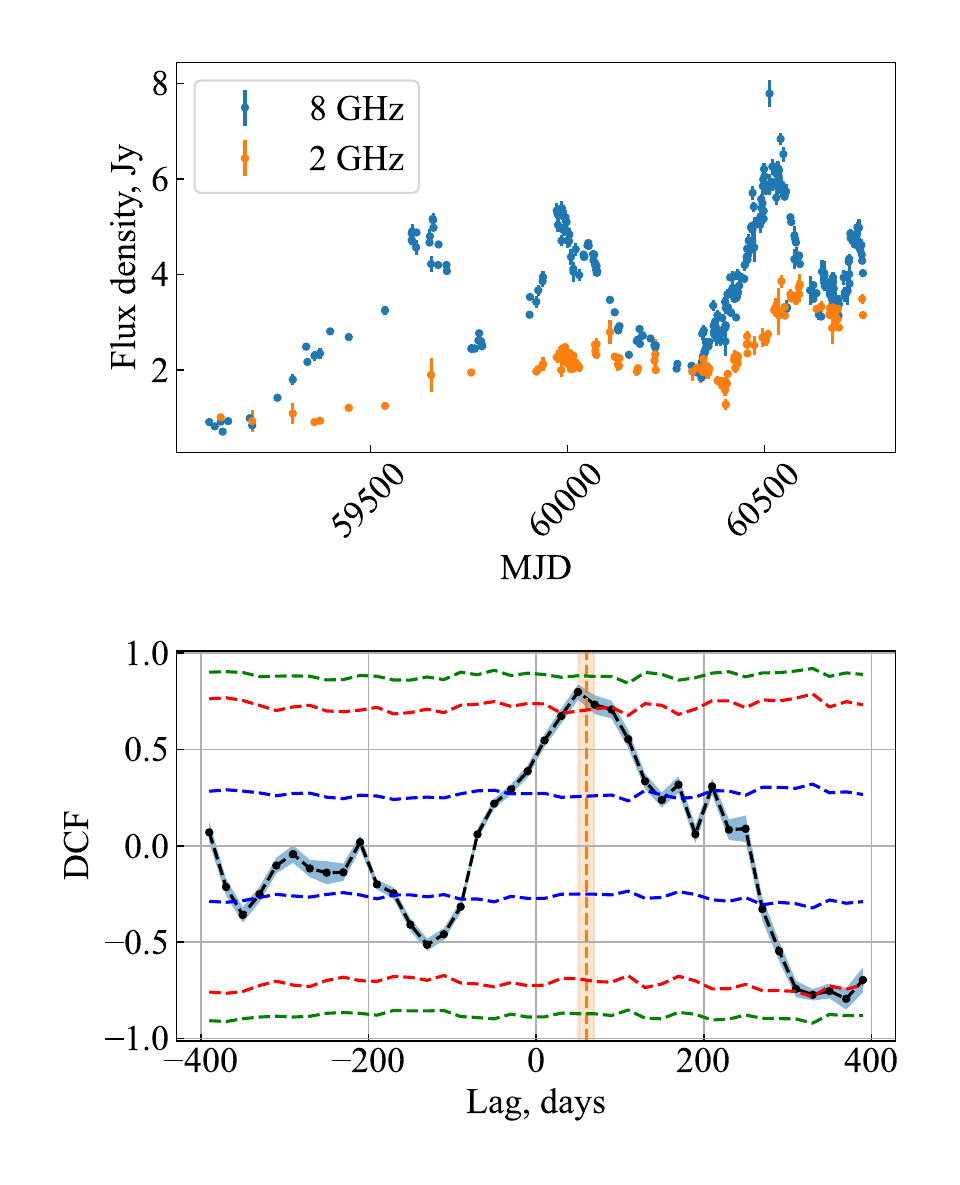}
}
\centerline{
\includegraphics[width=0.7\columnwidth]{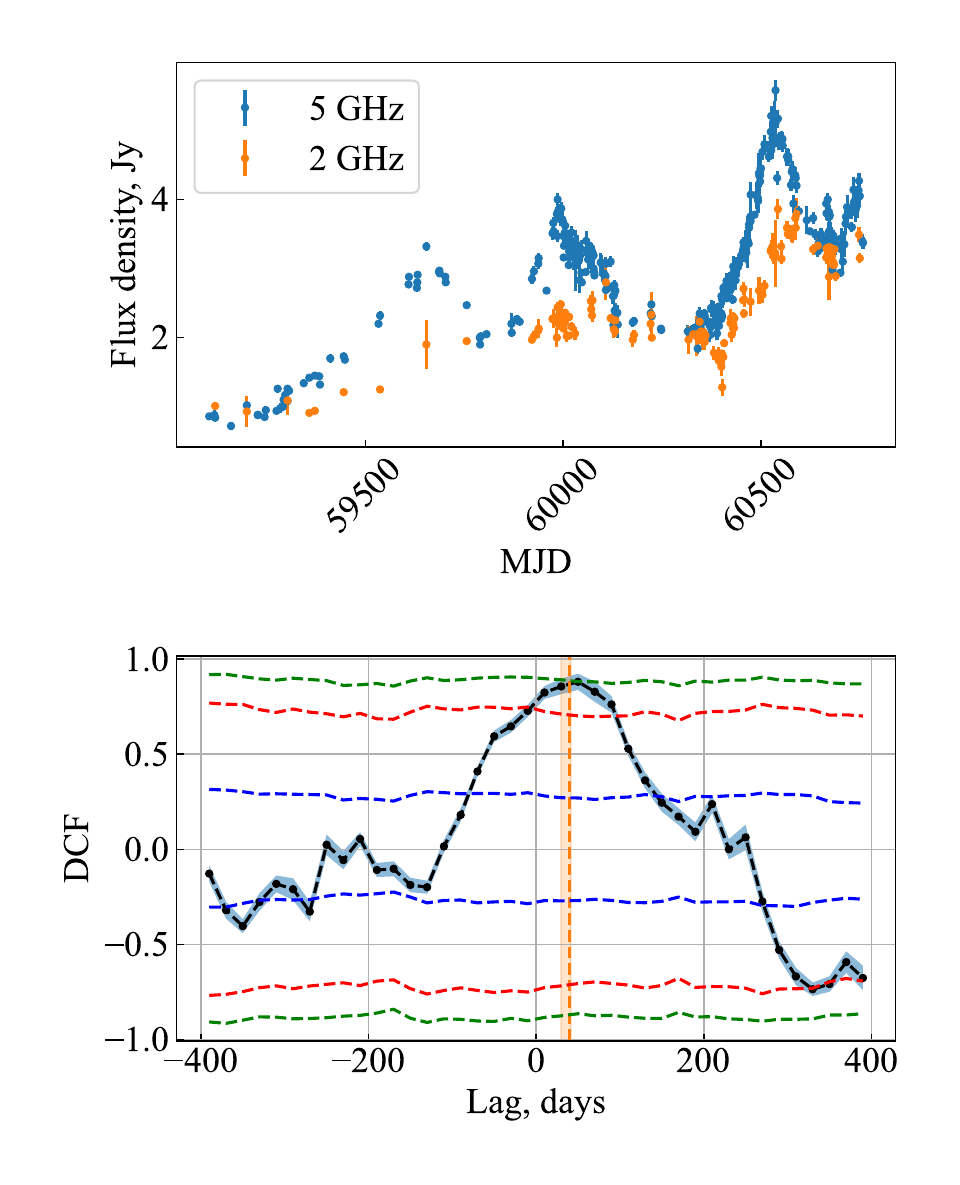}
}
\contcaption{The light curves and DCFs in epoch~4.} 
\end{figure*}

\clearpage
\newpage

\section{SMBH binary model}
Figures~\ref{fig:Ton599_model_2GHz(1.4-6)}--\ref{fig:Ton599_model_230GHz(1.3-7.7)} present the light curves of Ton\,599 at \mbox{2--230}~GHz, overlaid with the best-fitting binary SMBH model incorporating jet precession and orbital motion, as described in Section~\ref{sec:jet_precession_orbital}. Figure~\ref{fig:Ton599_compare_model} illustrates the improved agreement between the observed light curve and the combined orbital-precessional model compared to the precession-only scenario. Figures~\ref{fig:Ton599_model_2GHz_GARCH}--\ref{fig:Ton599_model_230GHz_GARCH} show the jet precession and orbital motion model plotted alongside the ARIMA and GARCH model fits to the 2--230 GHz light curves of Ton\,599. 

Table~\ref{tab:precession_orbit_results} presents the best-fitting parameters from the jet precession and orbital motion modelling. Table~\ref{tab:arma} lists the fit parameters of the ARIMA and GARCH models. Table~\ref{tab:compare} summarizes the binary SMBH and jet parameters for OJ\,287, 3C\,345, and Ton\,599, compiled from the literature and derived in the present study.

\begin{figure}
\centerline{\includegraphics[width=\columnwidth]{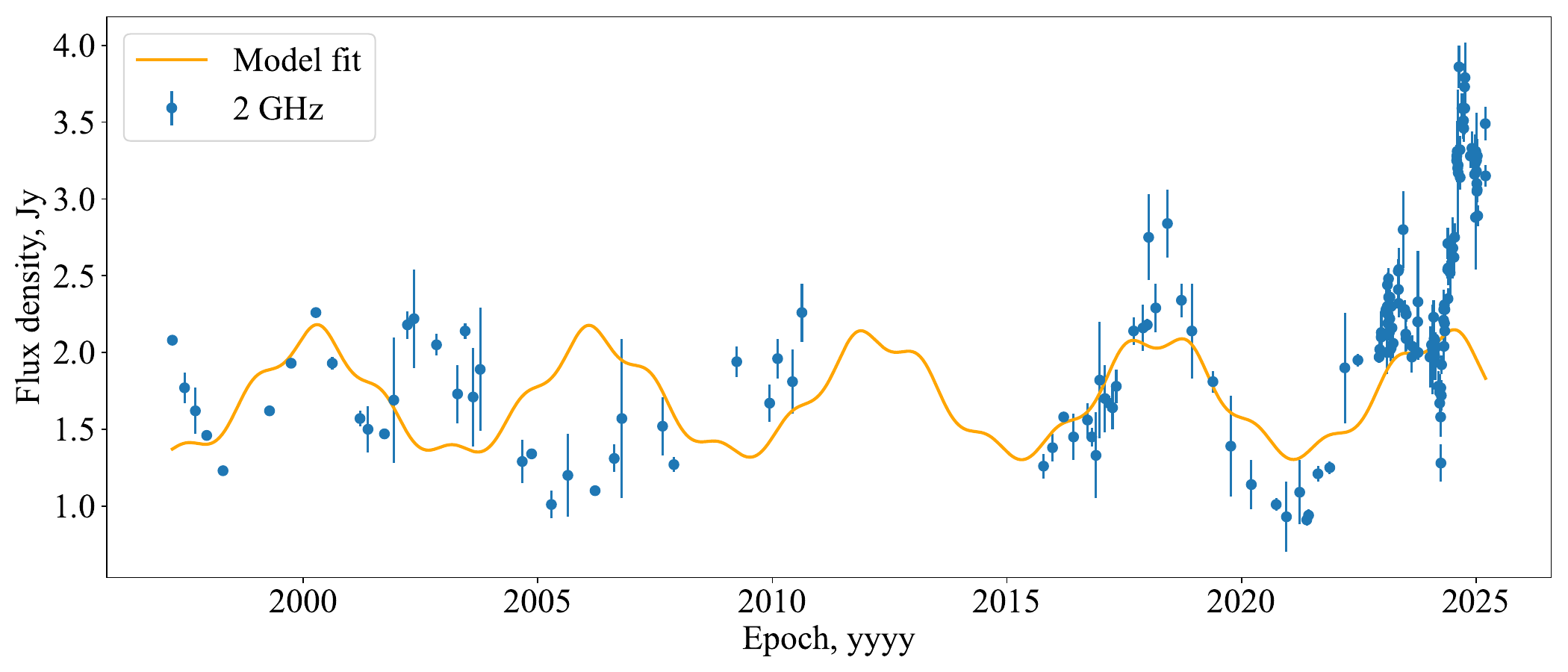}}
\caption{The jet orbital motion and precession model implemented at 
the 2 GHz light curve of Ton\,599 ($P_{\mathrm{pr}}=6$ yr, $P_{\mathrm{orb}}=1.4$ yr).} 
\label{fig:Ton599_model_2GHz(1.4-6)}
\end{figure}

\begin{figure}
\centerline{\includegraphics[width=\columnwidth]{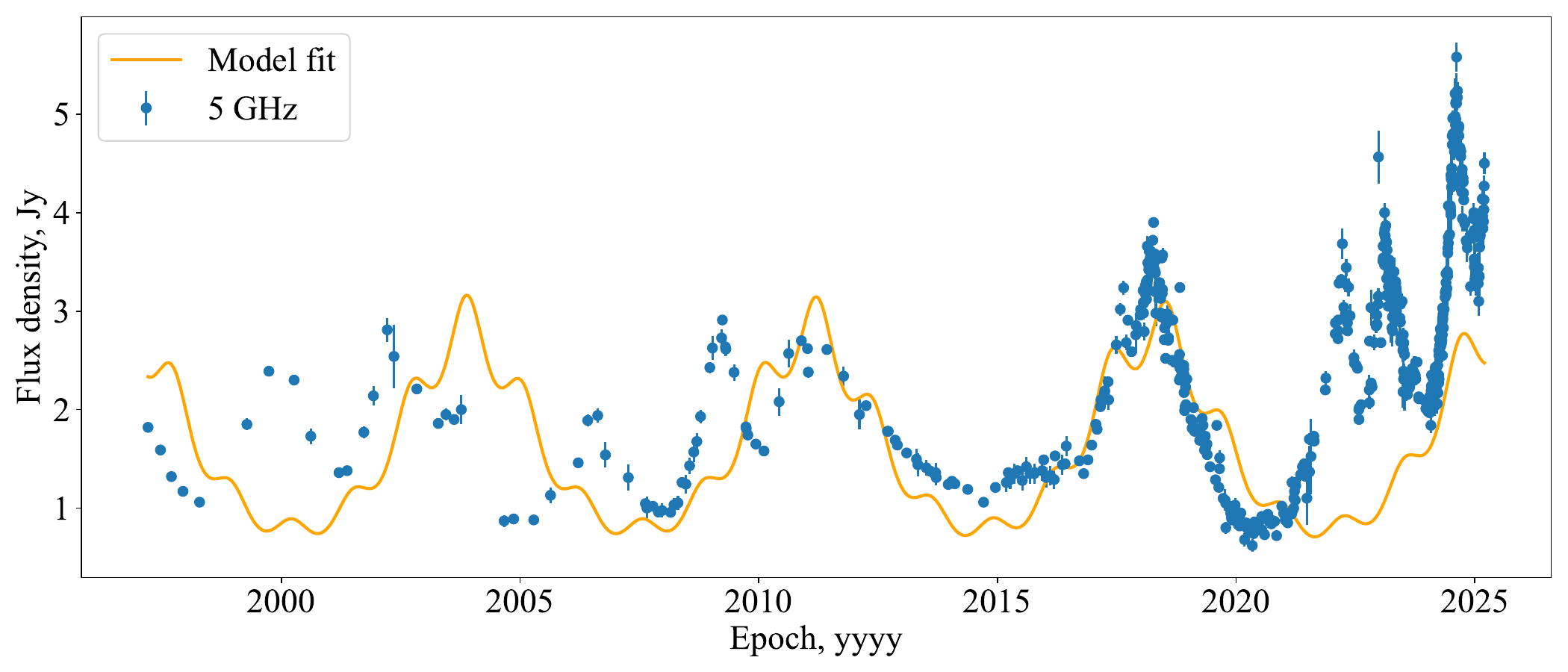}}
\caption{The jet orbital motion and precession model implemented at 
the 5 GHz light curve of Ton\,599 ($P_{\mathrm{pr}}=7.2$ yr, $P_{\mathrm{orb}}=1.2$ yr).} 
\label{fig:Ton599_model_5GHz(1.2-7.2)}
\end{figure}

\begin{figure}
\centerline{\includegraphics[width=\columnwidth]{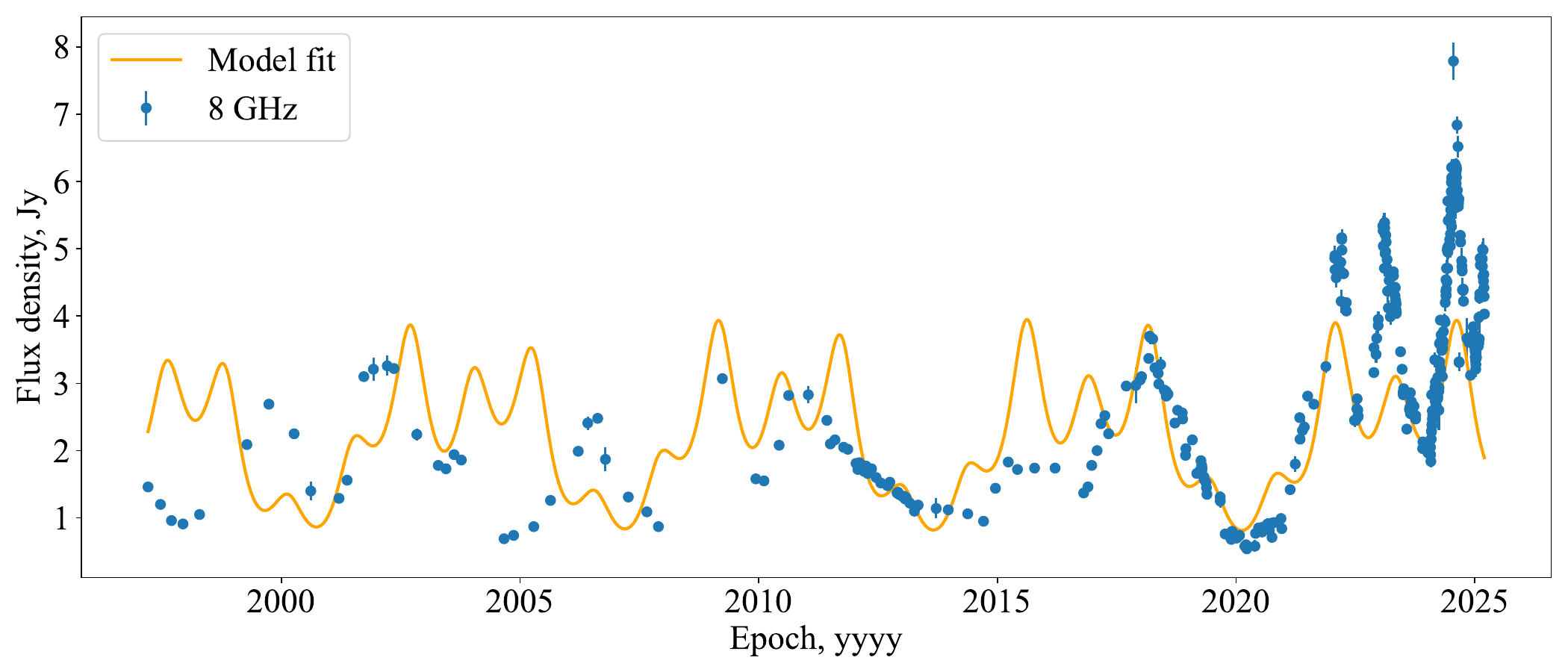}}
\caption{The jet orbital motion and precession model implemented at 
the 8 GHz light curve of Ton\,599 ($P_{\mathrm{pr}}=6.6$ yr, $P_{\mathrm{orb}}=1.3$ yr).} 
\label{fig:Ton599_model_8GHz(1.3-6.6)}
\end{figure}

\begin{figure}
\centerline{\includegraphics[width=\columnwidth]{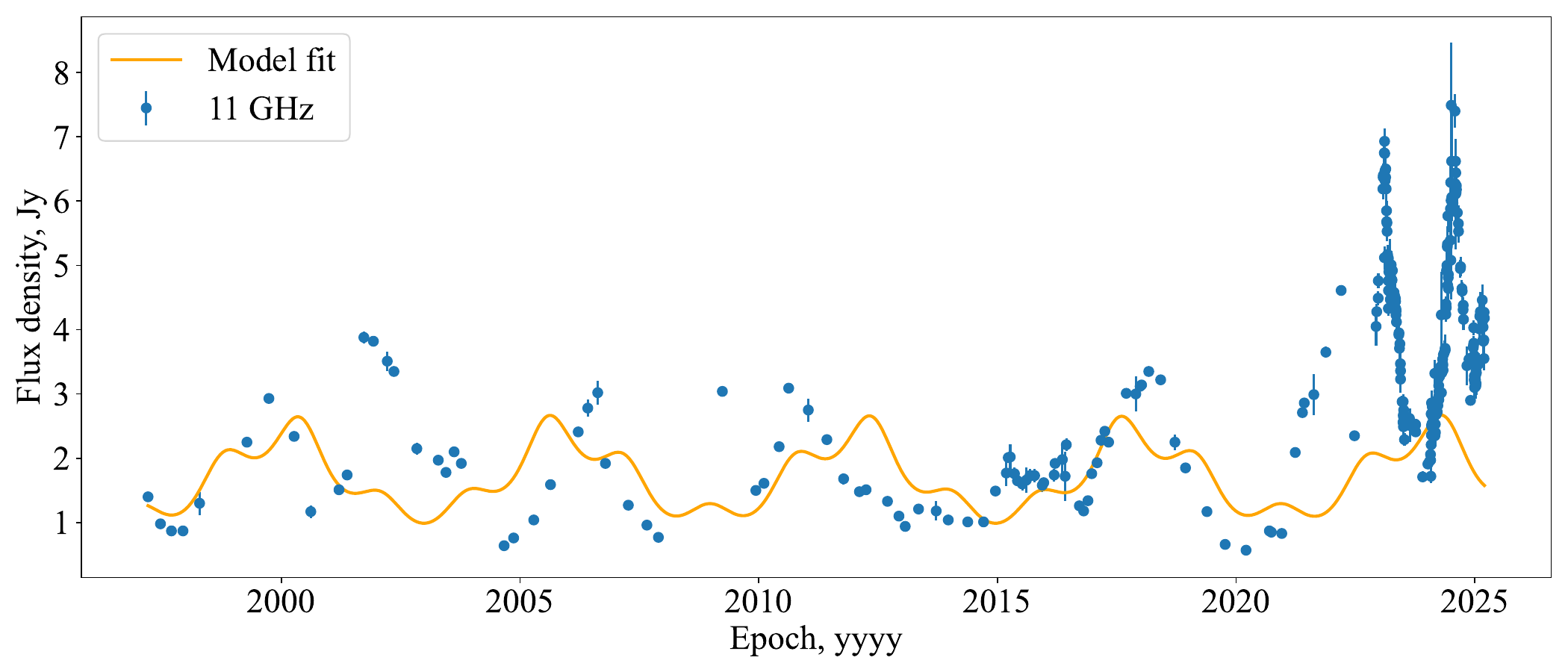}}
\caption{The jet orbital motion and precession model implemented at 
the 11 GHz light curve of Ton\,599 ($P_{\mathrm{pr}}=6$ yr, $P_{\mathrm{orb}}=1.7$ yr).} 
\label{fig:Ton599_model_11GHz(1.7-6.0)}
\end{figure}

\begin{figure}
\centerline{\includegraphics[width=\columnwidth]{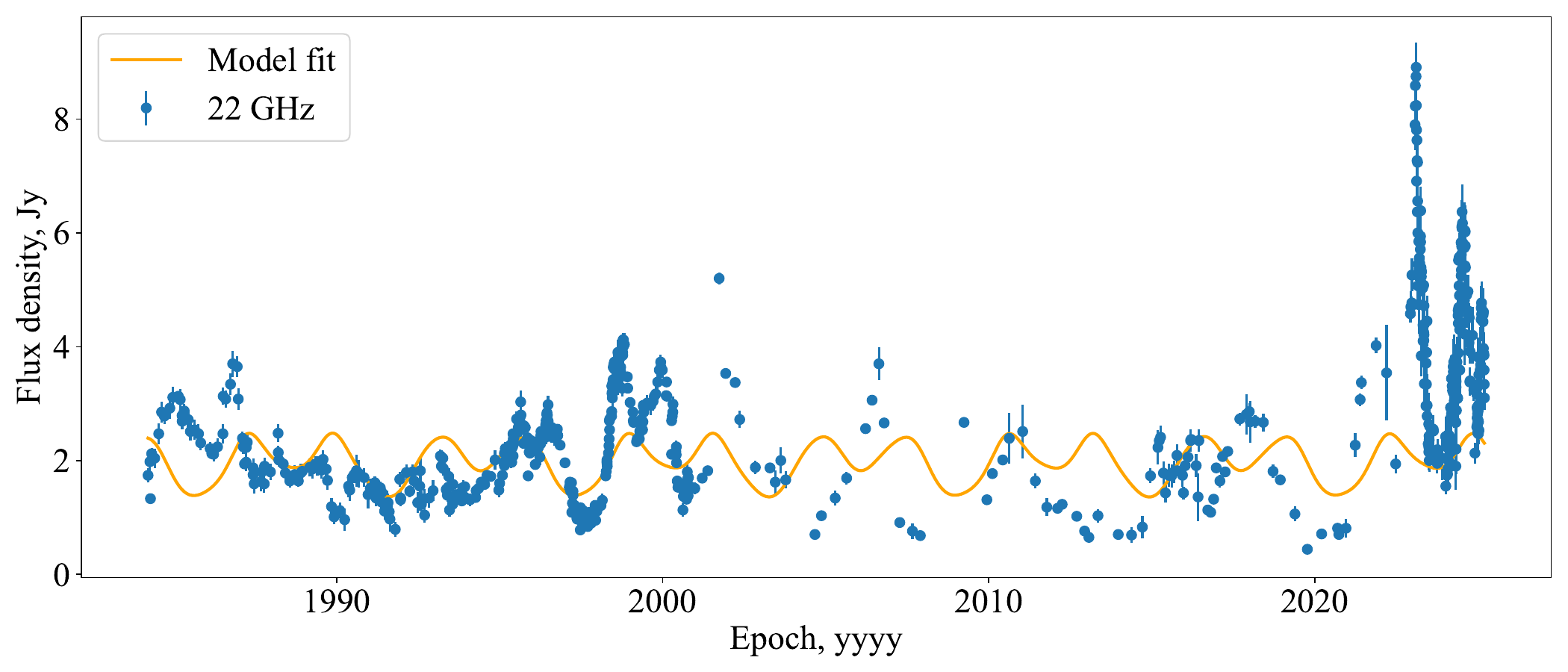}}
\caption{The jet orbital motion and precession model implemented at 
the 22 GHz light curve of Ton\,599 ($P_{\mathrm{pr}}=5.8$ yr, $P_{\mathrm{orb}}=1.3$ yr).} 
\label{fig:Ton599_model_22GHz(1.3-5.8)}
\end{figure}

\begin{figure}
\centerline{\includegraphics[width=\columnwidth]{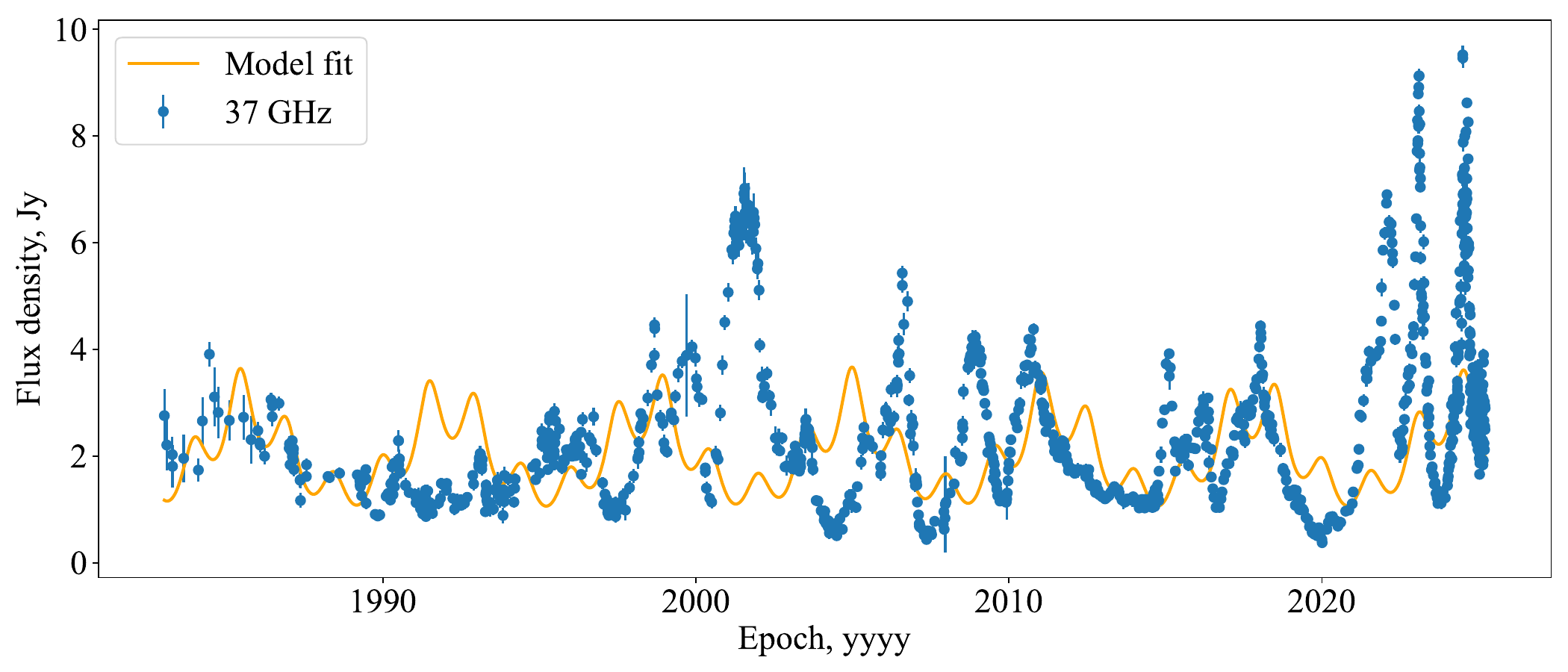}}
\caption{The jet orbital motion and precession model implemented at 
the 37 GHz light curve of Ton\,599 ($P_{\mathrm{pr}}=6.4$ yr, $P_{\mathrm{orb}}=1.5$ yr).} 
\label{fig:Ton599_model_37GHz(1.5-6.4)}
\end{figure}

\begin{figure}
\centerline{\includegraphics[width=\columnwidth]{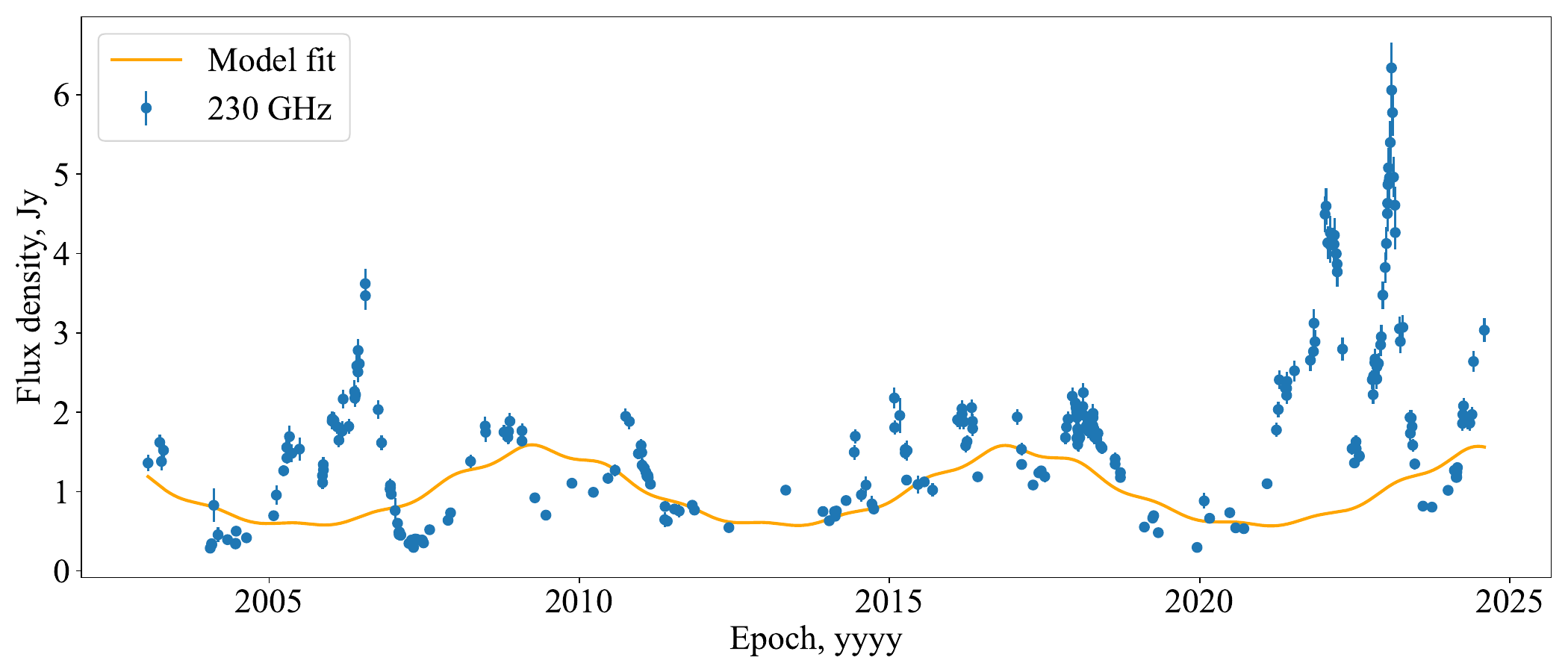}}
\caption{The jet orbital motion and precession model implemented at 
the 230 GHz light curve of Ton\,599 ($P_{\mathrm{pr}}=7.7$ yr, $P_{\mathrm{orb}}=1.3$ yr).} 
\label{fig:Ton599_model_230GHz(1.3-7.7)}
\end{figure}

\begin{figure}
\centerline{\includegraphics[width=\columnwidth]{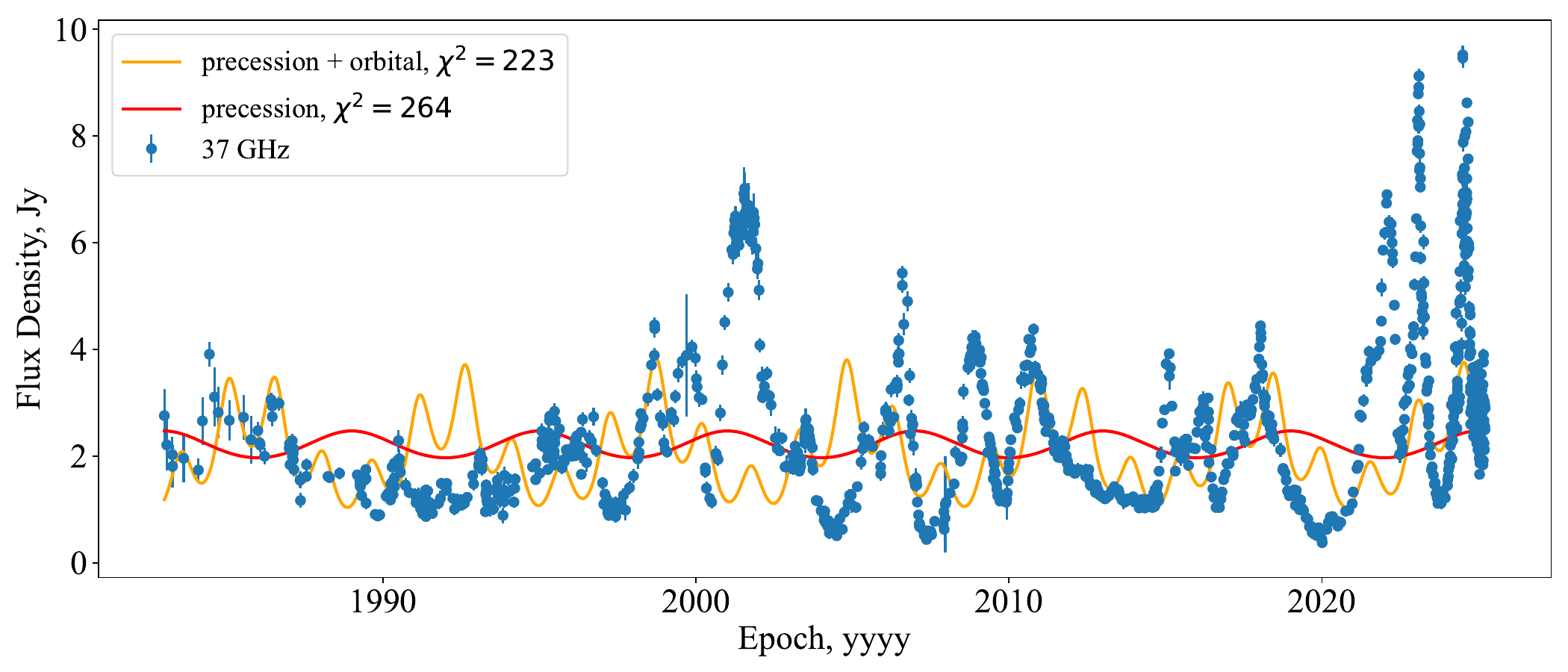}}
\caption{A comparison of the precession-only and combined orbital-precessional models,
shown based on the 37 GHz light curve of Ton\,599 ($P_{\mathrm{pr}}=6.4$ yr, $P_{\mathrm{orb}}=1.5$ yr). The reduced $\chi^2$ values are indicated for each 
of the fittings.} 
\label{fig:Ton599_compare_model}
\end{figure}

\begin{figure}
\centerline{\includegraphics[width=\columnwidth]{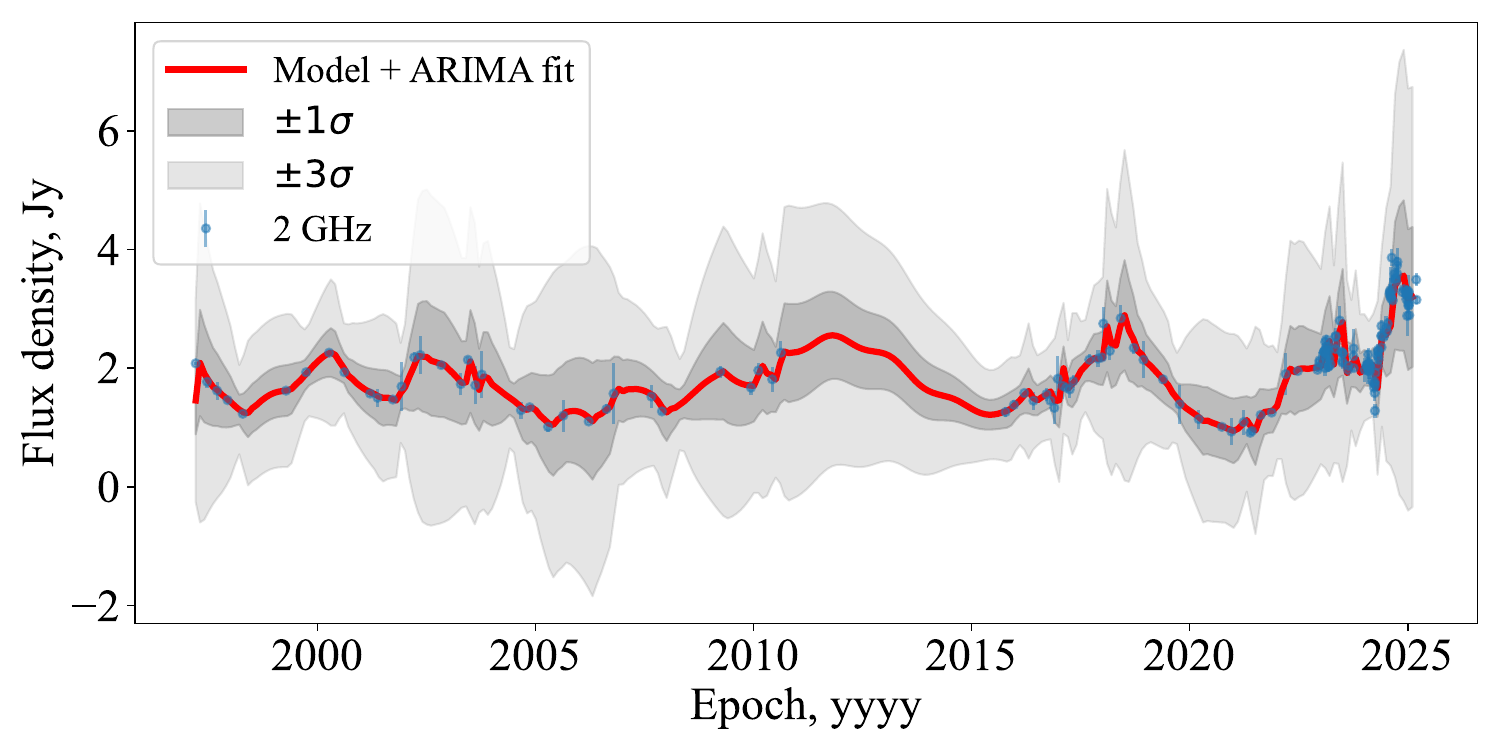}}
\caption{The jet orbital motion and precession model plotted together with the ARIMA model (red colour) based on the 2 GHz light curve of Ton\,599 (the observed data  are shown by blue dots). The shaded regions depict the calculated $\sigma$ and 3$\sigma$ uncertainties (GARCH method).} 
\label{fig:Ton599_model_2GHz_GARCH}
\end{figure}

\begin{figure}
\centerline{\includegraphics[width=\columnwidth]{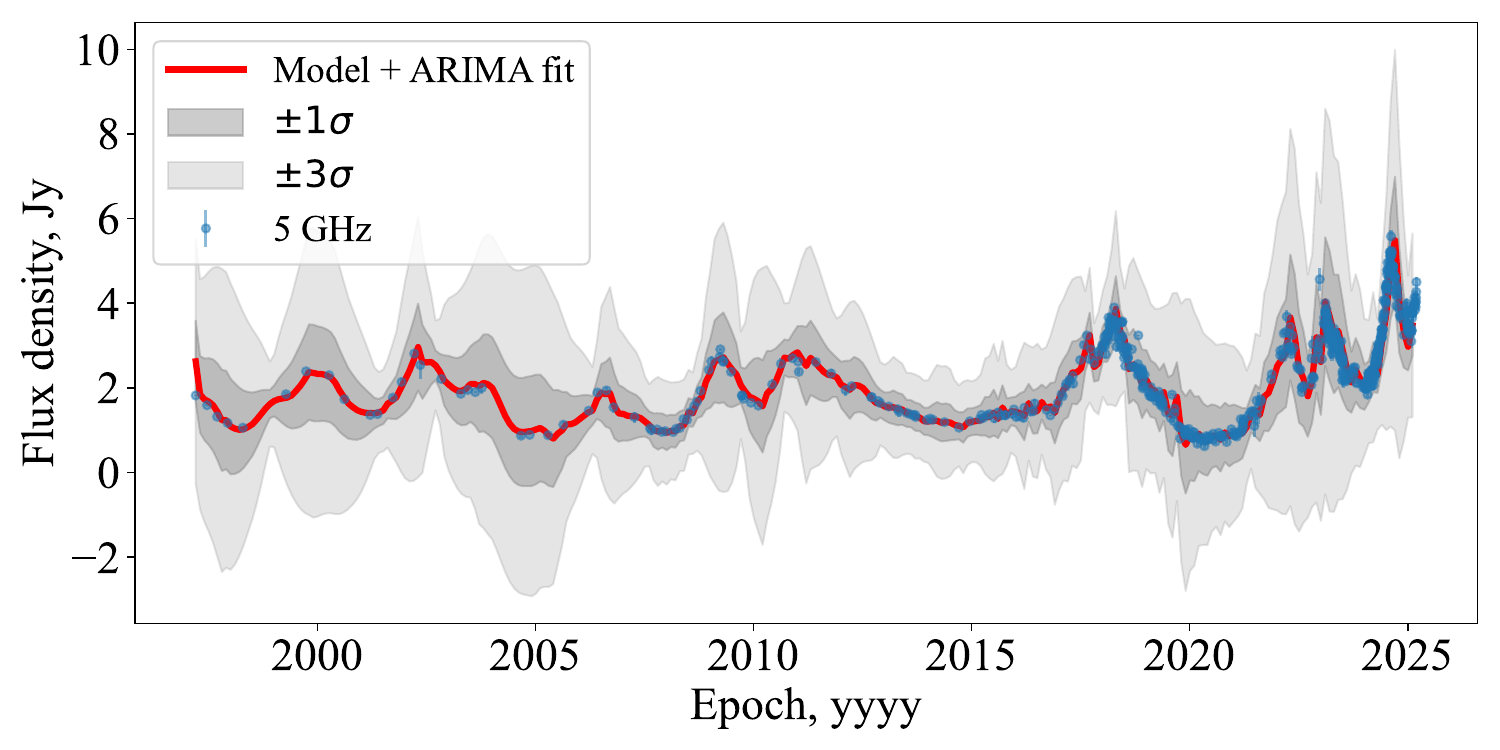}}
\caption{Same as in Fig.\ref{fig:Ton599_model_2GHz_GARCH} for 5 GHz.} 
\label{fig:Ton599_model_5GHz_GARCH}
\end{figure}

\begin{figure}
\centerline{\includegraphics[width=\columnwidth]{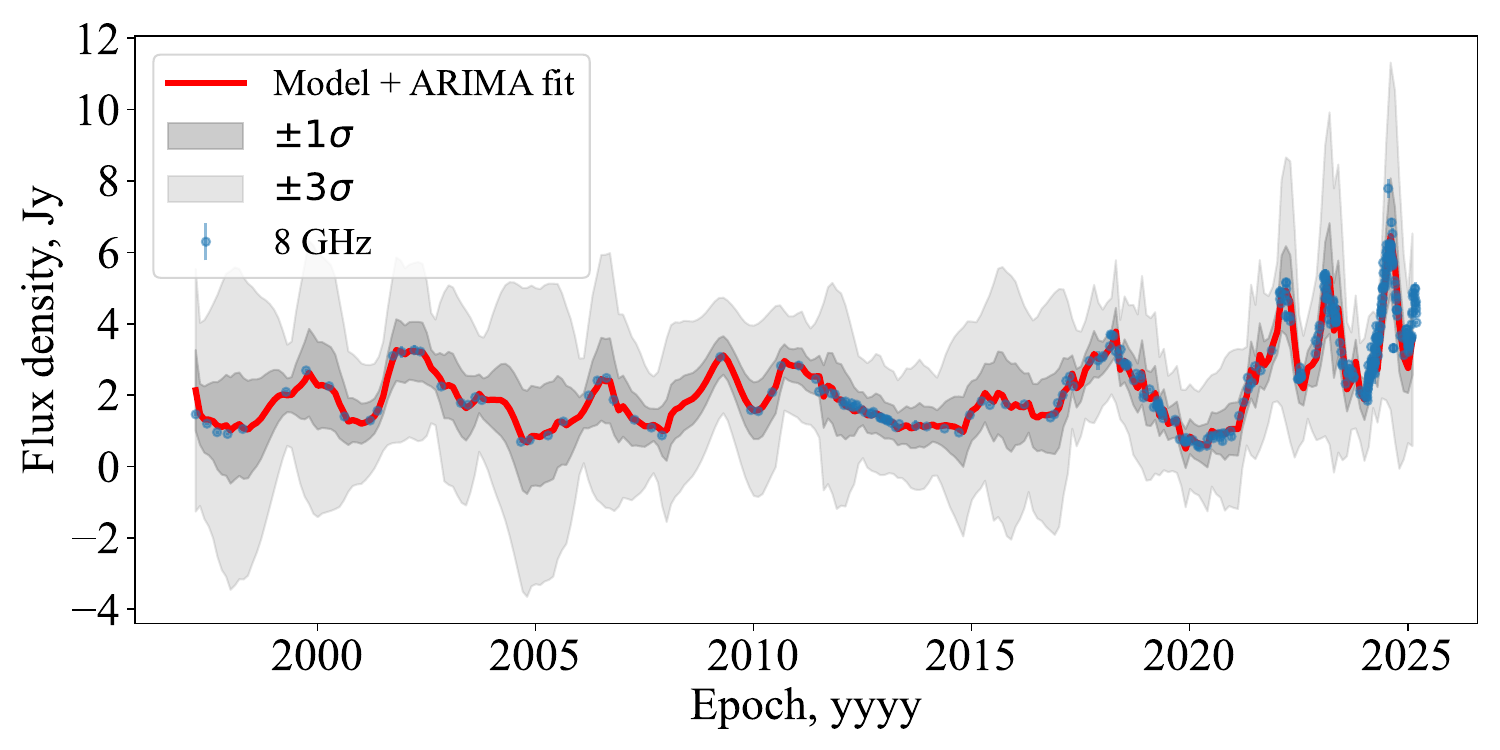}}
\caption{Same as in Fig.\ref{fig:Ton599_model_2GHz_GARCH} for 8 GHz.} 
\label{fig:Ton599_model_8GHz_GARCH}
\end{figure}

\begin{figure}
\centerline{\includegraphics[width=\columnwidth]{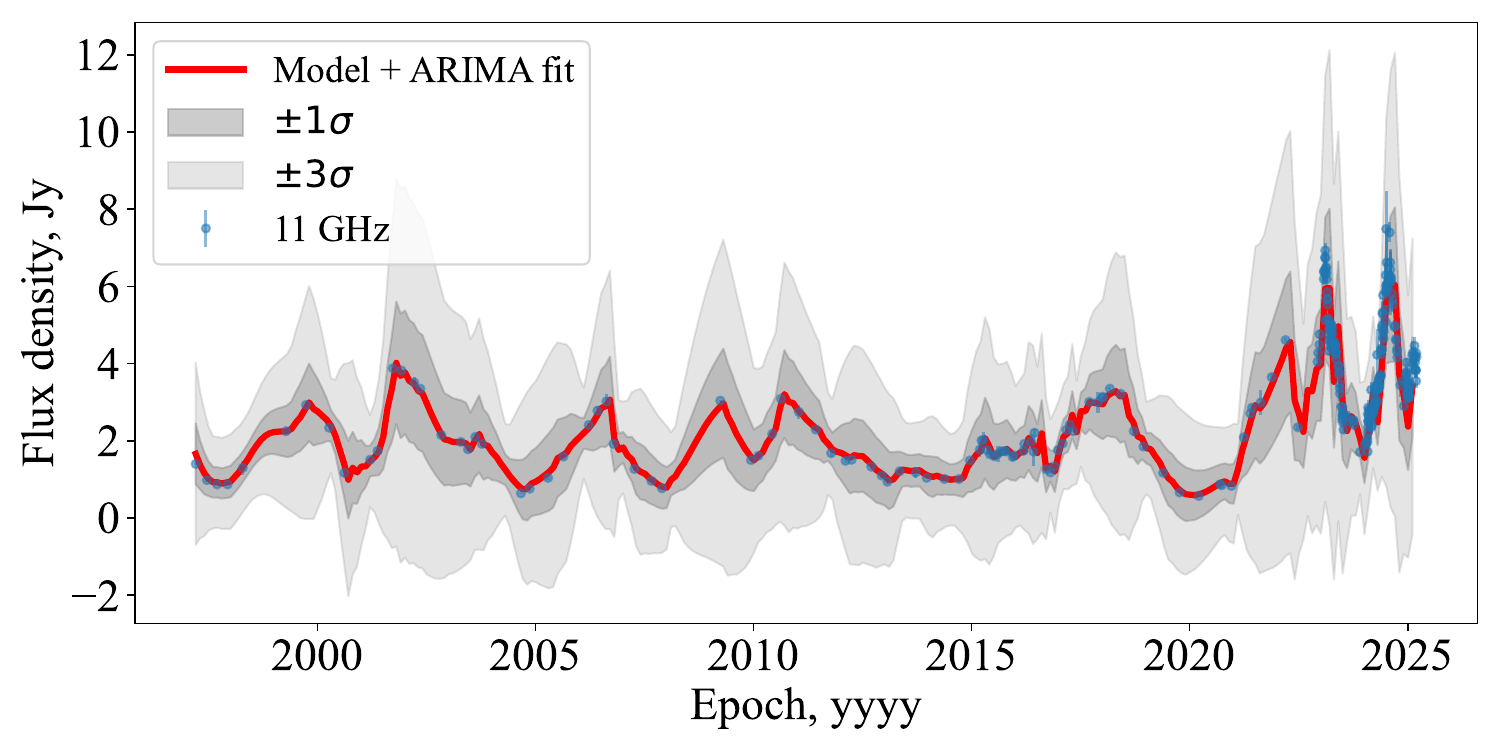}}
\caption{Same as in Fig.\ref{fig:Ton599_model_2GHz_GARCH} for 11 GHz.} 
\label{fig:Ton599_model_11GHz_GARCH}
\end{figure}

\begin{figure}
\centerline{\includegraphics[width=\columnwidth]{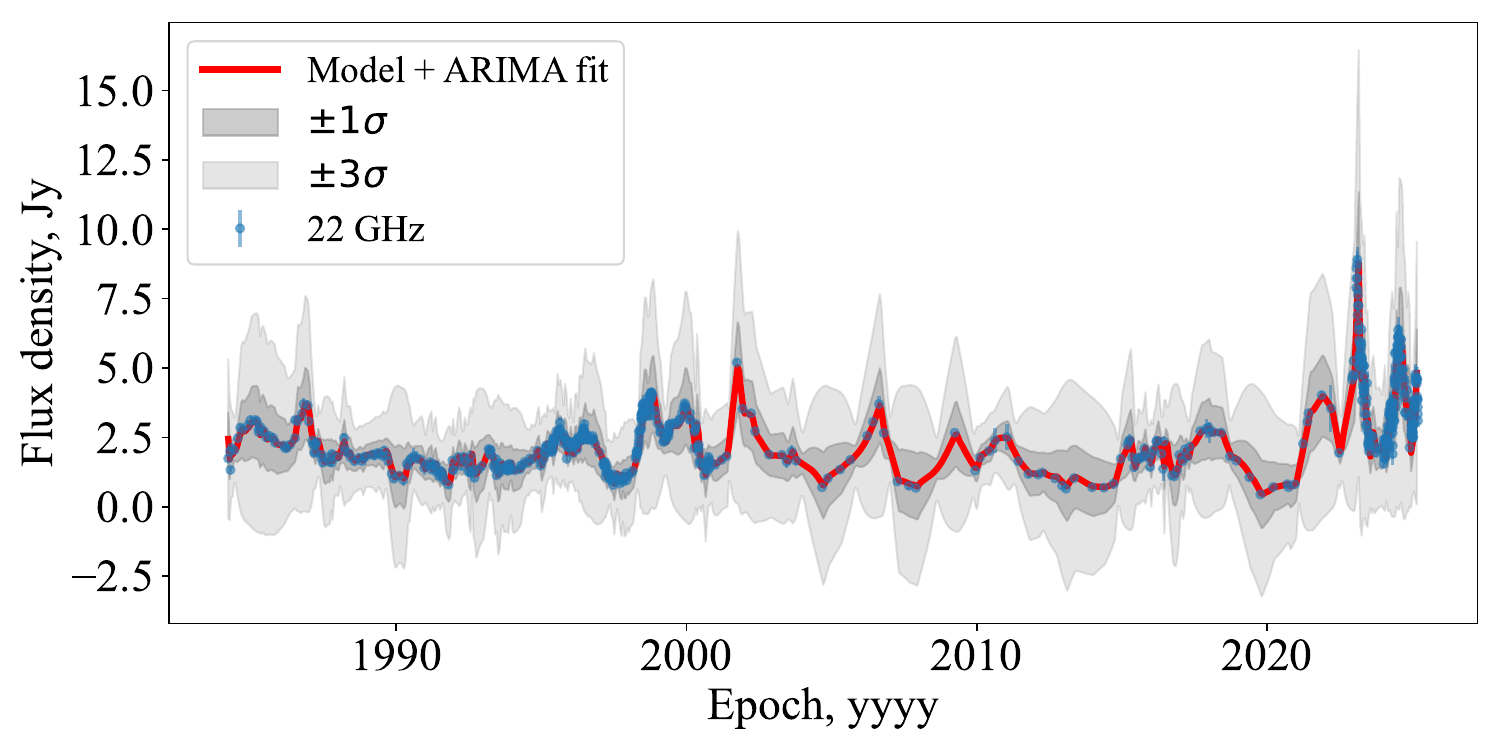}}
\caption{Same as in Fig.\ref{fig:Ton599_model_2GHz_GARCH} for 22 GHz.} 
\label{fig:Ton599_model_22GHz_GARCH}
\end{figure}

\begin{figure}
\centerline{\includegraphics[width=\columnwidth]{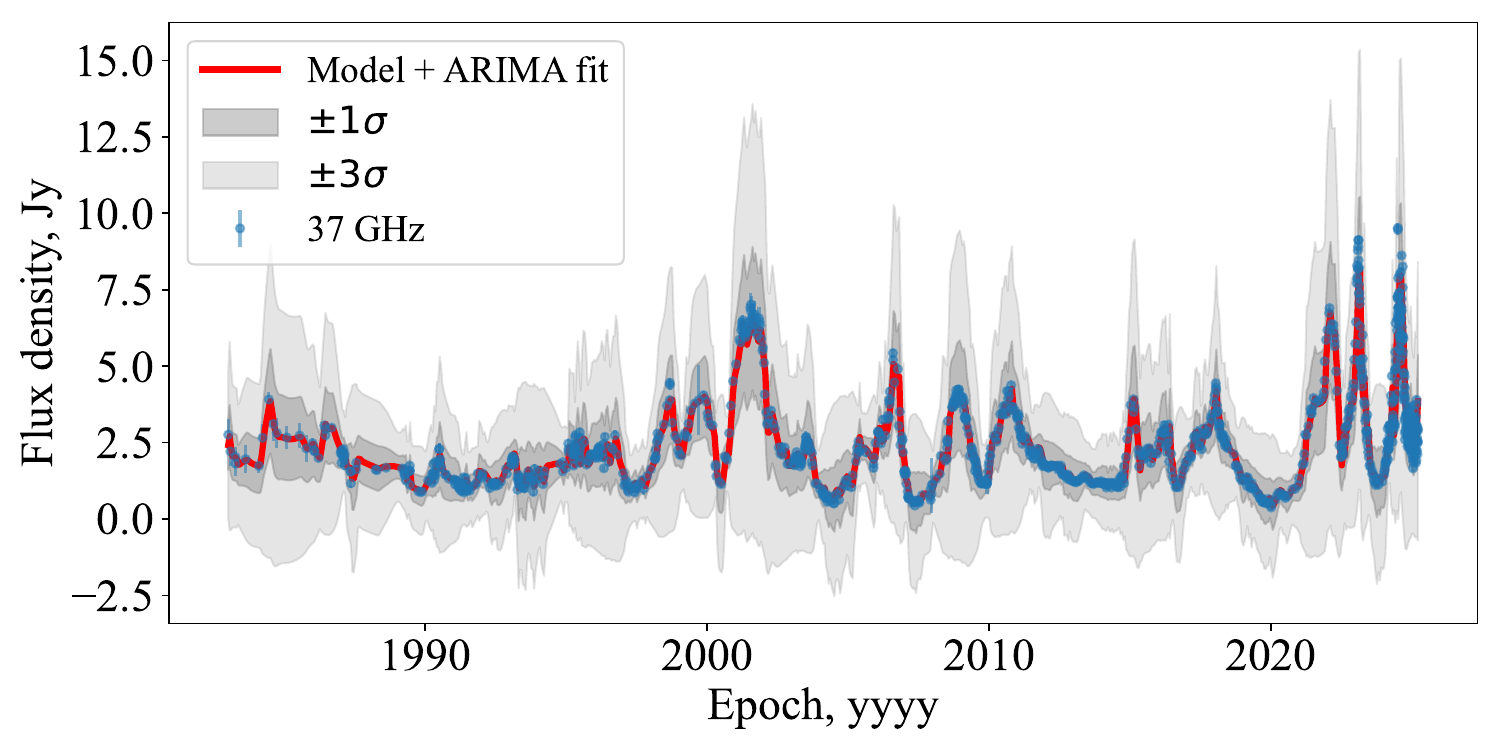}}
\caption{Same as in Fig.\ref{fig:Ton599_model_2GHz_GARCH} for 37 GHz.} 
\label{fig:Ton599_model_37GHz_GARCH}
\end{figure}

\begin{figure}
\centerline{\includegraphics[width=\columnwidth]{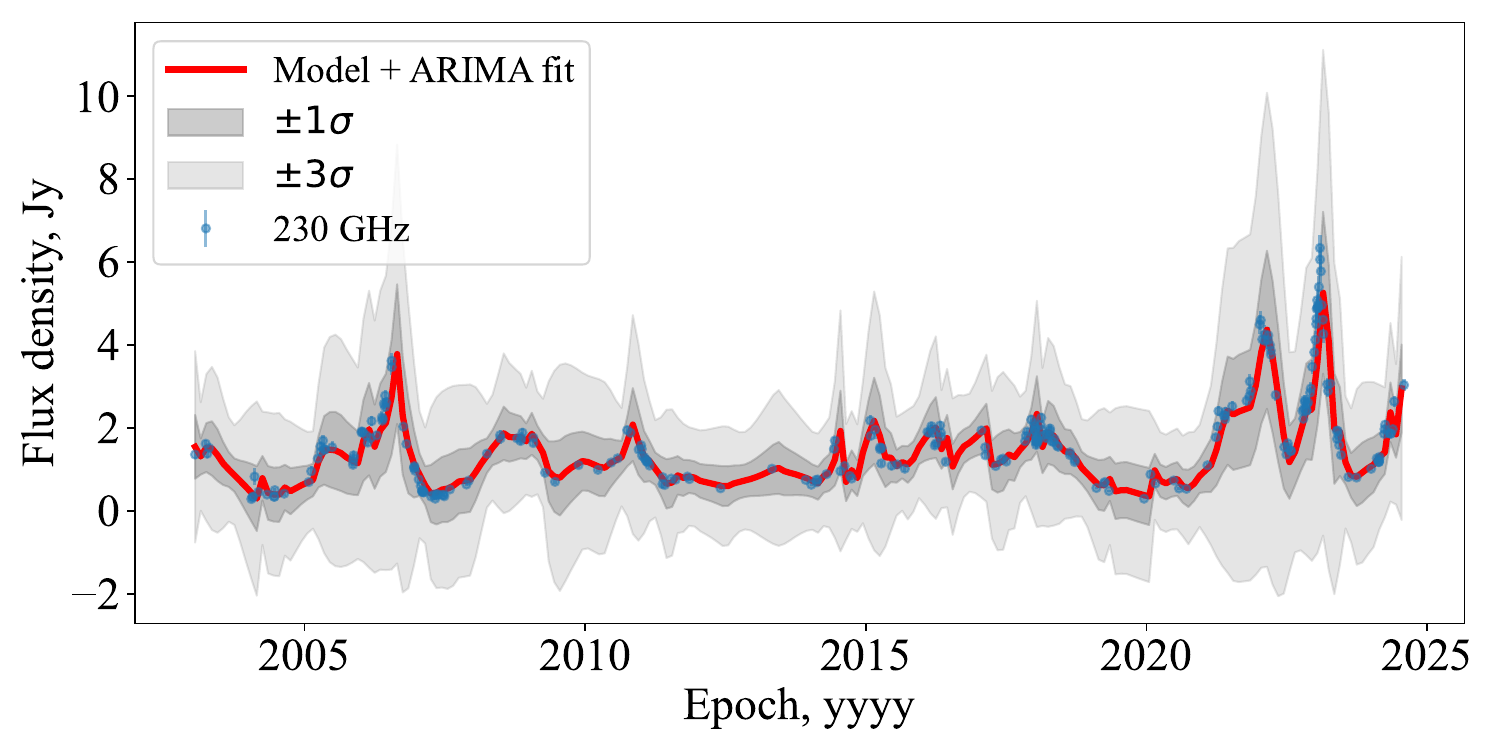}}
\caption{Same as in Fig.\ref{fig:Ton599_model_2GHz_GARCH} for 230 GHz.} 
\label{fig:Ton599_model_230GHz_GARCH}
\end{figure}

\begin{table*}
\centering
\caption{Best-fitting parameters of precession and orbital modeling in the radio bands. The parameter $\beta=v/c$ is determined through the Lorentz factor $\Gamma$ as in equation \ref{eq:beta}; the reduced $\chi^2$ and RMS residuals are listed in the last two columns; other parameters are described in Table~\ref{tab:precession_orbit_params}.}
\label{tab:precession_orbit_results}
\begin{tabular}{ccccccccccccc}
\hline
$\nu$ (GHz) & $P_{\rm pr}$ (yr) & $\beta_{\rm prec}$ & $\theta$ (\degr) & $\Omega$ (\degr) & $\phi_{\rm prec}$ (rad) & $\alpha$ & $\beta_{\rm orb}$ & $i$ (\degr) & $P_{\rm orb}$ (yr) & $\phi_{\rm orb}$ (rad) & $\chi^2$ & RMS (Jy) \\\\
(1) & (2) & (3) & (4) & (5) & (6) & (7) & (8) & (9) & (10) & (11) & (12) & (13) \\\\
\hline
2   & 6.0 & 0.999 & 4.8 & 0.4  & 2.9 & -0.3 & 0.997 & 1.3  & 1.4 & -1.3  & 77  & 0.65 \\
5   & 7.2 & 0.999 & 5.2 & 0.7  & -0.4& 1.0  & 0.991 & 2.3  & 1.2 & 0.4   & 897 & 1.20 \\
8   & 6.6 & 0.993 & 2.4 & 5.4  & 3.1 & -0.6 & 0.997 & 10.0 & 1.3 & -2.7  & 890 & 1.19 \\
11  & 6.0 & 0.998 & 5.1 & 0.9 & 2.7  & -0.4 & 0.986 & 5.0  & 1.7 & -1.2  & 339 & 1.99 \\
22  & 5.8 & 0.996 & 0.5 & 2.4 & -0.6 & 0.1  & 0.998 & 0.5  & 1.3 & -0.6  & 66  & 1.50 \\
37  & 6.4 & 0.999 & 6.8 & 1.2  & -3.1& -0.8 & 0.992 & 10.0  & 1.5 & 3.1   & 233 & 1.75 \\
230 & 7.7 & 0.997 & 4.7 & 1.4  & -1.1& -0.3 & 0.995 & 1.4  & 1.3 & 1.1   & 95  & 1.42 \\
\hline
\end{tabular}
\end{table*}

\begin{table*}
\centering
\caption{ARIMA (Const, AR, MA, $\sigma^2$) and GARCH ($\mu$, $\omega$, $\alpha$, $\beta$) model fit parameters: Const---the constant mean level of the process, AR---autoregressive coefficients at lags~1 and 2, MA---moving average coefficients at lags~1 and 2, $\sigma^2$---residual variance; $\mu$---mean of residuals, $\omega$---long-term volatility level, $\alpha$---short-term shock sensitivity, $\beta$---persistence.} 
\label{tab:arma}
\begin{tabular}{lrrrrrrrrrr}
\hline
 & Const & AR(1) & AR(2) & MA(1) & MA(2) & $\sigma^2$ & $\mu$ & $\omega$ & $\alpha$ & $\beta$ \\
\hline
\multicolumn{11}{c}{2 GHz}\\
\hline
Coeff.     & 0.08 & 1.95 & -0.96 & -0.54 & -0.37 & 0.01   & -0.15 & 0.01 & 1.00 & $<10^{-10}$ \\
Std. Error & 0.10  & 0.06 & 0.06  & 0.06 & 0.06 & 0.01  & 0.02   & 0.01  & 0.47  & 0.55 \\
z-value    & 0.81 & 33.43 &-16.64  &-9.05  & -6.11 & 25.97 & -6.30  & 0.82  & 2.12 & $<10^{-10}$ \\
p-value    & 0.42 & $<10^{-10}$ & $<10^{-10}$ &$<10^{-10}$ & $<10^{-10}$ & $<10^{-10}$ & $<10^{-10}$ & 0.41 & 0.03 & 1.00 \\   
\hline
\multicolumn{11}{c}{5 GHz}\\
\hline
Coeff.     & 0.29 & 1.14 & -0.45 & -0.09 & -0.05 & 0.03 & 0.03 & 0.01 & 0.97 & 0.03 \\
Std. Error & 0.29 & 0.19  & 0.19  & 0.19  & 0.12  & 0.001  & 0.02  & 0.01  & 0.08  & 0.13 \\
z-value    & 1.01 & 7.26 &-2.44  &0.48  & -0.40 & 21.72 & 1.44  & 1.21  & 12.30 & 0.26 \\
p-value    & 0.31 & $<10^{-10}$ & 0.02 & 0.63 & 0.69 & $<10^{-10}$ & 0.15 & 0.23 & $<10^{-10}$ & 0.80 \\   
\hline
\multicolumn{11}{c}{8 GHz}\\
\hline
Coeff.     & -0.13 & 1.08 & -0.17 & 0.56 & 0.38 & 0.04 & 0.14 & 0.04 & 1.00 & $<10^{-10}$ \\
Std. Error & 0.34 & 0.10  & 0.10   & 0.10   & 0.06  & 0.002  & 0.10   & 0.03  & 0.39  & 0.49 \\
z-value    & -0.39 & 11.17 & -1.70  & 6.16  & 6.20 & 20.83 & 1.47  & 1.18  & 2.59 & $<10^{-10}$ \\
p-value    & 0.69 & $<10^{-10}$ & 0.09 & $<10^{-10}$ & $<10^{-10}$ & $<10^{-10}$ & 0.14 & 0.24 &  0.01 & 1.00 \\   
\hline
\multicolumn{11}{c}{11 GHz}\\
\hline
Coeff.     & 0.40 & 1.54 & -0.60 & 0.16 & -0.30 & 0.06 & -0.56 & 0.02 & 1.00 & $<10^{-10}$ \\
Std. Error & 0.38 & 0.14 & 0.13  & 0.14 & 0.11  & 0.002 & 0.14   & 0.009  & 0.16  & 0.20 \\
z-value    & 1.04 & 11.23 &-4.59  &1.12  & -2.81 & 23.28 & -4.03  & 2.48  & 6.19 & $<10^{-10}$ \\
p-value    & 0.30 & $<10^{-10}$ & $<10^{-10}$ &  0.26 & 0.01 & $<10^{-10}$ & $<10^{-5}$  & 0.013 & $<10^{-10}$ & 1.00 \\   
\hline
\multicolumn{11}{c}{22 GHz}\\
\hline
Coeff.     & 0.06 & 0.94 & 0.01 & 0.12 & 0.23 & 0.06 & -0.20 & 0.02 & 0.92 & 0.08 \\
Std. Error & 0.32 & 0.10  & 0.10 & 0.09 & 0.02  & 0.01  & 0.04  & 0.004  & 0.04  & 0.04 \\
z-value    & 0.17 & 9.90 & 0.07  & 1.22  & 9.97 & 66.19 & 5.32  & 4.91  & 24.27 & 2.05 \\
p-value    & 0.86 & $<10^{-10}$ & 0.95 & 0.22 & $<10^{-10}$ & $<10^{-10}$ & $<10^{-6}$ & $<10^{-6}$ & $<10^{-10}$ & 0.04 \\   
\hline
\multicolumn{11}{c}{37 GHz}\\
\hline
Coeff.     & 0.02 & 1.66 & -0.69 & -0.62 & 0.16 & 0.10 & -0.72 & 0.02 & 0.93 & 0.07 \\
Std. Error & 0.32 & 0.05  & 0.05   & 0.05   & 0.01  & 0.002  & 0.05   & 0.007  & 0.05  & 0.04 \\
z-value    & 0.06 & 35.89 &-15.12  &-12.81  & 12.90 & 62.83 & -13.95  & 2.60  & 20.37 & 1.59 \\
p-value    & 0.95 & $<10^{-10}$ & $<10^{-10}$ & $<10^{-10}$ & $<10^{-10}$ & $<10^{-10}$ & $<10^{-10}$ & 0.009 &  $<10^{-10}$ & 0.11 \\   
\hline
\multicolumn{11}{c}{230 GHz}\\
\hline
Coeff.     & 0.36 & 1.27 & -0.40 & 0.07 & 0.12 & 0.08 & -0.28 & 0.02  & 1.00   & $<10^{-14}$ \\
Std. Error & 0.24 & 0.23 & 0.21  & 0.23 & 0.13 & 0.004 & 0.04  & 0.01  & 0.09  & 0.04 \\
z-value    & 1.51 & 5.42 & -1.90 & 0.30 & 0.90 & 18.13 & -8.04  & 3.50  & 10.61 & $<10^{-13}$ \\
p-value    & 0.13 & $<10^{-10}$ & 0.06 & 0.77 & 0.37  & $<10^{-10}$ &  $<10^{-10}$ & 0.001 & $<10^{-10}$ & 1.00 \\ 
\hline
\end{tabular}
\end{table*}

\begin{table*}
\centering
\caption{Binary SMBH and jet parameters in OJ\,287, 3C\,345, and Ton\,599, obtained from  \protect\cite{2005A&A...431..831L,2010CeMDA.106..235V,2012MNRAS.427...77V} and in the current study.}
\begin{tabular}{lccc}
\hline
Parameter & OJ\,287 & 3C\,345 & Ton\,599 \\
\hline
Primary mass $M_1$, $M_\odot$ & $1.83 \times 10^{10}$ & $7.1 \times 10^8$ & $3.6 \times 10^8$ \\
Secondary mass $M_2$, $M_\odot$ & $1.46 \times 10^8$ & $7.1 \times 10^8$ & $1.4 \times 10^8$ \\
Mass ratio, $q = M_2/M_1$ & $\sim 0.008$ & 1.0 & $\sim 0.39$ \\
Orbital period, yr & 12.0 & 480 & $\sim 1.4$ \\
Precession period, yr & $\sim$120 & 2570 & $\sim 6.5$ \\
Orbital separation, pc & $\sim$0.056 & 0.33 & 0.04--0.4 \\
Orbital eccentricity  & 0.65--0.70 & $\lesssim 0.1$ & 0 \\
Orbital inclination  & 9\degr & 5-10\degr & 0.5--10\degr \\
Jet viewing angle  & 1-2\degr & 1-2\degr & 0.5--5.2\degr \\
Jet precession cone angle  & & 1.5\degr & 0.4--5.4\degr \\
Flare periodicity, yr & 12 (double peaks) & Irregular & Quasi-periodic \\
\hline
\end{tabular}
\label{tab:compare}
\end{table*}

\clearpage
\onecolumn 
\bsp
\label{lastpage}

\end{document}